\documentclass[12pt]{article}

\pdfoutput = 1

\usepackage{times}
\usepackage{authblk}
\usepackage{amsmath}
\usepackage{amssymb}
\usepackage{amsthm}
\usepackage{bm}
\usepackage{braket}
\usepackage{slashed}
\usepackage[svgnames,psnames]{xcolor}
\usepackage[nottoc]{tocbibind}
\usepackage{cite}
\usepackage{here}
\usepackage[height = 8.8 in, width = 6.45 in]{geometry}
\usepackage{graphicx}
\usepackage{adjustbox}
\usepackage{tikz-cd}
\usepackage[titletoc]{appendix}
\usepackage{comment}
\usepackage[colorlinks, linktocpage, hypertexnames = false]{hyperref}
\usepackage{url}

\hypersetup{linkcolor = Crimson, citecolor = MediumBlue, urlcolor = MediumBlue}

\numberwithin{equation}{section}

\theoremstyle{definition}

\begin{document}

~\vspace{4cm}~
\begin{center}{\fontsize{17}{0}
\textbf{Fermionization of fusion category symmetries in 1+1 dimensions}}
\end{center}

\begin{center}
Kansei Inamura
\end{center}

\begin{center}
{\small Institute for Solid State Physics, University of Tokyo, Kashiwa, Chiba 277-8581, Japan}
\end{center}
~

\begin{abstract}
We discuss the fermionization of fusion category symmetries in two-dimensional topological quantum field theories (TQFTs).
When the symmetry of a bosonic TQFT is described by the representation category $\mathrm{Rep}(H)$ of a semisimple weak Hopf algebra $H$, the fermionized TQFT has a superfusion category symmetry $\mathrm{SRep}(\mathcal{H}^u)$, which is the supercategory of super representations of a weak Hopf superalgebra $\mathcal{H}^u$.
The weak Hopf superalgebra $\mathcal{H}^u$ depends not only on $H$ but also on a choice of a non-anomalous $\mathbb{Z}_2$ subgroup of $\mathrm{Rep}(H)$ that is used for the fermionization.
We derive a general formula for $\mathcal{H}^u$ by explicitly constructing fermionic TQFTs with $\mathrm{SRep}(\mathcal{H}^u)$ symmetry.
We also construct lattice Hamiltonians of fermionic gapped phases when the symmetry is non-anomalous.
As concrete examples, we compute the fermionization of finite group symmetries, the symmetries of finite gauge theories, and duality symmetries.
We find that the fermionization of duality symmetries depends crucially on $F$-symbols of the original fusion categories.
The computation of the above concrete examples suggests that our fermionization formula of fusion category symmetries can also be applied to non-topological QFTs.
\end{abstract}

\setcounter{page}{0}

\thispagestyle{empty}

\newpage

\tableofcontents

\flushbottom

\section{Introduction and summary}
Symmetry is a powerful tool to constrain the dynamics of quantum many-body systems and classify quantum phases of matter.
There has recently been much progress in understanding the general structures of symmetries. 
Specifically, it has been recognized that a large class of symmetries can be characterized by the algebraic structures of topological defects.
For example, an ordinary group symmetry is understood as the group of codimension one topological defects whose fusion rules are invertible.
Symmetries associated with invertible topological defects of higher codimensions are called higher form symmetries \cite{GKSW2015}, which are also described by groups.
On the other hand, when the fusion rules are non-invertible, topological defects no longer form a group.
Nevertheless, the algebraic structure of non-invertible topological defects should also be regarded as symmetry in a generalized sense.
Symmetries associated with non-invertible topological defects are called non-invertible symmetries, which are expected to be described by higher categories \cite{DR2018, GJF2019, Johnson-Freyd2022a, KLWZZ2020a, KLWZZ2020b}.
For recent reviews on generalized symmetries, see for example \cite{McGreevy2022, CDIS2022}.

In 1+1 dimensional bosonic systems, finite non-invertible symmetries are called fusion category symmetries \cite{TW2019}, for the algebraic structures of finitely many topological defect lines are described by unitary fusion categories \cite{BT2018, EGNO2015}.
Fusion category symmetries are ubiquitous in 2d conformal field theories (CFTs) \cite{Ver1988, PZ2001, FRS2002, FFRS2004, FFRS2007, FGRS2007, FFRS2010, CLSWY2019, CL2020, TW2021} and impose strong constraints on renormalization group flows \cite{CLSWY2019, TW2019, TW2021, GK2021}.
Fusion category symmetries also exist in 2d topological quantum field theories (TQFTs) \cite{DKR2011, CR2016, BCP2014a, BCP2014b, BCP2015, BT2018, TW2019, KORS2020, HL2022, Inamura2021, HLS2021, Inamura2022}.
In particular, 2d semisimple bosonic TQFTs with fusion category symmetry $\mathcal{C}$ are in one-to-one correspondence with finite semisimple $\mathcal{C}$-module categories \cite{TW2019, KORS2020}.
Physically, a $\mathcal{C}$-module category is the category of topological boundary conditions of a 2d TQFT with $\mathcal{C}$ symmetry \cite{KORS2020}.\footnote{A $\mathcal{C}$-module category that specifies a 2d TQFT with $\mathcal{C}$ symmetry can also be viewed as a single topological boundary condition of 3d Turaev-Viro TQFT constructed from $\mathcal{C}$ \cite{TV1992, BW1996}, see \cite{TW2019, GK2021, FT2018, FT2021, Freed2022} for this viewpoint.}
The explicit construction of $\mathcal{C}$-symmetric TQFTs classified by $\mathcal{C}$-module categories is given in \cite{HLS2021, Inamura2022}.
There are also concrete lattice models with these symmetries \cite{FTLTKWF2007, BG2017, AMF2016, AFM2020, HEHGV2016, GRLM2022, Inamura2022}, some of which have been used to search for novel quantum field theories with exotic fusion category symmetries \cite{HLOTT2021, VLDWOHV2021, LZR2022}.

In contrast to 1+1d bosonic systems, finite non-invertible symmetries of 1+1d fermionic systems cannot be described by ordinary fusion categories in general.
This is because some topological defect lines in 1+1d fermionic systems can have fermionic point-like defects on them, whose anti-commutation relation modifies the algebraic structures of topological defects.
A topological line that can have a fermionic point-like defect on it is called a q-type object, while a topological line that cannot have a fermion on it is called an m-type object \cite{ALW2019}.
The algebraic structures of these topological lines are generally described by superfusion categories \cite{BE2017, Ush2018, GWW2015, ALW2019}.
Thus, we call finite non-invertible symmetries of 1+1d fermionic systems superfusion category symmetries.
A comprehensive understanding of general 1+1d fermionic systems with superfusion category symmetries has not been achieved, although there are remarkable papers studying topological lines in 2d fermionic QFTs \cite{NR2020, RW2020, LSCH2020, KCXC2022} and particularly the relation between anomalous invertible symmetries of fermionic QFTs and non-invertible symmetries of bosonic QFTs \cite{Thorngren2020, JSW2020, LS2021, BKN2021}.\footnote{Other kinds of defects such as conformal defects and boundaries in 2d fermionic CFTs are studied in e.g. \cite{MW2017, Smith2021, FTZ2021, EW2021}.}

\begin{figure}
\begin{center}
\includegraphics[width = 14.5cm]{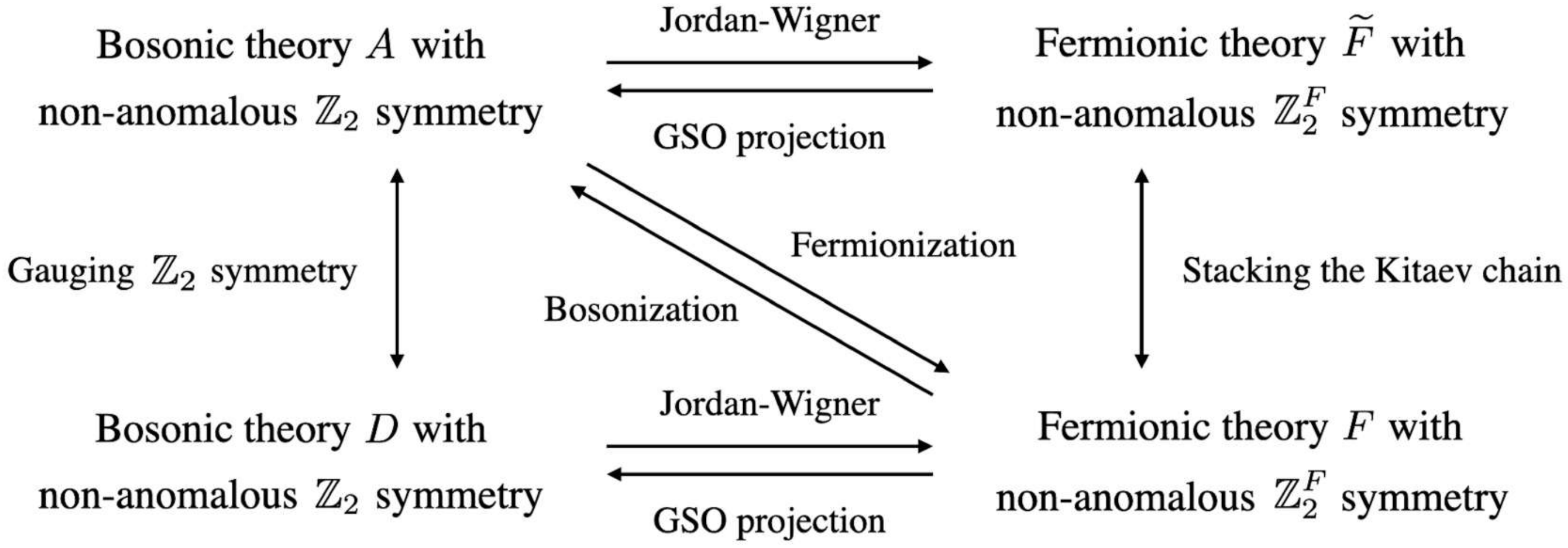}
\caption{Bosonization and fermionization in 1+1 dimensions can be summarized by the above diagram. The Jordan-Wigner transformation is the gauging of a $\mathbb{Z}_2$ symmetry coupled to a spin structure \cite{Tachikawa2019, GK2021}. The dual of the gauged $\mathbb{Z}_2$ symmetry is the fermion parity symmetry. The inverse of the Jordan-Wigner transformation is the summation over spin structures, which is called the Gliozzi-Scherk-Olive (GSO) projection \cite{GSO1977}. In this paper, fermionization refers to the map from bosonic theory $A$ to fermionic theory $F$. For example, the fermionization of a trivial TQFT is a trivial TQFT, while the fermionization of a $\mathbb{Z}_2$ symmetry broken TQFT is the Kitaev chain. The bosonization is the inverse map of the fermionization. We note that the fermionization in this paper is called the Jordan-Wigner transformation in \cite{HNT2021}, whereas the Jordan-Wigner transformation in this paper is called the fermionization in \cite{Thorngren2020, JSW2020}.}
\end{center}
\label{fig: fermionization}
\end{figure}

In this paper, we investigate superfusion category symmetries of 2d fermionic TQFTs by using the fermionization of 2d bosonic TQFTs.
Here, the fermionization refers to the Jordan-Wigner transformation followed by the operation of stacking (the low-energy limit of) the Kitaev chain \cite{Thorngren2020, GK2016, Tachikawa2019, KTT2019, JSW2020, HNT2021, Kulp2021}, see figure \ref{fig: fermionization} for details.
We give an explicit formula to determine the superfusion category symmetry $\mathcal{C}_f$ of the fermionized TQFT from the fusion category symmetry $\mathcal{C}_b$ of a bosonic TQFT.
Although the formula is derived in the context of topological field theories, we expect that our formula can also be applied to non-topological QFTs because the formula does not involve any data of TQFTs other than their symmetries.
Concrete examples also suggest that the fermionization formula of fusion category symmetries is applicable to general 1+1d systems.
Along with the derivation of the formula, we explicitly construct fermionic TQFTs with superfusion category symmetry $\mathcal{C}_f$ by fermionizing bosonic TQFTs with fusion category symmetry $\mathcal{C}_b$.\footnote{This generalizes the construction of 2d fermionic TQFTs with finite group symmetries given in \cite{KTY2018}.}
This construction gives a map between $\mathcal{C}_b$-module categories and $\mathcal{C}_f$-supermodule categories, which are the categories of topological boundary conditions of $\mathcal{C}_b$-symmetric bosonic TQFTs and $\mathcal{C}_f$-symmetric fermionic TQFTs in two dimensions. 
We will elaborate on the above results in the rest of this section.
A more detailed description will be given in the subsequent sections. 

\paragraph{Bosonic TQFTs with fusion category symmetries.}
Let us first recall the construction of 2d bosonic TQFTs with fusion category symmetries following \cite{Inamura2022}.
Bosonic TQFTs on oriented surfaces can be constructed from semisimple algebras via the state sum construction \cite{FHK94, BP1993}.
The bosonic state sum TQFT obtained from a semisimple algebra $K$ is denoted by $\mathcal{T}_b^K$.
In general, the bosonic TQFT $\mathcal{T}_b^K$ has a fusion category symmetry described by the category ${}_K \mathcal{M}_K$ of $(K, K)$-bimodules \cite{DKR2011}.
This is because $\mathcal{T}_b^K$ is the generalized gauging of a trivial TQFT by $K$ \cite{FRS2002, FFRS2010, CR2016, BCP2014a, BCP2014b, BCP2015, BT2018, CRS2019}.
When the input algebra $K$ is a left $H$-comodule algebra where $H$ is a weak Hopf algebra, we can pull back the ${}_K \mathcal{M}_K$ symmetry to the representation category $\mathrm{Rep}(H)$ by a tensor functor $F_K: \mathrm{Rep}(H) \rightarrow {}_K \mathcal{M}_K$.
Since the pullback of a fusion category symmetry does not change the underlying TQFT, $\mathcal{T}_b^K$ has $\mathrm{Rep}(H)$ symmetry when $K$ is a left $H$-comodule algebra.
The bosonic TQFT $\mathcal{T}_b^K$ with $\mathrm{Rep}(H)$ symmetry is indecomposable if $K$ is indecomposable as a left $H$-comodule algebra.
Conversely, any indecomposable semisimple bosonic TQFTs with $\mathrm{Rep}(H)$ symmetry can be obtained in this way \cite{Inamura2022}.\footnote{For example, a bosonic SPT phase with fusion category symmetry is the pullback of a trivial TQFT by a fiber functor, which is a tensor functor from a fusion category to the category of vector spaces. This is consistent with the fact that 1+1d bosonic SPT phases with fusion category symmetries are classified by fiber functors \cite{TW2019}.}
We note that $\mathrm{Rep}(H)$ symmetry is the most general fusion category symmetry because any unitary fusion category is equivalent to the representation category of some weak Hopf algebra \cite{Ostrik2003, Hayashi1999}. 
Hereafter, we only consider weak Hopf algebras whose representation categories are unitary fusion categories unless otherwise stated.

\paragraph{Fermionization of bosonic TQFTs.}
The bosonic TQFT $\mathcal{T}_b^K$ with $\mathrm{Rep}(H)$ symmetry can be fermionized if it has a non-anomalous $\mathbb{Z}_2$ subgroup symmetry.
In general, a bosonic state sum TQFT $\mathcal{T}_b^K$ has a non-anomalous $\mathbb{Z}_2$ symmetry if the input algebra $K$ is $\mathbb{Z}_2$-graded \cite{KTY2017, SR2017}.
In particular, a bosonic TQFT $\mathcal{T}_b^K$ with $\mathrm{Rep}(H)$ symmetry has a non-anomalous $\mathbb{Z}_2$ subgroup symmetry if the dual weak Hopf algebra $H^*$ has a $\mathbb{Z}_2$ group-like element.
This is because a $\mathbb{Z}_2$ group-like element $u \in H^*$ enables us to define a $\mathbb{Z}_2$-grading on a left $H$-comodule algebra $K$.
The algebra $K$ equipped with this $\mathbb{Z}_2$-grading is denoted by $K^u$, and the corresponding $\mathbb{Z}_2$ symmetry of $\mathcal{T}_b^K$ is denoted by $\mathbb{Z}^u_2$.
The relation between a $\mathbb{Z}_2$ group-like element in $H^*$ and a $\mathbb{Z}_2$ subgroup of $\mathrm{Rep}(H)$ can also be understood as follows: if $H^*$ has a $\mathbb{Z}_2$ group-like element $u \in H^*$, there is a representation $V_u \in \mathrm{Rep}(H)$ that obeys a $\mathbb{Z}_2$ group-like fusion rule \cite{Nikshych2002, Vecsernyes2003} and generates $\mathbb{Z}_2^u$ subgroup.
By using this $\mathbb{Z}_2^u$ symmetry, we can fermionize a bosonic TQFT $\mathcal{T}_b^K$ to obtain a fermionic TQFT $\mathcal{T}_f^{K^u}$, which is a fermionic state sum TQFT constructed from a semisimple superalgebra $K^u$ \cite{NR2015, GK2016, KTY2018}.

\paragraph{Symmetries of fermionized TQFTs.}
The fermionic state sum TQFT $\mathcal{T}_f^{K^u}$ has a superfusion category symmetry described by the supercategory ${}_{K^u} \mathcal{SM}_{K^u}$ of $(K^u, K^u)$-superbimodules.
This symmetry can be pulled back to the super representation category $\mathrm{SRep}(\mathcal{H}^u)$ by a supertensor functor $\mathcal{F}_{K^u}: \mathrm{SRep}(\mathcal{H}^u) \rightarrow {}_{K^u} \mathcal{SM}_{K^u}$, where $\mathcal{H}^u$ is a weak Hopf superalgebra defined in terms of $H$ and $u$, see section \ref{sec: Fermionization of fusion category symmetries} for the explicit definition of $\mathcal{H}^u$.
Thus, the fermionized TQFT $\mathcal{T}_f^{K^u}$ has $\mathrm{SRep}(\mathcal{H}^u)$ symmetry if the original bosonic TQFT $\mathcal{T}_b^K$ has $\mathrm{Rep}(H)$ symmetry whose $\mathbb{Z}_2^u$ subgroup is used for the fermionization.
This shows that the fermionization of a fusion category symmetry $\mathrm{Rep}(H)$ with respect to $\mathbb{Z}_2^u$ subgroup becomes a superfusion category symmetry $\mathrm{SRep}(\mathcal{H}^u)$.
Just as the category of topological boundary conditions of $\mathcal{T}_b^K$ is a $\mathrm{Rep}(H)$-module category ${}_K \mathcal{M}$ \cite{Inamura2022}, the category of topological boundary conditions of $\mathcal{T}_f^{K^u}$ is an $\mathrm{SRep}(\mathcal{H}^u)$-supermodule category ${}_{K^u} \mathcal{SM}$, where ${}_K \mathcal{M}$ and ${}_{K^u} \mathcal{SM}$ are the category of left $K$-modules and the supercategory of left $K^u$-supermodules respectively.

\paragraph{Examples of superfusion category symmetries.}
A basic example of $\mathrm{SRep}(\mathcal{H}^u)$ symmetry is the fermionization of non-anomalous finite group symmetry.
It turns out that the fermionization of a finite group symmetry $G$ crucially depends on whether $\mathbb{Z}_2^u$ subgroup of $G$ is central or not.
Specifically, when $\mathbb{Z}_2^u$ is a central subgroup of $G$, the fermionized symmetry is also a finite group $G$.
On the other hand, when $\mathbb{Z}_2^u$ is not a central subgroup, the fermionized symmetry becomes non-invertible.
This is similar to the fact that a $\mathbb{Z}_2$-gauging of a finite group symmetry $G$ becomes non-invertible if $\mathbb{Z}_2$ is not a central subgroup of $G$ \cite{BT2018}.
However, we note that the fermionization of $G$ is not quite the same as the $\mathbb{Z}_2$-gauging.
For example, if $G$ is a non-trivial central extension of $G_b$ by $\mathbb{Z}_2$, the fermionization of $G$ is also a finite group $G$, which is a non-trivial central extension of $G_b$ by the fermion parity symmetry $\mathbb{Z}_2^F$, whereas the $\mathbb{Z}_2$-gauging of $G$ is a finite group symmetry $\mathbb{Z}_2 \times G_b$ with a mixed 't Hooft anomaly \cite{BT2018, CW2022a}.

A more involved example is the fermionization of $\mathrm{Rep}(H_8)$ symmetry, where $H_8$ is a Hopf algebra known as the eight-dimensional Kac-Paljutkin algebra \cite{KP1966}.
The $\mathrm{Rep}(H_8)$ symmetry describes a self-duality under gauging $\mathbb{Z}_2 \times \mathbb{Z}_2$ symmetry \cite{BT2018, TW2019}.
The fusion category $\mathrm{Rep}(H_8)$ consists of four invertible defects corresponding to $\mathbb{Z}_2 \times \mathbb{Z}_2$ symmetry and a single non-invertible defect called a duality defect.
If we choose a $\mathbb{Z}_2^u$ subgroup of $\mathrm{Rep}(H_8)$ appropriately, the fermionized symmetry $\mathrm{SRep}(\mathcal{H}_8^u)$ consists of four invertible defects corresponding to $\mathbb{Z}_2 \times \mathbb{Z}_2^F$ symmetry and two non-invertible defects corresponding to a self-duality under the composite operation of gauging a $\mathbb{Z}_2$ subgroup and stacking the Kitaev chain.\footnote{The author thanks Ryohei Kobayashi for discussing the interpretation of $\mathrm{SRep}(\mathcal{H}_8^u)$ symmetry as a self-duality and pointing out the example of a self-dual Gu-Wen SPT phase mentioned below.}
These non-invertible defects turn out to be q-type objects, one of which is obtained by fusing the fermion parity defect with the other non-invertible defect.
The relation between $\mathrm{Rep}(H_8)$ and $\mathrm{SRep}(\mathcal{H}^u_8)$ is a non-anomalous analogue of the relation between the symmetries of the Ising CFT and a single massless Majorana fermion discussed in \cite{Thorngren2020, JSW2020}.
Physically, $\mathrm{Rep}(H_8)$ symmetry is realized by the stacking of two Ising CFTs \cite{TW2019}, whereas $\mathrm{SRep}(\mathcal{H}^u_8)$ symmetry is realized by the stacking of the Ising CFT and a single massless Majorana fermion.
More generally, since $\mathrm{Rep}(H_8)$ symmetry exists on the entire orbifold branch of $c=1$ CFTs \cite{TW2019, TW2021}, the fermionization of any $c=1$ orbifold CFT has $\mathrm{SRep}(\mathcal{H}_8^u)$ symmetry.
Another example exhibiting this self-duality is (the low-energy limit of) a Gu-Wen SPT phase with $\mathbb{Z}_2 \times \mathbb{Z}_2^F$ symmetry \cite{GW2014}.
This is the fermionization of a non-trivial bosonic SPT phase with $\mathbb{Z}_2 \times \mathbb{Z}_2$ symmetry, which is self-dual under gauging $\mathbb{Z}_2 \times \mathbb{Z}_2$ symmetry \cite{TW2019}.

The $\mathrm{Rep}(H_8)$ symmetry is one of the three non-anomalous self-dualities under $\mathbb{Z}_2 \times \mathbb{Z}_2$ gauging \cite{TW2019}.
The other two are described by $\mathrm{Rep}(D_8)$ and $\mathrm{Rep}(Q_8)$, where $D_8$ and $Q_8$ are the dihedral group of order $8$ and the quaternion group respectively.
The fusion categories $\mathrm{Rep}(D_8)$ and $\mathrm{Rep}(Q_8)$ have the same fusion rules as $\mathrm{Rep}(H_8)$ but have different $F$-symbols \cite{TY1998}.
In contrast to the fermionization of $\mathrm{Rep}(H_8)$, we will see that the fermionization of $\mathrm{Rep}(D_8)$ and $\mathrm{Rep}(Q_8)$ does not have q-type objects for any choice of $\mathbb{Z}_2^u$ subgroup.
The reason for this difference is that the self-dualities described by these fusion categories involve different pairings of background and dynamical $\mathbb{Z}_2 \times \mathbb{Z}_2$ gauge fields \cite{TW2019}, or equivalently different choices of an isomorphism between the dual and the original $\mathbb{Z}_2 \times \mathbb{Z}_2$ symmetries, see equations \eqref{eq: Rep(H8) duality} and \eqref{eq: Rep(D8) duality}.
This example illustrates that the existence of q-type objects depends on the $F$-symbols of fusion category symmetries.
In particular, a duality defect does not always become a q-type object after fermionization. 

Before proceeding, we mention that non-invertible symmetries have recently been studied in higher dimensions \cite{FSV2013, KS2010, CMS2020, CRS2018, CRS2020, CRS2019, Meusburger2022, NTU2021a, KNY2021, CCHLS2021, KOZ2022, RSS2022, BBSNT2022, ATRG2022, HT2022, CCHLS2022, KZZ2022, CLS2022, CO2022, AGR2022, BDZH2022}.
We leave the fermionization of these symmetries for future work.

The rest of the paper is organized as follows.
In section \ref{sec: Preliminaries}, we review the definitions and some basic properties of superfusion categories and (weak) Hopf superalgebras.
In section \ref{sec: Fermionization of fusion category symmetries}, we give the explicit formula for the fermionization of fusion category symmetries and discuss several concrete examples.
In section \ref{sec: Fermionic TQFTs with superfusion category symmetries}, we derive the fermionization formula of fusion category symmetries in the context of topological field theories.
Along the way, we explicitly construct fermionic TQFTs with superfusion category symmetries by fermionizing bosonic TQFTs with fusion category symmetries.
We also construct lattice models of fermionic gapped phases with non-anomalous superfusion category symmetries at the end of section \ref{sec: Fermionic TQFTs with superfusion category symmetries}.
Detailed derivations of several results described in the main text are relegated to three appendices.

\section{Preliminaries}
\label{sec: Preliminaries}
In this section, we briefly review some mathematical tools used in the later sections.
Throughout the paper, (super)algebras refer to finite dimensional semisimple (super)algebras over $\mathbb{C}$.

\subsection{Superfusion categories}
We first review superfusion categories and related notions such as supertensor functors and supermodule categories following \cite{BE2017, Ush2018}, see also \cite{GWW2015, ALW2019}.
Let us begin with the definitions of supercategories and superfunctors.
A supercategory is a category such that the set $\mathrm{Hom}(x, y)$ of morphisms between any objects $x$ and $y$ is a $\mathbb{Z}_2$-graded vector space and the composition of morphisms preserves the $\mathbb{Z}_2$-grading.
The $\mathbb{Z}_2$-grading of a homogeneous morphism $f$ is denoted by $|f|$, which is $0$ or $1$ depending on whether $f$ is even or odd.
A superfunctor between supercategories is a functor that preserves the $\mathbb{Z}_2$-grading of morphisms. 

A supercategory $\mathcal{C}$ is called a monoidal supercategory if it is equipped with a tensor product structure $\otimes: \mathcal{C} \times \mathcal{C} \rightarrow \mathcal{C}$, the unit object $1$, and the structure isomorphisms called associators and unit morphisms that satisfy the usual pentagon and triangle equations.
The tensor product of morphisms preserves the $\mathbb{Z}_2$-grading, which means $|f \otimes g| = |f| + |g| \bmod 2$ for any homogeneous morphisms $f$ and $g$.
The compatibility of the tensor product and the composition of morphisms is encoded in the following relation called the super interchange law:
\begin{equation}
(f \otimes g) \circ (h \otimes k) = (-1)^{|g| |h|} (f \circ h) \otimes (g \circ k).
\label{eq: super interchange law}
\end{equation}
This equation implies that $\mathbb{Z}_2$-odd morphisms obey the anti-commutation relation because $(f \otimes \mathrm{id}) \circ (\mathrm{id} \otimes g) = (-1)^{|f| |g|} (\mathrm{id} \otimes g) \circ (f \otimes \mathrm{id})$, which is represented  in terms of string diagrams as follows:
\begin{equation}
\adjincludegraphics[valign = c, width = 1.5cm]{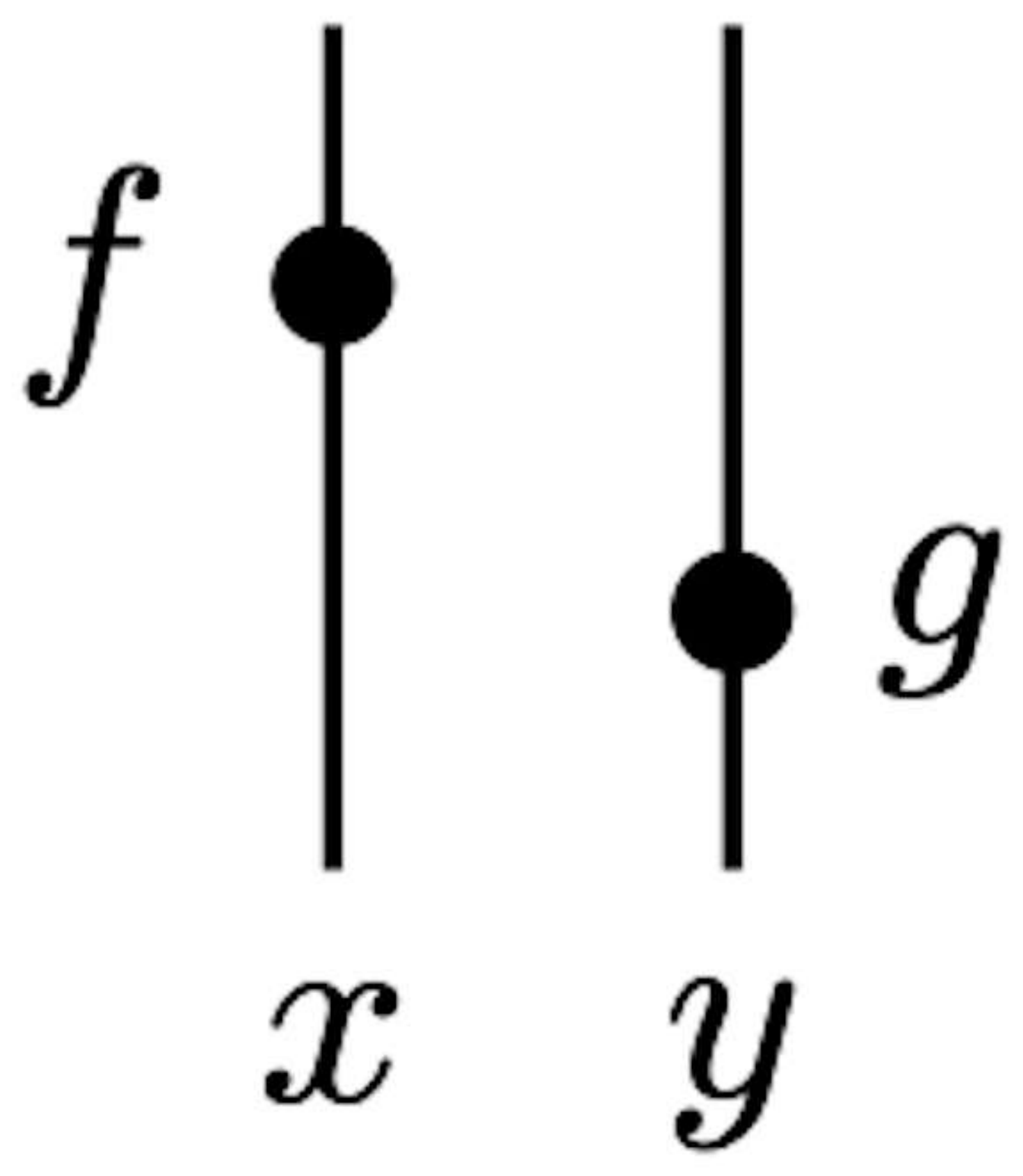} ~ = (-1)^{|f| |g|} ~ \adjincludegraphics[valign = c, width = 1.5cm]{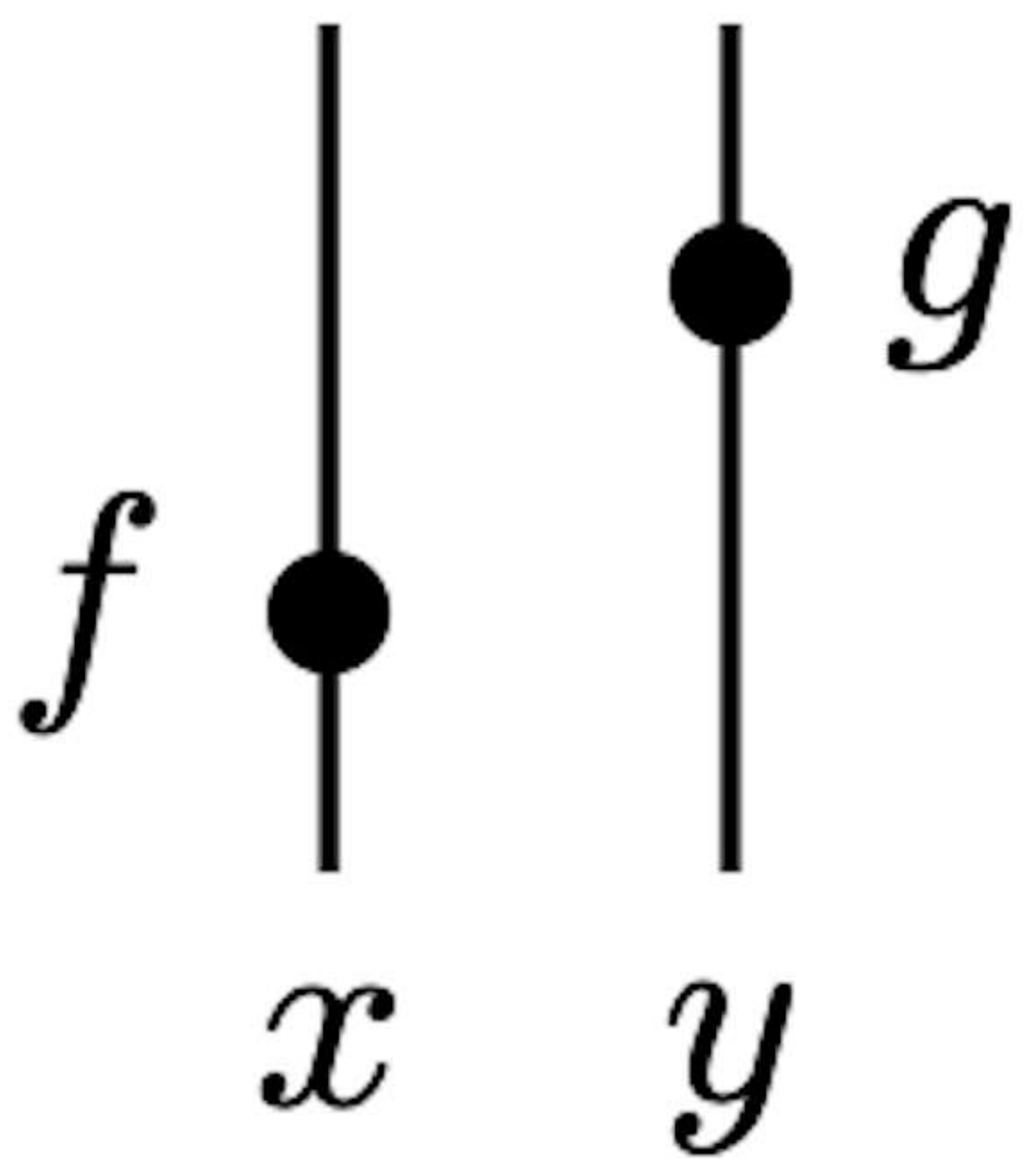}.
\label{eq: anti-commutation}
\end{equation}
Here, we read the string diagrams from the bottom to the top.
We note that a monoidal supercategory is not a monoidal category in the usual sense due to the extra minus sign on the right-hand side of the above equation.

A monoidal supercategory is called rigid if every object has both left dual and right dual.
Here, a left dual $x^{*}$ of an object $x$ is an object equipped with even morphisms $\mathrm{ev}_x^{L}: x^* \otimes x \rightarrow 1$ and $\mathrm{coev}_x^L: 1 \rightarrow x \otimes x^*$ that satisfy the following zigzag axiom:\footnote{We omit to write associators and unit morphisms in equations.}
\begin{equation}
(\mathrm{id}_x \otimes \mathrm{ev}_x^L) \circ (\mathrm{coev}_x^L \otimes \mathrm{id}_x) = \mathrm{id}_x, \quad 
(\mathrm{ev}_x^L \otimes \mathrm{id}_{x^*}) \circ (\mathrm{id}_{x^*} \otimes \mathrm{coev}_x^L) = \mathrm{id}_{x^*}.
\label{eq: zigzag}
\end{equation}
The morphisms $\mathrm{ev}_x^L$ and $\mathrm{coev}_x^L$ are called left evaluation and left coevaluation morphisms respectively.
A right dual ${}^* x$ is also defined similarly.

A superfusion category is a finite semisimple $\mathbb{C}$-linear rigid monoidal supercategory such that the unit object is simple.
In particular, any object $x$ of a superfusion category is isomorphic to a finite direct sum of simple objects, and the Hom space $\mathrm{Hom}(x, y)$ between any two objects $x$ and $y$ is a finite dimensional super vector space.
The Hom space $\mathrm{Hom}(x, x)$ for a simple object $x$ is isomorphic to either $\mathbb{C}^{1|0}$ or $\mathbb{C}^{1|1}$.\footnote{A vector space $\mathbb{C}^{p|q}$ denotes a super vector space of superdimension $(p, q)$.}
A simple object of the former type is called an m-type object, whereas a simple object of the latter type is called a q-type object \cite{ALW2019}.
We note that the unit object is an m-type object \cite{Ush2018}.

We assume that superfusion categories are pivotal. Namely, they are equipped with an even natural isomorphism $a_x: x \rightarrow x^{**}$ called a pivotal structure.
We use this natural isomorphism to identify the right dual ${}^* x$ with the left dual $x^{*}$.
Accordingly, we write the right evaluation and coevaluation morphisms in terms of $x^*$ rather than ${}^* x$.
Specifically, the right evaluation and coevaluation morphisms are defined as even morphisms $\mathrm{ev}_x^R: x \otimes x^* \rightarrow 1$ and $\mathrm{coev}_x^R: 1 \rightarrow x^* \otimes x$ that satisfy the zigzag axiom analogous to eq. \eqref{eq: zigzag}.

Since a superfusion category is rigid, we can define the left and right duals of a morphism.
Specifically, for a pivotal superfusion category, the left dual $f^*$ and the right dual ${}^* f$ of a morphism $f \in \mathrm{Hom}(x, y)$ are morphisms from $y^*$ to $x^*$ defined by 
\begin{equation}
\begin{aligned}
f^* & := (\mathrm{ev}_y^L \otimes \mathrm{id}_{x^*}) \circ (\mathrm{id}_{y^*} \otimes f \otimes \mathrm{id}_{x^*}) \circ (\mathrm{id}_{y^*} \otimes \mathrm{coev}_{x}^L), \\
{}^*f & := (\mathrm{id}_{x^*} \otimes \mathrm{ev}_{y}^R) \circ (\mathrm{id}_{x^*} \otimes f \otimes \mathrm{id}_{y^*}) \circ (\mathrm{coev}_{x}^R \otimes \mathrm{id}_{y^*}).
\end{aligned}
\label{eq: dual of a morphism}
\end{equation}
Following \cite{ALW2019}, we require that the left dual and the right dual of a morphism $f$ are related by
\begin{equation}
{}^*f = (-1)^{|f|} f^*,
\label{eq: fermion rotation}
\end{equation}
which physically corresponds to the fact that a fermion acquires a minus sign when rotated by $2\pi$.

Superfusion categories that describe the symmetries of 1+1d fermionic systems have a further structure known as $\Pi$-superfusion categories.
Here, a $\Pi$-superfusion category is a superfusion category equipped with a distinguished object $\pi$ and an odd isomorphism $\zeta: \pi \rightarrow 1$.\footnote{Physically, an odd isomorphism $\zeta$ corresponds to a probe local fermion, and $\pi$ corresponds to the worldline of it. Any superfusion category is equivalent to a $\Pi$-superfusion category \cite{BE2017}.}
In particular, every object $x$ of a $\Pi$-superfusion category has an oddly isomorphic object $\Pi x := \pi \otimes x$.
The odd isomorphism from $\Pi x$ to $x$ is denoted by $\zeta_x$. 
The evaluation and coevaluation morphisms of $\Pi x$ can be expressed in terms of those of $x$ as 
\begin{equation}
\begin{aligned}
\mathrm{ev}_{\Pi x}^L & = - \mathrm{ev}_{x}^L \circ ((\zeta^*_x)^{-1} \otimes \zeta_x), \quad 
\mathrm{coev}_{\Pi x}^L = (\zeta_x^{-1} \otimes \zeta_x^*) \circ \mathrm{coev}_x^L, \\
\mathrm{ev}_{\Pi x}^R & = \mathrm{ev}_x^R \circ (\zeta_x \otimes ({}^* \zeta_x)^{-1}), \quad
\mathrm{coev}_{\Pi x}^R = - ({}^* \zeta_x \otimes \zeta_x^{-1}) \circ \mathrm{coev}_x^R,
\end{aligned}
\label{eq: (co)evPix}
\end{equation}
which follow from equations \eqref{eq: anti-commutation}, \eqref{eq: zigzag}, and \eqref{eq: dual of a morphism}.
We note that in a pivotal $\Pi$-superfusion category, the quantum dimension of $\Pi x$ agrees with that of $x$ due to equations \eqref{eq: fermion rotation} and \eqref{eq: (co)evPix}:
\begin{equation}
\mathrm{qdim}(\Pi x) = \mathrm{ev}_{\Pi x}^R \circ \mathrm{coev}_{\Pi x}^L = \mathrm{ev}_{x}^R \circ \mathrm{coev}_{x}^L = \mathrm{qdim}(x).
\end{equation}

A supertensor functor $(\mathcal{F}, \mathcal{J}, \varphi)$ between superfusion categories $\mathcal{C}$ and $\mathcal{D}$ is a $\mathbb{C}$-linear superfunctor $\mathcal{F}: \mathcal{C} \rightarrow \mathcal{D}$ equipped with an even natural isomorphism $\mathcal{J}_{x, y}: \mathcal{F}(x) \otimes \mathcal{F}(y) \rightarrow \mathcal{F}(x \otimes y)$ and an even isomorphism $\varphi: 1_{\mathcal{D}} \rightarrow \mathcal{F}(1_{\mathcal{C}})$ that satisfy the same commutative diagrams as those for an ordinary tensor functor.
We often write a supertensor functor $(\mathcal{F}, \mathcal{J}, \varphi)$ simply as $\mathcal{F}$.

A supermodule category $\mathcal{M}$ over a superfusion category $\mathcal{C}$ is a supercategory equipped with a $\mathcal{C}$-action denoted by $\overline{\otimes}: \mathcal{C} \times \mathcal{M} \rightarrow \mathcal{M}$.
The structure of a $\mathcal{C}$-supermodule category on $\mathcal{M}$ is represented by a supertensor functor $\mathcal{F}: \mathcal{C} \rightarrow \mathrm{End}(\mathcal{M})$, where $\mathrm{End}(\mathcal{M})$ is the supercategory of superfunctors from $\mathcal{M}$ to itself. 

\paragraph{Example.}
The simplest example of a superfusion category is the supercategory $\mathrm{SVec}$ of super vector spaces.
Objects and morphisms of $\mathrm{SVec}$ are finite dimensional super vector spaces and (not necessarily even) linear maps between them.
The $\mathbb{Z}_2$-even part and the $\mathbb{Z}_2$-odd part of a super vector space $V$ are denoted by $V_0$ and $V_1$ respectively.
The direct sum and the tensor product of super vector spaces $V$ and $W$ are given by 
\begin{align}
(V \oplus W)_0 & = V_{0} \oplus W_0, \quad (V \oplus W)_1 = V_1 \oplus W_1,\\
(V \otimes W)_0 & = (V_0 \otimes W_0 ) \oplus (V_1 \otimes W_1), \quad (V \otimes W)_1 = (V_1 \otimes W_0) \oplus (V_0 \otimes W_1).
\end{align}
The tensor product of morphisms involves a non-trivial sign coming from the braiding of $\mathbb{Z}_2$-odd elements.
Specifically, for homogeneous morphisms $f \in \mathrm{Hom}(V, V^{\prime})$ and $g \in \mathrm{Hom}(W, W^{\prime})$, the tensor product $f \otimes g \in \mathrm{Hom}(V \otimes W, V^{\prime} \otimes W^{\prime})$ is defined by
\begin{equation}
f \otimes g (v \otimes w) = (-1)^{|g| |v|} f(v) \otimes g(w), \quad \forall v \in V, ~ \forall w \in W,
\label{eq: super tensor product of morphisms}
\end{equation}
which obeys the super interchange law \eqref{eq: super interchange law}.
The dual object of a super vector space $V$ is the dual vector space $V^* := \mathrm{Hom}(V, \mathbb{C})$ equipped with the obvious $\mathbb{Z}_2$-grading.
The evaluation and coevaluation morphisms are defined in the usual way:
\begin{align}
& \mathrm{ev}_V^L (\phi \otimes v) = \mathrm{ev}_V^R (v \otimes \phi) = \phi(v), \quad \forall \phi \in V^*, ~ \forall v \in V, 
\label{eq: ev of SVec} \\
& \mathrm{coev}_V^L(\lambda) = \lambda \sum_{i} v_i \otimes v^{i}, \quad \mathrm{coev}_V^R(\lambda) = \lambda \sum_i v^i \otimes v_i, \quad \forall \lambda \in \mathbb{C},
\label{eq: coev of SVec}
\end{align}
where $\{v_i\}$ and $\{v^i\}$ are dual bases of $V$ and $V^{*}$.
Based on the above definition, we can express the $\mathbb{Z}_2$-grading automorphism $p_V: V \rightarrow V$, which is defined by $p_V(v) = (-1)^{|v|} v$, in terms of the evaluation and coevaluation morphisms as follows:
\begin{equation}
p_V = (\mathrm{ev}_V^L \otimes \mathrm{id}_{V}) \circ (\mathrm{id}_{V^*} \otimes c_{\mathrm{super}}) \circ (\mathrm{coev}_{V}^R \otimes \mathrm{id}_V) = (\mathrm{id}_V \otimes \mathrm{ev}_V^R) \circ (c_{\mathrm{super}} \otimes \mathrm{id}_{V^*}) \circ (\mathrm{id}_V \otimes \mathrm{coev}_V^L),
\label{eq: p}
\end{equation}
where $c_{\mathrm{super}}$ is the symmetric braiding of super vector spaces defined by $c_{\mathrm{super}}(v \otimes w) = (-1)^{|v| |w|} w \otimes v$.
The subscript $V$ of the $\mathbb{Z}_2$-grading automorphism $p_V$ will be omitted if it is clear from the context. 
Combined with the axiom \eqref{eq: zigzag} and its counterpart for the right dual, equation \eqref{eq: p} implies that the left (co)evaluation morphism and the right (co)evaluation morphism are related to each other in the following way:
\begin{equation}
\begin{aligned}
\mathrm{ev}_V^L & = \mathrm{ev}_V^R \circ c_{\mathrm{super}} \circ (\mathrm{id} \otimes p) = \mathrm{ev}_V^R \circ c_{\mathrm{super}} \circ (p \otimes \mathrm{id}), \\
\mathrm{coev}_V^{L} & = (p \otimes \mathrm{id}) \circ c_{\mathrm{super}} \circ \mathrm{coev}_V^R = (\mathrm{id} \otimes p) \circ c_{\mathrm{super}} \circ \mathrm{coev}_V^R.
\end{aligned}
\label{eq: left-right relation}
\end{equation}
We can use this equation to show that $\mathrm{SVec}$ is a pivotal superfusion category satisfying the condition \eqref{eq: fermion rotation}.
Furthermore, $\mathrm{SVec}$ is a $\Pi$-superfusion category whose distinguished object $\pi$ is $\mathbb{C}^{0|1}$ and the odd isomorphism $\zeta: \mathbb{C}^{0|1} \rightarrow \mathbb{C}^{1|0}$ is the identity map of the underlying vector space.

\subsection{Semisimple superalgebras}
A superalgebra is an algebra equipped with a $\mathbb{Z}_2$-grading such that the multiplication and unit are even.\footnote{The unit is automatically even when the multiplication is even.}
A simple superalgebra is isomorphic to either $\mathrm{End}(\mathbb{C}^{p|q})$ or $\mathrm{End}(\mathbb{C}^{p|0}) \otimes \mathrm{Cl}(1)$, where $\mathrm{End}(\mathbb{C}^{p|q})$ is the endomorphism superalgebra of $\mathbb{C}^{p|q}$ and $\mathrm{Cl}(1)$ is the complex Clifford algebra with a single odd generator \cite{Wall1964}.
A semisimple superalgebra $K$ can be decomposed into a direct sum of simple superalgebras, namely we have \cite{Jozefiak1988}
\begin{equation}
K \cong (\bigoplus_i \mathrm{End}(\mathbb{C}^{p_i|q_i})) \oplus (\bigoplus_j \mathrm{End}(\mathbb{C}^{p_j|0}) \otimes \mathrm{Cl}(1)).
\label{eq: decomposition of a semisimple superalgebra}
\end{equation}

A left supermodule $M$ over a superalgebra $K$ is a $\mathbb{Z}_2$-graded vector space on which $K$ acts by an even linear map $\rho_M: K \otimes M \rightarrow M$.
Right supermodules and superbimodules are also defined similarly.
A linear map between $K$-supermodules is called a $K$-supermodule morphism if it commutes with the action of $K$.
More specifically, a $K$-supermodule morphism $f: M \rightarrow N$ is a linear map that satisfies $f \circ \rho_M = \rho_N \circ (\mathrm{id}_K \otimes f)$, where the tensor product on the right-hand side is defined by eq. \eqref{eq: super tensor product of morphisms}.
Simple supermodules over simple superalgebras $\mathrm{End}(\mathbb{C}^{p|q})$ and $\mathrm{End}(\mathbb{C}^{p|0}) \otimes \mathrm{Cl}(1)$ are $\mathbb{C}^{p|q}$ and $\mathbb{C}^{p|p}$ respectively, which are unique up to isomorphism of supermodules.
If a simple superalgebra $K_0$ is a direct summand of a semisimple superalgebra $K$, we can consider a $K_0$-supermodule $M_0$ as a $K$-supermodule by demanding that the other direct summands act trivially on $M_0$.
Any simple $K$-supermodule is of this form.
Therefore, simple supermodules over a semisimple superalgebra $K$ are in one-to-one correspondence with the direct summands on the right-hand side of eq. \eqref{eq: decomposition of a semisimple superalgebra} \cite{Jozefiak1988}.

Since a semisimple superalgebra $K$ is a semisimple algebra, it is a $\Delta$-separable symmetric Frobenius algebra \cite{FS2008, FRS2002}, which is also called a special symmetric Frobenius algebra in \cite{FRS2002}.
Namely, $K$ is equipped with a comultiplication $\Delta_K: K \rightarrow K \otimes K$ and a counit $\epsilon_K: K \rightarrow \mathbb{C}$ that satisfy the following $\Delta$-separability \eqref{eq: Delta separability}, symmetricity \eqref{eq: symmetricity}, and the Frobenius relation \eqref{eq: Frobenius rel}:
\begin{align}
m_K \circ \Delta_K & = \mathrm{id}_K,
\label{eq: Delta separability}\\
((\epsilon_K \circ m_K) \otimes \mathrm{id}_{K^*}) \circ (\mathrm{id}_K \otimes \mathrm{coev}_{K}^L) & = (\mathrm{id}_{K^*} \otimes (\epsilon_K \circ m_K)) \circ (\mathrm{coev}_K^R \otimes \mathrm{id}_K),
\label{eq: symmetricity}\\
(\mathrm{id}_K \otimes m_K) \circ (\Delta_K \otimes \mathrm{id}_K) & = \Delta_K \circ m_K = (m_K \otimes \mathrm{id}_K) \circ (\mathrm{id}_K \otimes \Delta_K).
\label{eq: Frobenius rel}
\end{align}
Here, $m_K: K \otimes K \rightarrow K$ and $\eta_K: \mathbb{C} \rightarrow K$ are the multiplication and unit of $K$ respectively.
The $\Delta$-separable symmetric Frobenius structure on $K$ is unique \cite{FRS2002} and is defined regardless of the $\mathbb{Z}_2$-grading on $K$.
Concretely, the comultiplication $\Delta_K$ and the counit $\epsilon_K$ are given by
\begin{align}
\Delta_K & = (\mathrm{id}_K \otimes m_K) \circ (\mathrm{id}_K \otimes \Phi^{-1} \otimes \mathrm{id}_K) \circ (\mathrm{coev}_K^L \otimes \mathrm{id}_K),
\label{eq: comultiplication on K}\\
\epsilon_K & = \mathrm{ev}_K^R \circ (m_K \otimes \mathrm{id}_{K^*}) \circ (\mathrm{id}_K \otimes \mathrm{coev}_K^L),
\label{eq: counit of K}
\end{align}
where $\Phi: K \rightarrow K^*$ is an even isomorphism of superalgebras defined by
\begin{equation}
\Phi = ((\epsilon_K \circ m_K) \otimes \mathrm{id}_{K^*}) \circ (\mathrm{id}_K \otimes \mathrm{coev}_K^L ).
\label{eq: Phi}
\end{equation}
We note that the symmetricity \eqref{eq: symmetricity} is equivalent to 
\begin{equation}
\begin{aligned}
\epsilon_K \circ m_K & = \epsilon_K \circ m_K \circ c_{\mathrm{triv}} = \epsilon_K \circ m_K \circ c_{\mathrm{super}} \circ (\mathrm{id}_K \otimes p) = \epsilon_K \circ m_K \circ c_{\mathrm{super}} \circ (p \otimes \mathrm{id}_K), \\
\Delta_K \circ \eta_K & = c_{\mathrm{triv}} \circ \Delta_K \circ \eta_K = (p \otimes \mathrm{id}_K) \circ c_{\mathrm{super}} \circ \Delta_K \circ \eta_K = (\mathrm{id}_K \otimes p) \circ c_{\mathrm{super}} \circ \Delta_K \circ \eta_K
\end{aligned}
\label{eq: left-right relation of K}
\end{equation}
due to eqs. \eqref{eq: zigzag} and \eqref{eq: p}, where $p: K \rightarrow K$ is the $\mathbb{Z}_2$-grading automorphism of $K$ and $c_{\mathrm{triv}}$ is the trivial braiding defined by $c_{\mathrm{triv}}(v \otimes w) = w \otimes v$ for all $v, w \in K$.

The supercategory of superbimodules over a semisimple superalgebra $K$ is a monoidal supercategory \cite{BE2017}, which is denoted by ${}_K \mathcal{SM}_K$.
More specifically, ${}_K \mathcal{SM}_K$ should be a multi-superfusion category, which is a superfusion category except that the unit object may not be simple.
This supercategory reduces to $\mathrm{SVec}$ when $K$ is a trivial superalgebra $\mathbb{C}^{1|0}$.
The monoidal structure on ${}_K \mathcal{SM}_K$ is given by the tensor product over $K$, that is, the tensor product $X_1 \otimes_K X_2$ of objects $X_1, X_2 \in {}_K \mathcal{SM}_K$ is given by the image of the projector $p_{X_1, X_2}: X_1 \otimes X_2 \rightarrow X_1 \otimes X_2$ defined by $p_{X_1, X_2} = (\rho_{X_1}^R \otimes \rho_{X_2}^L) \circ (\mathrm{id}_{X_1} \otimes (\Delta_K \circ \eta_K) \otimes \mathrm{id}_{X_2})$, where $\rho_{X}^{L}$ and $\rho_X^{R}$ denote the left and right $K$-actions on $X$.
The splitting maps of this projector are denoted by $\pi_{X_1, X_2}: X_1 \otimes X_2 \rightarrow X_1 \otimes_K X_2$ and $\iota_{X_1, X_2}: X_1 \otimes_K X_2 \rightarrow X_1 \otimes X_2$, which satisfy 
\begin{equation}
\iota_{X_1, X_2} \circ \pi_{X_1, X_2} = p_{X_1, X_2}, \quad \pi_{X_1, X_2} \circ \iota_{X_1, X_2} = \mathrm{id}_{X_1 \otimes_K X_2}.
\label{eq: splitting maps}
\end{equation}
The tensor product $f \otimes_K g$ of superbimodule morphisms $f: X_1 \rightarrow X_1^{\prime}$ and $g: X_2 \rightarrow X_2^{\prime}$ is given by $f \otimes_K g = \pi_{X_1^{\prime}, X_2^{\prime}} \circ (f \otimes g) \circ \iota_{X_1, X_2}$, where $f \otimes g$ is the tensor product of $\mathbb{Z}_2$-graded linear maps defined by eq. \eqref{eq: super tensor product of morphisms}.
The structure morphisms of ${}_K \mathcal{SM}_K$ are analogous to those of the category ${}_K \mathcal{M}_K$ of $(K, K)$-bimodules, which is an ordinary multi-fusion category \cite{EGNO2015}.

\subsection{Hopf algebras and Hopf superalgebras}
\paragraph{Hopf algebras.}
Let $H$ be an associative unital algebra that is also a coassociative counital coalgebra and is equipped with a linear map $S: H \rightarrow H$ called an antipode.
The multiplication and the unit of $H$ are denoted by $m: H \otimes H \rightarrow H$ and $\eta: \mathbb{C} \rightarrow H$.
Similarly, the comultiplication and the counit of $H$ are denoted by $\Delta: H \rightarrow H \otimes H$ and $\epsilon: H \rightarrow \mathbb{C}$.
The algebra $H$ is called a Hopf algebra if the structure maps $(m, \eta, \Delta, \epsilon, S)$ satisfy the conditions listed below \cite{Montgomery1993}:
\begin{itemize}
\item The comultiplication $\Delta$ is a unit-preserving algebra homomorphism, i.e.
\begin{equation}
\Delta \circ m = (m \otimes m) \circ (\mathrm{id} \otimes c_{\mathrm{triv}} \otimes \mathrm{id}) \circ (\Delta \otimes \Delta), \quad \Delta \circ \eta = \eta \otimes \eta,
\label{eq: Hopf comultiplication}
\end{equation}
where $c_{\mathrm{triv}}: H \otimes H \rightarrow H \otimes H$ is the trivial braiding.
\item The counit $\epsilon$ is a unit-preserving algebra homomorphism, i.e.
\begin{equation}
\epsilon \circ m = \epsilon \otimes \epsilon, \quad \epsilon \circ \eta(1) = 1.
\label{eq: Hopf counit}
\end{equation}
\item The antipode $S$ satisfies the following antipode axiom:
\begin{equation}
m \circ (\mathrm{id} \otimes S) \circ \Delta = m \circ (S \otimes \mathrm{id}) \circ \Delta = \eta \circ \epsilon.
\label{eq: Hopf antipode}
\end{equation}
\end{itemize}
The antipode $S$ becomes an algebra and coalgebra anti-homomorphism.
Furthermore, the antipode of a semisimple Hopf algebra squares to the identity, i.e. $S^2 = \mathrm{id}$.
We note that the dual vector space $H^*$ of a finite dimensional Hopf algebra $H$ is also a Hopf algebra, whose structure maps are given by the duals of the structure maps of the original Hopf algebra $H$.
Every semisimple (weak) Hopf algebra over $\mathbb{C}$ is finite dimensional \cite{Nikshych2004}.

\paragraph{Representation categories.}
The category of representations of a semisimple Hopf algebra $H$ is a fusion category, which is denoted by $\mathrm{Rep}(H)$.
The monoidal structure is given by the tensor product of representations.
Specifically, the tensor product of representations $V$ and $W$ is defined by the usual tensor product $V \otimes W$ with the $H$-action $\rho_{V \otimes W}: H \otimes (V \otimes W) \rightarrow V \otimes W$ given via the comultiplication $\Delta$ as follows:
\begin{equation}
\rho_{V \otimes W} = (\rho_V \otimes \rho_W) \circ (\mathrm{id}_H \otimes c_{\mathrm{triv}} \otimes \mathrm{id}_W) \circ (\Delta \otimes \mathrm{id}_{V \otimes W}),
\label{eq: tensor product representation}
\end{equation}
where $\rho_V$ and $\rho_W$ denote the $H$-actions on $V$ and $W$ respectively.
The action of $h \in H$ on $v \otimes w \in V \otimes W$ is sometimes written as $\Delta(h) \cdot v \otimes w$.
The unit object is a one-dimensional representation $\mathbb{C}$ on which $H$ acts by the formula $h \cdot v = \epsilon(h)v$ for any $h \in H$ and $v \in \mathbb{C}$.
The dual of a representation $V \in \mathrm{Rep}(H)$ is the dual vector space $V^{*}$.
The action of $H$ on the dual representation $V^*$ is defined by
\begin{equation}
\rho_{V^*}(h \otimes \phi)(v) = \phi(\rho_V(S(h) \otimes v)),
\label{eq: Hopf action on dual}
\end{equation}
for $h \in H$, $\phi \in V^*$, and $v \in V$. 
The evaluation and coevaluation morphisms are given by eqs. \eqref{eq: ev of SVec} and \eqref{eq: coev of SVec}.
The other structure isomorphisms of $\mathrm{Rep}(H)$ are trivial.

Any indecomposable module category over $\mathrm{Rep}(H)$ is equivalent to the category ${}_K \mathcal{M}$ of left $K$-modules, where $K$ is a left $H$-comodule algebra \cite{AM2007}.
Here, a left $H$-comodule algebra $K$ is an algebra equipped with a left $H$-comodule structure $\delta_K^H: K \rightarrow H \otimes K$ that is compatible with the multiplication $m_K: K \otimes K \rightarrow K$ and the unit $\eta_K: \mathbb{C} \rightarrow K$.
More specifically, an algebra $(K, m_K, \eta_K)$ equipped with an $H$-comodule action $\delta_K^H$ is called an $H$-comodule algebra if the linear maps $\delta_K^H$, $m_K$, and $\eta_K$ satisfy
\begin{equation}
\delta^H_K \circ m_K = (m \otimes m_K) \circ (\mathrm{id}_H \otimes c_{\mathrm{triv}} \otimes \mathrm{id}_K) \circ (\delta_K^H \otimes \delta_K^H), \quad \delta_K^H \circ \eta_K = \eta \otimes \eta_K,
\label{eq: comodule algebra}
\end{equation}
where $m$ and $\eta$ are the multiplication and the unit of a Hopf algebra $H$.
The action of a left $H$-module $V \in \mathrm{Rep}(H)$ on a left $K$-module $M \in {}_K \mathcal{M}$ is denoted by $V \overline{\otimes} M \in {}_K \mathcal{M}$.
The underlying vector space of $V \overline{\otimes} M$ is the usual tensor product $V \otimes M$ and the left $K$-action on $V \overline{\otimes} M$ is defined via the $H$-comodule action on $K$ in an analogous way to eq. \eqref{eq: tensor product representation}.
The $\mathrm{Rep}(H)$-module category structure on ${}_K \mathcal{M}$ is encoded in a tensor functor $F_K: \mathrm{Rep}(H) \rightarrow \mathrm{End}({}_K \mathcal{M}) \cong {}_K \mathcal{M}_K$, which maps a left $H$-module $V \in \mathrm{Rep}(H)$ to a $(K, K)$-bimodule $V \overline{\otimes} K$.\footnote{The right $K$-module action on $V \overline{\otimes} K$ is given by the multiplication of $K$ from the right.}
Here, we used the equivalence $\mathrm{End}({}_K \mathcal{M}) \cong {}_K \mathcal{M}_K$ \cite{Watts1960, Eilenberg1960}.
We note that $F_K(V) \otimes_K M \cong V \overline{\otimes} M$.

\paragraph{Hopf superalgebras.}
Hopf superalgebras are defined as Hopf algebra objects in the symmetric fusion category $\underline{\mathrm{SVec}}$ whose objects are super vector spaces and whose morphisms are even linear maps.\footnote{The category $\underline{\mathrm{SVec}}$ is the underlying fusion category of the superfusion category $\mathrm{SVec}$ \cite{BE2017}.}
More specifically, a Hopf superalgebra $\mathcal{H}$ is an associative unital superalgebra $(\mathcal{H}, m, \eta)$ equipped with a structure of a coassociative counital supercoalgebra $(\mathcal{H}, \Delta, \epsilon)$ and an even linear map $S: \mathcal{H} \rightarrow \mathcal{H}$ called an antipode such that the structure maps $(m, \eta, \Delta, \epsilon, S)$ satisfy the conditions \eqref{eq: Hopf comultiplication}, \eqref{eq: Hopf counit}, and \eqref{eq: Hopf antipode} in which the trivial braiding $c_{\mathrm{triv}}$ is replaced by the symmetric braiding $c_{\mathrm{super}}$.
As in the case of Hopf algebras, the antipode of a Hopf superalgebra is a superalgebra and supercoalgebra anti-homomorphism.
Furthermore, when a Hopf superalgebra $\mathcal{H}$ is semisimple, the antipode $S$ squares to the $\mathbb{Z}_2$-grading automorphism of $\mathcal{H}$ \cite{AEG2001}. 
In particular, the antipode $S$ satisfies $S^4 = \mathrm{id}$.
The dual of a Hopf superalgebra $\mathcal{H}$ is also a Hopf superalgebra.

\paragraph{Representation supercategories.}
The supercategory of super representations of a semisimple Hopf superalgebra $\mathcal{H}$ is a superfusion category, which we write as $\mathrm{SRep}(\mathcal{H})$.
Objects and morphisms of $\mathrm{SRep}(\mathcal{H})$ are left $\mathcal{H}$-supermodules (i.e., super representations of $\mathcal{H}$) and $\mathcal{H}$-supermodule morphisms respectively.
The monoidal structure on $\mathrm{SRep}(\mathcal{H})$ is given by the tensor product of super representations.
The underlying super vector space of the tensor product representation is the usual tensor product of super vector spaces.
The $\mathcal{H}$-action on the tensor product representation is given by eq. \eqref{eq: tensor product representation} where the trivial braiding $c_{\mathrm{triv}}$ is replaced by the symmetric braiding $c_{\mathrm{super}}$.
The unit object is a one-dimensional super representation $\mathbb{C}^{1|0}$ on which $\mathcal{H}$ acts as $h \cdot v = \epsilon(h)v$ for all $h \in \mathcal{H}$ and $v \in \mathbb{C}^{1|0}$.
The dual object of a super representation $V \in \mathrm{SRep}(\mathcal{H})$ is the dual super vector space $V^*$.
The action of $\mathcal{H}$ on $V^*$ is defined by
\begin{equation}
\rho_{V^*}(h \otimes \phi) (v) = (-1)^{|h| |\phi|} \phi (\rho_V(S(h) \otimes v))
\end{equation}
for homogeneous elements $h \in \mathcal{H}$, $\phi \in V^*$, and $v \in V$.
This action is linearly extended to inhomogeneous elements.
The evaluation and coevaluation morphisms and the other structure isomorphisms of $\mathrm{SRep}(\mathcal{H})$ are induced from those of $\mathrm{SVec}$.

The supercategory ${}_K \mathcal{SM}$ of left $K$-supermodules is an $\mathrm{SRep}(\mathcal{H})$-supermodule category when $K$ is a left $\mathcal{H}$-supercomodule algebra.
Here, a left $\mathcal{H}$-supercomodule algebra $K$ is a superalgebra equipped with a left $\mathcal{H}$-supercomodule structure that satisfies the compatibility condition \eqref{eq: comodule algebra} where the trivial braiding $c_{\mathrm{triv}}$ is replaced by the symmetric braiding $c_{\mathrm{super}}$.
The supertensor functor $\mathcal{F}_K: \mathrm{SRep}(\mathcal{H}) \rightarrow \mathrm{End}({}_K \mathcal{SM}) \cong {}_K \mathcal{SM}_K$ that represents an $\mathrm{SRep}(\mathcal{H})$-supermodule structure on ${}_K \mathcal{SM}$ is analogous to the tensor functor $F_K: \mathrm{Rep}(H) \rightarrow {}_K \mathcal{M}_K$ mentioned above.

\subsection{Weak Hopf algebras and weak Hopf superalgebras}
\label{sec: Weak Hopf algebras and weak Hopf superalgebras}
\paragraph{Weak Hopf algebras.}
Weak Hopf algebras are defined by relaxing the defining properties of ordinary Hopf algebras.
Specifically, a weak Hopf algebra $H$ is an associative unital algebra $(H, m, \eta)$ equipped with a structure of a coassociative counital coalgebra $(H, \Delta, \epsilon)$ and an antipode $S: H \rightarrow H$ that satisfy the following properties \cite{BNS1999}:
\begin{itemize}
\item The comultiplication $\Delta$ is multiplicative, i.e.
\begin{equation}
\Delta \circ m = (m \otimes m) \circ (\mathrm{id} \otimes c_{\mathrm{triv}} \otimes \mathrm{id}) \circ (\Delta \otimes \Delta).
\label{eq: weak Hopf comultiplication}
\end{equation}
We note that the comultiplication need not preserve the unit.
\item The counit $\epsilon$ satisfies
\begin{equation}
\epsilon \circ m \circ (m \otimes \mathrm{id}) = (\epsilon \otimes \epsilon) \circ (m \otimes m) \circ (\mathrm{id} \otimes \Delta \otimes \mathrm{id}) = (\epsilon \otimes \epsilon) \circ (m \otimes m) \circ (\mathrm{id} \otimes (c_{\mathrm{triv}} \circ \Delta) \otimes \mathrm{id}).
\label{eq: weak Hopf counit}
\end{equation}
\item The unit $\eta$ satisfies
\begin{equation}
(\Delta \otimes \mathrm{id}) \circ \Delta \circ \eta = (\mathrm{id} \otimes m \otimes \mathrm{id}) \circ (\Delta \otimes \Delta) \circ (\eta \otimes \eta) = (\mathrm{id} \otimes (m \circ c_{\mathrm{triv}}) \otimes \mathrm{id}) \circ (\Delta \otimes \Delta) \circ (\eta \otimes \eta).
\label{eq: weak Hopf unit}
\end{equation}
\item The antipode $S$ satisfies
\begin{equation}
\begin{aligned}
m \circ (\mathrm{id} \otimes S) \circ \Delta & = ((\epsilon \circ m) \otimes \mathrm{id}) \circ (\mathrm{id} \otimes c_{\mathrm{triv}}) \circ ((\Delta \circ \eta) \otimes \mathrm{id}), \\
m \circ (S \otimes \mathrm{id}) \circ \Delta & = (\mathrm{id} \otimes (\epsilon \circ m)) \circ (c_{\mathrm{triv}} \otimes \mathrm{id}) \circ (\mathrm{id} \otimes (\Delta \circ \eta)), \\
S & = m \circ (m \otimes \mathrm{id}) \circ (S \otimes \mathrm{id} \otimes S) \circ (\Delta \otimes \mathrm{id}) \circ \Delta.
\end{aligned}
\label{eq: weak Hopf antipode}
\end{equation}
\end{itemize}
The antipode $S$ of a weak Hopf algebra is an algebra and coalgebra homomorphism.
The linear maps defined by the first and the second equations in eq. \eqref{eq: weak Hopf antipode} are called the target counital map $\epsilon_t$ and the source counital map $\epsilon_s$ respectively.
These counital maps are idempotents, i.e. we have $\epsilon_t \circ \epsilon_t = \epsilon_t$ and $\epsilon_s \circ \epsilon_s = \epsilon_s$.
The images of these idempotents are called the target and source counital subalgebras, which are denoted by $H_t$ and $H_s$ respectively.

\paragraph{Representation categories.}
Representations of a semisimple weak Hopf algebra $H$ form a multi-fusion category $\mathrm{Rep}(H)$ \cite{NTV2003}, which is a finite semisimple tensor category whose unit object is not necessarily simple.
Conversely, any multi-fusion category is equivalent to the representation category of an appropriate weak Hopf algebra $H$ \cite{Hayashi1999, Ostrik2003}.
The tensor product $V \boxtimes W$ of representations $V, W \in \mathrm{Rep}(H)$ is defined so that the unit element $\eta(1) \in H$ acts as the identity.
More specifically, the underlying vector space of $V \boxtimes W$ is the image of the action of $\eta(1)$ on $V \otimes W$, i.e. $V \boxtimes W = \Delta(\eta(1)) \cdot V \otimes W$, where the action of $H$ on $V \otimes W$ is defined by eq. \eqref{eq: tensor product representation}.\footnote{The vector space $V \boxtimes W$ can also be viewed as the tensor product of $V$ and $W$ over $H_t$.}
The action of $H$ on $V \boxtimes W$ is given by the restriction of the $H$-action on $V \otimes W$.
The unit object of $\mathrm{Rep}(H)$ is the target counital subalgebra $H_t$, where $H$ acts on $H_t$ by the multiplication followed by the projection to $H_t$, i.e. $h \cdot v = \epsilon_t(hv)$ for all $h \in H$ and $v \in H_t$.
The dual of a representation $V \in \mathrm{Rep}(H)$ is the dual vector space $V^*$ on which $H$ acts by eq. \eqref{eq: Hopf action on dual}.
The details of evaluation and coevaluation morphisms and the other structure isomorphisms of $\mathrm{Rep}(H)$ can be found in \cite{NTV2003}.

Any indecomposable module category over a unitary (multi-)fusion category $\mathrm{Rep}(H)$ can be written as the category ${}_K \mathcal{M}$ of left $K$-modules, where $K$ is a left $H$-comodule algebra \cite{Henker2011}.
Here, a left $H$-comodule algebra $K$ for a weak Hopf algebra $(H, m, \eta, \Delta, \epsilon, S)$ is an algebra equipped with a left $H$-comodule structure $\delta_K^H: K \rightarrow H \otimes K$ that is compatible with the algebra structure $(K, m_K, \eta_K)$ in the following sense:
\begin{equation}
\delta_K^H \circ m_K = (m \otimes m_K) \circ (\mathrm{id}_H \otimes c_{\mathrm{triv}} \otimes \mathrm{id}_K) \circ (\delta_K^H \otimes \delta_K^H), \quad
\delta_K^H \circ \eta_K = (\epsilon_s \otimes \mathrm{id}_K) \circ \delta_K^H \circ \eta_K.
\label{eq: weak comodule algebra}
\end{equation}
We note that this equation reduces to eq. \eqref{eq: comodule algebra} when $H$ is a Hopf algebra. 
The action of $V \in \mathrm{Rep}(H)$ on $M \in {}_K \mathcal{M}$ is denoted by $V \overline{\otimes} M \in {}_K \mathcal{M}$, which is the subspace of $V \otimes M$ on which the unit element $\eta_K(1) \in K$ acts as the identity.
The left $K$-action on $V \overline{\otimes} M$ is the restriction of the left $K$-action on $V \otimes M$ defined via the left $H$-comodule structure on $K$.
The corresponding tensor functor $F_K: \mathrm{Rep}(H) \rightarrow {}_K \mathcal{M}_K$ maps a left $H$-module $V$ to a $(K, K)$-bimodule $V \overline{\otimes} K$.
In particular, we have an isomorphism of $(K, K)$-bimodules $F_K(V) \otimes_K M \cong V \overline{\otimes} M$.

\paragraph{Weak Hopf superalgebras.}
Weak Hopf superalgebras are defined as weak Hopf algebra objects in the symmetric fusion category $\underline{\mathrm{SVec}}$.\footnote{Weak Hopf algebra objects in more general braided tensor categories are discussed in \cite{AAFVGR2008, PS2009}.}
More specifically, a weak Hopf superalgebra $\mathcal{H}$ is an associative unital superalgebra $(\mathcal{H}, m, \eta)$ equipped with a structure of a coassociative counital supercoalgebra $(\mathcal{H}, \Delta, \epsilon)$ and an antipode $S: \mathcal{H} \rightarrow \mathcal{H}$ that satisfy the properties \eqref{eq: weak Hopf comultiplication}--\eqref{eq: weak Hopf antipode} where the trivial braiding $c_{\mathrm{triv}}$ is replaced by the symmetric braiding $c_{\mathrm{super}}$.
The first and the second equations in eq. \eqref{eq: weak Hopf antipode} define the target counital map $\epsilon_t$ and the source counital map $\epsilon_s$, whose images are the target and source counital subalgebras denoted by $\mathcal{H}_t$ and $\mathcal{H}_s$ respectively.

\paragraph{Representation supercategories.}
The supercategory $\mathrm{SRep}(\mathcal{H})$ of super representations of a weak Hopf superalgebra $\mathcal{H}$ is defined in a similar way to the representation category $\mathrm{Rep}(H)$ of a weak Hopf algebra $H$.
Supermodule categories over $\mathrm{SRep}(\mathcal{H})$ and the corresponding supertensor functors are similar to module categories over $\mathrm{Rep}(H)$ and the corresponding tensor functors.
In particular, the supercategory ${}_K \mathcal{SM}$ of left $K$-supermodules is an $\mathrm{SRep}(\mathcal{H})$-supermodule category if $K$ is a left $\mathcal{H}$-supercomodule algebra, which is a superalgebra equipped with a left $\mathcal{H}$-supercomodule structure satisfying the compatibility condition \eqref{eq: weak comodule algebra} with $c_{\mathrm{triv}}$ replaced by $c_\mathrm{super}$.
Accordingly, we have a supertensor functor $\mathcal{F}_K: \mathrm{SRep}(\mathcal{H}) \rightarrow {}_K \mathcal{SM}_K$ when $K$ is a left $\mathcal{H}$-supercomodule algebra.

\section{Fermionization of fusion category symmetries}
\label{sec: Fermionization of fusion category symmetries}
In this section, we discuss the symmetries of 1+1d fermionic systems obtained by the fermionization of bosonic systems with fusion category symmetries.
We give a general formula to determine the symmetry of the fermionized theory from the symmetry of a bosonic theory.
The derivation of this formula will be given in section \ref{sec: Fermionic TQFTs with superfusion category symmetries} in the context of topological field theories.

\subsection{Fermionization of non-anomalous symmetries}
\label{sec: Fermionization of non-anomalous symmetries}
A fusion category symmetry of a 1+1d bosonic system is said to be non-anomalous if there exist gapped phases with unique ground states preserving the symmetry.
Mathematically, non-anomalous symmetries are described by fusion categories that admit fiber functors \cite{TW2019}.
Such fusion categories are equivalent to the representation categories $\mathrm{Rep}(H)$ of semisimple Hopf algebras $H$.
A $\mathrm{Rep}(H)$ symmetry can be fermionized if it has a non-anomalous $\mathbb{Z}_2$ subgroup symmetry.
Henceforth, a non-anomalous $\mathbb{Z}_2$ subgroup will be simply called a $\mathbb{Z}_2$ subgroup because any $\mathbb{Z}_2$ subgroup of a non-anomalous symmetry $\mathrm{Rep}(H)$ is non-anomalous.

Let us describe a $\mathbb{Z}_2$ subgroup of $\mathrm{Rep}(H)$ symmetry in terms of a group-like element in the dual Hopf algebra $H^*$.
In general, a subgroup of $\mathrm{Rep}(H)$ symmetry consists of one-dimensional representations of $H$, which are denoted by $V_g$.
Since $V_g$ is a one-dimensional representation, the action of $a \in H$ on $ v \in V_g$ can be written as $a \cdot v = g(a) v$ by using a linear functional $g \in H^*$ that satisfies 
\begin{equation}
g(ab) = g(a) g(b), \quad g(\eta(1)) = 1, \quad \forall a, b \in H.
\label{eq: group-like rep}
\end{equation}
The first equality of the above equation is equivalent to 
\begin{equation}
\Delta_{H^*} (g) = g \otimes g,
\label{eq: group-like element}
\end{equation}
where $\Delta_{H^{*}}: H^* \rightarrow H^* \otimes H^*$ is the comultiplication on $H^*$.
A non-zero element $g \in H^*$ satisfying eq. \eqref{eq: group-like element} is called a group-like element in $H^*$.
When $g \in H^*$ is a group-like element, the second equality of eq. \eqref{eq: group-like rep} is automatically satisfied due to the defining property \eqref{eq: Hopf comultiplication} of Hopf algebra $H$.
An immediate consequence of eq. \eqref{eq: group-like element} is that group-like elements of $H^{*}$ form a group $G(H^*)$, whose multiplication is the multiplication on $H^*$ and whose unit element is the counit $\epsilon: H \rightarrow \mathbb{C}$.\footnote{The inverse of $g \in G(H^*)$ is given by $S_{H^*}(g)$, where $S_{H^*}: H^* \rightarrow H^*$ is the antipode of $H^*$.}
Accordingly, one-dimensional representations of $H$ also form a group, which is isomorphic to $G(H^*)$.
The multiplication of one-dimensional representations $V_g$ and $V_h$ is given by the tensor product of representations $V_g \otimes V_h \cong V_{gh}$.
In particular, if $G(H^*)$ has a $\mathbb{Z}_2$ subgroup generated by $u \in G(H^*)$, a fusion category $\mathrm{Rep}(H)$ has a $\mathbb{Z}_2$ subgroup generated by a one-dimensional representation $V_u \in \mathrm{Rep}(H)$.
The generator $u$ of a $\mathbb{Z}_2$ subgroup of $G(H^*)$ is called a $\mathbb{Z}_2$ group-like element in $H^*$, which satisfies $u \otimes u (\Delta (a)) = \epsilon (a)$ for all $a \in H$ in addition to eq. \eqref{eq: group-like rep}.

Let $u \in G(H^*)$ be a $\mathbb{Z}_2$ group-like element in $H^*$ and let $\mathbb{Z}^u_2$ be the $\mathbb{Z}_2$ subgroup of $\mathrm{Rep}(H)$ generated by $V_u \in \mathrm{Rep}(H)$.
We propose that the fermionization of $\mathrm{Rep}(H)$ symmetry with respect to $\mathbb{Z}^u_2$ subgroup is a superfusion category symmetry $\mathrm{SRep}(\mathcal{H}^u)$, where $\mathcal{H}^u$ is a Hopf superalgebra defined as follows:
\begin{itemize}
\item The underlying vector space of $\mathcal{H}^u$ is $H$.
\item The $\mathbb{Z}_2$-grading automorphism $p_u: \mathcal{H}^u \rightarrow \mathcal{H}^u$ is defined by the adjoint action of $u$, i.e.
\begin{equation}
p_u := (u \otimes \mathrm{id} \otimes u) \circ (\mathrm{id} \otimes \Delta) \circ \Delta.
\label{eq: grading of Hopf superalgebra}
\end{equation}
The $\mathbb{Z}_2$-grading of a homogeneous element $a \in \mathcal{H}^u$ is denoted by $p_u(a) = (-1)^{|a|} a$.
\item The multiplication $m_u: \mathcal{H}^u \otimes \mathcal{H}^u \rightarrow \mathcal{H}^u$ of homogeneous elements $a$ and $b$ is given by
\begin{equation}
m_u(a \otimes b) = m \circ (\mathrm{id} \otimes u^{|b|} \otimes \mathrm{id}) \circ (\Delta \otimes \mathrm{id}) (a \otimes b),
\label{eq: multiplication of Hopf superalgebra}
\end{equation}
where $m$ on the right-hand side is the multiplication on the original Hopf algebra $H$.
The above multiplication is extended linearly to the multiplication of inhomogeneous elements.
\item The antipode $S_u: \mathcal{H}^u \rightarrow \mathcal{H}^u$ of a homogeneous element $a$ is given by
\begin{equation}
S_u (a) = (u^{|a|} \otimes S) \circ \Delta (a),
\label{eq: antipode of Hopf superalgebra}
\end{equation}
where $S$ on the right-hand side is the antipode of the original Hopf algebra $H$.
The above antipode is extended linearly to inhomogeneous elements.
\item The comultiplication $\Delta: \mathcal{H}^u \rightarrow \mathcal{H}^u \otimes \mathcal{H}^u$, the unit $\eta: \mathbb{C} \rightarrow \mathcal{H}^u$, and the counit $\epsilon: \mathcal{H}^u \rightarrow \mathbb{C}$ are the same as those of $H$.
\end{itemize}
The same relation between a Hopf algebra and a Hopf superalgebra is discussed in Theorem 3.1.1 of \cite{AEG2001}.

We note that the inverse of the map $(H, m, \eta, \Delta, \epsilon, S) \mapsto (\mathcal{H}^u, m_u, \eta, \Delta, \epsilon, S_u)$ is given by
\begin{equation}
m(a \otimes b) = m_u \circ (\mathrm{id} \otimes u^{|b|} \otimes \mathrm{id}) \circ (\Delta \otimes \mathrm{id}) (a \otimes b), \quad S (a) = (u^{|a|} \otimes S_u) \circ \Delta (a).
\label{eq: inverse map}
\end{equation}
This inverse map describes the bosonization of a superfusion category symmetry.\footnote{This is different from the bosonization of Hopf superalgebra discussed in \cite{Majid1994, AAY2011}.}
We also note that the fermion parity symmetry $\mathbb{Z}_2^F \subset \mathrm{SRep}(\mathcal{H}^u)$ is generated by a one-dimensional super representation $V_u$ on which $a \in \mathcal{H}^u$ acts as the multiplication by a complex number $u(a)$.

In the rest of this subsection, we show that $\mathcal{H}^u$ defined above is a Hopf superalgebra.
To this end, we first notice that the structure maps defined above preserve the $\mathbb{Z}_2$-grading because the $\mathbb{Z}_2$ group-like element $u \in G(H^*)$ is $\mathbb{Z}_2$-even and the structure maps of the original Hopf algebra $H$ preserve the $\mathbb{Z}_2$-grading defined by eq. \eqref{eq: grading of Hopf superalgebra}.
Moreover, a straightforward computation shows that the multiplication $m_u$ defined by eq. \eqref{eq: multiplication of Hopf superalgebra} is associative and the multiplicative unit with respect to $m_u$ is given by $\eta$.
This shows that $(\mathcal{H}^u, m_u, \eta)$ is an associative unital superalgebra.
It is also clear that $(\mathcal{H}^u, \Delta, \epsilon)$ is a coassociative counital supercoalgebra.
Therefore, it remains to show eqs. \eqref{eq: Hopf comultiplication}, \eqref{eq: Hopf counit}, and \eqref{eq: Hopf antipode} in which the trivial braiding $c_{\mathrm{triv}}$ is replaced by the symmetric braiding $c_{\mathrm{super}}$.
Among these equations, the second equality of eq. \eqref{eq: Hopf comultiplication} and the second equality of eq. \eqref{eq: Hopf counit} are obvious because they do not involve the modified structure maps $m_u$ and $S_u$.
An efficient way to show the other equations is to use string diagrams.
The sting diagrams for the structure maps of the original Hopf algebra $H$ are written as
\begin{equation}
\adjincludegraphics[valign = c, width = 1.2cm]{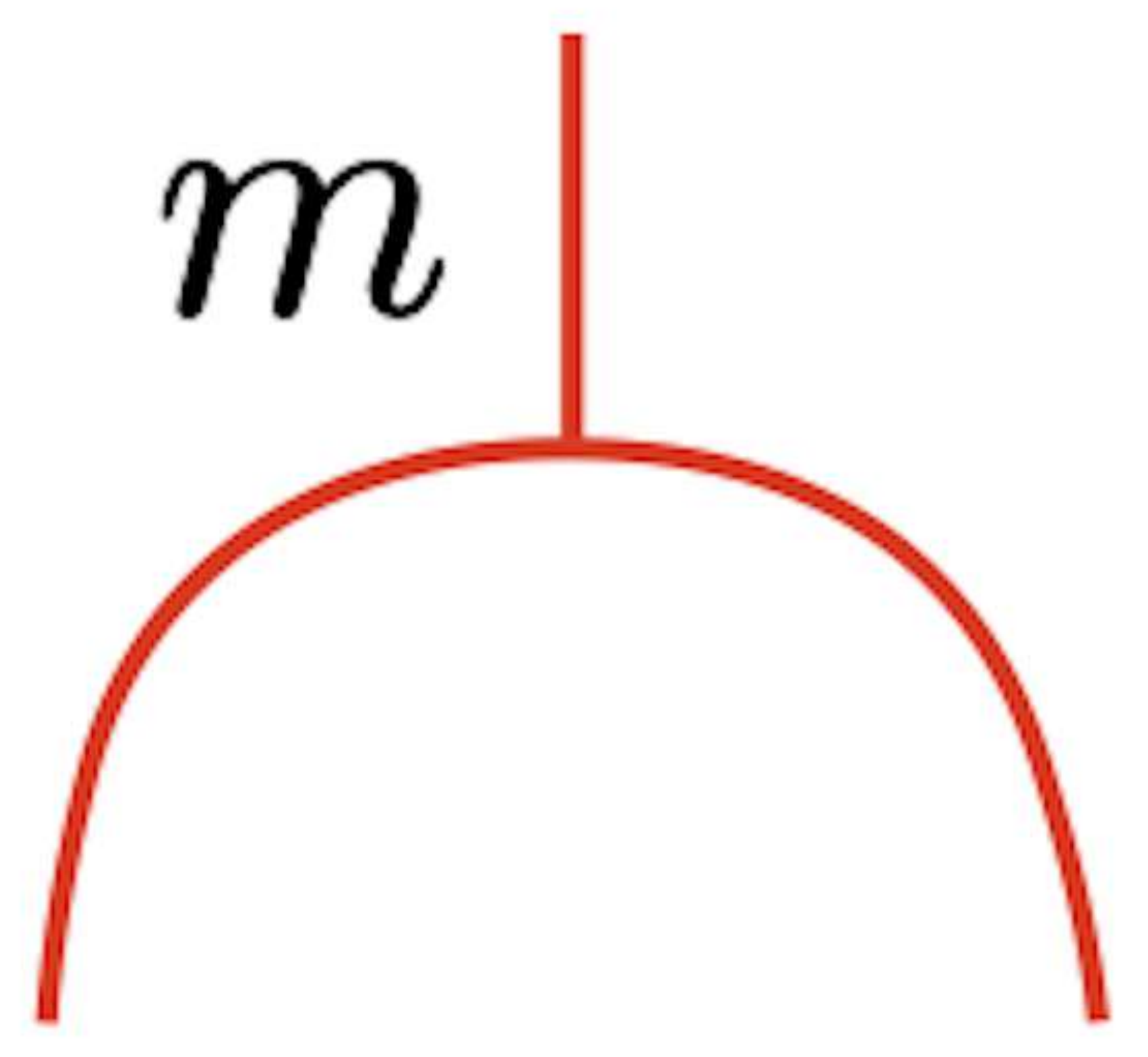}, \quad \quad
\adjincludegraphics[valign = c, width = 1.2cm]{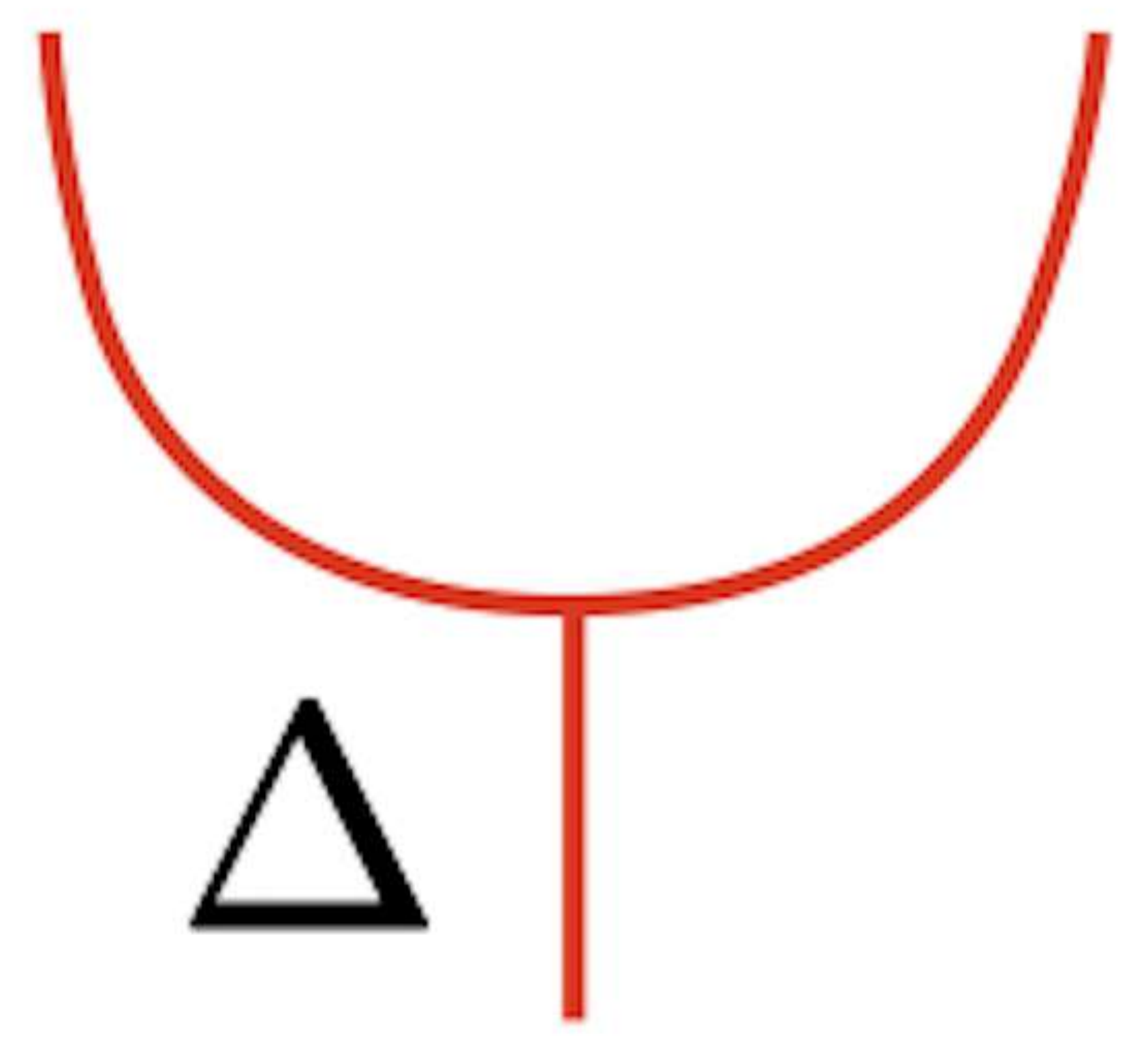}, \quad \quad
\adjincludegraphics[valign = c, width = 0.3cm]{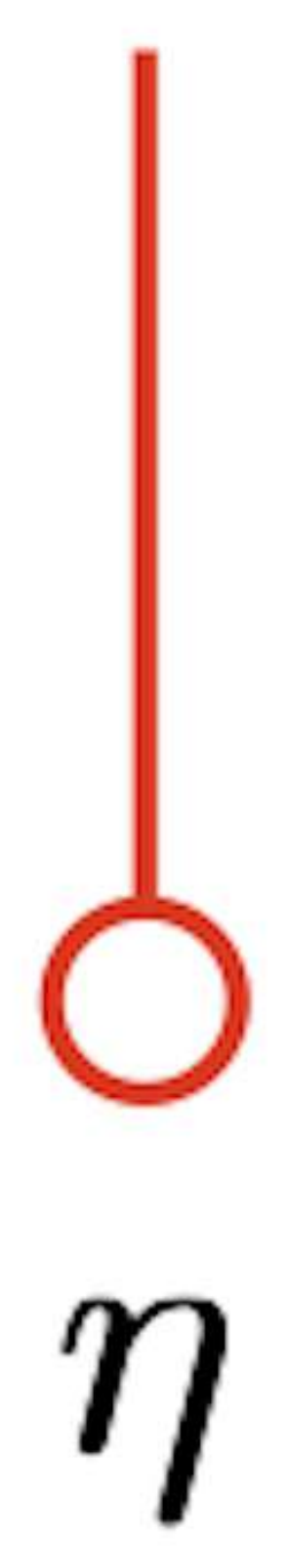}, \quad \quad
\adjincludegraphics[valign = c, width = 0.3cm]{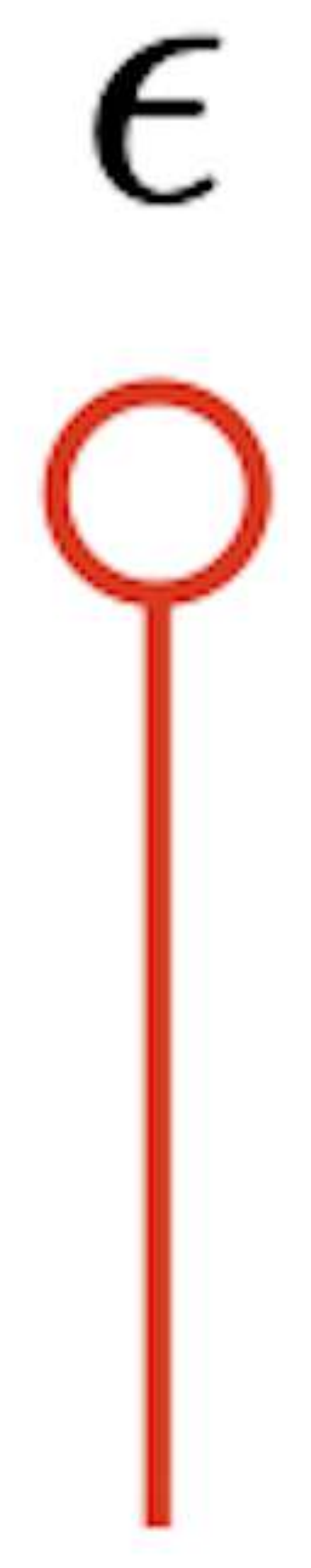}, \quad \quad
\adjincludegraphics[valign = c, width = 0.55cm]{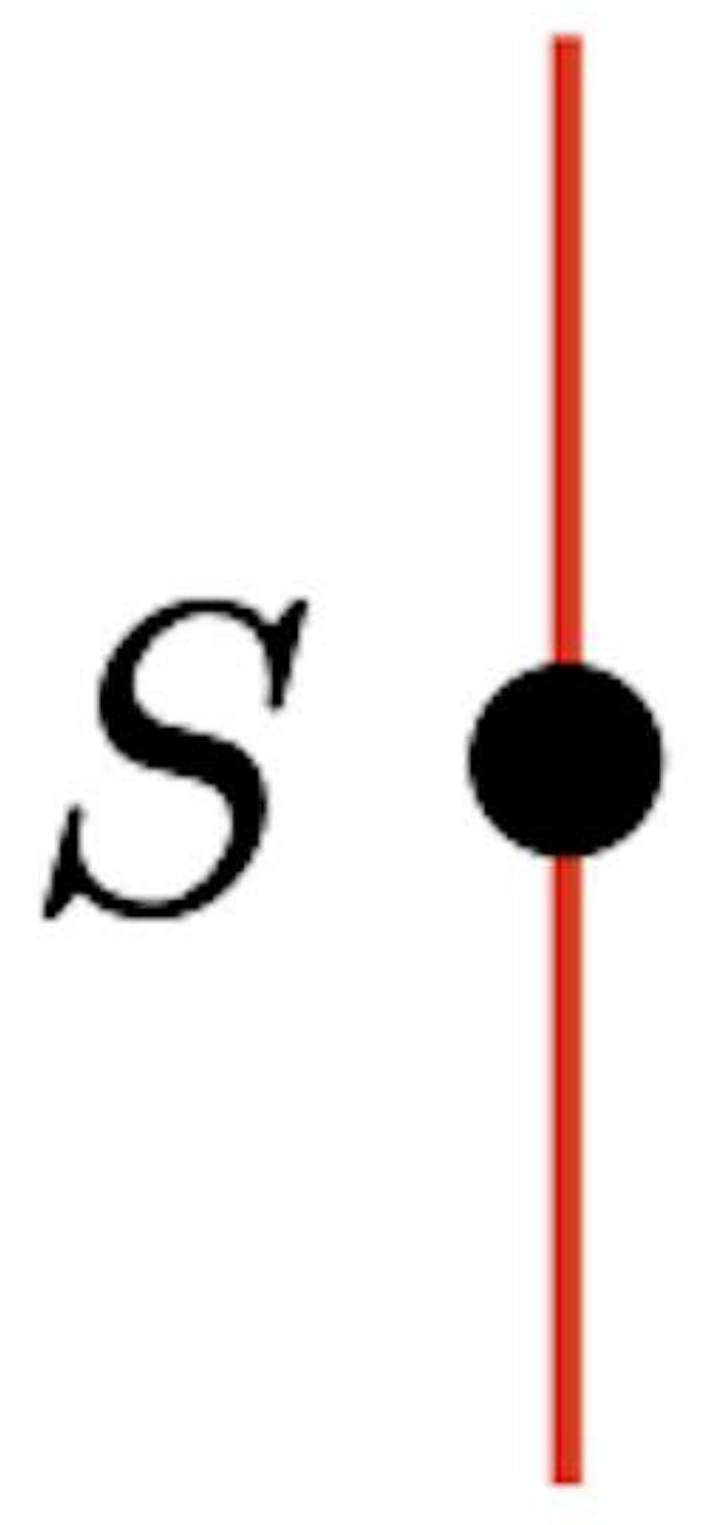}.
\end{equation}
By using these string diagrams, we can show the first equality of eq. \eqref{eq: Hopf comultiplication} as follows:
\begin{equation}
\mathrm{LHS} ~ = ~
\adjincludegraphics[valign = c, width = 1.5cm]{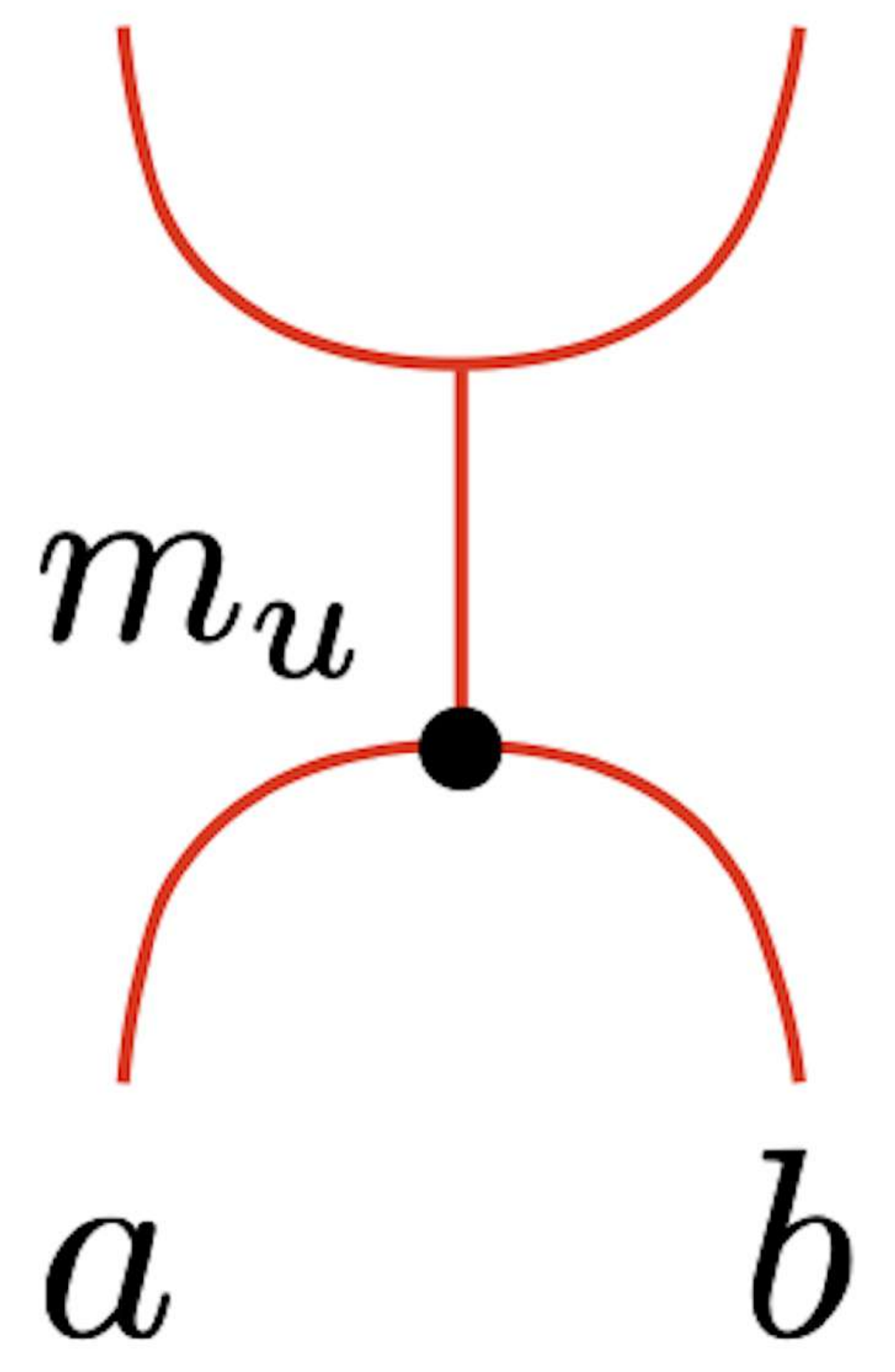} ~ = ~
\adjincludegraphics[valign = c, width = 3.0cm]{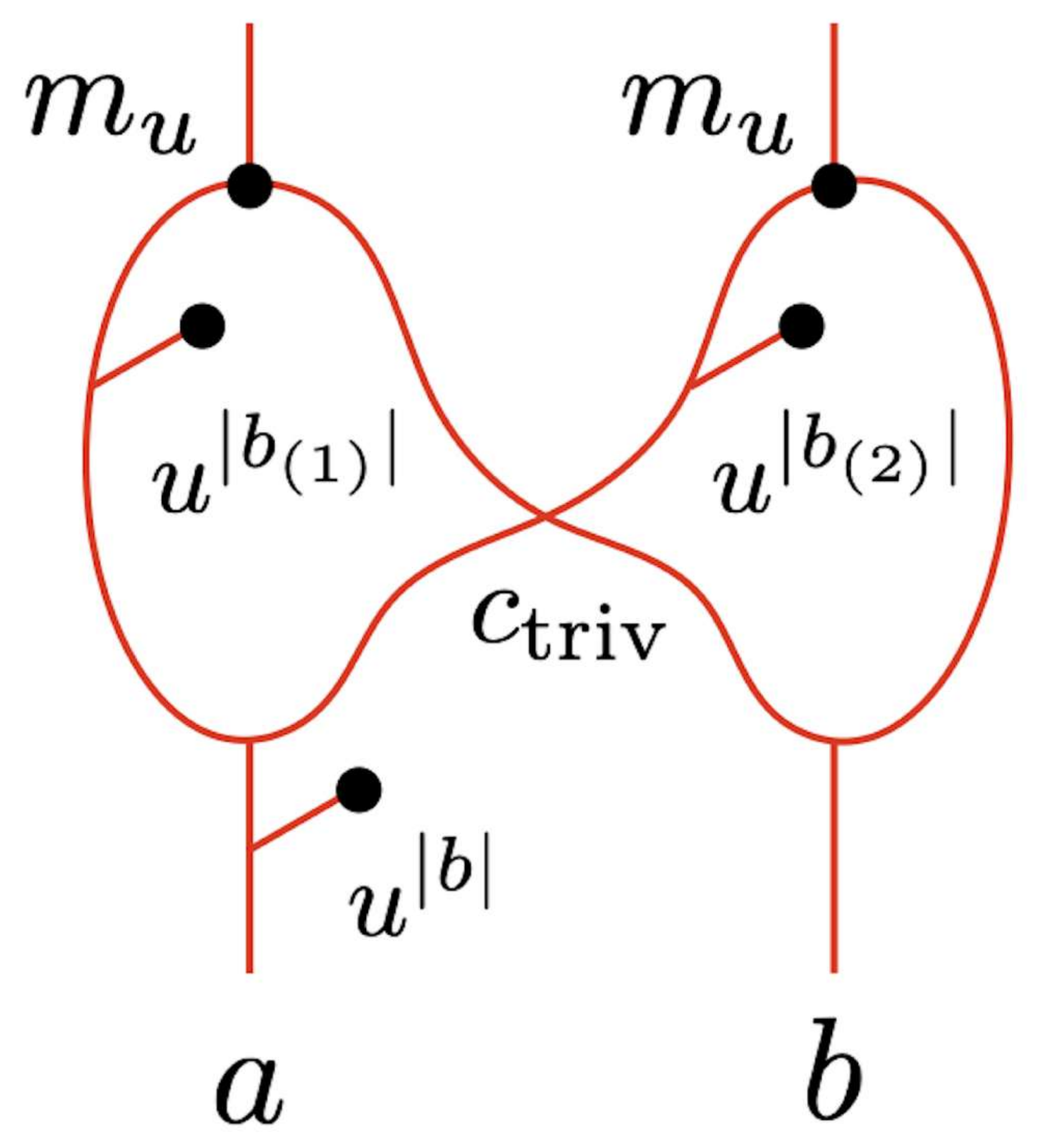} ~ = ~
\adjincludegraphics[valign = c, width = 2.8cm]{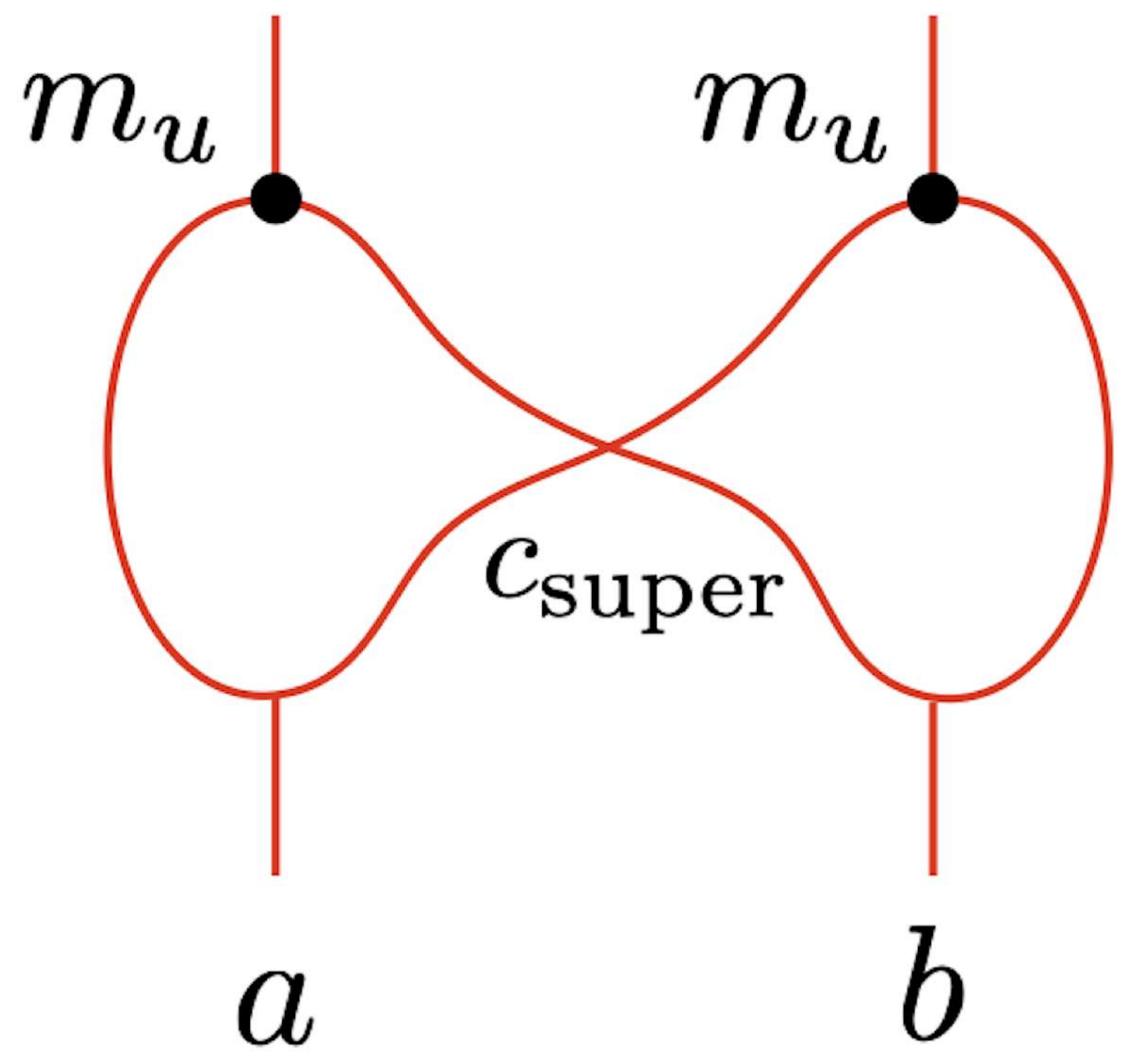} ~ = ~
\mathrm{RHS},
\label{eq: multiplicative Delta of Hopf superalgebra}
\end{equation}
where we used the Sweedler notation $\Delta(b) = b_{(1)} \otimes b_{(2)}$ and the identity $c_{\mathrm{triv}}(a_{(2)} \otimes b_{(1)}) = c_{\mathrm{super}} \circ (p_u^{|b_{(1)}|} \otimes \mathrm{id}) (a_{(2)} \otimes b_{(1)})$ between the trivial braiding $c_{\mathrm{triv}}$ and the symmetric braiding $c_{\mathrm{super}}$.
When $b$ is homogeneous, we can assume without loss of generality that $b_{(1)}$ and $b_{(2)}$ are homogeneous because the comultiplication $\Delta$ is $\mathbb{Z}_2$-even, and therefore the notations $|b_{(1)}|$ and $|b_{(2)}|$ in eq. \eqref{eq: multiplicative Delta of Hopf superalgebra} make sense.
The first equality of eq. \eqref{eq: Hopf counit} also holds because
\begin{equation}
\mathrm{LHS} ~ = ~
\adjincludegraphics[valign = c, width = 1.5cm]{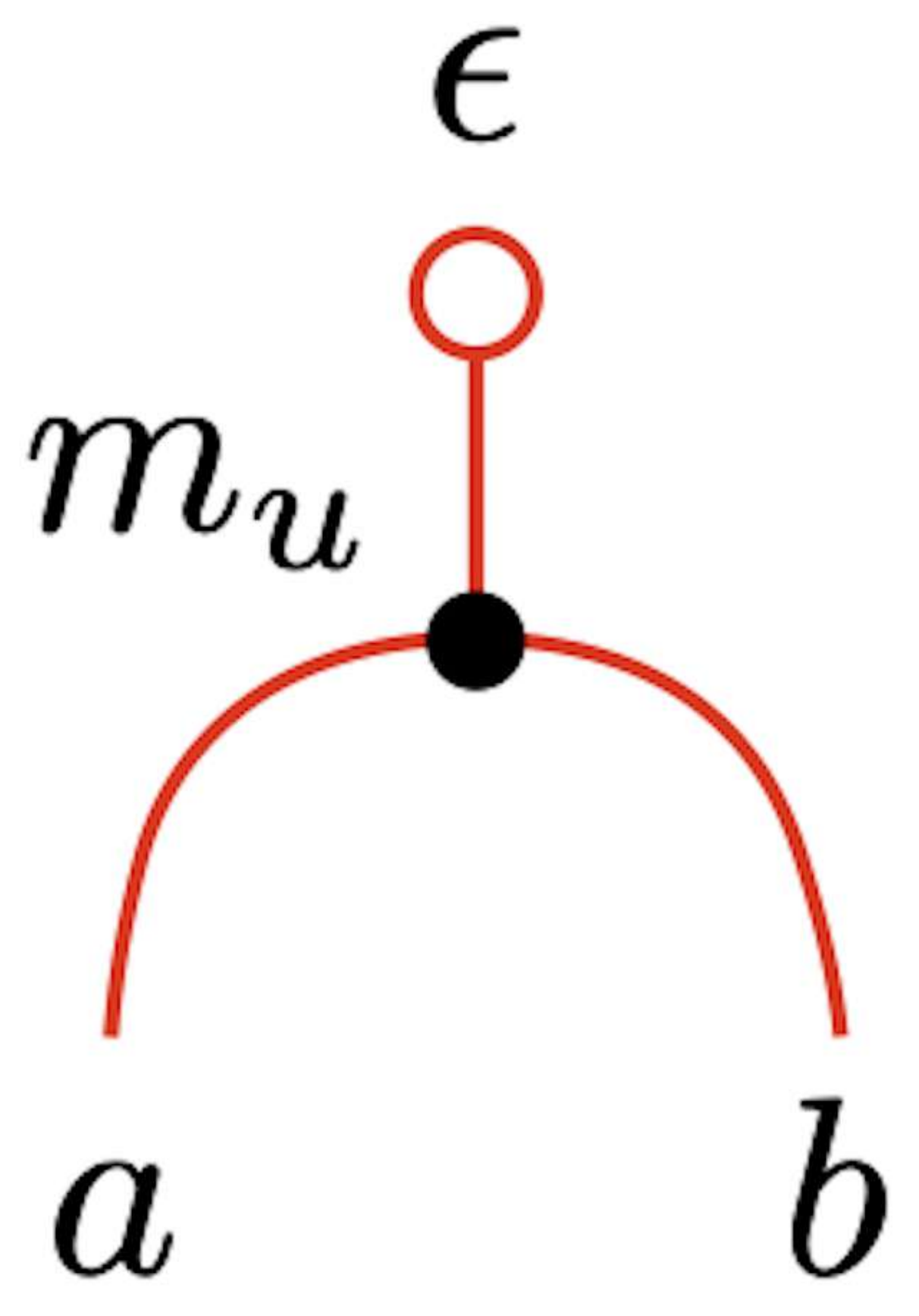} ~ = ~
\adjincludegraphics[valign = c, width = 1.8cm]{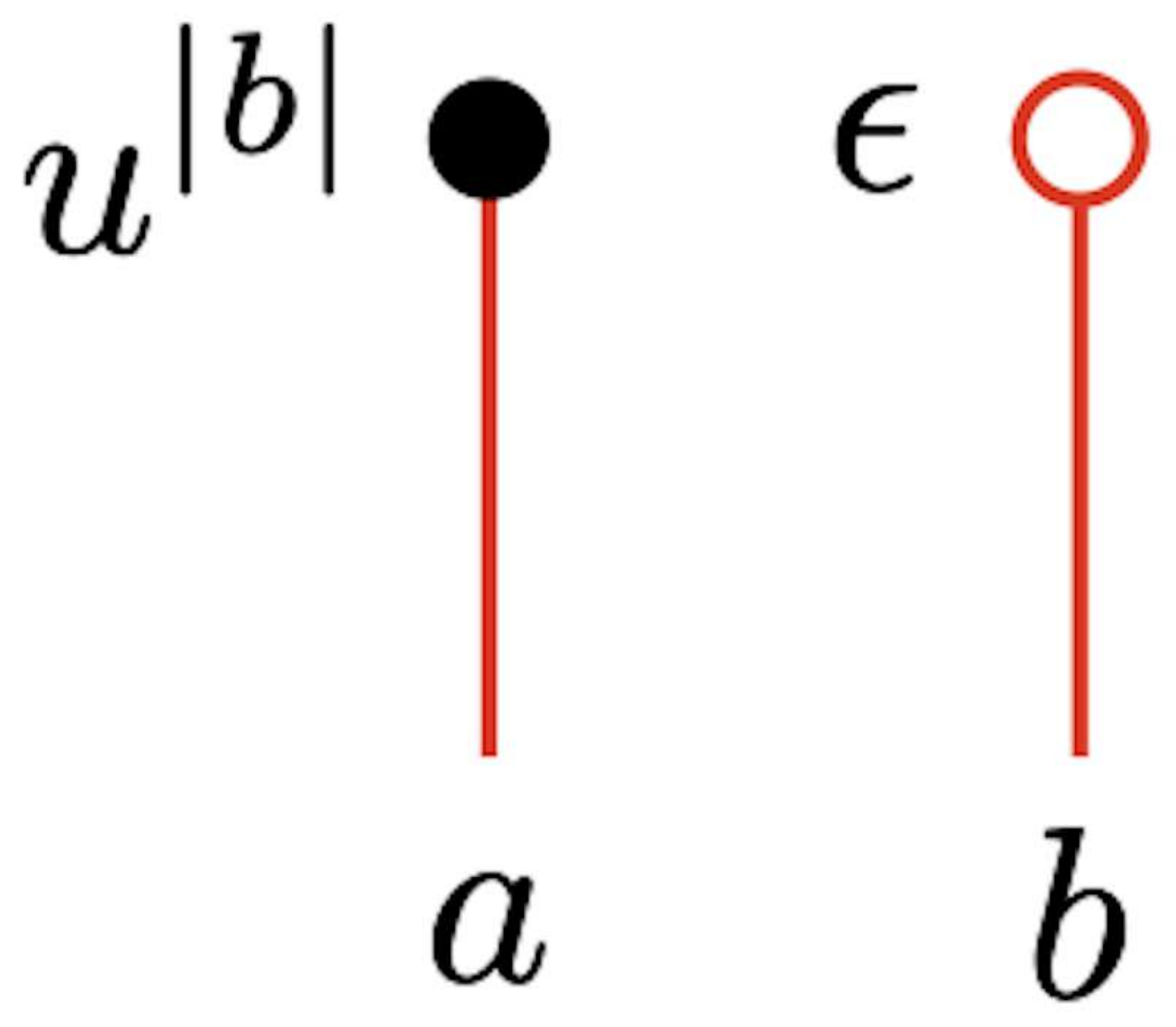} ~ = ~
\adjincludegraphics[valign = c, width = 1.5cm]{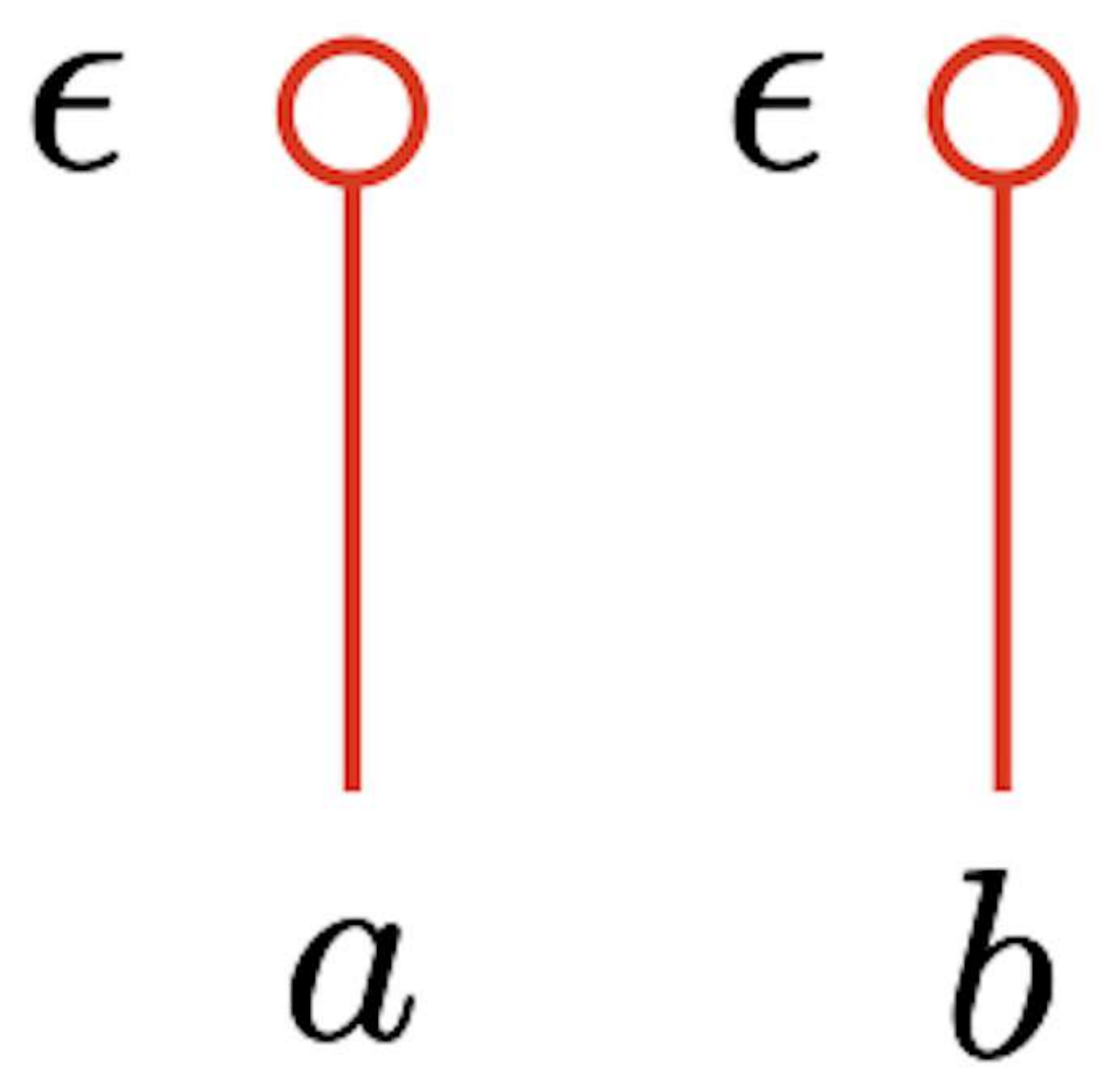} ~ = ~
\mathrm{RHS},
\label{eq: unit-preserving epsilon of Hopf superalgebra}
\end{equation}
where the third equality follows from the fact that $\epsilon(b) = 0$ unless $|b| = 0$.
Equation \eqref{eq: Hopf antipode} can also be checked similarly.
Thus, we find that the super vector space $\mathcal{H}^u$ equipped with the structure maps $(m_u, \eta, \Delta, \epsilon, S_u)$ is a Hopf superalgebra.

\subsection{Fermionization of anomalous symmetries}
\label{sec: Fermionization of anomalous symmetries}
A fusion category symmetry $\mathcal{C}$ is said to be anomalous if it does not admit gapped phases with unique ground states that preserve the symmetry $\mathcal{C}$.
Any anomalous fusion category symmetry $\mathcal{C}$ is equivalent to the representation category $\mathrm{Rep}(H)$ of a semisimple weak Hopf algebra $H$.

In order to perform the fermionization of $\mathrm{Rep}(H)$ symmetry, we first specify a non-anomalous $\mathbb{Z}_2$ subgroup symmetry of $\mathrm{Rep}(H)$.
As in the case of non-anomalous symmetries, the existence of a $\mathbb{Z}_2$ group-like element in the dual weak Hopf algebra $H^*$ implies that $\mathrm{Rep}(H)$ symmetry has a non-anomalous $\mathbb{Z}_2$ subgroup symmetry.
Here, a group-like element $g \in H^*$ is defined as an invertible element that satisfies
\begin{equation}
\Delta_{H^*}(g) = \Delta_{H^*}(\epsilon) \cdot g \otimes g = g \otimes g \cdot \Delta_{H^*}(\epsilon),
\label{eq: group-like of weak Hopf}
\end{equation}
where $\epsilon \in H^*$ is the counit of $H$.
In particular, we call $u \in H^*$ a $\mathbb{Z}_2$ group-like element if it satisfies $u \cdot u = \epsilon$ in addition to eq. \eqref{eq: group-like of weak Hopf}.
The representation $V_u \in \mathrm{Rep}(H)$ associated with a $\mathbb{Z}_2$ group-like element $u \in H^*$ is defined as follows \cite{Nikshych2002}: the underlying vector space of $V_u$ is the target counital subalgebra $H_t$ and the $H$-action $\rho_{V_u}: H \otimes V_u \rightarrow V_u$ is given by $\rho_{V_u} = (\epsilon_t \otimes u) \circ \Delta \circ m$, where $m: H \otimes V_u \rightarrow H$ and $\Delta: V_u \rightarrow H \otimes H$ is the multiplication and the comultiplication on $H$ restricted to $H \otimes V_u$ and $V_u$ respectively.
The representation $V_u$ generates a $\mathbb{Z}_2$ subgroup of $\mathrm{Rep}(H)$ \cite{Vecsernyes2003}, which is denoted by $\mathbb{Z}_2^u$.

We propose that the fermionization of an anomalous fusion category symmetry $\mathrm{Rep}(H)$ with respect to $\mathbb{Z}^u_2$ subgroup symmetry is a superfusion category symmetry $\mathrm{SRep}(\mathcal{H}^u)$, where $\mathcal{H}^u$ is a weak Hopf superalgebra defined as follows:
\begin{itemize}
\item The underlying vector space of $\mathcal{H}^u$ is $H$.
\item The $\mathbb{Z}_2$-grading on $\mathcal{H}^u$ is defined by eq. \eqref{eq: grading of Hopf superalgebra}.
\item The multiplication $m_u: \mathcal{H}^u \otimes \mathcal{H}^u \rightarrow \mathcal{H}^u$ and the antipode $S_u: \mathcal{H}^u \rightarrow \mathcal{H}^u$ are given by eqs. \eqref{eq: multiplication of Hopf superalgebra} and \eqref{eq: antipode of Hopf superalgebra} respectively.
\item The other structure maps are the same as those of $H$.
\end{itemize}
A direct computation shows that super vector space $\mathcal{H}^u$ equipped with the above structure maps becomes a weak Hopf superalgebra, see appendix \ref{sec: weak Hopf superalgebra} for details.
As in the case of non-anomalous symmetries, the inverse of the map $(H, m, \eta, \Delta, \epsilon, S) \mapsto (\mathcal{H}^u, m_u, \eta, \Delta, \epsilon, S_u)$ given by eq. \eqref{eq: inverse map} describes the bosonization of a superfusion category symmetry.

The above proposal will be verified in section \ref{sec: Fermionic TQFTs with superfusion category symmetries} when the underlying QFT is topological.
Although the discussion in section \ref{sec: Fermionic TQFTs with superfusion category symmetries} is restricted to TQFTs, the concrete examples discussed below suggest that our proposal for the fermionization of fusion category symmetries can also be applied to non-topological QFTs.

\subsection{Examples}
\label{sec: Examples of fermionization}
We explicitly compute the fermionization of fusion category symmetries for several examples including both non-anomalous and anomalous symmetries.

\subsubsection{Fermionization of finite group symmetries}
\label{sec: Fermionization of finite group symmetries}
The first example is the fermionization of a non-anomalous finite group symmetry $G$.
As a fusion category symmetry, a finite group symmetry $G$ is described by the representation category $\mathrm{Rep}(\mathbb{C}[G]^*)$ of the dual group algebra $\mathbb{C}[G]^*$.
The basis of $\mathbb{C}[G]^*$ is denoted by $\{\widehat{g} \mid g \in G\}$, which are dual to the standard basis $\{g \in G\}$ of $\mathbb{C}[G]$, i.e. we have $\widehat{g}(h) = \delta_{g, h}$.
The Hopf algebra structure on the dual group algebra $\mathbb{C}[G]^*$ is given by
\begin{equation}
m(\widehat{g} \otimes \widehat{h}) = \delta_{g, h} \widehat{g}, \quad \eta(1) = \sum_{g \in G} \widehat{g}, \quad \Delta(\widehat{g}) = \sum_{h \in G} \widehat{gh^{-1}} \otimes \widehat{h}, \quad \epsilon(\widehat{g}) = \delta_{g, e}, \quad S(\widehat{g}) = \widehat{g^{-1}}.
\label{eq: Hopf str. on group algebra}
\end{equation}
We note that the basis $\widehat{g}$ is an idempotent and thus $\mathbb{C}[G]^*$ is isomorphic to the direct sum of one-dimensional algebras $\mathbb{C} \widehat{g}$ as a semisimple algebra: 
\begin{equation}
\mathbb{C}[G]^* = \bigoplus_{g \in G} \mathbb{C} \widehat{g} \cong \bigoplus_{g \in G} \mathrm{End}(V_g).
\label{eq: decomposition of dual group algebra}
\end{equation}
Here, $V_g$ is a one-dimensional representation of $\mathbb{C}[G]^*$ defined by
\begin{equation}
\widehat{h} \cdot v_g = \delta_{h, g} v_g, \quad \forall v_g \in V_g.
\label{eq: one-dim rep Vg}
\end{equation}
The above decomposition \eqref{eq: decomposition of dual group algebra} indicates that simple objects of $\mathrm{Rep}(\mathbb{C}[G]^*)$ are one-dimensional representations $V_g$.
These simple objects form a group under the tensor product of representations because we have an isomorphism $V_g \otimes V_h \cong V_{gh}$.
Thus, the representation category $\mathrm{Rep}(\mathbb{C}[G]^*)$ describes an ordinary finite group symmetry $G$.

To perform the fermionization of a finite group symmetry, we specify a $\mathbb{Z}_2$ subgroup of a fusion category $\mathrm{Rep}(\mathbb{C}[G]^*)$.
A $\mathbb{Z}_2$ subgroup associated with a $\mathbb{Z}_2$ group-like element $u \in \mathbb{C}[G]$ is denoted by $\mathbb{Z}^u_2$, which is generated by $V_u \in \mathrm{Rep}(\mathbb{C}[G]^*)$.
If we use $\mathbb{Z}_2^u$ subgroup for fermionization, the fermionized symmetry becomes $\mathrm{SRep}(\mathcal{H}^u_{\mathbb{C}[G]^*})$ where $\mathcal{H}^u_{\mathbb{C}[G]^*}$ is a Hopf superalgebra obtained by the prescription given in section \ref{sec: Fermionization of non-anomalous symmetries}.

\paragraph{Hopf superalgebra $\mathcal{H}_{\mathbb{C}[G]^*}^u$.}
Let us investigate the Hopf superalgebra structure on $\mathcal{H}^u_{\mathbb{C}[G]^*}$.
The $\mathbb{Z}_2$-grading automorphism $p_u: \mathcal{H}^u_{\mathbb{C}[G]^*} \rightarrow \mathcal{H}^u_{\mathbb{C}[G]^*}$ defined by the adjoint action \eqref{eq: grading of Hopf superalgebra} of the $\mathbb{Z}_2$ group-like element $u \in \mathbb{C}[G]$ is computed as
\begin{equation}
p_u(\widehat{g}) = \widehat{ugu}.
\end{equation}
This equation implies that $\widehat{g} \in \mathbb{C}[G]^*$ is a $\mathbb{Z}_2$-even element when $g \in G$ commutes with $u$.
On the other hand, when $g \in G$ does not commute with $u$, $\widehat{g}$ is not a homogeneous element, but the linear combinations $\widehat{g}_{\pm} = \widehat{g} \pm \widehat{ugu}$ are homogeneous because $p(\widehat{g}_{\pm}) = \pm \widehat{g}_{\pm}$.
Therefore, the $\mathbb{Z}_2$-even sector and the $\mathbb{Z}_2$-odd sector of $\mathcal{H}^u_{\mathbb{C}[G]^*}$ are given by
\begin{equation}
(\mathcal{H}^u_{\mathbb{C}[G]^*})_0 = ( \bigoplus_{g \in C_G(u)} \mathbb{C}\widehat{g} ) \oplus ( \bigoplus_{\substack{[g] \in G/\sim \\ \text{s.t. } g \notin C_G(u)}} \mathbb{C} \widehat{g}_{+} ), \quad
(\mathcal{H}^u_{\mathbb{C}[G]^*})_1 = \bigoplus_{\substack{[g] \in G/\sim \\ \text{s.t. } g \notin C_G(u)}} \mathbb{C} \widehat{g}_{-}.
\end{equation}
Here, $C_G(u) := \{g \in G \mid gu = ug\}$ is the centralizer of $u \in G$ and $G/\sim$ is the quotient of $G$ by the equivalence relation $g \sim ugu$.
Given the above $\mathbb{Z}_2$-grading on $\mathcal{H}^u_{\mathbb{C}[G]^*}$, we can compute the multiplication $m_u$ and the antipode $S_u$ defined by eqs. \eqref{eq: multiplication of Hopf superalgebra} and \eqref{eq: antipode of Hopf superalgebra} as follows:
\begin{align}
m_u (\widehat{g} \otimes \widehat{h}) & = \frac{1}{2} (\delta_{g, h} + \delta_{g, uhu}) \widehat{g} + \frac{1}{2} (\delta_{g, hu} - \delta_{g, uh}) \widehat{gu}, 
\label{eq: multiplication of fermionized group} \\
S_u(\widehat{g}) & = \frac{1}{2} \left( \widehat{g^{-1}} + \widehat{ug^{-1}u} + \widehat{g^{-1}u} - \widehat{ug^{-1}} \right).
\label{eq: antipode of fermionized group}
\end{align}

It turns out that the superalgebra $\mathcal{H}^u_{\mathbb{C}[G]^*}$ with multiplication \eqref{eq: multiplication of fermionized group} can be decomposed into the direct sum of simple superalgebras as follows:
\begin{equation}
\mathcal{H}^u_{\mathbb{C}[G]^*} \cong ( \bigoplus_{g \in C_G(u)} \mathbb{C} \widehat{g} ) \oplus ( \bigoplus_{\substack{[g] \in G/ \sim^{\prime} \\ \text{s.t. } g \notin C_G(u)}} \mathrm{Span}\{\widehat{g}_{\pm}, \widehat{gu}_{\pm}\} ),
\label{eq: decomposition}
\end{equation}
where $G/ \sim^{\prime}$ is the quotient of $G$ by the equivalence relation $g \sim^{\prime} ug \sim^{\prime} gu \sim^{\prime} ugu$.
The direct summand $\mathbb{C}\widehat{g}$ is isomorphic to the endomorphism superalgebra $\mathrm{End}(V_g)$ of a one-dimensional super representation $V_g \cong \mathbb{C}^{1|0}$ on which $\mathcal{H}^u_{\mathbb{C}[G]^*}$ acts by eq. \eqref{eq: one-dim rep Vg}.
On the other hand, the direct summand $\mathrm{Span}\{\widehat{g}_{\pm}, \widehat{gu}_{\pm}\}$ is isomorphic to the complex Clifford algebra $\mathrm{Cl}(2)$ with two odd generators.
This isomorphism can be seen from the fact that the linear combinations 
\begin{equation}
1_g := \widehat{g}_{+} + \widehat{gu}_{+}, \quad \Gamma_g := \widehat{g}_{-} + \widehat{gu}_{-}, \quad \Gamma_g^{\prime} := \widehat{g}_{-} - \widehat{gu}_{-}
\end{equation}
satisfy the same algebra as $\mathrm{Cl}(2)$, i.e. $(\Gamma_g)^2 = - (\Gamma_g^{\prime})^{2} = 1_g$ and $\Gamma_g \Gamma_g^{\prime} = - \Gamma_g^{\prime} \Gamma_g$.
Since the Clifford algebra $\mathrm{Cl}(2)$ is isomorphic to $\mathrm{End}(\mathbb{C}^{1|1})$ \cite{Jozefiak1988}, we have an isomorphism
\begin{equation}
\mathrm{Span}\{\widehat{g}_{\pm}, \widehat{gu}_{\pm}\} \cong \mathrm{End}(W_{g})
\end{equation}
for a two-dimensional super representation $W_{g} \cong \mathbb{C}^{1|1}$.
The action of $\mathcal{H}^u_{\mathbb{C}[G]^*}$ on $W_g$ is given by
\begin{equation}
\Gamma_g \cdot (w_g)_+  = \Gamma_g^{\prime} \cdot (w_g)_+ = (w_g)_-, \quad \Gamma_g \cdot (w_g)_- = - \Gamma_g^{\prime} \cdot (w_g)_- = (w_g)_+,
\label{eq: two-dim rep}
\end{equation}
where $(w_g)_+$ and $(w_g)_-$ are a $\mathbb{Z}_2$-even basis and a $\mathbb{Z}_2$-odd basis of $W_g$ respectively.\footnote{We can equally define the action of $\mathcal{H}^u_{\mathbb{C}[G]^*}$ on $W_g$ as $\Gamma_g \cdot (w_g)_+ = - \Gamma^{\prime}_g \cdot (w_g)_+ = (w_g)_-$ and $\Gamma_g \cdot (w_g)_- = \Gamma^{\prime}_g \cdot (w_g)_- = (w_g)_+$. The super representation defined by this action is oddly isomorphic to the super representation defined by eq. \eqref{eq: two-dim rep}}
A direct sum component other than $\mathrm{Span}\{\widehat{g}_{\pm}, \widehat{gu}_{\pm}\}$ acts trivially on $W_{g}$, namely we have $a \cdot (w_g)_{\pm} = 0$ for any $a \notin \mathrm{Span}\{\widehat{g}_{\pm}, \widehat{gu}_{\pm} \}$.
We note that $W_g$ is evenly isomorphic to $W_{ugu}$ and oddly isomorphic to $W_{gu}$ and $W_{ug}$. In particular, the isomorphism class of a two-dimensional super representation $W_g$ is labeled by $[g] \in G/ \sim^{\prime}$ where $g \notin C_G(u)$.
In terms of irreducible representations $V_g$ and $W_{g}$, we can write the direct sum decomposition \eqref{eq: decomposition} as
\begin{equation}
\mathcal{H}^u_{\mathbb{C}[G]^*} \cong (\bigoplus_{g \in C_G(u)} \mathrm{End}(V_g)) \oplus (\bigoplus_{\substack{[g] \in G/\sim^{\prime} \\ \text{s.t. } g \notin C_G(u)}} \mathrm{End}(W_{g})).
\label{eq: direct sum decompostion of HCG*u}
\end{equation}

\paragraph{Superfusion category $\mathrm{SRep}(\mathcal{H}_{\mathbb{C}[G]^*}^u)$.}
The direct sum decomposition \eqref{eq: direct sum decompostion of HCG*u} indicates that irreducible super representations of $\mathcal{H}^u_{\mathbb{C}[G]^*}$, or equivalently simple objects of $\mathrm{SRep}(\mathcal{H}^u_{\mathbb{C}[G]^*})$, are isomorphic to either one-dimensional super representations $V_g$ for $g \in C_G(u)$ or two-dimensional super representations $W_{g}$ for $g \notin C_G(u)$.
All of these simple objects are m-type objects because irreducible super representations of superalgebras $\mathrm{End}(\mathbb{C}^{p|q})$ do not have odd automorphisms for any $p$ and $q$ \cite{Jozefiak1988}.
We can also compute the fusion rules as follows:
\begin{equation}
\begin{aligned}
& V_g \otimes V_h \cong V_{gh}, \quad V_g \otimes W_{h} \cong W_{gh}, \quad W_{g} \otimes V_h \cong W_{gh}, \\
& W_{g} \otimes W_{h} \cong 
\begin{cases}
V_{gh} \oplus V_{uguh} \oplus \Pi V_{ghu} \oplus \Pi V_{guh}, \quad & \text{if } gh, uguh \in C_G(u),\\
V_{gh} \oplus \Pi V_{ghu} \oplus W_{uguh}, \quad & \text{if } gh \in C_G(u), uguh \notin C_G(u), \\
W_{gh} \oplus V_{uguh} \oplus \Pi V_{guh}, \quad & \text{if } gh \notin C_G(u), uguh \in C_G(u), \\
W_{gh} \oplus W_{uguh}, \quad & \text{if } gh, uguh \notin C_G(u).
\end{cases}
\end{aligned}
\end{equation}
These fusion rules will be derived in appendix \ref{sec: Fusion rules of superfusion category symmetries}.
The isomorphisms in the above equation are even isomorphisms and $\Pi V_g$ denotes a one-dimensional super representation oddly isomorphic to $V_g$.
We note that the fermionization of a finite group symmetry $G$ is also $G$ if $\mathbb{Z}^u_2$ is a central subgroup of $G$.
In particular, if $G$ is a non-trivial central extension of $G_b$ by $\mathbb{Z}_2^u$, the fermionized symmetry is a non-trivial central extension of $G_b$ by the fermion parity symmetry $\mathbb{Z}_2^F$.
On the other hand, if $\mathbb{Z}_2^u$ is not a central subgroup of $G$, the symmetry of the fermionized theory is no longer a group.
This is reminiscent of the gauging of a non-central $\mathbb{Z}_2$ subgroup, which turns a finite group symmetry into a non-invertible symmetry \cite{BT2018}.
However, the fermionization is not quite the same as the $\mathbb{Z}_2$ gauging.
For example, the fermionization of a non-anomalous finite group symmetry is always non-anomalous, while the $\mathbb{Z}_2$ gauging of a non-anomalous finite group symmetry can become anomalous \cite{BT2018, CW2022a}.

\subsubsection{Fermionization of $\mathrm{Rep}(G)$ symmetry}
Another example of fusion category symmetries related to a finite group $G$ is the representation category of a group algebra $\mathbb{C}[G]$.
This fusion category is denoted by $\mathrm{Rep}(G)$ rather than $\mathrm{Rep}(\mathbb{C}[G])$.
Physically, a fusion category $\mathrm{Rep}(G)$ describes the symmetry of $G$ gauge theories.

The fermionization of $\mathrm{Rep}(G)$ symmetry is denoted by $\mathrm{SRep}(\mathcal{H}^u_{G})$, where $u$ is a $\mathbb{Z}_2$ group-like element in $\mathbb{C}[G]^*$.
We note that a $\mathbb{Z}_2$ group-like element in $\mathbb{C}[G]^*$ is nothing but an algebra homomorphism from $\mathbb{C}[G]$ to $\mathbb{C}$ that satisfies $u(g)^2 = 1$ for all $g \in G$.
The $\mathbb{Z}_2$-grading automorphism $p_u: \mathcal{H}^u_{G} \rightarrow \mathcal{H}^u_G$ associated with $u \in \mathbb{C}[G]^*$ can be computed as
\begin{equation}
p_u(g) = (u \otimes \mathrm{id} \otimes u) \circ (\mathrm{id} \otimes \Delta) \circ \Delta (g) = u(g)^2 g = g,
\label{eq: grading of C(G)}
\end{equation}
where we used the equality $\Delta(g) = g \otimes g$ for $g \in \mathbb{C}[G]$.
The above equation shows that the Hopf superalgebra $\mathcal{H}^u_G$ is purely even and hence is just the group algebra $\mathbb{C}[G]$ equipped with the trivial $\mathbb{Z}_2$ grading.
Therefore, simple objects of $\mathrm{SRep}(\mathcal{H}^u_G)$ are irreducible $\mathbb{C}[G]$-modules equipped with purely even or purely odd $\mathbb{Z}_2$-gradings.
In particular, $\mathrm{SRep}(\mathcal{H}_G^u)$ has the same set of (isomorphism classes of) simple objects as $\mathrm{Rep}(G)$. 
The fusion rules of $\mathrm{SRep}(\mathcal{H}^u_G)$ are also the same as those of $\mathrm{Rep}(G)$.
We note that all simple objects of $\mathrm{SRep}(\mathcal{H}_G^u)$ are m-type objects.

In general, the Hopf superalgebra $\mathcal{H}^u$ becomes purely even if the $\mathbb{Z}_2$ group-like element $u$ is in the center of $H^*$. 
In this case, the fermionized symmetry is described by essentially the same category as the original symmetry.
More specifically, the superfusion category $\mathrm{SRep}(\mathcal{H}^u)$ is equivalent to the Deligne tensor product $\mathrm{Rep}(H) \boxtimes \mathrm{SVec}$ of a fusion category $\mathrm{Rep}(H)$ and a superfusion category $\mathrm{SVec}$.
The fermionization of $\mathrm{Rep}(G)$ symmetry is a special example of this.

\subsubsection{Fermionization of $\mathrm{Rep}(H_8)$ symmetry}
The next example is the fermionization of the representation category $\mathrm{Rep}(H_8)$ of a Hopf algebra $H_8$ known as the eight-dimensional Kac-Paljutkin algebra, which is the smallest non-commutative and non-cocommutative semisimple Hopf algebra \cite{KP1966}. 
A simple example of a physical system that realizes $\mathrm{Rep}(H_8)$ symmetry is the stacking of two Ising CFTs \cite{TW2019}.
The category $\mathrm{Rep}(H_8)$ is equivalent to one of the three non-anomalous $\mathbb{Z}_2 \times \mathbb{Z}_2$ Tambara-Yamagami categories,\footnote{There are four $\mathbb{Z}_2 \times \mathbb{Z}_2$ Tambara-Yamagami categories \cite{TY1998}, one of which is anomalous and the others are non-anomalous \cite{TW2019, Tam2000}.} which describe self-dualities under gauging $\mathbb{Z}_2 \times \mathbb{Z}_2$ symmetry.
Specifically, $\mathrm{Rep}(H_8)$ consists of four one-dimensional representations forming $\mathbb{Z}_2 \times \mathbb{Z}_2$ and a two-dimensional representation corresponding to a duality defect.
The other two non-anomalous $\mathbb{Z}_2 \times \mathbb{Z}_2$ Tambara-Yamagami categories are $\mathrm{Rep}(D_8)$ and $\mathrm{Rep}(Q_8)$, where $D_8$ and $Q_8$ are the dihedral group of order eight and the quaternion group respectively.
Although these three fusion categories have the same set of simple objects and the same fusion rules,  the fermionization of $\mathrm{Rep}(H_8)$ turns out to be qualitatively different from the fermionization of $\mathrm{Rep}(D_8)$ and $\mathrm{Rep}(Q_8)$.
In particular, we will see that the fermionization of $\mathrm{Rep}(H_8)$ has q-type objects in contrast to the fact that the fermionization of $\mathrm{Rep}(D_8)$ and $\mathrm{Rep}(Q_8)$ does not have q-type objects as we discussed in the previous example.

Let us first recall the definition of the Kac-Paljutkin algebra $H_8$.
The Kac-Paljutkin algebra $H_8$ is generated by three elements $x, y, z \in H_8$ that obey the following multiplication rules:
\begin{equation}
x^2 = y^2 = 1, \quad xy = yx, \quad xz = zy, \quad zx = yz, \quad z^2 = \frac{1}{2} (1 + x + y - xy).
\end{equation}
The other structure maps of $H_8$ are given by
\begin{equation}
\begin{aligned}
\Delta(x) & = x \otimes x, \quad \Delta (y) = y \otimes y, \quad \Delta(z) = \frac{1}{2} (1 \otimes 1 + 1 \otimes x + y \otimes 1 - y \otimes x) (z \otimes z) \\
\epsilon (x) & = \epsilon (y) = \epsilon (z) = 1, \quad
S(x) = x, \quad S(y) = y, \quad S(z) = z.
\end{aligned}
\end{equation}
These structure maps are extended to the entire Hopf algebra $H_8$ by demanding that $\Delta$ and $\epsilon$ are algebra homomorphisms and $S$ is an algebra anti-homomorphism.

A $\mathbb{Z}_2$ subgroup of $\mathrm{Rep}(H_8)$ is associated with a $\mathbb{Z}_2$ group-like element $u \in H_8^{*}$.
Since a $\mathbb{Z}_2$ group-like element  $u \in H_8^*$ is multiplicative, i.e. it satisfies $u(ab) = u(a) u(b)$ for any $a, b \in H_8$, the element $u$ is uniquely determined by its values evaluated at the generators $x, y, z \in H_8$.
A straightforward calculation shows that there are three $\mathbb{Z}_2$ group-like elements in $H_8^*$ except for the trivial one $u = \epsilon$:
\begin{equation}
(u(x), u(y), u(z)) = (1, 1, -1), (-1, -1, i), (-1, -1, -i).
\end{equation}
These correspond to three $\mathbb{Z}_2$ subgroups of $\mathbb{Z}_2 \times \mathbb{Z}_2$ group-like symmetry of $\mathrm{Rep}(H_8)$.
If we choose $u$ as $(u(x), u(y), u(z)) = (1, 1, -1)$, the $\mathbb{Z}_2$-grading defined by the adjoint action of $u$ becomes purely even, which means that the fermionization of $\mathrm{Rep}(H_8)$ symmetry with respect to the $\mathbb{Z}_2$ subgroup corresponding to this choice of $u$ is the Deligne tensor product $\mathrm{Rep}(H_8) \boxtimes \mathrm{SVec}$.
In particular, the fermionized symmetry does not have q-type objects in this case.
On the other hand, if we choose $u$ as $(u(x), u(y), u(z)) = (-1, -1, \pm i)$, the adjoint action of $u$ endows $H_8$ with a non-trivial $\mathbb{Z}_2$-grading, which consequently makes the fermionization of $\mathrm{Rep}(H_8)$ qualitatively different from the original fusion category $\mathrm{Rep}(H_8)$.
For this reason, we will focus on the latter case below.
The fermionization of $\mathrm{Rep}(H_8)$ symmetry will be denoted by $\mathrm{SRep}(\mathcal{H}^u_8)$, which is realized by, for example, the stacking of the Ising CFT and a single massless Majorana fermion.

\paragraph{Hopf superalgebra $\mathcal{H}_8^u$.}
The $\mathbb{Z}_2$-grading automorphism $p_u: \mathcal{H}^u_8 \rightarrow \mathcal{H}^u_8$ defined by eq. \eqref{eq: grading of Hopf superalgebra} is computed as 
\begin{equation}
p_u(x) = x, \quad p_u(y) = y, \quad p_u(z) = xyz.
\end{equation}
Since the $\mathbb{Z}_2$-grading $p_u$ is multiplicative, we also have $p_u(xy) = xy$ and $p_u(xz) = zx$.
The above equation implies that the $\mathbb{Z}_2$-even part and the $\mathbb{Z}_2$-odd part of $\mathcal{H}^u_8$ are given by
\begin{equation}
(\mathcal{H}^u_8)_0 = \mathrm{Span}\{1, x, y, xy, z+xyz, zx+xz\}, \quad (\mathcal{H}^u_8)_1 = \mathrm{Span}\{z-xyz, zx-xz\}.
\end{equation}
Based on this $\mathbb{Z}_2$-grading, we can compute the Hopf superalgebra structure on $\mathcal{H}^u_8$ following the definitions \eqref{eq: multiplication of Hopf superalgebra} and \eqref{eq: antipode of Hopf superalgebra}.
Specifically, if we define the $\mathbb{Z}_2$-even basis $\{e_i \mid 1 \leq i \leq 6\}$ and the $\mathbb{Z}_2$-odd basis $\{e_7, e_8\}$ of $\mathcal{H}^u_8$ as
\begin{equation}
\begin{aligned}
e_1 & = \frac{1}{8} [ 1 + x + y + xy + (z + xyz) + (zx + xz) ], \\
e_2 &= \frac{1}{8} [ 1 + x + y + xy - (z + xyz) - (zx + xz) ], \\
e_3 & = \frac{1}{8} [ 1 - x - y + xy + i (z + xyz) - i (zx + xz) ], \\
e_4 &= \frac{1}{8} [ 1 - x - y + xy - i (z + xyz) + i (zx + xz) ], \\
e_5 & = \frac{1}{4}(1 + x - y - xy), \quad
e_6 = \frac{1}{4} (1 - x + y - xy), \\
e_7 & = \frac{i}{4 \sqrt{u(z)}} [(z - xyz) + (zx - xz)], \quad
e_8 = \frac{1}{4 \sqrt{u(z)}} [(z - xyz) - (zx - xz)],
\end{aligned}
\label{eq: basis of H8}
\end{equation}
the multiplication $m_u$ and the antipode $S_u$ can be computed as follows:
\begin{equation}
m_u (e_i \otimes e_j) = 
\begin{cases}
e_i \quad & \text{for } 1 \leq i = j \leq 6, \\
e_5 \quad & \text{for } i = j = 7, \\
e_6 \quad & \text{for } i = j = 8, \\
e_7 \quad & \text{for } (i, j) = (5, 7), (7, 5), \\
e_8 \quad & \text{for } (i, j) = (6, 8), (8, 6), \\
0 \quad & \text{otherwise}, 
\end{cases}
\quad 
S_u(e_i) = 
\begin{cases}
e_i \quad & \text{for } 1 \leq i \leq 6, \\
u(z) e_7 \quad & \text{for } i = 7, \\
-u(z) e_8 \quad & \text{for } i = 8.
\end{cases}
\end{equation}
This implies that the Hopf superalgebra $\mathcal{H}^u_8$ is decomposed into the direct sum of simple superalgebras as
\begin{equation}
\mathcal{H}^u_8 \cong (\bigoplus_{1 \leq i \leq 4} \mathbb{C} e_i) \oplus \mathrm{Span}\{e_5, e_7\} \oplus \mathrm{Span}\{e_6, e_8\}.
\label{eq: direct sum decomposition of H8}
\end{equation}
The direct summand $\mathbb{C} e_i$ for $1 \leq i \leq 4$ is isomorphic to the endomorphism superalgebra $\mathrm{End}(\mathbb{C}^{1|0})$, whereas the other two direct summands $\mathrm{Span}\{e_5, e_7\}$ and $\mathrm{Span}\{e_6, e_8\}$ are isomorphic to the Clifford algebra $\mathrm{Cl}(1)$ with one odd generator.

\paragraph{Superfusion category $\mathrm{SRep}(\mathcal{H}_8^u)$.}
The direct sum decomposition \eqref{eq: direct sum decomposition of H8} indicates that the superfusion category $\mathrm{SRep}(\mathcal{H}^u_8)$ consists of the following simple objects:
\begin{itemize}
\item a one-dimensional super representation $V_i \cong \mathbb{C}^{1|0}$ for each $i \in \{1, 2, 3, 4\}$, which is a unique (up to isomorphism) irreducible super representation of $\mathbb{C} e_i \cong \mathrm{End}(\mathbb{C}^{1|0})$,
\item a two-dimensional super representation $W_1 \cong \mathbb{C}^{1|1}$, which is a unique (up to isomorphism) irreducible super representation of $\mathrm{Span}\{e_5, e_7\} \cong \mathrm{Cl}(1)$,
\item a two-dimensional super representation $W_2 \cong \mathbb{C}^{1|1}$, which is a unique (up to isomorphism) irreducible super representation of $\mathrm{Span}\{e_6, e_8\} \cong \mathrm{Cl}(1)$.
\end{itemize}
We note that the two-dimensional super representations $W_1$ and $W_2$ are q-type objects because they are irreducible super representations of $\mathrm{Cl}(1)$ \cite{Jozefiak1988}.
Therefore, the superfusion category $\mathrm{SRep}(\mathcal{H}_8^u)$ has q-type objects in contrast to the fermionization of the other non-anomalous $\mathbb{Z}_2 \times \mathbb{Z}_2$ Tambara-Yamagami categories $\mathrm{Rep}(D_8)$ and $\mathrm{Rep}(Q_8)$.
This illustrates that the existence of q-type objects depends on $F$-symbols of the original fusion category symmetry.

The fusion rules of $\mathrm{SRep}(\mathcal{H}^u_8)$ can be computed explicitly, see appendix \ref{sec: Fusion rules of superfusion category symmetries} for details.
The one-dimensional super representations $V_i$ for $1 \leq i \leq 4$ obey a group-like fusion rule $\mathbb{Z}_2 \times \mathbb{Z}_2^F$, whose unit element is given by $V_1$:
\begin{equation}
V_1 \otimes V_i \cong V_i \otimes V_1 \cong V_i, \quad V_2 \otimes V_2 \cong V_3 \otimes V_3 \cong V_4 \otimes V_4 \cong V_1, \quad V_2 \otimes V_3 \cong V_3 \otimes V_2 \cong V_4.
\end{equation}
As we discussed in section \ref{sec: Fermionization of non-anomalous symmetries}, the generator of the fermion parity symmetry $\mathbb{Z}_2^F$ is a one-dimensional super representation on which $e_i \in \mathcal{H}^u_8$ acts as a scalar multiplication by $u(e_i)$.
If we choose a $\mathbb{Z}_2$ group-like element $u$ as $(u(x), u(y), u(z)) = (-1, -1, i)$, the generator of $\mathbb{Z}_2^F$ is given by $V_4$ because $u(e_i) = \delta_{i, 4}$.
On the other hand, if we choose $u$ as $(u(x), u(y), u(z)) = (-1, -1, -i)$, the generator of $\mathbb{Z}_2^F$ is given by $V_3$ because $u(e_i) = \delta_{i, 3}$.
The fusion rules of one-dimensional super representations $V_i$ and two-dimensional super representations $W_j$ are given by
\begin{equation}
\begin{aligned}
V_1 \otimes W_j & \cong V_2 \otimes W_j \cong W_j \cong W_j \otimes V_1 \cong W_j \otimes V_2, \\ V_3 \otimes W_1 & \cong V_4 \otimes W_1 \cong W_2 \cong W_1 \otimes V_3 \cong W_1 \otimes V_4.
\end{aligned}
\end{equation}
In particular, the second equation implies that the generator of the fermion parity symmetry $\mathbb{Z}_2^F$ exchanges two q-type objects $W_1$ and $W_2$ by the fusion.
Finally, the fusion rule of two-dimensional super representations is given by
\begin{equation}
W_1 \otimes W_1 \cong V_1 \oplus V_2 \oplus \Pi V_1 \oplus \Pi V_2,
\label{eq: fermionic duality}
\end{equation}
where $\Pi V_i$ is a one-dimensional super representation oddly isomorphic to $V_i$.
The other fusion rules are determined by the associativity of the fusion rules.
The above fusion rule \eqref{eq: fermionic duality} implies that $W_1$ is a duality defect for the condensation of $V_1 \oplus V_2 \oplus \Pi V_1 \oplus \Pi V_2$.
In particular, a fermionic system with $\mathrm{SRep}(\mathcal{H}^u_8)$ symmetry should be invariant under condensing the sum of topological lines $V_1 \oplus V_2 \oplus \Pi V_1 \oplus \Pi V_2$.
Since we have an isomorphism $V_1 \oplus V_2 \oplus \Pi V_1 \oplus \Pi V_2 \cong (V_1 \oplus V_2) \otimes (V_1 \oplus \Pi V_1)$, the condensation of $V_1 \oplus V_2 \oplus \Pi V_1 \oplus \Pi V_2$ is implemented by condensing $V_1 \oplus V_2$ and $V_1 \oplus \Pi V_1$ successively.
The condensation of $V_1 \oplus V_2$ gives rise to gauging a $\mathbb{Z}_2$ symmetry generated by $V_2$, while the condensation of $V_1 \oplus \Pi V_1$ gives rise to stacking the Kitaev chain.\footnote{In particular, condensing $V_1 \oplus \Pi V_1$ on a closed spin surface amounts to multiplying the partition function by the Arf invariant, which can be checked by a direct computation.}
Therefore, $\mathrm{SRep}(\mathcal{H}^u_8)$ symmetry can be understood as a self-duality under gauging a $\mathbb{Z}_2$ subgroup and stacking the Kitaev chain,\footnote{The author thanks Ryohei Kobayashi for a discussion on this point.} see below for more details.
We note that the relation between the fusion category symmetry $\mathrm{Rep}(H_8)$ and its fermionization $\mathrm{SRep}(\mathcal{H}^u_8)$ is a non-anomalous analogue of the relation between the symmetry of the Ising CFT and that of a single massless Majorana fermion, which will be discussed in section \ref{sec: Fermionization of Z2 Tambara-Yamagami symmetries}.

\paragraph{Self-duality.}
We can confirm that the partition function of a fermionic theory $\mathcal{T}_f$ is indeed invariant under gauging $\mathbb{Z}_2$ symmetry and stacking the Kitaev chain if it is the fermionization of a bosonic theory $\mathcal{T}_b$ with $\mathrm{Rep}(H_8)$ symmetry.
Let us write the partition function of $\mathcal{T}_b$ and $\mathcal{T}_f$ as $Z_b(\Sigma; \alpha_1, \alpha_2)$ and $Z_f(\Sigma; \alpha_1, \eta)$ respectively, where $\Sigma$ is a closed oriented surface, $\alpha_1$ and $\alpha_2$ are background $\mathbb{Z}_2$ gauge fields, and $\eta$ is a spin structure on $\Sigma$.
Since $\mathcal{T}_f$ is the fermionization of $\mathcal{T}_b$, the partition functions $Z_b$ and $Z_f$ are related by \cite{GK2016}
\begin{equation}
Z_f(\Sigma; \alpha_1, \eta) = \frac{1}{\sqrt{|H^1(\Sigma, \mathbb{Z}_2)|}} \sum_{\alpha_2 \in H^1(\Sigma, \mathbb{Z}_2)} Z_b(\Sigma; \alpha_1, \alpha_2) (-1)^{q_{\eta}(\alpha_2)} \mathrm{Arf}(\eta),
\label{eq: fermionization of RepH8 partition function}
\end{equation}
where $q_{\eta}$ is a quadratic refinement of the intersection form and $\mathrm{Arf}(\eta) = \pm 1$ is the partition function of the Kitaev chain known as the Arf invariant.
When a bosonic theory $\mathcal{T}_b$ has $\mathrm{Rep}(H_8)$ symmetry, the partition function $Z_b$ is invariant under gauging $\mathbb{Z}_2 \times \mathbb{Z}_2$ symmetry with the diagonal pairing of background gauge fields \cite{TW2019}:
\begin{equation}
Z_b(\Sigma; \alpha_1, \alpha_2) = \frac{1}{|H^1(\Sigma, \mathbb{Z}_2)|} \sum_{\beta_1, \beta_2} Z_b(\Sigma; \beta_1, \beta_2) (-1)^{\int \alpha_1 \cup \beta_1 + \alpha_2 \cup \beta_2}.
\label{eq: Rep(H8) duality}
\end{equation}
This equation implies that the fermionic partition function \eqref{eq: fermionization of RepH8 partition function} satisfies 
\begin{equation}
Z_f(\Sigma; \alpha_1, \eta) = \frac{1}{\sqrt{|H^1(\Sigma, \mathbb{Z}_2)|}} \sum_{\beta_1} Z_f(\Sigma; \beta_1, \eta) (-1)^{\int \alpha_1 \cup \beta_1} \mathrm{Arf}(\eta),
\label{eq: self-duality for SRepH8}
\end{equation}
which shows that the fermionized theory $\mathcal{T}_f$ is self-dual under gauging $\mathbb{Z}_2$ symmetry and stacking the Kitaev chain.

Let us compare this self-duality with the fermionization of $\mathrm{Rep}(D_8)$ and $\mathrm{Rep}(Q_8)$ symmetries.
If the original bosonic theory $\mathcal{T}_b$ has $\mathrm{Rep}(D_8)$ or $\mathrm{Rep}(Q_8)$ symmetry, the bosonic partition function $Z_b$ is invariant under gauging $\mathbb{Z}_2 \times \mathbb{Z}_2$ symmetry with the off-diagonal pairing of background gauge fields \cite{TW2019}:
\begin{equation}
Z_b(\Sigma; \alpha_1, \alpha_2) = \frac{1}{|H^1(\Sigma, \mathbb{Z}_2)|} \sum_{\beta_1, \beta_2} Z_b(\Sigma; \beta_1, \beta_2) (-1)^{\int \alpha_1 \cup \beta_2 + \alpha_2 \cup \beta_1}.
\label{eq: Rep(D8) duality}
\end{equation}
Correspondingly, the fermionic partition function $Z_f$ is invariant under doing the GSO projection and the Jordan-Wigner transformation simultaneously:
\begin{equation}
Z_f(\Sigma; \alpha_1, \eta) = \frac{1}{|H^1(\Sigma, \mathbb{Z}_2)|} \sum_{\beta_1} \sum_{\xi} Z_f(\Sigma; \beta_1, \xi) (-1)^{q_{\xi}(\alpha_1)} (-1)^{q_{\eta}(\beta_1)}.
\label{eq: self-duality for SRepD8}
\end{equation}
The difference between \eqref{eq: self-duality for SRepH8} and \eqref{eq: self-duality for SRepD8} is attributed to the different fusion rules of superfusion categories $\mathrm{SRep}(\mathcal{H}^u_8)$ and $\mathrm{SRep}(\mathcal{H}_{D_8 \text{ or } Q_8}^u)$.
More specifically, eq. \eqref{eq: self-duality for SRepD8} is a consequence of the following fusion rule of a duality line $D \in \mathrm{SRep}(\mathcal{H}_{D_8 \text{ or } Q_8}^u)$:
\begin{equation}
D \otimes D \cong 1 \oplus V \oplus (-1)^F \oplus V(-1)^F \cong (1 \oplus V) \otimes (1 \oplus (-1)^F),
\end{equation}
where $V$ and $(-1)^F$ are the generators of $\mathbb{Z}_2 \times \mathbb{Z}_2^F$ symmetry.
The above fusion rule implies that a fermionic system with $\mathrm{SRep}(\mathcal{H}_{D_8 \text{ or } Q_8}^u)$ symmetry is invariant under condensing $1 \oplus V$ and $1 \oplus (-1)^F$ successively.
Here, the condensation of $1 \oplus V$ should be understood as the gauging of a $\mathbb{Z}_2$ symmetry with a coupling $(-1)^{q_{\eta}(\alpha)}$ between a $\mathbb{Z}_2$ gauge field $\alpha$ and a spin structure $\eta$.
Therefore, the condensation of $1 \oplus V$ is the Jordan-Wigner transformation.
On the other hand, the condensation of $1 \oplus (-1)^F$ is the GSO projection.

\subsubsection{Fermionization of $\mathbb{Z}_2$ Tambara-Yamagami symmetries}
\label{sec: Fermionization of Z2 Tambara-Yamagami symmetries}
As an example of anomalous superfusion category symmetry, we consider the fermionization of self-dualities under gauging $\mathbb{Z}_2$ symmetry.
This example was studied in detail in \cite{JSW2020} by using a different method from ours.
These self-dualities are described by $\mathbb{Z}_2$ Tambara-Yamagami categories $\mathrm{TY}(\mathbb{Z}_2, \chi, \pm 1)$, where $\chi: \mathbb{Z}_2 \times \mathbb{Z}_2 \rightarrow \mathrm{U}(1)$ is the unique symmetric non-degenerate bicharacter on $\mathbb{Z}_2$ \cite{TY1998}.
Fusion category symmetries $\mathrm{TY}(\mathbb{Z}_2, \chi, +1)$ and $\mathrm{TY}(\mathbb{Z}_2, \chi, -1)$ are realized by, e.g., the Ising CFT and the $\mathrm{SU}(2)_2$ WZW model respectively \cite{TW2019, JSW2020}.
The fermionization of the Ising CFT is a single massless Majorana fermion, which has $\mathbb{Z}_2 \times \mathbb{Z}_2^F$ symmetry with an 't Hooft anomaly $1 \bmod 8$ \cite{RZ2012, Thorngren2020, JSW2020}.
Similarly, the fermionization of the $\mathrm{SU}(2)_2$ WZW model is three massless Majorana fermions, which have $\mathbb{Z}_2 \times \mathbb{Z}_2^F$ symmetry with an 't Hooft anomaly $3 \bmod 8$ \cite{JSW2020}.
We note that the generator of the chiral $\mathbb{Z}_2$ symmetry of a massless Majorana fermion is a q-type object \cite{KORS2020}.
Therefore, the above CFT examples suggest that the fermionization of $\mathrm{TY}(\mathbb{Z}_2, \chi, \pm 1)$ has two q-type objects, one of which is obtained by fusing the fermion parity defect with the other. 
Furthermore, since $\mathbb{Z}_2$ Tambara-Yamagami categories describe self-dualities under gauging $\mathbb{Z}_2$ symmetry, the fermionization of them should imply the invariance under stacking the Kitaev chain, cf. figure \ref{fig: fermionization}.
In the following, we will see that the fermionization of $\mathrm{TY}(\mathbb{Z}_2, \chi, \pm1)$ indeed has these properties.
This suggests that the fermionization formula of fusion category symmetries is also applicable to non-topological QFTs.

We first recall the definition of the Tambara-Yamagami category $\mathrm{TY}(A, \chi, \epsilon)$, where $A$ is a finite abelian group, $\chi: A \times A \rightarrow \mathrm{U}(1)$ is a symmetric non-degenerate bicharacter on $A$, and $\epsilon \in \{\pm 1\}$ is a sign.
The fusion category $\mathrm{TY}(A, \chi, \epsilon)$ consists of simple objects labeled by group elements $g \in A$ and an additional simple object $m$ called a duality object.
The set of simple objects of $\mathrm{TY}(A, \chi, \epsilon)$ will be denoted by $\Omega := A \sqcup \{m\}$.
The fusion rules are given by $g \otimes h \cong gh, g \otimes m \cong m \otimes g \cong m, m \otimes m \cong \bigoplus_{g \in A} g$ for $g, h \in A$.
In particular, $A$ is the group-like symmetry of $\mathrm{TY}(A, \chi, \epsilon)$.
The $F$-symbols can be found in \cite{TY1998}.

The Tambara-Yamagami category $\mathrm{TY}(A, \chi, \epsilon)$ is equivalent to the representation category of a weak Hopf algebra $H_{A, \chi, \epsilon}$, whose structure maps are spelled out in \cite{Mevel2010}.
As a semisimple algebra, $H_{A, \chi, \epsilon}$ is isomorphic to the direct sum of endomorphism algebras $\mathrm{End}(H^x)$ for $x \in \Omega$, i.e.
\begin{equation}
H_{A, \chi, \epsilon} \cong \bigoplus_{x \in \Omega} \mathrm{End}(H^x) = (\bigoplus_{g \in A} \mathrm{End}(H^g)) \oplus \mathrm{End}(H^m),
\end{equation}
where $H^g \cong \mathbb{C}^{|A|+1}$ and $H^m \cong \mathbb{C}^{2|A|}$.
The bases of the subalgebras $\mathrm{End}(H^g)$ and $\mathrm{End}(H^m)$ are denoted by $\{e^g_{\alpha, \beta} \mid \alpha, \beta \in \Omega\}$ and $\{e^{m}_{\alpha, \beta} \mid \alpha, \beta \in A \sqcup \overline{A}\}$ where $\overline{A} := \{\overline{a} \mid a \in A\}$ is a copy of $A$.\footnote{The bar is just a notation, which allows us to distinguish, e.g., $e^m_{a, b}$ and $e^m_{a, \overline{b}}$ for $a, b \in A$.}
The multiplication on $H_{A, \chi, \epsilon}$ is given by the usual multiplication of matrices: $m(e^x_{\alpha, \beta} \otimes e^y_{\gamma, \delta}) = \delta_{x, y} \delta_{\beta, \gamma} e^x_{\alpha, \delta}$.
The comultiplication of each basis element $e^x_{\alpha, \beta}$ is given by
\begin{equation*}
\begin{aligned}
\Delta(e^g_{a, b}) & = \sum_{h \in A} e^{gh^{-1}}_{h^{-1}a, h^{-1}b} \otimes e^h_{a, b} + e^m_{g^{-1}a, g^{-1}b} \otimes e^m_{\overline{a}, \overline{b}}, \\
\Delta(e^g_{a, m}) & = \sum_{h \in A} e^{gh^{-1}}_{h^{-1}a, m} \otimes e^h_{a, m} + \frac{\epsilon}{\sqrt{|A|}} \sum_{b \in A} \chi(g, b^{-1}) e^m_{g^{-1}a, \overline{b}} \otimes e^m_{\overline{a}, b}, \\
\Delta(e^g_{m ,a}) & = \sum_{h \in A} e^{gh^{-1}}_{m, h^{-1}a} \otimes e^h_{m, a} + \frac{\epsilon}{\sqrt{|A|}} \sum_{b \in A} \chi(g, b) e^m_{\overline{b}, g^{-1}a} \otimes e^m_{b, \overline{a}}, \\
\Delta(e^g_{m, m}) & = \sum_{h \in A} e^{gh^{-1}}_{m, m} \otimes e^h_{m, m} + \frac{1}{|A|} \sum_{a, b \in A} \chi(g, ab^{-1}) e^m_{\overline{a}, \overline{b}} \otimes e^m_{a, b},\\
\Delta(e^m_{a, b}) & = \sum_{g \in A} \chi(g, a^{-1}b) e^m_{a, b} \otimes e^g_{m, m} + \sum_{g \in A} e^g_{ag, bg} \otimes e^m_{ag, bg}, \\
\Delta(e^m_{a, \overline{b}}) & = \sum_{g \in A} \chi(g, a^{-1}) e^m_{a, \overline{g^{-1}b}} \otimes e^g_{m, b} + \sum_{g \in A} \chi(g, b) e^g_{ag, m} \otimes e^m_{ag, \overline{b}},\\
\Delta(e^m_{\overline{a}, b}) & = \sum_{g \in A} \chi(g, b) e^m_{\overline{g^{-1}a}, b} \otimes e^g_{a, m} + \sum_{g \in A} \chi(g, a^{-1}) e^g_{m, bg} \otimes e^m_{\overline{a}, bg},
\end{aligned}
\end{equation*}
\begin{equation*}
\begin{aligned}
\Delta(e^m_{\overline{a}, \overline{b}}) & = \sum_{g \in A} e^m_{\overline{g^{-1}a}, \overline{g^{-1}b}} \otimes e^g_{a, b} + \sum_{g \in A} \chi(g, a^{-1}b) e^g_{m, m} \otimes e^m_{\overline{a}, \overline{b}}.
\end{aligned}
\end{equation*}
The unit and the counit of $H_{A, \chi, \epsilon}$ are defined by $\eta(1) = \sum_{x \in \Omega} \sum_{\alpha} e^x_{\alpha, \alpha}$ and $\epsilon (e^x_{\alpha, \beta}) = \delta_{x, 1}$ respectively.
The antipode is also given in \cite{Mevel2010}, although we will not use it in the following discussion.

A non-anomalous $\mathbb{Z}_2$ subgroup symmetry of $\mathrm{TY}(A, \chi, \epsilon)$ generated by $u \in A$ is associated with a $\mathbb{Z}_2$ group-like element $\eta_u \in H_{A, \chi, \epsilon}^*$ defined by $\eta_u(e^x_{\alpha, \beta}) = \delta_{x, u}$.
The adjoint action of $\eta_{u}$ gives a $\mathbb{Z}_2$-grading $p_u$ on $H_{A, \chi, \epsilon}$ as follows:
\begin{equation}
\begin{aligned}
p_u(e^g_{a, b}) & = e^g_{ua, ub}, \quad p_u(e^g_{a, m}) = e^g_{ua, m}, \quad p_u(e^g_{m, a}) = e^g_{m, ua}, \quad p_u(e^g_{m, m}) = e^g_{m, m}, \\
p_u(e^m_{a, b}) & = \chi(u, a^{-1}b) e^m_{ua, ub}, \quad p_u(e^m_{a, \overline{b}}) = \chi(u, u) \chi(u, a^{-1}b) e^m_{ua, \overline{ub}}, \\
p_u(e^m_{\overline{a}, b}) & = \chi(u, u) \chi(u, a^{-1}b) e^m_{\overline{ua}, ub}, \quad p_u(e^m_{\overline{a}, \overline{b}}) = \chi(u, a^{-1}b) e^m_{\overline{ua}, \overline{ub}}.
\end{aligned}
\label{eq: grading on Z2 TY}
\end{equation}
A weak Hopf algebra $H_{A, \chi, \epsilon}$ equipped with the above $\mathbb{Z}_2$-grading becomes a weak Hopf superalgebra $\mathcal{H}_{A, \chi, \epsilon}^u$ if we modify the multiplication and the antipode as discussed in section \ref{sec: Fermionization of anomalous symmetries}.

\paragraph{Weak Hopf superalgebra $\mathcal{H}_{\mathbb{Z}_2, \chi, \epsilon}^u$.}
We now restrict our attention to the case $A = \mathbb{Z}_2$.
In this case, the symmetric non-degenerate bicharacter $\chi$ is given by
\begin{equation}
\chi(1, 1) = \chi(1, u) = \chi(u, 1) = 1, \quad \chi(u, u) = -1.
\end{equation}
Thus, eq. \eqref{eq: grading on Z2 TY} implies that the $\mathbb{Z}_2$-even part and the $\mathbb{Z}_2$-odd part of $\mathcal{H}_{\mathbb{Z}_2, \chi, \epsilon}^u$ are given by
\begin{equation}
\begin{aligned}
(\mathcal{H}^u_{\mathbb{Z}_2, \chi, \epsilon})_0 & = (\bigoplus_{g \in \mathbb{Z}_2} \mathrm{Span}\{(e^g_{11})_+, (e^g_{1u})_+, (e^g_{1m})_+, (e^g_{m1})_+, e^g_{mm}\}) \\
& \quad \quad \oplus (\bigoplus_{a \in \mathbb{Z}_2} \mathrm{Span}\{(e^m_{1a})_+, (e^m_{1\overline{a}})_+, (e^m_{\overline{1}a})_+, (e^m_{\overline{1} \overline{a}})_+\}), \\
(\mathcal{H}^u_{\mathbb{Z}_2, \chi, \epsilon})_1 & = (\bigoplus_{g \in \mathbb{Z}_2} \mathrm{Span}\{(e^g_{11})_-, (e^g_{1u})_-, (e^g_{1m})_-, (e^g_{m1})_-\}) \\
& \quad \quad \oplus (\bigoplus_{a \in \mathbb{Z}_2} \mathrm{Span}\{(e^m_{1a})_-, (e^m_{1\overline{a}})_-, (e^m_{\overline{1}a})_-, (e^m_{\overline{1} \overline{a}})_-\}),
\end{aligned}
\end{equation}
where the homogeneous elements $(e^x_{\alpha, \beta})_{\pm}$ are defined as follows:
\begin{equation}
\begin{aligned}
(e^g_{11})_{\pm} & = e^g_{11} \pm e^g_{uu}, \quad (e^g_{1u})_{\pm} = e^g_{1u} \pm e^g_{u1}, \quad (e^g_{1m})_{\pm} = e^g_{1m} \pm e^g_{um}, \quad (e^g_{m1})_{\pm} = e^g_{m1} \pm e^g_{mu}, \\
(e^m_{11})_{\pm} & = e^m_{11} \pm e^m_{uu}, \quad (e^m_{1u})_{\pm} = e^m_{1u} \mp e^m_{u1}, \quad (e^m_{1\overline{1}})_{\pm} = e^m_{1\overline{1}} \mp e^m_{u \overline{u}}, \quad (e^m_{1 \overline{u}})_{\pm} = e^m_{1 \overline{u}} \pm e^m_{u \overline{1}}, \\
(e^m_{\overline{1} 1})_{\pm} & = e^m_{\overline{1} 1} \mp e^m_{\overline{u} u}, \quad (e^m_{\overline{1} u})_{\pm} = e^m_{\overline{1} u} \pm e^m_{\overline{u} 1}, \quad (e^m_{\overline{1} \overline{1}})_{\pm} = e^m_{\overline{1} \overline{1}} \pm e^m_{\overline{u} \overline{u}}, \quad (e^m_{\overline{1} \overline{u}})_{\pm} = e^m_{\overline{1} \overline{u}} \mp e^m_{\overline{u} \overline{1}}.
\end{aligned}
\end{equation}
Let us compute the multiplication on $\mathcal{H}^u_{\mathbb{Z}_2, \chi, \epsilon}$ based on the above $\mathbb{Z}_2$-grading and determine the direct sum decomposition of $\mathcal{H}^u_{\mathbb{Z}_2, \chi, \epsilon}$.
To this end, we first notice that $\mathcal{H}^u_{\mathbb{Z}_2, \chi, \epsilon}$ is decomposed into the direct sum of a subalgebra spanned by \{$e^g_{\alpha, \beta} \mid g \in \mathbb{Z}_2, \alpha, \beta \in \Omega \}$ and its complement spanned by $\{ e^m_{\gamma, \delta} \mid \gamma, \delta \in \mathbb{Z}_2 \sqcup \overline{\mathbb{Z}_2} \}$.
This is because the multiplication of $e^g_{\alpha, \beta}$ and $e^m_{\gamma, \delta}$ vanishes for any choice of $\alpha, \beta, \gamma$, and $\delta$.
As we will see below, each of these subalgebras is further decomposed into the direct sum of two simple superalgebras.
Specifically, the former subalgebra is isomorphic to the direct sum of two copies of the endomorphism superalgebra $\mathrm{End}(\mathbb{C}^{2|1})$, whereas the latter subalgebra is isomorphic to the direct sum of two copies of $\mathrm{End}(\mathbb{C}^{2|0}) \otimes \mathrm{Cl}(1)$.
In order to see an isomorphism $\mathrm{Span}\{e^g_{\alpha, \beta} \mid g \in \mathbb{Z}_2, \alpha, \beta \in \Omega\} \cong \mathrm{End}(\mathbb{C}^{2|0}) \oplus \mathrm{End}(\mathbb{C}^{2|0})$, we define a new basis of the algebra spanned by $\{e^g_{\alpha, \beta} \mid g \in \mathbb{Z}_2, \alpha, \beta \in \Omega\}$ as follows:
\begin{equation}
\begin{aligned}
x^g_{11} & = \frac{1}{2} [(e^g_{11})_+ + (e^g_{1u})_+], \quad x^g_{1u} = - \frac{1}{2} [(e^{ug}_{11})_- - (e^{ug}_{1u})_-], \quad x^g_{1m} = \frac{1}{2} (e^g_{1m})_+, \\
x^g_{u1} & = \frac{1}{2} [(e^g_{11})_- + (e^g_{1u})_-], \quad x^g_{uu} = \frac{1}{2} [(e^{ug}_{11})_+ - (e^{ug}_{1u})_+], \quad x^g_{um} = \frac{1}{2} (e^g_{1m})_-, \\
x^g_{m1} & = (e^g_{m1})_+, \quad x^g_{mu} = - (e^{ug}_{m1})_-, \quad x^g_{mm} = e^g_{mm}.
\end{aligned}
\label{eq: new basis}
\end{equation}
For this basis, the multiplication on $\mathcal{H}^u_{\mathbb{Z}_2, \chi, \epsilon}$ can be expressed as
\begin{equation}
m_u (x^g_{\alpha, \beta} \otimes x^h_{\gamma, \delta}) = \delta_{g, h} \delta_{\beta, \gamma} x^g_{\alpha, \delta},
\end{equation}
which shows that the subalgebra spanned by $\{x^g_{\alpha, \beta} \mid \alpha, \beta \in \Omega\}$ is a full matrix algebra for each $g \in \mathbb{Z}_2$.
Since this subalgebra has superdimension $(5, 4)$ as can be seen from eq. \eqref{eq: new basis}, there is an even isomorphism between $\mathrm{Span}\{x^g_{\alpha, \beta} \mid \alpha, \beta \in \Omega\}$ and $\mathrm{End}(\mathbb{C}^{2|1})$.
Therefore, we have the following direct sum decomposition of a superalgebra:
\begin{equation}
\mathrm{Span}\{e^g_{\alpha, \beta} \mid g \in \mathbb{Z}_2, \alpha, \beta \in \Omega \} \cong \mathrm{End}(\mathbb{C}^{2|1}) \oplus \mathrm{End}(\mathbb{C}^{2|1}).
\label{eq: direct sum decomposition of TY 1}
\end{equation}
Similarly, an isomorphism $\mathrm{Span}\{e^m_{\alpha, \beta} \mid \alpha, \beta \in \mathbb{Z}_2 \sqcup \overline{\mathbb{Z}_2}\} \cong (\mathrm{End}(\mathbb{C}^{2|0}) \otimes \mathrm{Cl}(1)) \oplus (\mathrm{End}(\mathbb{C}^{2|0}) \otimes \mathrm{Cl}(1))$ also becomes clear if we define a new basis of the algebra spanned by $\{e^m_{\alpha, \beta} \mid \alpha, \beta \in \mathbb{Z}_2 \sqcup \overline{\mathbb{Z}_2} \}$ as 
\begin{equation}
\begin{aligned}
(x^{m, s}_{11})_{+} & = \frac{1}{2} [(e^m_{11})_{+} + i s (e^m_{1u})_{+}], \quad (x^{m, s}_{11})_{-} = \frac{1}{2} [(e^m_{11})_{-} + i s (e^m_{1u})_{-}], \\
(x^{m, s}_{1\overline{1}})_{+} & = \frac{1}{2} [(e^m_{1\overline{1}})_{+} - i s (e^m_{1 \overline{u}})_{+}], \quad (x^{m, s}_{1\overline{1}})_{-} = \frac{1}{2} [(e^m_{1\overline{1}})_{-} - i s (e^m_{1 \overline{u}})_{-}], \\
(x^{m, s}_{\overline{1} 1})_{+} & = \frac{1}{2} [(e^m_{\overline{1} 1})_+ + i s (e^m_{\overline{1} u})_+], \quad (x^{m, s}_{\overline{1} 1})_-  = \frac{is}{2} [(e^m_{\overline{1} 1})_- + i s (e^m_{\overline{1} u})_-], \\
(x^{m, s}_{\overline{1} \overline{1}})_+ & = \frac{1}{2} [(e^m_{\overline{1} \overline{1}})_+ - i s (e^m_{\overline{1} \overline{u}})_+], \quad (x^{m, s}_{\overline{1} \overline{1}})_- = \frac{is}{2} [(e^m_{\overline{1} \overline{1}})_- - i s (e^m_{\overline{1} \overline{u}})_-],
\end{aligned}
\end{equation}
where the superscript $s$ takes values in $\{\pm1\}$.
The multiplication on $\mathcal{H}^u_{\mathbb{Z}_2, \chi, \epsilon}$ for this basis can be written as
\begin{equation}
m_u ((x^{m, s}_{\alpha, \beta})_{p} \otimes (x^{m, t}_{\gamma, \delta})_{q}) = \delta_{s, t} \delta_{\beta, \gamma} (x^{m, s}_{\alpha, \delta})_{pq}.
\end{equation}
This implies that the subalgebra spanned by $\{(x^{m, s}_{\alpha, \beta})_{\pm} \mid \alpha, \beta = 1, \overline{1} \}$ for each $s \in \{\pm 1\}$ is isomorphic to a simple superalgebra $\mathrm{End}(\mathbb{C}^{2|0}) \otimes \mathrm{Cl}(1)$, where the odd generator of the Clifford algebra $\mathrm{Cl}(1)$ corresponds to $(x^{m, s}_{1 1})_{-} + (x^{m, s}_{\overline{1} \overline{1}})_{-}$.
Thus, we find the following direct sum decomposition:
\begin{equation}
\mathrm{Span}\{e^m_{\alpha, \beta} \mid \alpha, \beta \in \mathbb{Z}_2 \sqcup \overline{\mathbb{Z}_2}\} \cong (\mathrm{End}(\mathbb{C}^{2|0}) \otimes \mathrm{Cl}(1)) \oplus (\mathrm{End}(\mathbb{C}^{2|0}) \otimes \mathrm{Cl}(1)).
\label{eq: direct sum decomposition of TY 2}
\end{equation}
Equations \eqref{eq: direct sum decomposition of TY 1} and \eqref{eq: direct sum decomposition of TY 2} show that the weak Hopf superalgebra $\mathcal{H}_{\mathbb{Z}_2, \chi, \epsilon}^u$ is decomposed into the direct sum of two copies of $\mathrm{End}(\mathbb{C}^{2|1})$ and two copies of $\mathrm{End}(\mathbb{C}^{2|0}) \otimes \mathrm{Cl}(1)$.

\paragraph{Superfusion category $\mathrm{SRep}(\mathcal{H}_{\mathbb{Z}_2, \chi, \epsilon}^u)$.}
The direct sum decomposition of $\mathcal{H}^u_{\mathbb{Z}_2, \chi, \epsilon}$ indicates that $\mathcal{H}_{\mathbb{Z}_2, \chi, \epsilon}^u$ has three-dimensional irreducible super representations $V_g \cong \mathbb{C}^{2|1}$ for $g \in \mathbb{Z}_2$ and four-dimensional irreducible super representations $W_s \cong \mathbb{C}^{2|2}$ for $s = \pm 1$.
The actions of $\mathcal{H}_{\mathbb{Z}_2, \chi, \epsilon}^u$ on $V_g$ and $W_s$ are given by the standard actions of the direct summands $\mathrm{End}(\mathbb{C}^{2|1})$ and $\mathrm{End}(\mathbb{C}^{2|0}) \otimes \mathrm{Cl}(1)$.
More specifically, if we write the bases of $V_g$ and $W_s$ as $\{v^g_{\alpha} \mid \alpha \in \Omega \}$ and $\{(w^s_{\alpha})_{\pm} \mid \alpha = 1, \overline{1}\}$ respectively, the actions of $\mathcal{H}_{\mathbb{Z}_2, \chi, \epsilon}^u$ on $V_g$ and $W_s$ are given by
\begin{align}
x^h_{\alpha, \beta} \cdot v^g_{\gamma} & = \delta_{g, h} \delta_{\beta, \gamma} v^g_{\alpha}, \quad (x^{m, t}_{\alpha, \beta})_q \cdot v^g_{\gamma} = 0,
\label{eq: Z2 TY action on Vg}\\
x^h_{\alpha, \beta} \cdot (w^s_{\gamma})_p & = 0, \quad (x^{m, t}_{\alpha, \beta})_q \cdot (w^s_{\gamma})_p = \delta_{s, t} \delta_{\beta, \gamma} (w^s_{\alpha})_{pq},
\label{eq: Z2 TY action on Ws}
\end{align}
where $v^g_{1}, v^g_{m} \in V_g$ and $(w^s_1)_+, (w^s_{\overline{1}})_+ \in W_s$ are $\mathbb{Z}_2$-even elements and the others are $\mathbb{Z}_2$-odd elements.
We note that $W_+$ and $W_-$ are q-type objects because they are irreducible super representations of $\mathrm{End}(\mathbb{C}^{2|0}) \otimes \mathrm{Cl}(1)$ \cite{Jozefiak1988}.

As we will see in appendix \ref{sec: Fusion rules of superfusion category symmetries}, simple objects $V_1, V_u, W_+, W_- \in \mathrm{SRep}(\mathcal{H}^u_{\mathbb{Z}_2, \chi, \epsilon})$ obey the following fusion rules:
\begin{equation}
V_u \boxtimes V_u \cong V_1, \quad V_u \boxtimes W_{+} \cong W_{-} \cong W_{+} \boxtimes V_u, \quad W_{+} \boxtimes W_{+} \cong V_1 \oplus \Pi V_1,
\label{eq: fusion rules of fermionized Z2 TY}
\end{equation}
where $\Pi V_1$ is oddly isomorphic to the trivial defect $V_1$.
The associativity of the fusion rules uniquely determines the other fusion rules.
Equation \eqref{eq: fusion rules of fermionized Z2 TY} implies that a q-type object $W_+$ and an m-type object $V_u$ generate $\mathbb{Z}_2 \times \mathbb{Z}_2^F$ symmetry with an odd 't Hooft anomaly.\footnote{In general, when $\mathbb{Z}_2 \times \mathbb{Z}_2^F$ symmetry has an odd 't Hooft anomaly, the generator $\eta$ of the $\mathbb{Z}_2$ subgroup does not satisfy the ordinary $\mathbb{Z}_2$ group-like fusion rule because $\eta$ is a q-type object and hence $\eta \otimes \eta$ has an odd automorphism. More specifically, $\eta \otimes \eta$ is not a trivial defect $1$ but rather the direct sum of a trivial defect $1$ and another defect $\Pi$ that is oddly isomorphic to $1$. If normalized appropriately, the action of $\eta$ on the $\mathrm{NS}$ sector satisfies the ordinary $\mathbb{Z}_2$ fusion rule because both $1$ and $\Pi$ act as the identity operator on the $\mathrm{NS}$ sector, cf. eq. \eqref{eq: action of PiX}.}
In particular, a fermionic system with $\mathrm{SRep}(\mathcal{H}^u_{\mathbb{Z}_2, \chi, \epsilon})$ symmetry is invariant under stacking the Kitaev chain \cite{JSW2020}, which is a consequence of the fusion rule $W_+ \boxtimes W_+ \cong V_1 \oplus \Pi V_1$.
This is consistent with the fact that the fermionization of the Ising CFT and the $\mathrm{SU}(2)_2$ WZW model has $\mathbb{Z}_2 \times \mathbb{Z}_2^F$ symmetries with 't Hooft anomalies $1$ and $3 \bmod 8$.
The determination of the anomaly would require further analysis on the $F$-symbols of $\mathrm{SRep}(\mathcal{H}_{\mathbb{Z}_2, \chi, \epsilon}^u)$.

\section{Fermionic TQFTs with superfusion category symmetries}
\label{sec: Fermionic TQFTs with superfusion category symmetries}
In this section, we explicitly construct fermionic state sum TQFTs with super fusion category symmetries, which can be thought of as the fermionization of bosonic state sum TQFTs with fusion category symmetries.
In particular, we derive the fermionization formula of fusion category symmetries by comparing the symmetries of fermionic and bosonic state sum TQFTs.
We will also write down gapped Hamiltonians with non-anomalous superfusion category symmetries on the lattice.
The content of this section is a straightforward generalization of the state sum construction of bosonic TQFTs with fusion category symmetries discussed in \cite{DKR2011, Inamura2022}.

\subsection{State sum construction of fermionic TQFTs and fermionization}
\label{sec: State sum construction of fermionic TQFTs as fermionization}
In this subsection, we review the state sum construction of 2d fermionic TQFTs on triangulated spin surfaces \cite{NR2015, GK2016}.
We will see that fermionic state sum TQFTs can be understood as the fermionization of bosonic state sum TQFTs \cite{GK2016, KTY2018}.

We first recall the combinatorial description of spin structures on triangulated surfaces following \cite{NR2015}.
Let $\Sigma$ be an oriented surface. 
We choose a triangulation of $\Sigma$.
The triangulated surface is also denoted by $\Sigma$.
A spin structure on a triangulated surface $\Sigma$ is specified by the following set of data, which is called a marking:
\begin{itemize}
\item an orientation of each edge of $\Sigma$,
\item an edge index $s(e) \in \{ 0, 1 \}$ of each edge $e$,
\item a choice of a marked edge on the boundary of each triangle, and 
\item a choice of a marked vertex on each connected component of the boundary $\partial \Sigma$.
\end{itemize}
A connected component of the boundary $\partial \Sigma$ belongs to either the in-boundary $\partial_{\mathrm{in}} \Sigma$ or the out-boundary $\partial_{\mathrm{out}} \Sigma$.
The orientations of edges on $\partial_{\mathrm{in}} \Sigma$ are induced by the orientation of $\Sigma$.
On the other hand, the orientations of edges on $\partial_{\mathrm{out}} \Sigma$ are opposite to the induced orientation.

\begin{figure}
\begin{minipage}{0.25 \hsize}
\begin{center}
\includegraphics[width = 3.5cm]{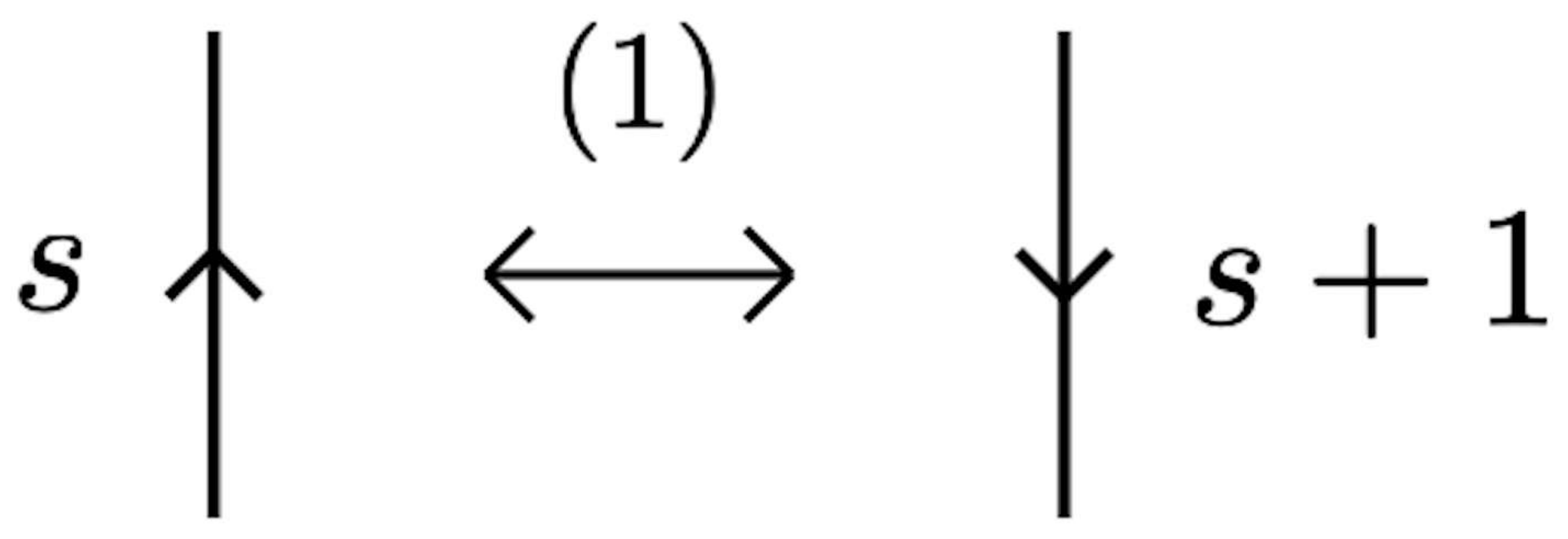}
\end{center}
\end{minipage}%
\begin{minipage}{0.03 \hsize}
\begin{center}
\end{center}
\end{minipage}%
\begin{minipage}{0.33 \hsize}
\begin{center}
\includegraphics[width = 5.4cm]{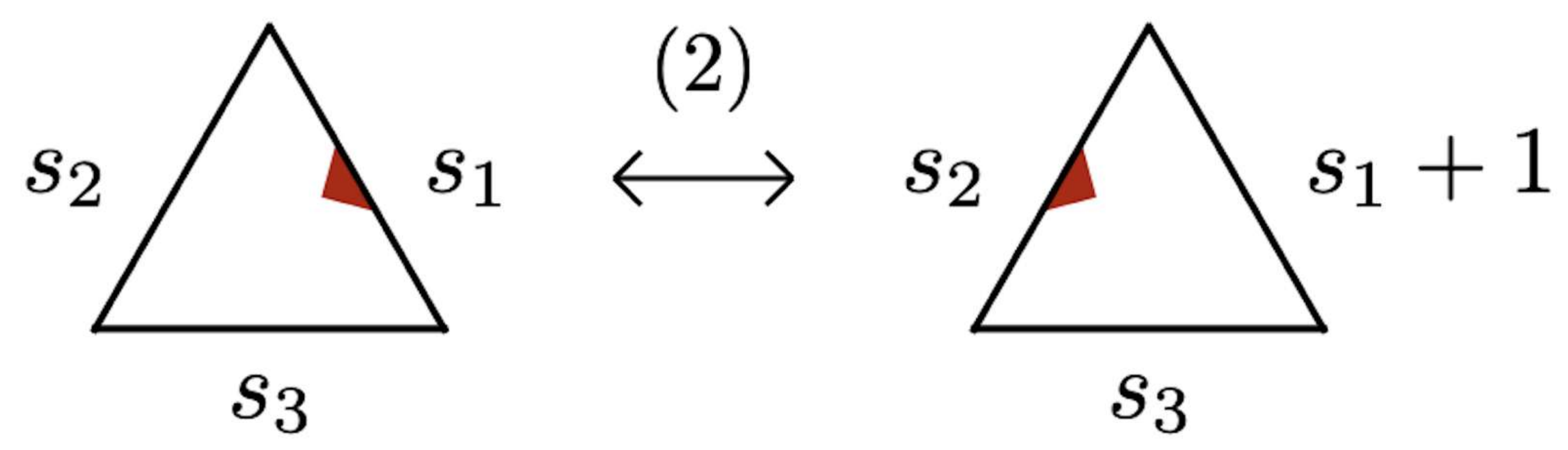}
\end{center}
\end{minipage}%
\begin{minipage}{0.03 \hsize}
\begin{center}
\end{center}
\end{minipage}%
\begin{minipage}{0.36 \hsize}
\begin{center}
\includegraphics[width = 5.7cm]{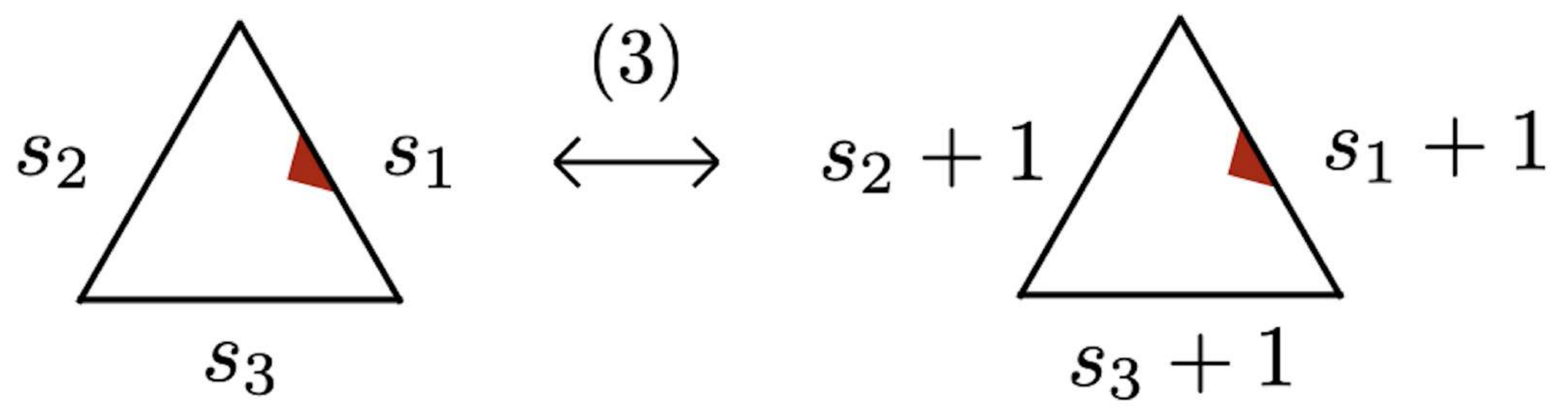}
\end{center}
\end{minipage}
\caption{Markings related by the above local moves give the same spin structure. (1) Change of an edge orientation: we reverse the orientation of an edge and shift the edge index by 1. (2) Change of a marked edge: we choose the edge next to the original marked edge as the new marked edge and shift the edge index accordingly. The marked edge is represented by an edge with a small red triangle attached. (3) Leaf exchange on a triangle: we shift the edge indices of all edges on the boundary of a triangle simultaneously. We note that the leaf exchange is equivalent to changing the marked edge three times.}
\label{fig: local moves}
\end{figure}

A marking on a triangulated surface $\Sigma$ gives a spin structure on $\Sigma$ if it satisfies the following admissibility condition for each vertex $v$:
\begin{equation}
\sum_{\substack{e: \text{ edges} \\ \text{s.t. } v \in \partial e}} s(e) =
\begin{cases}
D_v + E_v + \lambda \bmod 2 \quad & \text{when } v \text{ is a marked vertex on } \partial \Sigma, \\
D_v + E_v + 1 \bmod 2 \quad & \text{otherwise}.
\end{cases}
\label{eq: admissibility}
\end{equation}
Here, $D_v$ is the number of triangles $t$ such that a small counterclockwise loop around $v$ enters $t$ through the marked edge of $t$, $E_v$ is the number of edges whose initial vertex is $v$, and $\lambda$ is an index defined by
\begin{equation}
\lambda =
\begin{cases}
0 \quad & \text{if } v \text{ is on an }\mathrm{NS}\text{ boundary}, \\
1 \quad & \text{if } v \text{ is on an }\mathrm{R}\text{ boundary}.
\end{cases}
\end{equation}
Here, $\mathrm{NS}$ and $\mathrm{R}$ denote the Neveu-Schwarz (i.e., bounding) and Ramond (i.e., non-bounding) spin structures on a circle, which correspond to the anti-periodic and periodic boundary conditions respectively.
Two markings on $\Sigma$ correspond to the same spin structure if they are related by a sequence of the following local moves: (1) change of an edge orientation, (2) change of a marked edge, (3) leaf exchange on a triangle, see figure \ref{fig: local moves}.
These local moves define an equivalence relation between markings.
The quotient of the set of admissible markings on $\Sigma$ by this equivalence relation is in bijective correspondence with the set of spin structures on $\Sigma$ \cite{NR2015}.

\begin{figure}
\begin{minipage}{0.45 \hsize}
\begin{center}
\includegraphics[width = 7cm]{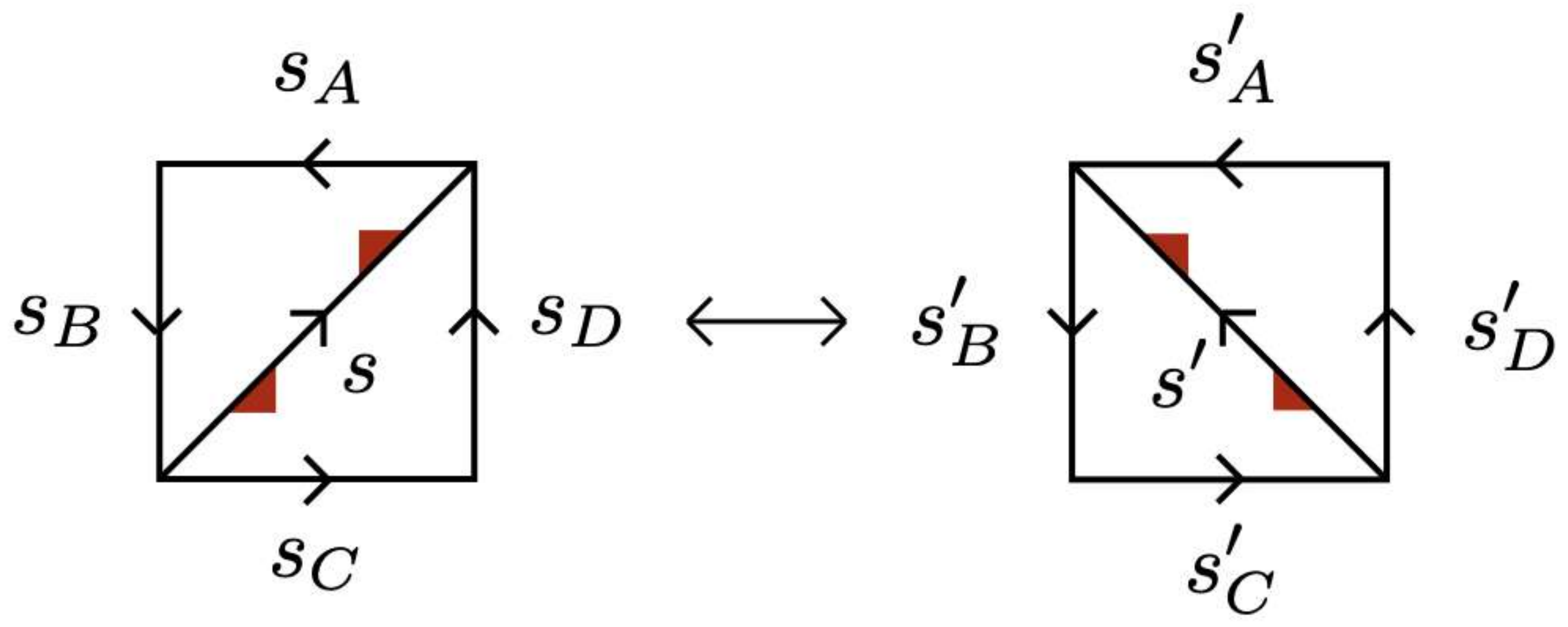}
\end{center}
\end{minipage}%
\begin{minipage}{0.05 \hsize}
\begin{center}
\end{center}
\end{minipage}%
\begin{minipage}{0.45 \hsize}
\begin{center}
\includegraphics[width = 6.5cm]{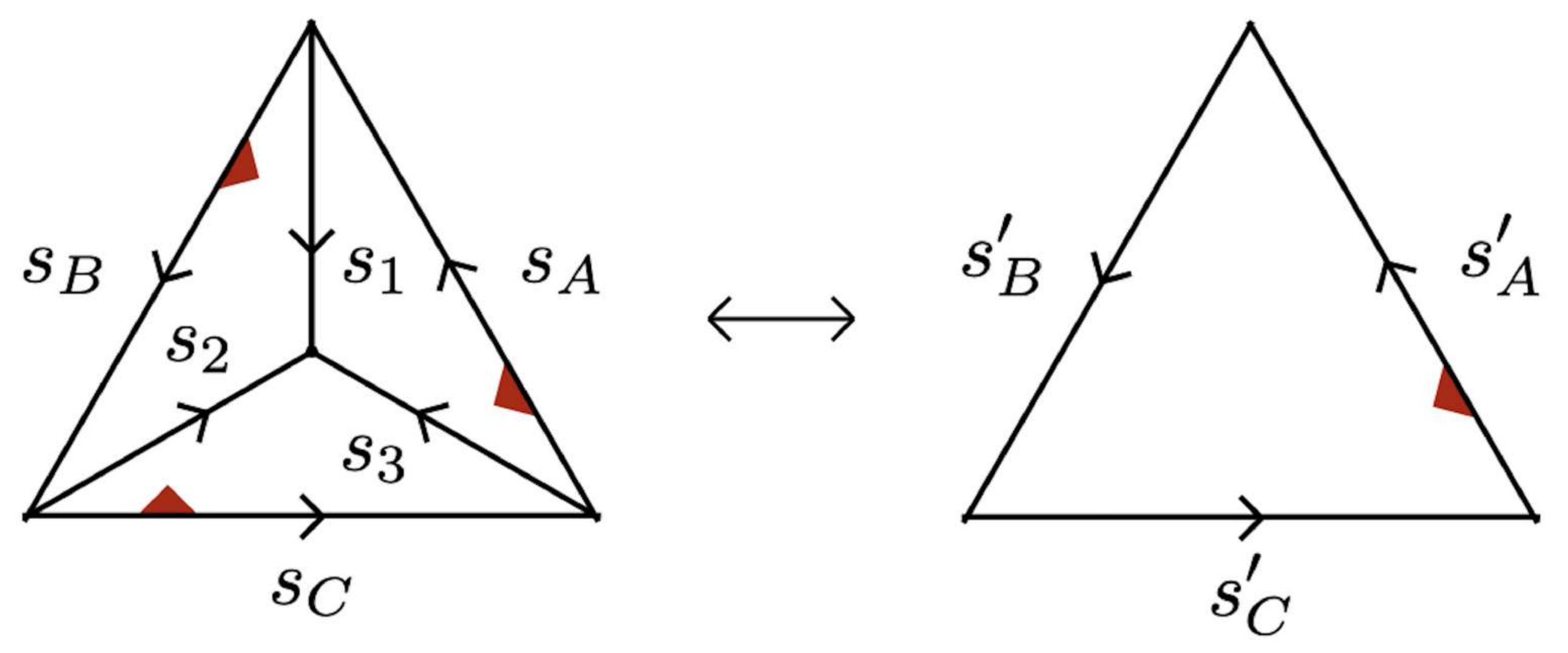}
\end{center}
\end{minipage}
\caption{The Pachner 2-2 move (left) and the Pachner 3-1 move (right). The edge indices are determined uniquely up to local moves. For the Pachner 2-2 move, we have $s^{\prime} = s, s^{\prime}_A = s_A, s^{\prime}_B = s_B + s + 1, s^{\prime}_C = s_C + 1, s^{\prime}_D = s_D + s + 1$. For the Pachner 3-1 move, we have $s^{\prime}_A = s_A, s^{\prime}_B = s_B + s_1, s^{\prime}_C = s_C + s_1 + s_2$, $s_3 = s_1 + s_2 + 1$.}
\label{fig: Pachner}
\end{figure}

Any two triangulations of an oriented surface are related by a finite sequence of Pachner moves. 
For a triangulated spin surface $\Sigma$ equipped with a marking, the Pachner moves also change the marking on affected triangles, see figure \ref{fig: Pachner}.
The change of a marking is determined by the admissibility condition \eqref{eq: admissibility} uniquely up to local moves shown in figure \ref{fig: local moves}.

Based on the above combinatorial description of spin structures, we can construct fermionic TQFTs from semisimple superalgebras by the state sum construction \cite{NR2015, GK2016}. 
The fermionic TQFT obtained from a semisimple superalgebra $K$ is denoted by $\mathcal{T}_f^K$.
The construction of $\mathcal{T}_f^K$ goes as follows: we first define the transition amplitude $Z_T(\Sigma)$ on a triangulated spin surface $\Sigma$ equipped with a marking and then restrict the domain and the codomain of $Z_T(\Sigma)$ to the images of the cylinder amplitudes $Z_T(\partial_{\mathrm{in}} \Sigma \times [0, 1])$ and $Z_T(\partial_{\mathrm{out}} \Sigma \times [0, 1])$ respectively.
We will generalize this construction to fermionic TQFTs on spin surfaces with defects in the following subsections.
The detailed definition of the transition amplitude $Z_T(\Sigma)$ will be given there.

\paragraph{Relation to fermionization.}
The fermionic TQFT $\mathcal{T}_f^K$ is the fermionization of a bosonic state sum TQFT $\mathcal{T}_b^K$ constructed from $K$ by the state sum construction of \cite{FHK94}.\footnote{Even though $\mathcal{T}_b^K$ and $\mathcal{T}_f^K$ are constructed from the same algebra $K$, these TQFTs are different from each other because the constructions are different.}
To see this, we notice that the partition functions of $\mathcal{T}_f^K$ and $\mathcal{T}_b^K$ on a closed surface $\Sigma$ can be written as \cite{GK2016, KTY2018}
\begin{align}
Z_f^K (\Sigma; \eta) & = \frac{1}{\sqrt{|H^1(\Sigma, \mathbb{Z}_2)|}} \sum_{\beta \in H^1(\Sigma, \mathbb{Z}_2)} Z(\Sigma; \beta) (-1)^{q_{\eta}(\beta)},
\label{eq: fermionic state sum partition function} \\
Z_b^K (\Sigma) & = \frac{1}{\sqrt{|H^1(\Sigma, \mathbb{Z}_2)|}} \sum_{\beta \in H^1(\Sigma, \mathbb{Z}_2)} Z(\Sigma; \beta),
\label{eq: bosonic state sum partition function}
\end{align}
where $\eta$ is a spin structure on $\Sigma$, $Z(\Sigma; \beta)$ is the partition function of some bosonic TQFT $\mathcal{T}$ on a surface $\Sigma$ with a background $\mathbb{Z}_2$ gauge field $\beta$, and $q_{\eta}$ is a quadratic refinement of the intersection form.
Equation \eqref{eq: fermionic state sum partition function} implies that the fermionic state sum TQFT $\mathcal{T}_f^K$ is the Jordan-Wigner transformation of $\mathcal{T}$, or equivalently a bosonic TQFT $\mathcal{T}$ is the GSO projection of $\mathcal{T}_f^{K}$ because summing over spin structures in eq. \eqref{eq: fermionic state sum partition function} leads to
\begin{equation}
Z(\Sigma; \beta = 0) = \frac{1}{\sqrt{|H^1(\Sigma, \mathbb{Z}_2)|}} \sum_{\eta} Z_f^K (\Sigma; \eta),
\end{equation}
where the left-hand side is the partition function of $\mathcal{T}$ on $\Sigma$ without a background $\mathbb{Z}_2$ gauge field.
On the other hand, eq. \eqref{eq: bosonic state sum partition function} implies that the bosonic state sum TQFT $\mathcal{T}_b^K$ is the $\mathbb{Z}_2$-gauging of $\mathcal{T}$.
Therefore, we find that $\mathcal{T}_b^K$ is obtained by the GSO projection of $\mathcal{T}_f^K$ followed by the $\mathbb{Z}_2$ gauging.
This shows that the bosonic state sum TQFT $\mathcal{T}_b^K$ and the fermionic state sum TQFT $\mathcal{T}_f^K$ are related by the bosonization and the fermionization, cf. figure \ref{fig: fermionization}.
More precisely, $\mathcal{T}_f^K$ is the fermionization of $\mathcal{T}_b^K$ with respect to the $\mathbb{Z}_2$ symmetry arising from the $\mathbb{Z}_2$-grading on $K$.
This fact allows us to derive the fermionization formula of fusion category symmetries later by comparing the symmetry of $\mathcal{T}_b^K$ and that of $\mathcal{T}_f^K$.

\subsection{Fermionic TQFTs with ${}_K \mathcal{SM}_{K}$ symmetry}
\label{sec: Fermionic TQFTs with KSMK symmetry}
In this subsection, we define the transition amplitudes of a fermionic TQFT $\mathcal{T}_f^K$ on spin surfaces with defects.
The construction of fermionic TQFTs with defects in this subsection is parallel to the construction of bosonic TQFTs with defects in \cite{DKR2011, Inamura2022}.
As we will see below, $\mathcal{T}_f^K$ has non-zero transition amplitudes on spin surfaces with topological defects labeled by $(K, K)$-superbimodules.
This implies that the fermionic state sum TQFT $\mathcal{T}_f^K$ has a superfusion category symmetry ${}_K \mathcal{SM}_K$.

We begin with triangulations of spin surfaces with defects.
We assume that every junction of topological defects is trivalent.
A triangulation of a spin surface with defects is a triangulation of the underlying spin surface such that each edge intersects a topological defect at most once and each triangle contains at most one trivalent junction.
Specifically, possible configurations of topological defects on a single triangle up to local moves are listed as follows:
\begin{equation}
(1) ~ \adjincludegraphics[valign = c, width = 1.9cm]{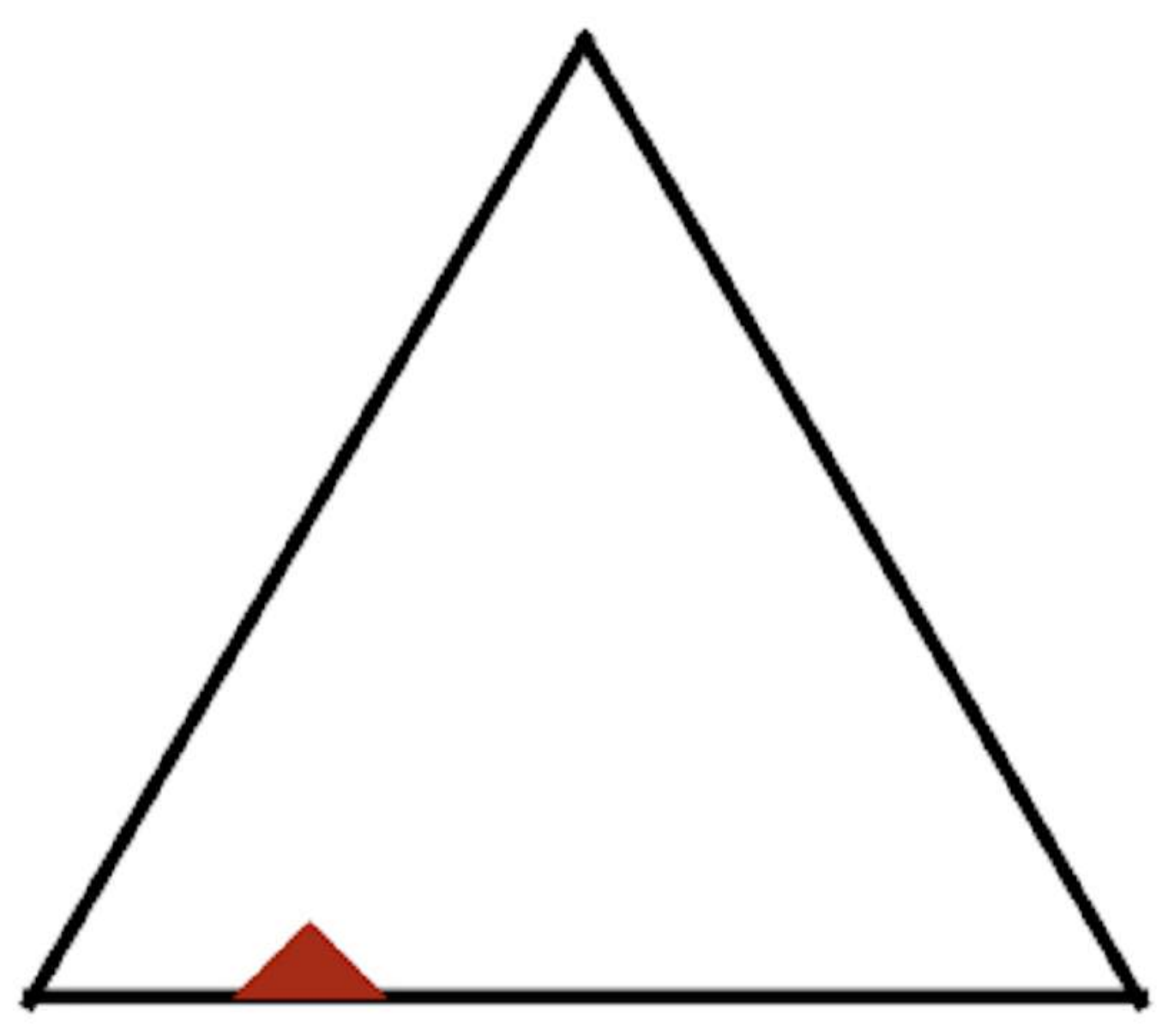}, ~
(2) ~ \adjincludegraphics[valign = c, width = 2cm]{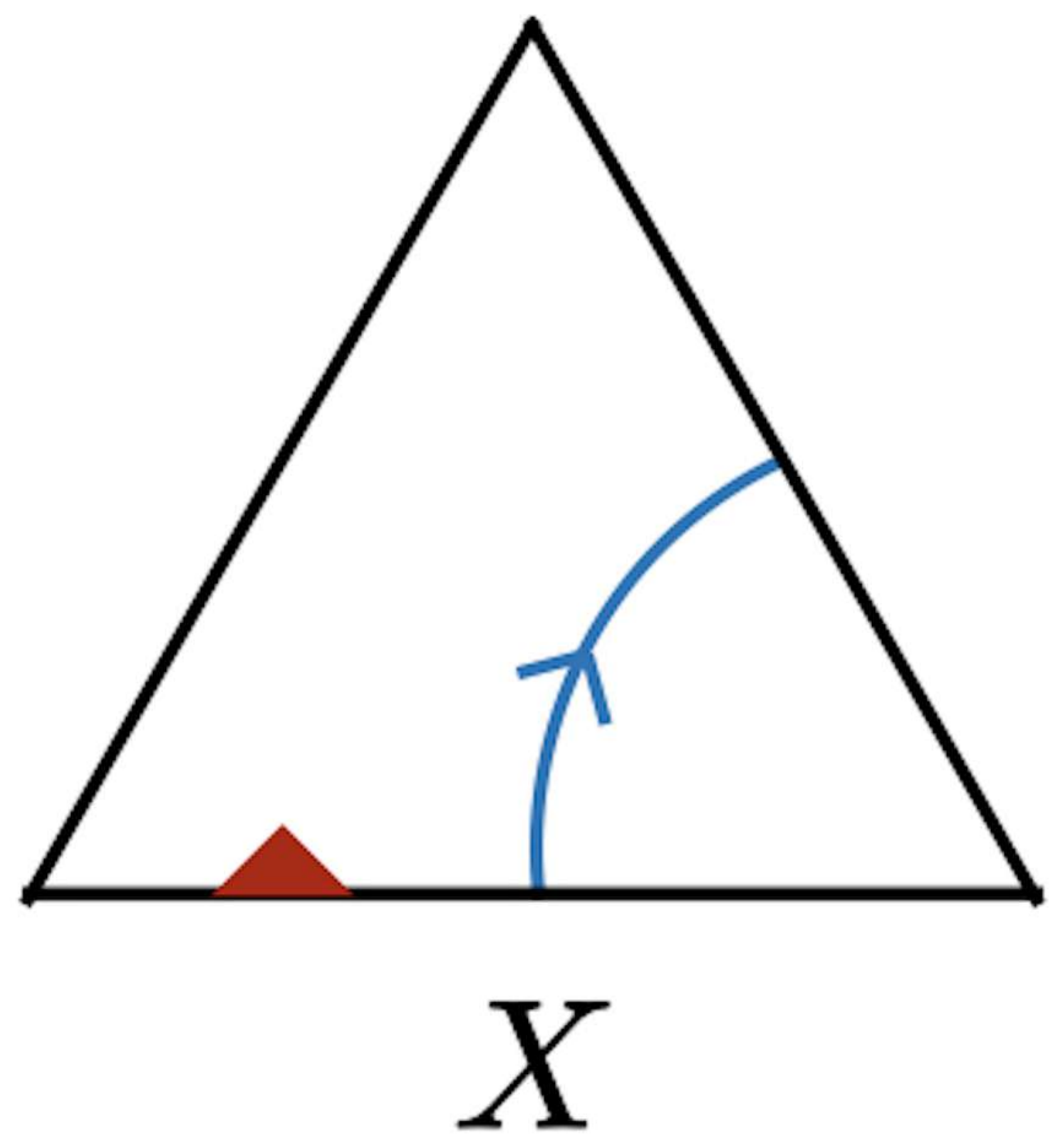}, ~
(3) ~ \adjincludegraphics[valign = c, width = 2cm]{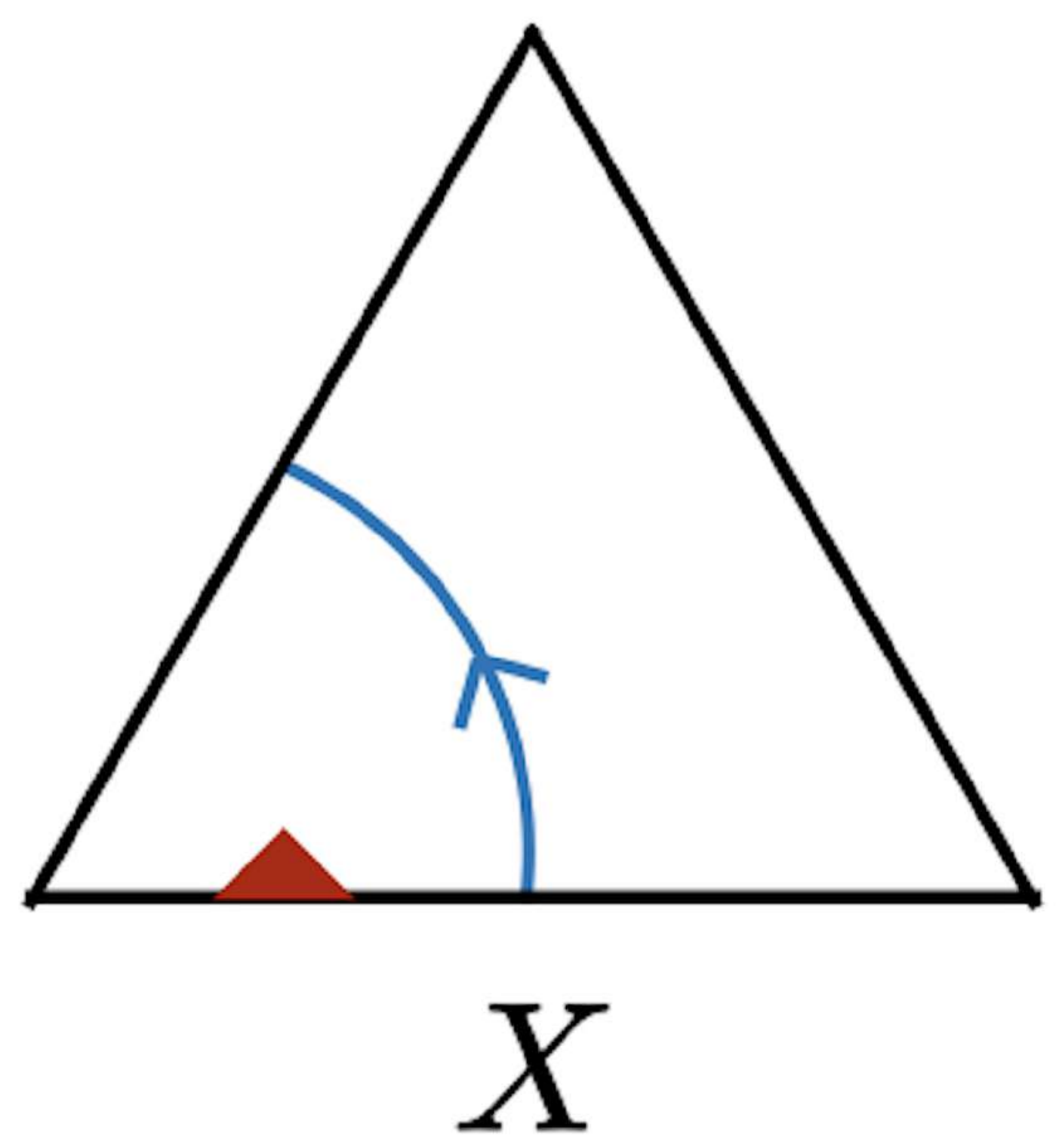}, ~
(4) ~ \adjincludegraphics[valign = c, width = 2.2cm]{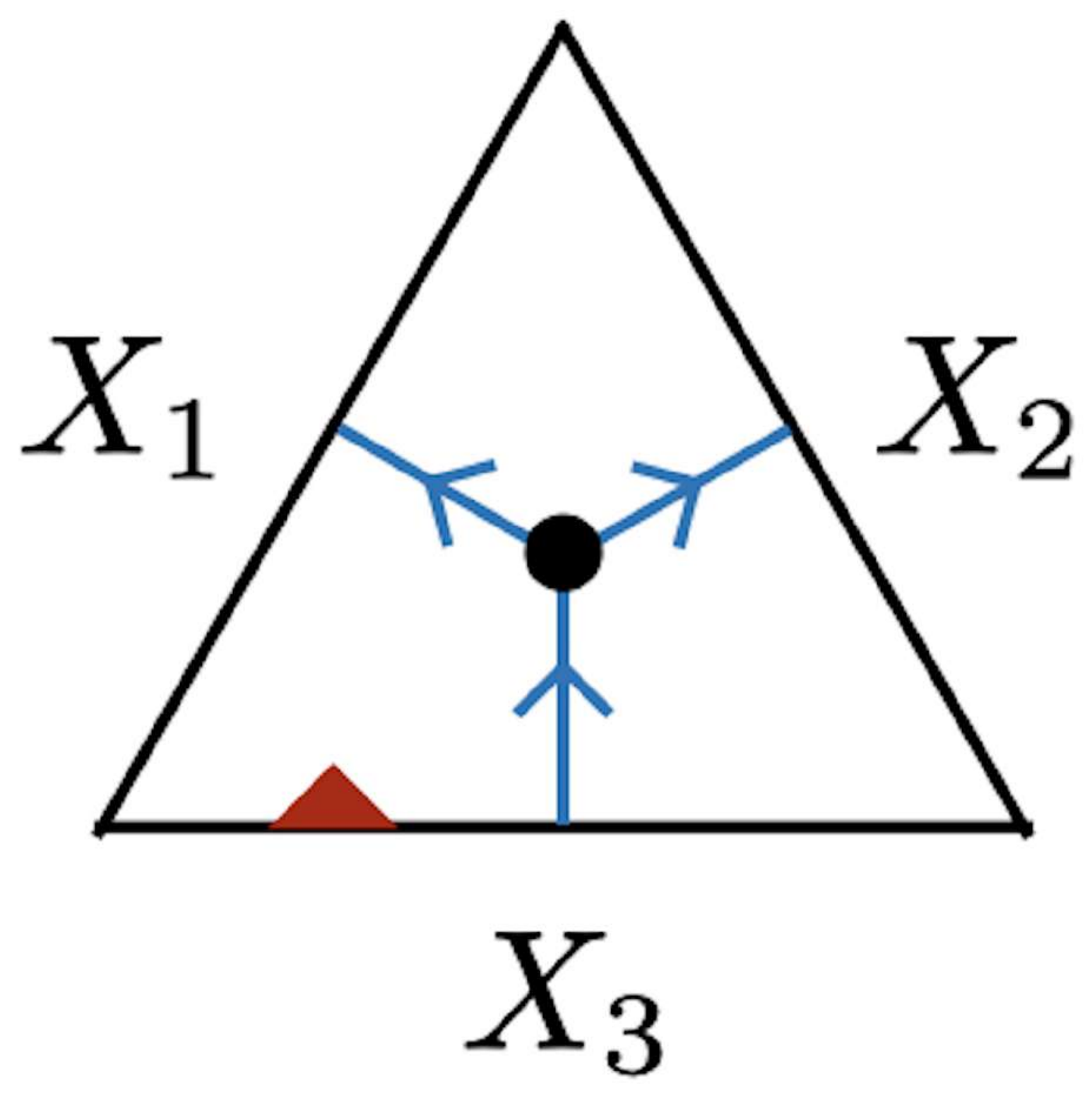}, ~
(5) ~ \adjincludegraphics[valign = c, width = 2.2cm]{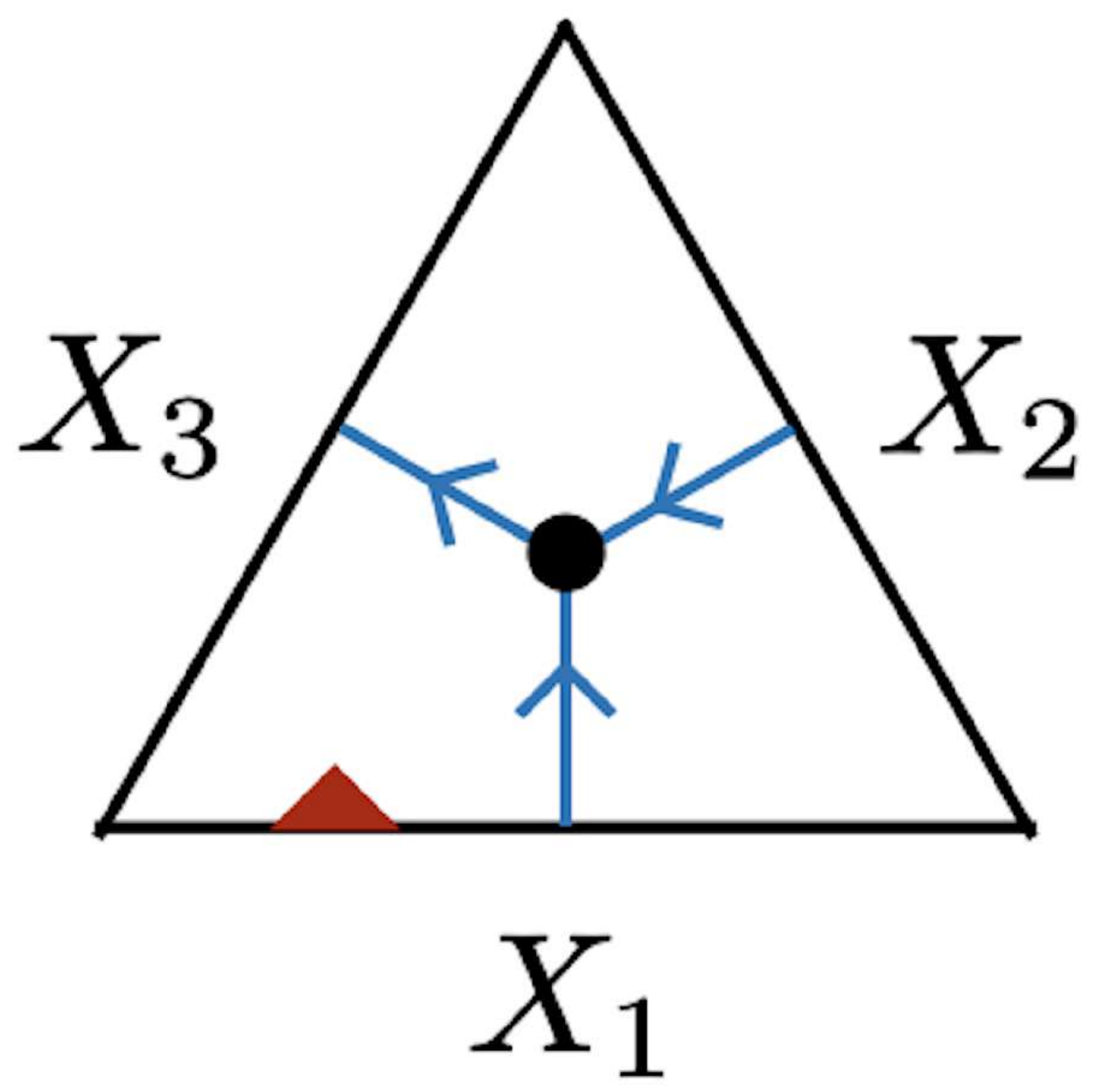}, 
\label{eq: configurations}
\end{equation}
We note that every configuration listed above can be regarded as a special case of configuration (5) because configurations (1)--(3) are obtained by taking some of the topological defects to be trivial defects and configuration (4) is obtained by replacing one of the topological defects by its dual.
Therefore, we need not consider configurations (1)--(4) independently. 
However, it is convenient to deal with these configurations separately when we compute transition amplitudes. 

In order to define the transition amplitude on $\Sigma$, we first define the vector spaces $Z_T(\partial_{\mathrm{in}} \Sigma)$ and $Z_T(\partial_{\mathrm{out}} \Sigma)$ on the in-boundary and the out-boundary.
To this end, we fix the order of the connected components of $\partial_{a} \Sigma$ for $a = \mathrm{in}, \mathrm{out}$.
The $i$th component of $\partial_{a} \Sigma$ is denoted by $(\partial_{a} \Sigma)_i$.
The vector space $Z_T(\partial_{a} \Sigma)$ is given by the tensor product of the vector spaces $Z_T((\partial_a \Sigma)_i)$ on the connected components $(\partial_a \Sigma)_i$, namely we have
\begin{equation}
Z_T(\partial_{a} \Sigma) = Z_T((\partial_a \Sigma)_1) \otimes Z_T((\partial_a \Sigma)_2) \otimes \cdots \otimes Z_T((\partial_a \Sigma)_{n_a}),
\label{eq: vector space on disjoint union}
\end{equation}
where $n_a$ denotes the number of connected components of $\partial_a \Sigma$.
The vector space $Z_T((\partial_a \Sigma)_i)$ is defined as the tensor product of vector spaces $X^a_e$ assigned to edges $e \in (\partial_{a} \Sigma)_i$:\footnote{The state space $Z((\partial_a \Sigma)_i)$ of the TQFT is obtained by restricting the vector space \eqref{eq: vector space on connected component} to the image of the transition amplitude \eqref{eq: transition amplitude} on a cylinder $(\partial_a \Sigma)_i \times [0, 1]$.}
\begin{equation}
Z_T((\partial_a \Sigma)_i) = \bigotimes_{e \in (\partial_a \Sigma)_i} X_e^a.
\label{eq: vector space on connected component}
\end{equation}
The order of the tensor product on the right-hand side is determined by the order of boundary edges $e \in (\partial_a \Sigma)_i$ induced by the edge orientations on the boundary $(\partial_a \Sigma)_i$, where the first edge is the one whose initial vertex is the marked vertex on $(\partial_a \Sigma)_i$.
Here, we recall that all edges on a boundary $(\partial_a \Sigma)_i$ are oriented in the same direction, which enables us to define the ordering of boundary edges.
The vector space $X_e^a$ assigned to an edge $e$ is a $(K, K)$-superbimodule that labels the topological defect intersecting the edge $e$.
More specifically, when a topological defect labeled by $X_e$ is oriented from the right of $e$ to the left of $e$, we assign the vector space $X_e$ to the edge $e$, i.e. $X_e^a = X_e$.
On the other hand, when a topological defect labeled by $X_e$ is oriented from the left of $e$ to the right of $e$, we assign the dual vector space $X_e^*$ to the edge $e$, i.e. $X_e^a = X_e^*$.
When a boundary edge $e$ does not intersect a topological defect, we assign a regular $(K, K)$-superbimodule $K$ to $e$, i.e. $X_e^a = K$.

Let us now define the transition amplitude $Z_T(\Sigma): Z_T(\partial_{\mathrm{in}} \Sigma) \rightarrow Z_T(\partial_{\mathrm{out}} \Sigma)$ on a triangulated spin surface $\Sigma$ with defects.
We first consider the case where all trivalent junctions on $\Sigma$ are bosonic. 
In this case, the transition amplitude is defined as the composition\footnote{Precisely, as mentioned in section \ref{sec: State sum construction of fermionic TQFTs as fermionization}, the transition amplitude $Z(\Sigma)$ of the TQFT is obtained by restricting the domain and codomain of eq. \eqref{eq: transition amplitude} to the images of the cylinder amplitudes.}
\begin{equation}
Z_T(\Sigma) = E(\Sigma) \circ c(\Sigma) \circ P(\Sigma): Z_T(\partial_{\mathrm{in}} \Sigma) \rightarrow Z_T(\partial_{\mathrm{out}} \Sigma),
\label{eq: transition amplitude}
\end{equation}
where the linear maps $P(\Sigma)$, $c(\Sigma)$, and $E(\Sigma)$ are defined below.
Although each of the linear maps $P(\Sigma)$, $c(\Sigma)$, and $E(\Sigma)$ depends on the orders of edges and triangles, their composition $Z_T(\Sigma)$ does not depend on these data.
Thus, we fix orders of edges and triangles arbitrarily in what follows.

\paragraph{The linear map $P(\Sigma): Z_T(\partial_{\mathrm{in}} \Sigma) \rightarrow Z_T(\partial_{\mathrm{in}} \Sigma) \otimes (\bigotimes_{e \in \Sigma \setminus \partial_{\mathrm{in}} \Sigma} X_e^L \otimes X_e^R)$.}
The vector spaces $X_e^L$ and $X_e^R$ are $(K, K)$-superbimodules assigned to the left and the right of an edge $e \in \Sigma \setminus \partial_{\mathrm{in}} \Sigma $.
Specifically, when a topological defect labeled by $X_e$ goes across $e$ from the right, we set $X_e^L = X_e$ and $X_e^R = X_e^*$.
On the other hand, when a topological defect labeled by $X_e$ goes across $e$ from the left, we set $X_e^L = X_e^*$ and $X_e^R = X_e$.
When $e$ does not intersect a topological defect, we assign a regular $(K, K)$-superbimodule $K$ to both left and right of $e$ because we have an even isomorphism \eqref{eq: Phi} between $K$ and $K^*$.
The linear map $P(\Sigma)$ is given by the tensor product 
\begin{equation}
P(\Sigma) = (\bigotimes_{e \in \partial_{\mathrm{in}} \Sigma} p_{X_e}^{s(e) + 1}) \otimes (\bigotimes_{e \in \Sigma \setminus \partial_{\mathrm{in}} \Sigma} (\mathrm{id}_{X_e^L} \otimes p_{X_e^R}^{s(e) + 1}) \circ P_e), 
\label{eq: P}
\end{equation}
where $p_{X}: X \rightarrow X$ is the $\mathbb{Z}_2$-grading automorphism of a $(K, K)$-superbimodule $X$ and $P_e: \mathbb{C} \rightarrow X_e^L \otimes X_e^R$ is either the left coevaluation morphism $\mathrm{coev}_{X_e}^L$ or the right coevaluation morphism $\mathrm{coev}_{X_e}^R$ depending on the orientation of the topological defect $X_e$.
Specifically, we define $P_e = \mathrm{coev}_{X_e}^L$ when $X_e^L = X_e$ and $X_e^R = X_e^*$, whereas we define $P_e = \mathrm{coev}_{X_e}^R$ when $X_e^L = X_e^*$ and $X_e^R = X_e$.
When $e$ does not intersect a topological defect, the linear map $P_e$ reduces to $\Delta_K \circ \eta_K: \mathbb{C} \rightarrow K \otimes K$, where $\Delta_K: K \rightarrow K \otimes K$ and $\eta_K: \mathbb{C} \rightarrow K$ are the comultiplication \eqref{eq: comultiplication on K} and the unit of $K$.
In the following, we will omit the subscript $X$ of the $\mathbb{Z}_2$-grading automorphism $p_X$ when it is clear from the context.

\paragraph{The linear map $c(\Sigma): Z_T(\partial_{\mathrm{in}} \Sigma) \otimes (\bigotimes_{e \in \Sigma \setminus \partial_{\mathrm{in}} \Sigma} X_e^L \otimes X_e^R) \rightarrow (\bigotimes_{t \in \Sigma} X_t^1 \otimes X_t^2 \otimes X_t^3) \otimes Z_T(\partial_{\mathrm{out}} \Sigma)$.}
The vector space $X_t^i$ is a $(K, K)$-superbimodule assigned to the $i$th edge on $\partial t$, where the order of the edges on $\partial t$ is given by the counterclockwise order starting from the marked edge.
More specifically, when a triangle $t$ is on the left side of the $i$th edge $e \in \partial t$, we define $X_t^i = X_e^L$.
On the other hand, when $t$ is on the right side of the $i$th edge $e \in \partial t$, we define $X_t^i = X_e^R$.
The linear map $c(\Sigma)$, which only changes the order of the tensor product, is given by the composition of the symmetric braiding $c_{\mathrm{super}}$ of super vector spaces assigned to edges.

\paragraph{The linear map $E(\Sigma): (\bigotimes_{t \in \Sigma} X_t^1 \otimes X_t^2 \otimes X_t^3) \otimes Z_T(\partial_{\mathrm{out}} \Sigma) \rightarrow Z_T(\partial_{\mathrm{out}} \Sigma)$.}
The linear map $E(\Sigma)$ is given by the tensor product of linear maps $E_t: X_t^1 \otimes X_t^2 \otimes X_t^3 \rightarrow \mathbb{C}$ for all triangles $t \in \Sigma$:
\begin{equation}
E(\Sigma) = (\bigotimes_{t \in \Sigma} E_t) \otimes \mathrm{id}_{Z_T(\partial_{\mathrm{out}} \Sigma)}.
\label{eq: E}
\end{equation}
The explicit form of $E_t$ depends on the configuration of topological defects on $t$.
Specifically, the linear map $E_t$ for each configuration in eq. \eqref{eq: configurations} is given by
\begin{align}
E_t = 
\begin{cases}
(1) & \epsilon_K \circ m_K \circ (m_K \otimes \mathrm{id}_K): K \otimes K \otimes K \rightarrow \mathbb{C}, \\
(2) & \mathrm{ev}_X^{R} \circ (\mathrm{id}_{X} \otimes \rho_{X^*}^R): X \otimes X^* \otimes K \rightarrow \mathbb{C}, \\
(3) & \mathrm{ev}_X^{R} \circ (\rho_{X}^R \otimes \mathrm{id}_{X^*}): X \otimes K \otimes X^* \rightarrow \mathbb{C}, \\
(4) & \mathrm{ev}_{X_1 \otimes X_2}^R \circ ((\iota_{X_1, X_2} \circ b_{3}^{12}) \otimes \mathrm{id}_{X_2^* \otimes X_1^*}): X_3 \otimes X_2^* \otimes X_1^* \rightarrow \mathbb{C}, \\
(5) & \mathrm{ev}_{X_3}^R \circ ((b_{12}^3 \circ \pi_{X_1, X_2}) \otimes \mathrm{id}_{X^*_3}): X_1 \otimes X_2 \otimes X_3^* \rightarrow \mathbb{C},
\end{cases}
\label{eq: Et}
\end{align}
where $\epsilon_K$ and $m_K$ are the counit \eqref{eq: counit of K} and the multiplication on $K$, $\rho_X^R: X \otimes K \rightarrow X$ is the right $K$-supermodule action on $X$, $\iota_{X_1, X_2}: X_1 \otimes_K X_2 \rightarrow X_1 \otimes X_2$ and $\pi_{X_1, X_2}: X_1 \otimes X_2 \rightarrow X_1 \otimes_K X_2$ are the splitting maps \eqref{eq: splitting maps}, and $b_{3}^{12}: X_3 \rightarrow X_1 \otimes_K X_2$ and $b_{12}^3: X_1 \otimes_K X_2 \rightarrow X_3$ are $\mathbb{Z}_2$-even $(K, K)$-superbimodule morphisms that represent bosonic trivalent junctions.
When the marked edge on a triangle $t$ is not the bottom edge in eq. \eqref{eq: configurations}, we first move the marked edge to the bottom by the local move depicted in figure \ref{fig: local moves} and define the transition amplitude as in eq. \eqref{eq: Et}.
Since applying the change of a marked edge three times gives rise to the leaf exchange, the above definition makes sense only when the transition amplitude is invariant under the leaf exchange.
The invariance under the leaf exchange is guaranteed by the assumption that the trivalent junctions $b_3^{12}$ and $b_{12}^3$ are $\mathbb{Z}_2$-even, i.e. bosonic.

The transition amplitude $Z_T(\Sigma)$ is invariant under the change of an edge orientation due to eqs. \eqref{eq: left-right relation} and \eqref{eq: left-right relation of K}.
The invariance under the other local moves in figure \ref{fig: local moves} also follows immediately from the definition of $Z_T(\Sigma)$.
Furthermore, a straightforward calculation shows that $Z_T(\Sigma)$ is invariant under the Pachner moves as well.
Thus, the transition amplitude $Z_T(\Sigma)$ is a topological invariant of a spin surface $\Sigma$ with defects.
We note that $Z_T(\Sigma)$ reduces to the transition amplitude given in \cite{NR2015} when there are no topological defects on $\Sigma$.
In particular, the transition amplitudes defined above are not identically zero.

We now incorporate fermionic trivalent junctions.
Let $\Sigma$ be a triangulated spin surface $\Sigma$ with fermionic junctions.
We fix the order of the fermionic junctions. 
The $i$th fermionic junction is denoted by $f_i$, which is a $\mathbb{Z}_2$-odd morphism of $(K, K)$-superbimodules.
In order to define the transition amplitude $Z_T(\Sigma)$, we first consider the transition amplitude on the punctured spin surface $\Sigma \setminus \bigsqcup_i D_i$, where $D_i$ is a small disk around $f_i$.
We note that the boundary of a disk $D_i$ is an $\mathrm{NS}$ circle, which we regard as an out-boundary.
Since the punctured surface $\Sigma \setminus \bigsqcup_i D_i$ only contains bosonic trivalent junctions, the transition amplitude on $\Sigma \setminus \bigsqcup_i D_i$ is defined by eq. \eqref{eq: transition amplitude}:
\begin{equation}
Z_T(\Sigma \setminus \bigsqcup_i D_i): Z_T(\partial_{\mathrm{in}} \Sigma) \rightarrow (\bigotimes_i Z_T(\partial D_i)) \otimes Z_T(\partial_{\mathrm{out}} \Sigma).
\end{equation}
The order of the tensor product on the right-hand side is determined by the order of fermionic junctions.
If we choose a triangulation of $\Sigma \setminus \bigsqcup_i D_i$ so that the boundary of each disk $D_i$ consists of three edges, the topological defect network on a disk $D_i$ looks like configurations (4) or (5) in eq. \eqref{eq: configurations}, where the edges of a triangle are boundary edges on $\partial D_i$ and the left bottom vertex is chosen as the marked vertex.
With this choice of a triangulation, we define the transition amplitude $Z_T(\Sigma)$ as follows:
\begin{equation}
Z_T(\Sigma) = ((\bigotimes_i E_{D_i}) \otimes \mathrm{id}_{Z_T(\partial_{\mathrm{out}} \Sigma)}) \circ Z_T(\Sigma \setminus \bigsqcup_i D_i).
\label{eq: fermionic transition amplitude}
\end{equation}
Here, the linear map $E_{D_i}$ is defined by the last two equations of eq. \eqref{eq: Et} in which bosonic junctions $b_3^{12}$ and $b_{12}^3$ are replaced by a fermionic junction $f_i$.
The topological invariance of the transition amplitude \eqref{eq: fermionic transition amplitude} follows from the topological invariance of the transition amplitude on the punctured surface.
We note that $Z_T(\Sigma)$ depends on the order of fermionic junctions because of the anti-commutation relation \eqref{eq: anti-commutation}.
We also note that $Z_T(\Sigma)$ depends on a spin structure on the punctured surface $\Sigma \setminus \bigsqcup_i D_i$ rather than a spin structure on $\Sigma$.
More specifically, the transition amplitude acquires an extra minus sign if we apply the leaf exchange on a disk $D_i$, which does not change a spin structure on $\Sigma$ but changes a spin structure on the punctured surface $\Sigma \setminus \bigsqcup_i D_i$.
This corresponds to the fact that the leaf exchange on $D_i$ is equivalent to winding a fermion parity defect around a fermionic junction $f_i$.

In this way, we can construct fermionic quantum field theories on spin surfaces with defects.
These are fermionic TQFTs with ${}_K \mathcal{SM}_K$ symmetry because the transition amplitudes are independent of triangulations of spin surfaces and invariant under continuous deformations of defects labeled by $(K, K)$-superbimodules.
It would be possible to formulate these TQFTs as some sort of superfunctors from the supercategory of spin manifolds with defects to the supercategory of super vector spaces.
Rigorously formulating fermionic TQFTs with superfusion category symmetries from this point of view is beyond the scope of this manuscript.

Before proceeding, we note that the above definition of the transition amplitudes can be generalized to spin surfaces with interfaces between different state sum TQFTs.
In general, a topological interface between state sum TQFTs $\mathcal{T}_f^K$ and $\mathcal{T}_f^{K^{\prime}}$ is labeled by a $(K, K^{\prime})$-superbimodule.
In particular, a topological defect labeled by a $(K, K)$-superbimodule can be thought of as a topological interface between the same state sum TQFTs $\mathcal{T}_f^K$.
The transition amplitudes on spin surfaces with interfaces are defined just by replacing $(K, K)$-superbimodules by $(K, K^{\prime})$-superbimodules in the above definition.

\subsection{Action of ${}_K \mathcal{SM}_K$ symmetry}
\label{sec: Action of symmetry}
Let $X$ be a simple object of ${}_K \mathcal{SM}_K$.
The action of $X$ on the state space on a circle $S^1_{\lambda}$ is defined by the transition amplitude on a cylinder $S^1_{\lambda} \times [0, 1]$ with a topological defect $X$ wrapping around a spatial cycle, where $S^1_{\lambda = 0}$ and $S^1_{\lambda = 1}$ are $\mathrm{NS}$ and $\mathrm{R}$ circles respectively.
When $X \in {}_K \mathcal{SM}_K$ is a q-type object, we can modify this action by putting a fermionic point-like defect $f$ on $X$. 
The above unmodified and modified actions are denoted by $U_{X; \mathrm{id}}^{\lambda}$ and $U_{X; f}^{\lambda}$ respectively.
In this subsection, we will explicitly compute these actions.

\begin{figure}
\begin{center}
\includegraphics[width = 5cm]{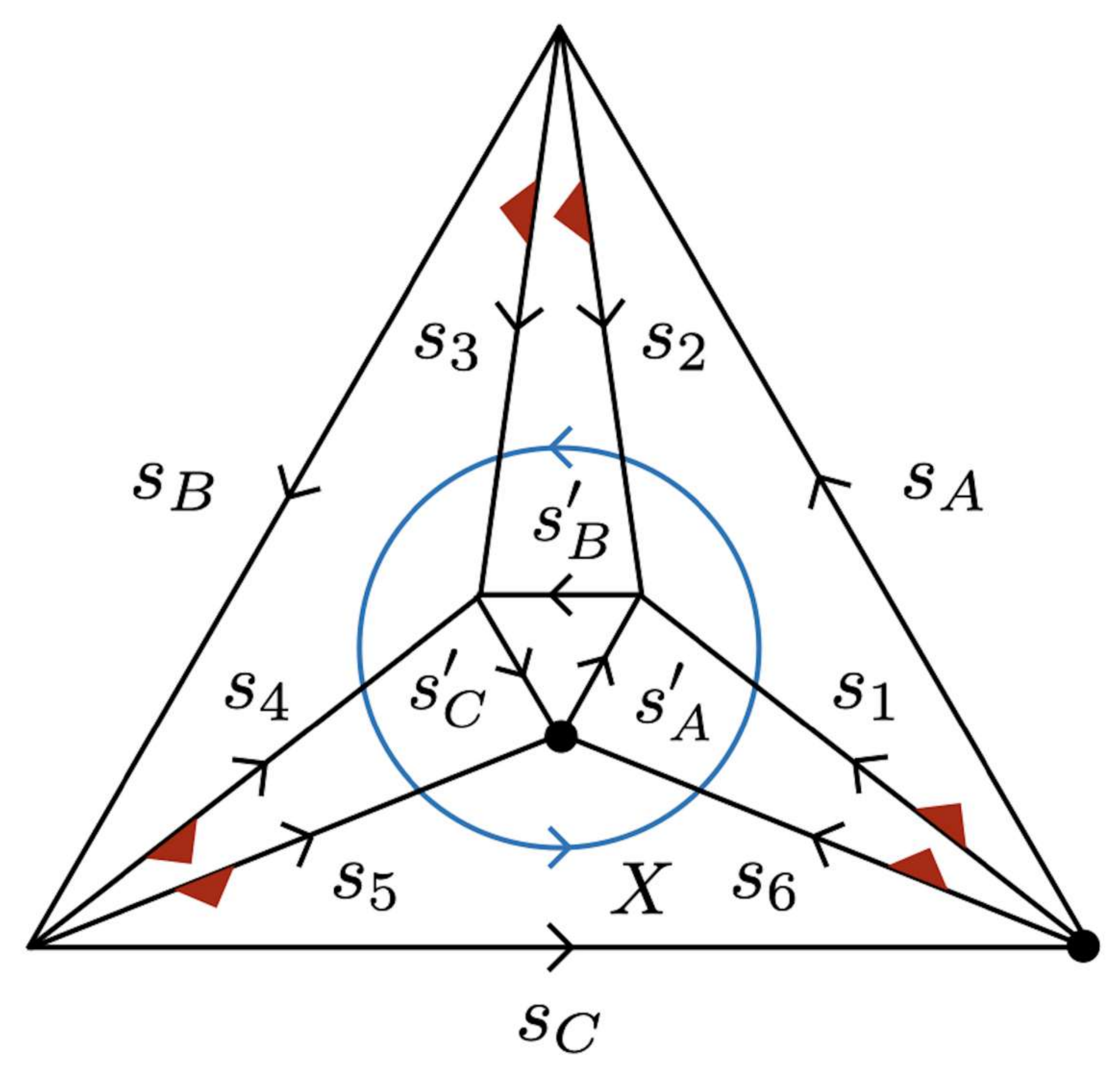}
\end{center}
\caption{A triangulation of a cylinder with topological defect $X$ wrapping around a spatial cycle. The outer triangle is the in-boundary, and the inner triangle is the out-boundary. The black dots represent marked vertices. The edge indices are determined by the admissibility condition uniquely up to local moves as eq. \eqref{cylinder indices}.}
\label{fig: spin triangulation of a cylinder}
\end{figure}

Let us first compute the unmodified action $U_{X; \mathrm{id}}^{\lambda}$.
We can choose a triangulation of a cylinder as shown in figure \ref{fig: spin triangulation of a cylinder}.
The admissibility condition \eqref{eq: admissibility} determines the edge indices up to local moves as
\begin{equation}
s_1 = s_2 = s_3 = s_4 = s_5 = s^{\prime}_A = s^{\prime}_B = s^{\prime}_C = 0, \quad s_6 = 1+\lambda, \quad s_A = s_B = s_C =: s.
\label{cylinder indices}
\end{equation}
The two solutions labeled by $s \in \{0, 1\}$ correspond to spin cylinders with and without a sheet exchange.
In order to identify the solution corresponding to a spin cylinder without a sheet exchange, we consider the transition amplitude on a cylinder without a topological defect, or equivalently the action of a trivial defect $X = K$.
This cylinder amplitude is denoted by $P_{\lambda}^s$.
The string diagram representation of the linear map $P_{\lambda}^s$ is given by \cite{NR2015}
\begin{equation}
P_{\lambda}^s ~ = ~ \adjincludegraphics[valign = c, width = 4.0cm]{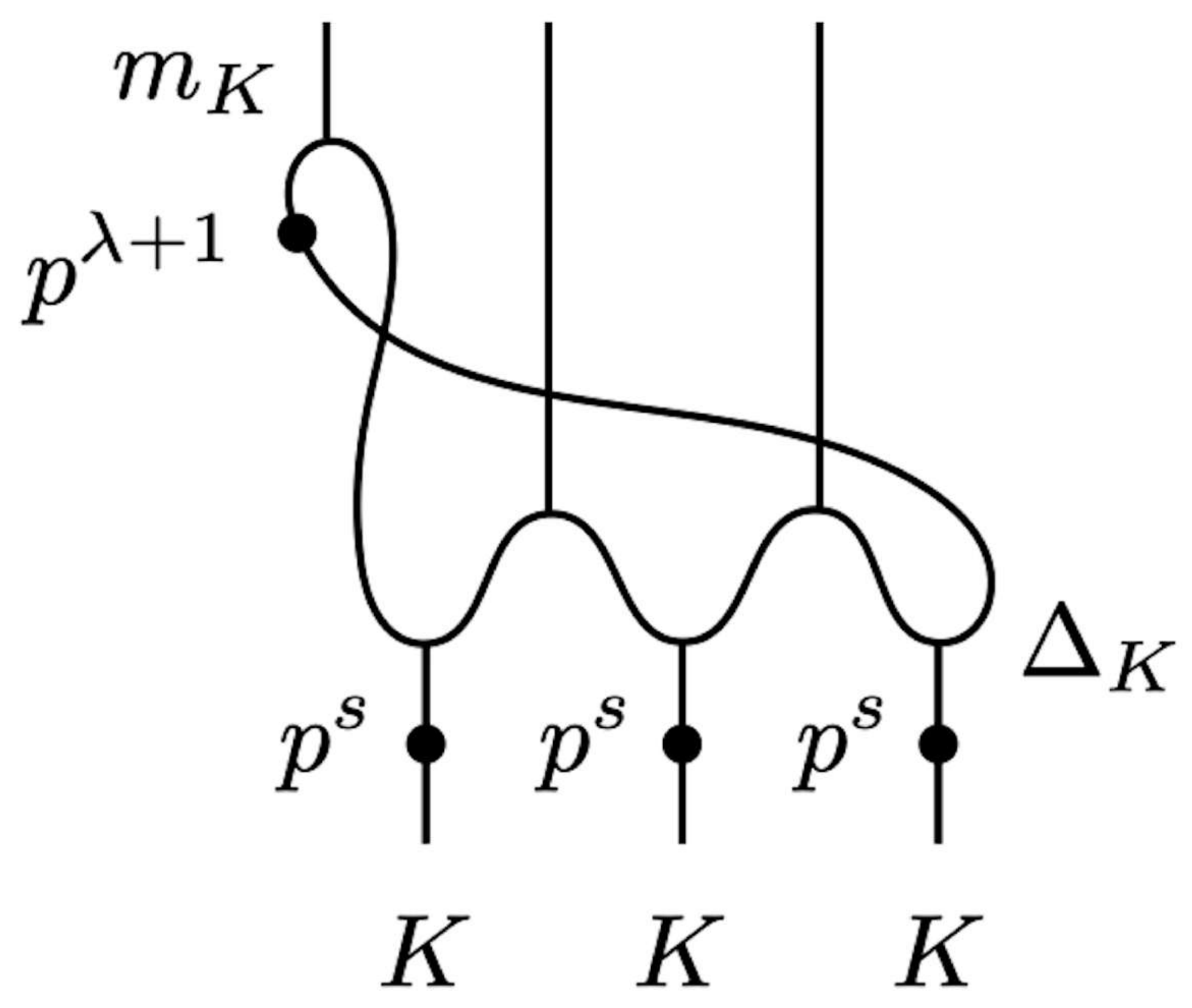},
\label{eq: cylinder projector}
\end{equation}
where the trivalent junctions represent the multiplication $m_K$ and the comultiplication $\Delta_K$, and the braiding of two strands represents the symmetric braiding isomorphism $c_{\mathrm{super}}$.\footnote{The braiding in the string diagrams appearing in the rest of the paper is always the symmetric braiding $c_{\mathrm{super}}$.}
We note that the superscript $s$ is additive under the composition of linear maps, i.e. we have $P_{\lambda}^{s_2} \circ P_{\lambda}^{s_1} = P_{\lambda}^{s_1 + s_2}$.
In particular,  the linear map $P_{\lambda}^{0}$ is an idempotent, which indicates that the solution $s = 0$ corresponds to a spin cylinder without a sheet exchange.
We can compute the unmodified action $U_{X; \mathrm{id}}^{\lambda}$ by wrapping a topological defect $X \in {}_K \mathcal{SM}_K$ around a spatial cycle of a spin cylinder without a sheet exchange.
The linear map $U_{X; \mathrm{id}}^{\lambda}$ is expressed by the following string diagram:
\begin{equation}
U_{X; \mathrm{id}}^{\lambda} ~ = ~ \adjincludegraphics[valign = c, width = 3.75cm]{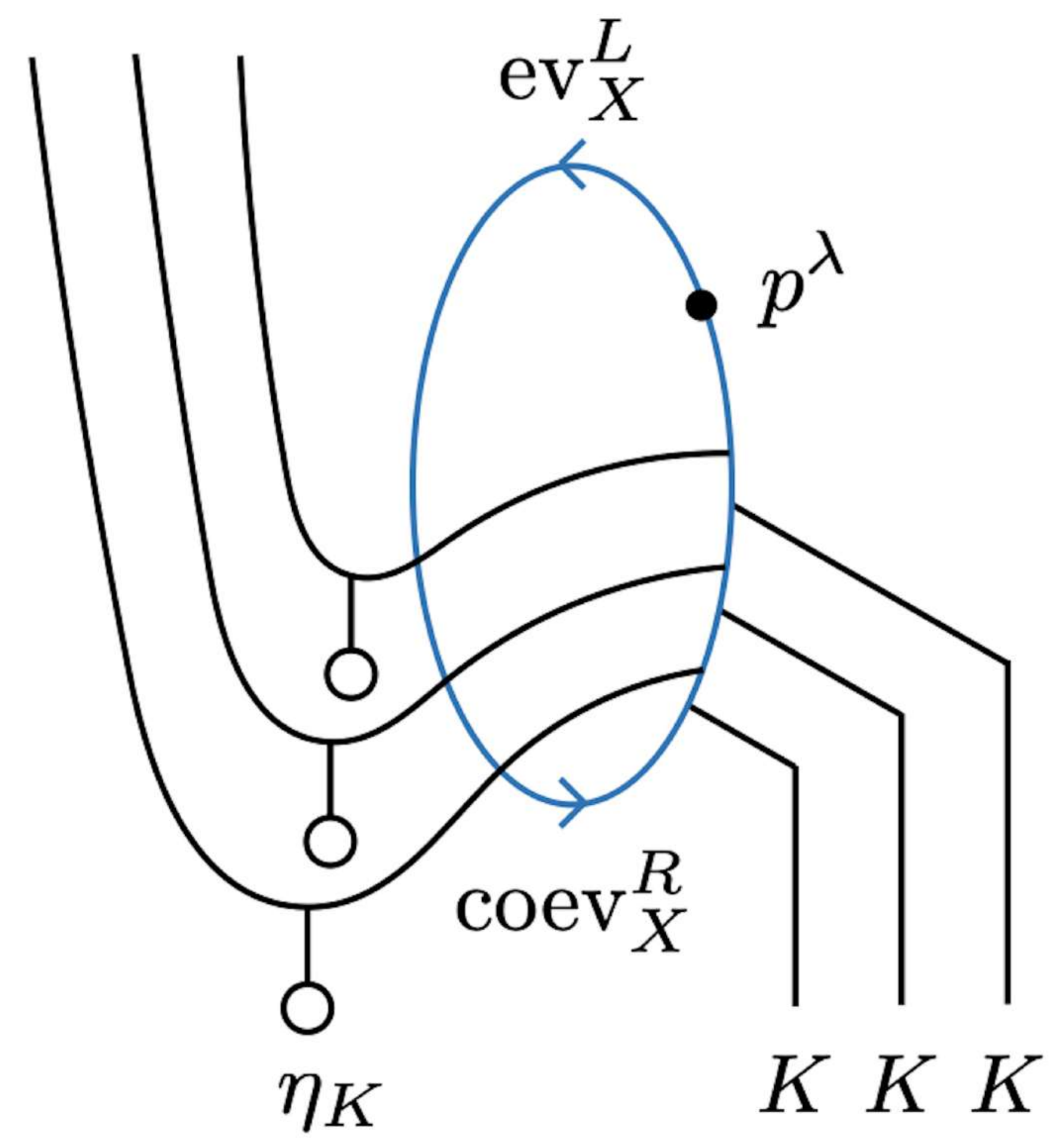} ~ = ~ \adjincludegraphics[valign = c, width = 3.5cm]{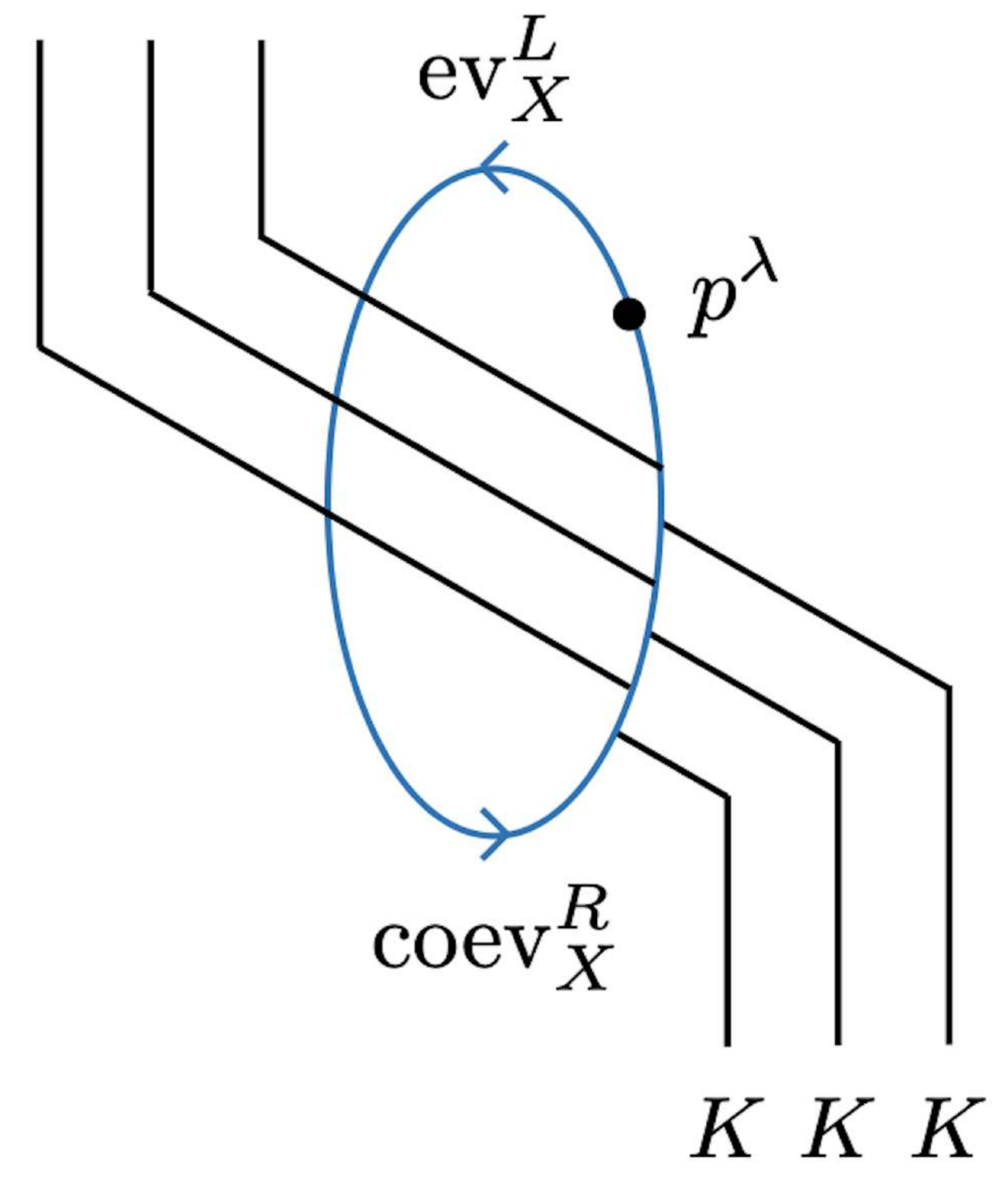}.
\label{eq: action of X}
\end{equation}
The trivalent junctions in the left diagram represent the left and right $K$-supermodule actions on $X$.
The right-hand side is a string diagram written in terms of the left $K$-supercomodule action $\delta^L_X: X \rightarrow K \otimes X$ defined by $\delta^L_X = (\mathrm{id}_K \otimes \rho^L_X) \circ ((\Delta_K \circ \eta_K) \otimes \mathrm{id}_X)$, where $\rho^L_X: K \otimes X \rightarrow X$ is the left $K$-supermodule action on $X$.
We note that eq. \eqref{eq: action of X} reduces to the cylinder amplitude $P_{\lambda}^0$ when $X$ is a trivial defect $K$.

For any $(K, K)$-superbimodule $X$, we can define another $(K, K)$-superbimodule ${}_p X$ by twisting the left $K$-action on $X$ by the $\mathbb{Z}_2$ grading automorphism $p$ of $K$.
More specifically, the left $K$-action on ${}_p X$ is given by $\rho_{{}_p X}^L(a \otimes x) := \rho_{X}^L(p(a) \otimes x)$ for $a \in K$ and $x \in X$.
It follows from eq. \eqref{eq: action of X} that the action of ${}_p X$ is related to that of $X$ as 
\begin{equation}
U_{{}_p X; \mathrm{id}}^{\lambda} = (-1)^F \circ U_{X; \mathrm{id}}^{\lambda} = U_{X; \mathrm{id}}^{\lambda} \circ (-1)^F.
\label{eq: action of pX}
\end{equation}
Here, $(-1)^F := P_{\lambda}^{1}$ is the transition amplitude on a spin cylinder with a sheet exchange, which physically describes the action of the fermion parity symmetry.
When $X$ is a trivial defect $K$, eq. \eqref{eq: action of pX} reduces to $U_{{}_p K; \mathrm{id}}^{\lambda} = (-1)^F$.
This implies that ${}_p K$ is the generator of the fermion parity symmetry $\mathbb{Z}_2^F$.
Equation \eqref{eq: action of pX} shows that the fermion parity symmetry is central, i.e., it commutes with the unmodified action $U_{X; \mathrm{id}}^{\lambda}$ for any $X \in {}_K \mathcal{SM}_K$.

By equipping ${}_p X$ with the opposite $\mathbb{Z}_2$-grading, we can define another $(K, K)$-superbimodule $\Pi X$, which is oddly isomorphic to $X$.
The $(K, K)$-superbimodule isomorphism $\zeta_X: \Pi X \rightarrow X$ is given by the identity map of the underlying vector space.
Equations \eqref{eq: action of X} and \eqref{eq: (co)evPix} imply that the action of $\Pi X$ is related to that of X as follows:
\begin{equation}
U_{\Pi X; \mathrm{id}}^{\lambda} = (-1)^{\lambda} U_{X; \mathrm{id}}^{\lambda}.
\label{eq: action of PiX}
\end{equation}
In particular, when $X$ is a trivial defect $K$, the oddly isomorphic defect $\Pi K$ acts as $P_{\lambda}^0$ on the $\mathrm{NS}$ sector and $-P_{\lambda}^0$ on the $\mathrm{R}$ sector.
This suggests that the topological defect $\Pi K$ detects the spin structure on the circle that it winds around.
Equation \eqref{eq: action of PiX} implies that the action of a q-type object $X$ on the $\mathrm{R}$ sector vanishes because $U_{X; \mathrm{id}}^{\lambda} = U_{\Pi X; \mathrm{id}}^{\lambda} = (-1)^{\lambda} U_{X; \mathrm{id}}^{\lambda}$, where the first equality follows from the existence of an even isomorphism $\Pi X \cong X$ for a q-type object $X$.
This particularly means that the torus partition function vanishes if a q-type object $X$ is winding around a spatial cycle equipped with an $\mathrm{R}$ spin structure.
Thus, due to the modular invariance, the $X$-twisted sector has the same number of bosonic states and fermionic states when $X \in {}_K \mathcal{SM}_K$ is a q-type object.
This can also be seen from the fact that the $X$-twisted sector has an odd automorphism when $X$ is a q-type object.

We next compute the modified action $U_{X; f}^{\lambda}$ of a q-type object $X$.
Since we have a fermionic point-like defect $f$ on $X$, the transition amplitude on a cylinder with $X$ wrapping around a spatial cycle depends on the spin structure on the punctured cylinder.
We can choose a spin structure so that the action $U_{X; f}^{\lambda}$ is represented by the string diagram in eq. \eqref{eq: action of X} where $p^{\lambda}$ is replaced by $f \circ p^{\lambda}$.
This modified action $U_{X; f}^{\lambda}$ satisfies $U_{X; f}^{\lambda} = (-1)^{\lambda + 1} U_{X; f}^{\lambda}$ because moving a fermionic point-like defect $f$ around a topological defect $X$ by using eq. \eqref{eq: dual of a morphism} produces a sign $(-1)^{\lambda + 1}$.
Therefore, the modified action $U_{X; f}^{\lambda}$ on the $\mathrm{NS}$ sector vanishes in contrast to the fact that the unmodified action $U_{X; \mathrm{id}}^{\lambda}$ vanishes on the $\mathrm{R}$ sector.
Thus, we find that the action of a q-type object is given by $U_{X; \mathrm{id}}^{\lambda}$ on the $\mathrm{NS}$ sector and $U_{X; f}^{\lambda}$ on the $\mathrm{R}$ sector.
We note that the action of a q-type object on the $\mathrm{R}$ sector anti-commutes with the fermion parity symmetry,\footnote{The anti-commutation relation between the fermion parity defect and a q-type object is observed in the example of massless Majorana fermions \cite{DGG2021, Thorngren2020, KTT2019}.} i.e.
\begin{equation}
U_{X; f}^{\lambda} \circ (-1)^F = - (-1)^F \circ U_{X; f}^{\lambda},
\end{equation}
which implies that the $\mathrm{R}$ sector is at least two-fold degenerate if the superfusion category symmetry ${}_K \mathcal{SM}_K$ has a q-type object whose action on the $\mathrm{R}$ sector is non-zero.
However, we emphasize that the existence of a q-type object in a general superfusion category symmetry does not imply degenerate ground states.
Indeed, the fermionization $\mathrm{SRep}(\mathcal{H}^u_8)$ of a non-anomalous fusion category symmetry $\mathrm{Rep}(H_8)$ is an example of a superfusion category symmetry with q-type objects that admits a non-degenerate ground state.

Let us apply the above linear maps $U_{X; \mathrm{id}}^{\lambda}$ and $U_{X; f}^{\lambda}$ to the ground states of TQFT $\mathcal{T}_f^K$.
We first write down the ground states of $\mathcal{T}_f^K$, which are in one-to-one correspondence with simple left $K$-supermodules \cite{KTY2018}.
The ground state corresponding to a simple left $K$-supermodule $M$ is obtained by evaluating the transition amplitude on a cylinder with a topological boundary condition $M$ imposed on one end of the cylinder.
The other end of the cylinder is regarded as an out-boundary.
Since a topological boundary is an interface between the state sum TQFT $\mathcal{T}_f^K$ and the trivial TQFT, this transition amplitude is a linear map from the state space of the trivial TQFT to the state space of $\mathcal{T}_f^K$.
This linear map can be canonically identified with a state of $\mathcal{T}_f^K$ because the state space of the trivial TQFT is $\mathbb{C}$.
The state obtained by this canonical identification is denoted by $\ket{M; \mathrm{id}}_{\lambda}$, where $\lambda$ labels the spin structure on a circle.
When a simple $K$-supermodule $M$ has an odd automorphism $f: M \rightarrow M$, which is unique up to rescaling, we can modify the boundary condition by putting the fermionic point-like defect $f$ on the topological boundary $M$.
The state corresponding to this modified boundary condition is denoted by $\ket{M; f}_{\lambda}$.
The string diagram representations of these states are given as follows:
\begin{equation}
\ket{M; \beta}_{\lambda} = \adjincludegraphics[valign = c, width = 3cm]{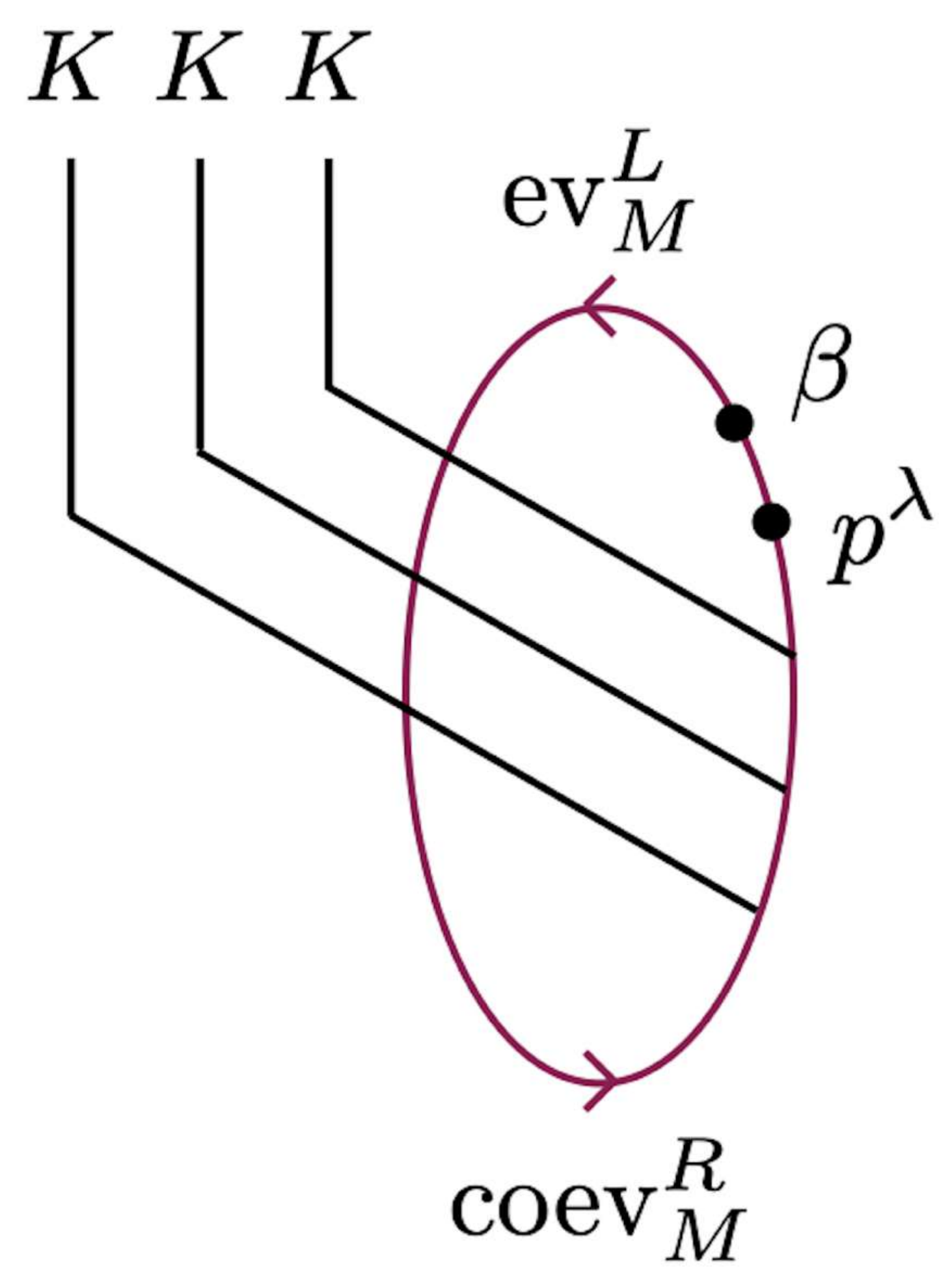}, 
\label{eq: ground states}
\end{equation}
where $\beta: M \rightarrow M$ is either the identity morphism $\mathrm{id}$ or the fermionic point-like defect $f$.
The trivalent junctions in the above diagram represent the left $K$-supercomodule action on $M$, which is defined in the same way as the second equality in eq. \eqref{eq: action of X}.
We note that the fermion parity of $\ket{M; \beta}_{\lambda}$ is given by the $\mathbb{Z}_2$-grading of $\beta$, i.e. we have $(-1)^F \ket{M; \beta}_{\lambda} = (-1)^{|\beta|} \ket{M; \beta}_{\lambda}$.
When $M$ has an odd automorphism, the $\mathrm{NS}$ sector state $\ket{M; f}_{\lambda = 0}$ and the $\mathrm{R}$ sector state $\ket{M; \mathrm{id}}_{\lambda = 1}$ vanish due to the equalities $\ket{M; f}_{\lambda} = (-1)^{\lambda + 1} \ket{M; f}_{\lambda}$ and $\ket{M; \mathrm{id}}_{\lambda} = (-1)^{\lambda} \ket{M; \mathrm{id}}_{\lambda}$.\footnote{In particular, the $\mathrm{NS}$ sector does not have fermionic states \cite{MS2006, KTY2018}.}
Therefore, states in the $\mathrm{NS}$ and $\mathrm{R}$ sectors are both in one-to-one correspondence with simple $K$-supermodules.
Although we can also define a state $\ket{N; \beta}$ for a non-simple $K$-supermodule $N$ and a general supermodule morphism $\beta: N \rightarrow N$ by replacing $M$ by $N$ in eq. \eqref{eq: ground states}, such a state is not linearly independent of the ground states labeled by simple $K$-supermodules.

The action of a simple topological defect $X \in {}_K \mathcal{SM}_K$ on the ground states \eqref{eq: ground states} can be computed as follows:
\begin{equation}
U_{X; \alpha}^{\lambda} \ket{M; \beta}_{\lambda} = \ket{X \otimes_K M; \alpha \otimes_K \beta}_{\lambda}.
\label{eq: action of X on M}
\end{equation}
Here, $\alpha: X \rightarrow X$ is a point-like defect on $X$ and $\beta: M \rightarrow M$ is a point-like defect on $M$.
Since $X$ and $M$ are a simple $(K, K)$-superbimodule and a simple $K$-supermodule respectively, the point-like defects $\alpha$ and $\beta$ are either the identity morphism or an odd automorphism $f$ unique up to scalar multiplication.
The expression \eqref{eq: action of X on M} is also valid for non-simple topological defects $X \in {}_K \mathcal{SM}_K$ and a general point-like defect $\alpha: X \rightarrow X$.

\subsection{Fermionic TQFTs with $\mathrm{SRep}(\mathcal{H}^u)$ symmetry}
\label{sec: Fermionic TQFTs with SRep symmetry}
In this subsection, we show that the fermionization of a bosonic state sum TQFT with $\mathrm{Rep}(H)$ symmetry is a fermionic state sum TQFT with $\mathrm{SRep}(\mathcal{H}^u)$ symmetry, where $H$ is a weak Hopf algebra and $\mathcal{H}^u$ is a weak Hopf superalgebra defined in section \ref{sec: Fermionization of fusion category symmetries}.
This verifies the fermionization formula of fusion category symmetries proposed in section \ref{sec: Fermionization of fusion category symmetries}.

We first recall the state sum construction of bosonic TQFTs with $\mathrm{Rep}(H)$ symmetry.
As shown in \cite{DKR2011}, the bosonic state sum TQFT $\mathcal{T}_b^K$ constructed from a semisimple algebra $K$ has a (multi-)fusion category symmetry ${}_K \mathcal{M}_K$ described by the category of $(K, K)$-bimodules.
When the input algebra $K$ is a left $H$-comodule algebra, the ${}_K \mathcal{M}_K$ symmetry can be pulled back to $\mathrm{Rep}(H)$ by a tensor functor $F_K: \mathrm{Rep}(H) \rightarrow {}_K \mathcal{M}_K$ \cite{Inamura2022}. 
Since the pullback of a fusion category symmetry leaves the underlying TQFT unchanged, the state sum TQFT $\mathcal{T}_b^K$ has $\mathrm{Rep}(H)$ symmetry if $K$ is a left $H$-comodule algebra.
We note that any indecomposable semisimple bosonic TQFTs with $\mathrm{Rep}(H)$ symmetry can be constructed in this way \cite{Inamura2022}.\footnote{Semisimple 2d TQFTs include unitary 2d TQFTs \cite{KTY2017, HLS2021}.}

In order to fermionize a bosonic TQFT $\mathcal{T}_b^K$ with $\mathrm{Rep}(H)$ symmetry, we specify a non-anomalous $\mathbb{Z}_2$ subgroup that we use for the fermionization. 
A bosonic TQFT $\mathcal{T}_b^K$ with $\mathrm{Rep}(H)$ symmetry has a non-anomalous $\mathbb{Z}_2$ subgroup symmetry when the input algebra $K$ is equipped with a $\mathbb{Z}_2$-grading $p: K \rightarrow K$ \cite{KTY2017, SR2017}.
The generator of this $\mathbb{Z}_2$ subgroup is a representation $V \in \mathrm{Rep}(H)$ such that $F_K(V) \cong {}_p K$, namely the generator $V$ is the pullback of a $(K, K)$-bimodule ${}_p K \in {}_K \mathcal{M}_K$.
On the other hand, as we discussed in section \ref{sec: Fermionization of fusion category symmetries}, a fusion category symmetry $\mathrm{Rep}(H)$ has a non-anomalous $\mathbb{Z}^u_2$ subgroup generated by $V_u \in \mathrm{Rep}(H)$ if the dual weak Hopf algebra $H^*$ has a $\mathbb{Z}_2$ group-like element $u \in H^*$.
This $\mathbb{Z}_2^u$ subgroup is associated with the following $\mathbb{Z}_2$-grading $p$ on the input left $H$-comodule algebra $K$:
\begin{equation}
p = (u \otimes \mathrm{id}_K) \circ \delta_K^H,
\label{eq: grading on H-comodule algebra K}
\end{equation}
where $\delta_K^H: K \rightarrow H \otimes K$ is the $H$-comodule action on $K$.
Indeed, the generator $V_u \in \mathrm{Rep}(H)$ can be viewed as the pullback of a $(K, K)$-bimodule ${}_p K \in {}_K \mathcal{M}_K$ because we have a $(K, K)$-bimodule isomorphism $\epsilon \overline{\otimes} \mathrm{id}_K: F_K(V_u) = V_u \overline{\otimes} K \rightarrow {}_p K$.
A semisimple algebra $K$ equipped with the above $\mathbb{Z}_2$-grading is denoted by $K^u$, which is a semisimple superalgebra because the multiplication and unit of $K$ are even with respect to the $\mathbb{Z}_2$-grading \eqref{eq: grading on H-comodule algebra K}.
The fermionization of a bosonic TQFT $\mathcal{T}_b^K$ with respect to $\mathbb{Z}_2^u$ subgroup symmetry is a fermionic TQFT $\mathcal{T}_f^{K^u}$ as we discussed at the end of section \ref{sec: State sum construction of fermionic TQFTs as fermionization}.

The symmetry of $\mathcal{T}_f^{K^u}$ can be understood in terms of pullback just as in the bosonic case.
When $K$ is a left $H$-comodule algebra, as we will see in appendix \ref{sec: supercomodule algebra}, a semisimple superalgebra $K^u$ becomes a left $\mathcal{H}^u$-supercomodule algebra, where the left $\mathcal{H}^u$-supercomodule action on $K^u$ is given by $\delta_K^H$.
Therefore, we have a supertensor functor $\mathcal{F}_{K^u}: \mathrm{SRep}(\mathcal{H}^u) \rightarrow {}_{K^u} \mathcal{SM}_{K^u}$, which enables us to pull back the ${}_{K^u} \mathcal{SM}_{K^u}$ symmetry that we discussed in detail in section \ref{sec: Fermionic TQFTs with KSMK symmetry}.
Since the pullback of the symmetry leaves the underlying TQFT unaffected, the fermionic TQFT $\mathcal{T}_f^{K^u}$ has a superfusion category symmetry $\mathrm{SRep}(\mathcal{H}^u)$.
This shows the fermionization formula of fusion category symmetries.

Let us explicitly compute the action of a topological defect $V \in \mathrm{SRep}(\mathcal{H}^u)$ on the state space on a circle $S^1_{\lambda}$.
The action of $V$ is denoted by $\mathcal{U}_{V; \alpha}^{\lambda}$, where $\alpha: V \rightarrow V$ is a point-like defect on $V$.
Since the $\mathrm{SRep}(\mathcal{H}^u)$ symmetry of a fermionic TQFT $\mathcal{T}_f^{K^u}$ is obtained as the pullback of the ${}_{K^u} \mathcal{SM}_{K^u}$ symmetry by a supertensor functor $\mathcal{F}_{K^u}$, the action of $V \in \mathrm{SRep}(\mathcal{H}^u)$ reduces to the action of $\mathcal{F}_{K^u}(V) \in {}_{K^u} \mathcal{SM}_{K^u}$.
Therefore, a topological defect $V$ acts on the ground states \eqref{eq: ground states} as
\begin{equation}
\mathcal{U}_{V; \alpha}^{\lambda} \ket{M; \beta}_{\lambda} = \ket{\mathcal{F}_{K^u}(V) \otimes_{K^u} M; \mathcal{F}_{K^u}(\alpha) \otimes_{K^u} \beta}_{\lambda} = \ket{V \overline{\otimes} M; \alpha \overline{\otimes} \beta}_{\lambda},
\label{eq: action of V on M}
\end{equation}
where $\overline{\otimes}: \mathrm{SRep}(\mathcal{H}^u) \times {}_{K^u} \mathcal{SM} \rightarrow {}_{K^u} \mathcal{SM}$ is the $\mathrm{SRep}(\mathcal{H}^u)$-supermodule action on ${}_{K^u} \mathcal{SM}$.

Since a left $K^u$-supermodule $M$ labelling a ground state \eqref{eq: ground states} is a topological boundary condition of $\mathcal{T}_f^{K^u}$, eq. \eqref{eq: action of V on M} implies that the category of boundary conditions of $\mathcal{T}_f^{K^u}$ is an $\mathrm{SRep}(\mathcal{H}^u)$-supermodule category ${}_{K^u} \mathcal{SM}$.
On the other hand, the category of boundary conditions of the original bosonic TQFT $\mathcal{T}_b^K$ is a $\mathrm{Rep}(H)$-module category ${}_K \mathcal{M}$ \cite{Inamura2022}.
Therefore, we find that the fermionization of a $\mathrm{Rep}(H)$-symmetric bosonic TQFT whose category of boundary conditions is a $\mathrm{Rep}(H)$-module category ${}_K \mathcal{M}$ is an $\mathrm{SRep}(\mathcal{H}^u)$-symmetric fermionic TQFT whose category of boundary conditions is an $\mathrm{SRep}(\mathcal{H}^u)$-supermodule category ${}_{K^u} \mathcal{SM}$.
This gives a map between $\mathrm{Rep}(H)$-module categories and $\mathrm{SRep}(\mathcal{H}^u)$-supermodule categories.

\subsection{Hamiltonians with superfusion category symmetries on the lattice}
We can construct lattice models of gapped phases described by fermionic TQFTs $\mathcal{T}_f^{K^u}$ in the low-energy limit.
In this subsection, we explicitly write down the Hamiltonians of these lattice models and discuss the superfusion category symmetries on the lattice.
For simplicity, we will restrict our attention to non-anomalous symmetries $\mathrm{SRep}(\mathcal{H}^u)$ where $\mathcal{H}^u$ is a Hopf superalgebra.

Let $N$ be the number of lattice sites.
The state space $K^u_i$ on the $i$th site is given by a left $\mathcal{H}^u$-supercomodule algebra $K^u$ defined in the previous subsection, namely $K^u_i$ is a left $H$-comodule algebra $K$ equipped with a $\mathbb{Z}_2$-grading \eqref{eq: grading on H-comodule algebra K}.
We suppose that $K^u$ is chosen so that the action of $\mathrm{SRep}(\mathcal{H}^u)$ symmetry defined below becomes faithful on the lattice.\footnote{We can always choose $K^u$ so that the symmetry acts faithfully as in the bosonic case \cite{Inamura2022}.}
The Hamiltonian $H_{\lambda}$ on a circular lattice with a spin structure $\lambda$ is of the form
\begin{equation}
H_{\lambda} = \sum_{i} (1 - h_{i, i+1}^{\lambda}),
\label{eq: Hamiltonian}
\end{equation}
where $\lambda = 0$ on an $\mathrm{NS}$ circle and $\lambda = 1$ on an $\mathrm{R}$ circle.
The interaction terms $h_{i, i+1}^{\lambda}$ between the neighboring sites are given by
\begin{equation}
h_{i, i+1}^{\lambda} = 
\begin{cases}
\Delta_K \circ m_K: K^u_i \otimes K^u_{i+1} \rightarrow K^u_i \otimes K^u_{i+1} \quad & \text{for } i \neq N, \\
(\mathrm{id}_{K} \otimes p^{\lambda + 1}) \circ \Delta_K \circ m_K \circ (\mathrm{id}_{K} \otimes p^{\lambda + 1}): K^u_N \otimes K^u_1 \rightarrow K^u_N \otimes K^u_1 \quad & \text{for } i = N,
\end{cases}
\label{eq: local interaction}
\end{equation}
where $p: K^u \rightarrow K^u$ is the $\mathbb{Z}_2$-grading automorphism \eqref{eq: grading on H-comodule algebra K}.
The above Hamiltonian is a fermionic analogue of the bosonic Hamiltonian in \cite{Inamura2022}.\footnote{The Hamiltonians for bosonic state sum TQFTs in general dimensions are given in \cite{WW2017, Bullivant2018}.}
The interaction terms $h^{\lambda}_{i, i+1}$ are commuting projectors because a semisimple (super)algebra $K^u$ is a Frobenius algebra satisfying eq. \eqref{eq: Delta separability}.
Thus, the ground states of the Hamiltonian \eqref{eq: Hamiltonian} form a subspace of the state space invariant under each projector.
This ground state subspace agrees with the state space of fermionic TQFT $\mathcal{T}_f^{K^u}$ on a circle $S^1_{\lambda}$ because the cylinder amplitude $P_{\lambda}^{0}$ given by eq. \eqref{eq: cylinder projector} is the composition of commuting projectors $h^{\lambda}_{i, i+1}$ for all $i$. 
Therefore, the ground states of the Hamiltonian \eqref{eq: Hamiltonian} are given by eq. \eqref{eq: ground states}.
As shown in \cite{NR2015}, the number of ground states is equal to the dimension of the $\mathbb{Z}_2$-graded center of $K$, which is also equal to the number of simple $K$-supermodules.
This implies that equation \eqref{eq: ground states} exhausts all ground states.\footnote{We recall that eq. \eqref{eq: ground states} is non-zero for either $\beta = \mathrm{id}$ or $\beta = f$.}
The one-to-one correspondence between the ground states and simple $K$-supermodules are also discussed in \cite{KTY2018}.

We define the symmetry operator $\widehat{\mathcal{U}}_{V; \alpha}^{\lambda}: \bigotimes_i K_i^u \rightarrow \bigotimes_i K_i^u$ for $V \in \mathrm{SRep}(\mathcal{H}^u)$ and $\alpha: V \rightarrow V$ by the following string diagram:
\begin{equation}
\widehat{\mathcal{U}}_{V; \alpha}^{\lambda} = ~
\adjincludegraphics[valign = c, width = 4.5cm]{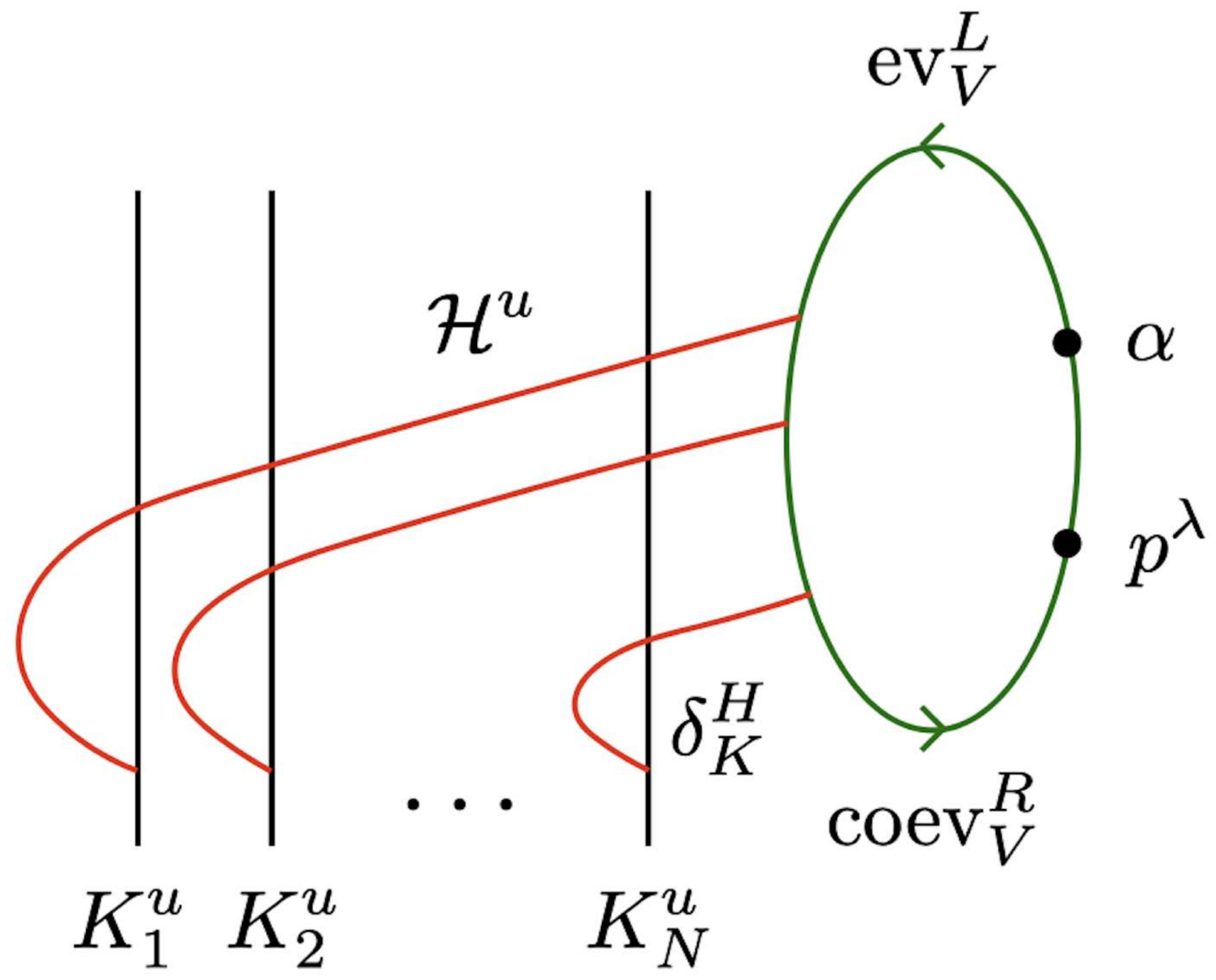} ~ = ~
\adjincludegraphics[valign = c, width = 4.2cm]{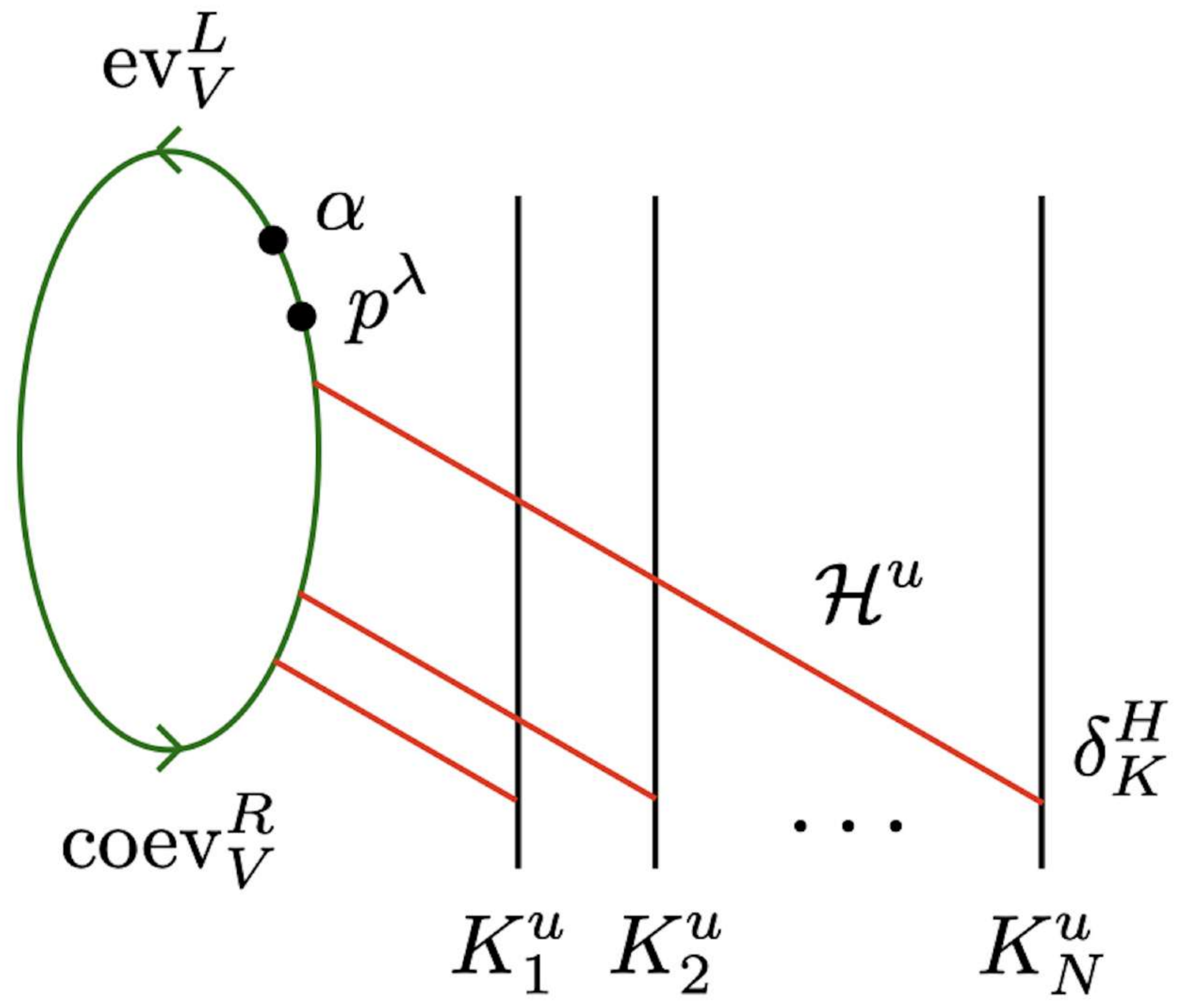}.
\label{eq: action of V on the lattice}
\end{equation}
The second equality follows from the definition of the right $\mathcal{H}^u$-action on $V$ induced by the left $\mathcal{H}^u$-action on $V^*$.
Specifically, the right $\mathcal{H}^u$-action $\rho_V^R: V \otimes \mathcal{H}^u \rightarrow V$ is defined by
\begin{equation}
\rho_V^R = (\mathrm{ev}_V^R \otimes \mathrm{id}_V) \circ (\mathrm{id}_V \otimes \rho_{V^*}^L \otimes \mathrm{id}_V) \circ (\mathrm{id}_{V} \otimes \mathrm{id}_{\mathcal{H}^u} \otimes \mathrm{coev}_V^R),
\label{eq: induced right action}
\end{equation}
where $\rho^L_{V^*}: \mathcal{H}^u \otimes V^* \rightarrow V^*$ is the left $\mathcal{H}^u$-action on $V^*$.
We note that a morphism $\alpha: V \rightarrow V$ commutes with the right $\mathcal{H}^u$-action \eqref{eq: induced right action} because the dual of $\alpha$ commutes with the left $\mathcal{H}^u$-action on $V^*$.
For a simple object $V \in \mathrm{SRep}(\mathcal{H}^u)$, a point-like defect $\alpha$ in eq. \eqref{eq: action of V on the lattice} is either the identity morphism $\mathrm{id}$ or a $\mathbb{Z}_2$-odd automorphism $f$.
When $V$ is a q-type object, the symmetry action on the lattice satisfies $\widehat{\mathcal{U}}_{V; f}^{\lambda = 0} = \widehat{\mathcal{U}}_{V; \mathrm{id}}^{\lambda = 1} = 0$ just as the corresponding operators of TQFT $\mathcal{T}_f^{K^u}$ satisfy $\mathcal{U}_{V; f}^{\lambda = 0} = \mathcal{U}_{V; \mathrm{id}}^{\lambda = 1} = 0$.
The composition of operators defined by eq. \eqref{eq: action of V on the lattice} are compatible with the monoidal structure on $\mathrm{SRep}(\mathcal{H}^u)$, namely we have $\widehat{\mathcal{U}}_{V; \alpha}^{\lambda} \circ \widehat{\mathcal{U}}_{V^{\prime}; \alpha^{\prime}}^{\lambda} = \widehat{\mathcal{U}}_{V \otimes V^{\prime}; \alpha \otimes \alpha^{\prime}}^{\lambda}$.

The symmetry action \eqref{eq: action of V on the lattice} commutes with the Hamiltonian \eqref{eq: Hamiltonian}.
The commutativity of the symmetry action $\widehat{\mathcal{U}}_{V; \alpha}^{\lambda}$ and the commuting projector $h_{i, i+1}^{\lambda}$ for $i \neq N$ follows from the equality
\begin{equation}
\adjincludegraphics[valign = c, width = 3cm]{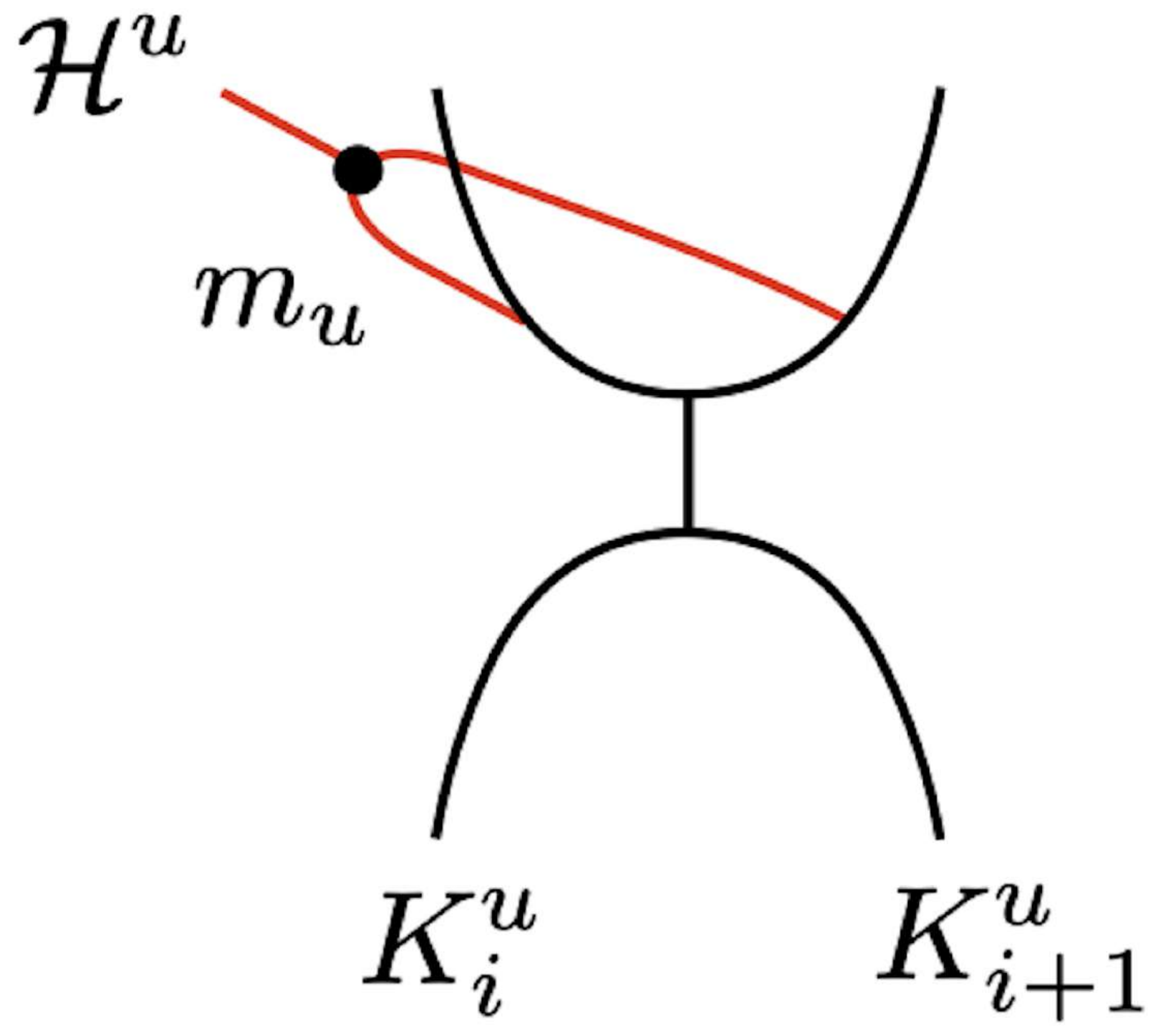} ~ = ~ \adjincludegraphics[valign = c, width = 3cm]{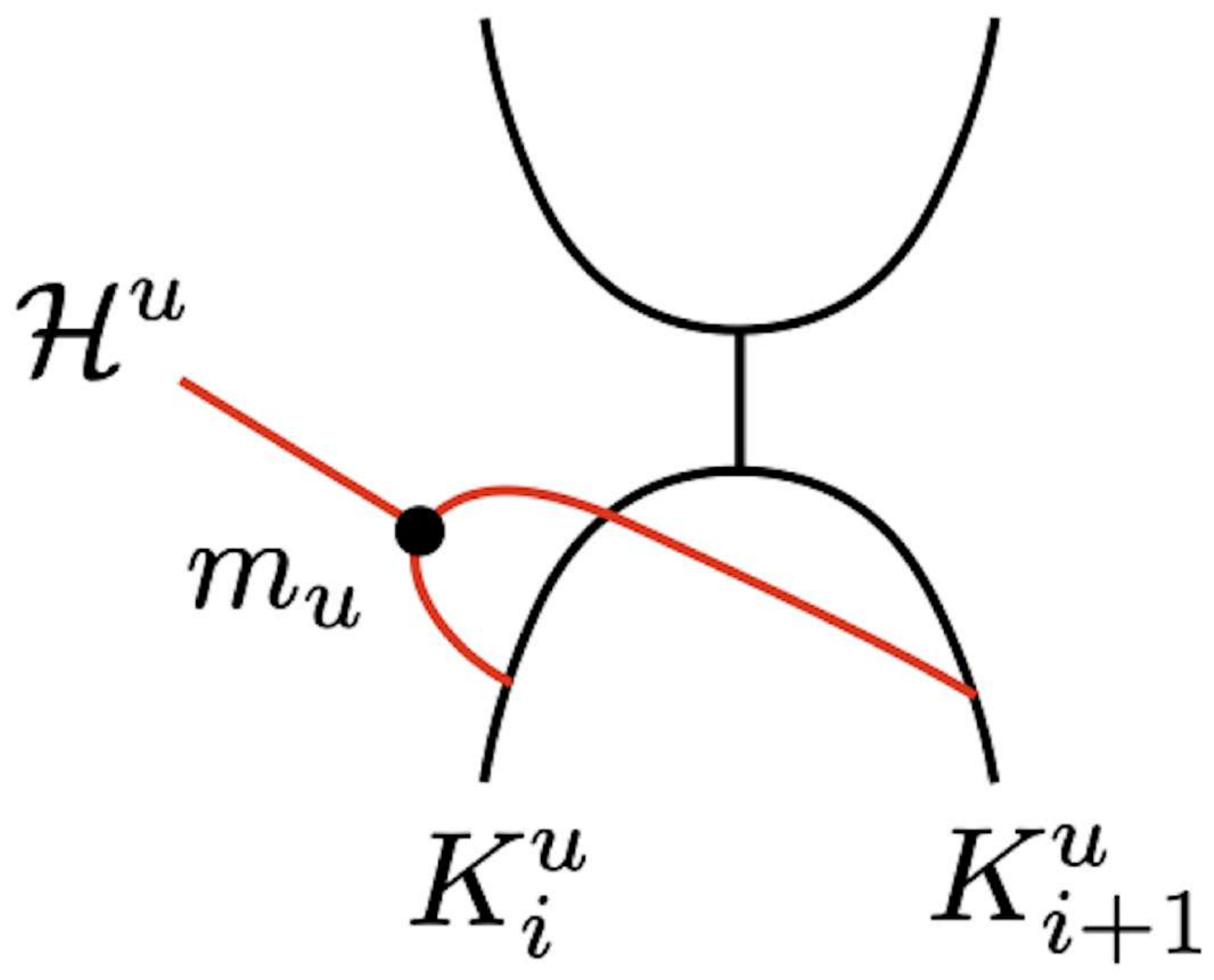},
\label{eq: commutativity on the lattice}
\end{equation}
which we can show in the same way as the bosonic case \cite{Inamura2022}.
In order to show that $\widehat{\mathcal{U}}_{V; \alpha}^{\lambda}$ also commutes with $h^{\lambda}_{N, 1}$, we write the equation $\widehat{\mathcal{U}}_{V; \alpha}^{\lambda} h^{\lambda}_{N, 1} = h^{\lambda}_{N, 1} \widehat{\mathcal{U}}_{V; \alpha}^{\lambda}$ in terms of string diagrams as
\begin{equation}
\adjincludegraphics[valign = c, width = 3.6cm]{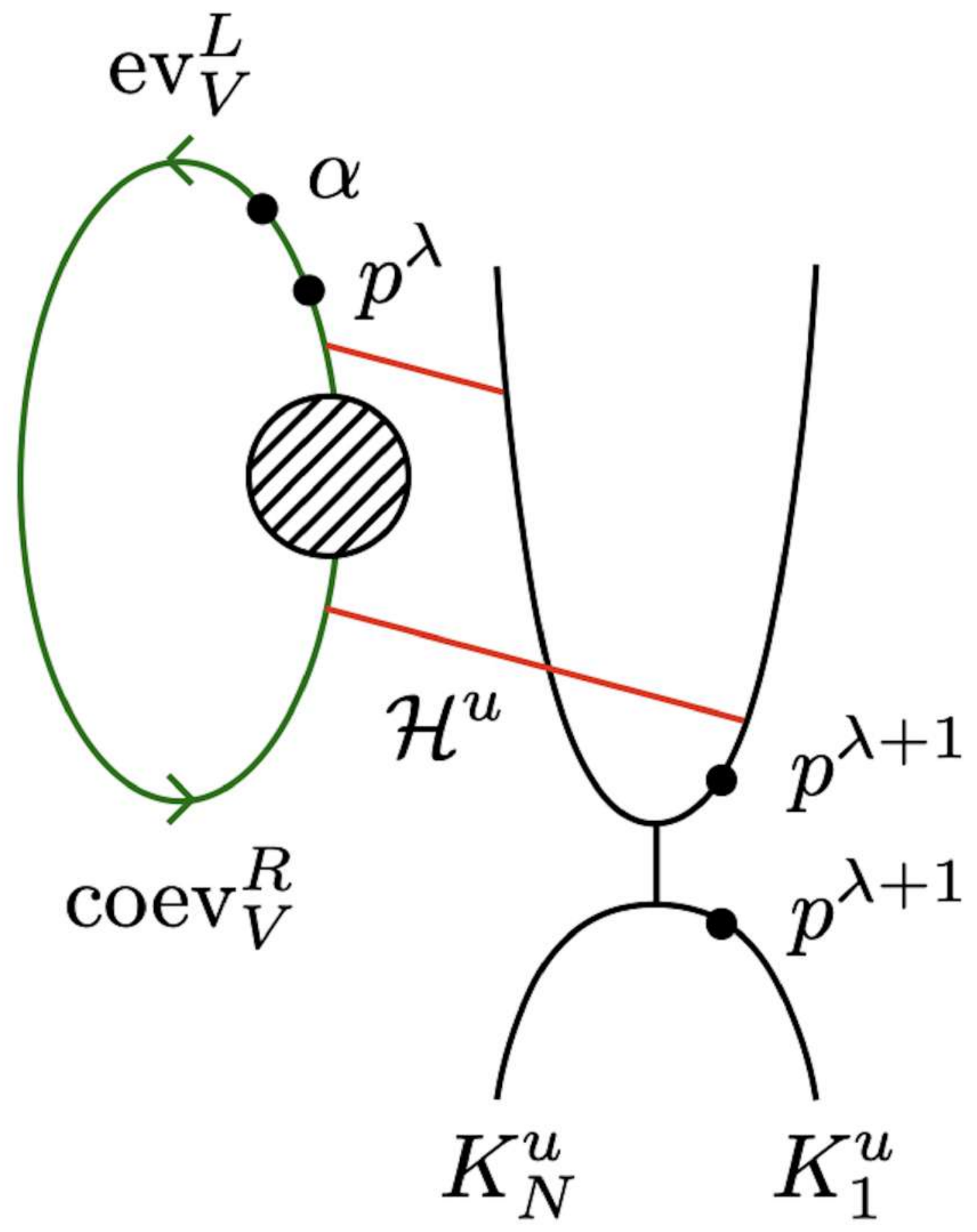} \quad = \quad \adjincludegraphics[valign = c, width = 4cm]{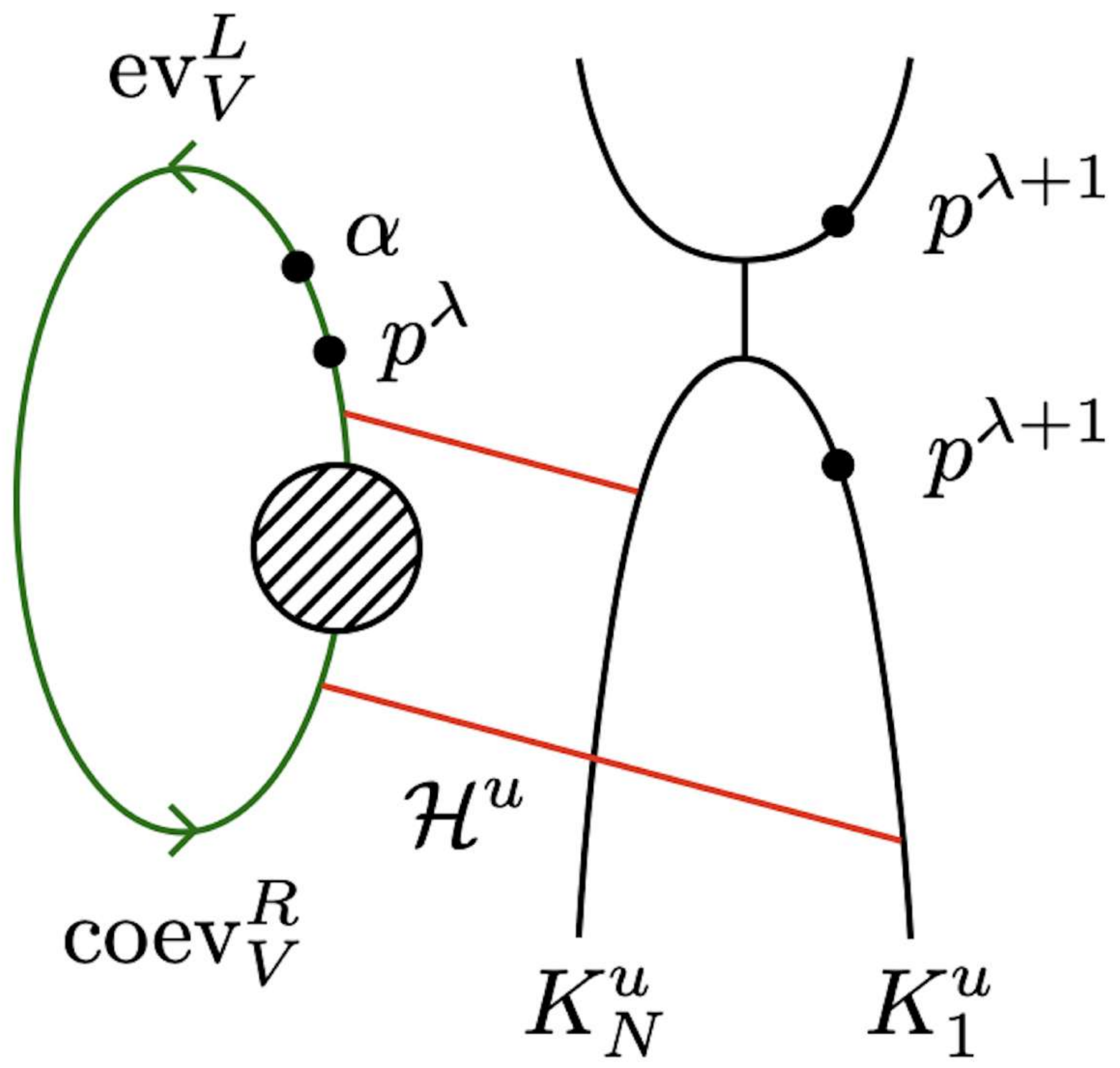},
\end{equation}
where the blob represents the action of $\mathcal{H}^u$ coacting on $K_{i \neq 1, N}^u$.
The above equation reduces to eq. \eqref{eq: commutativity on the lattice} due to the equality
\begin{equation}
\adjincludegraphics[valign = c, width = 3cm]{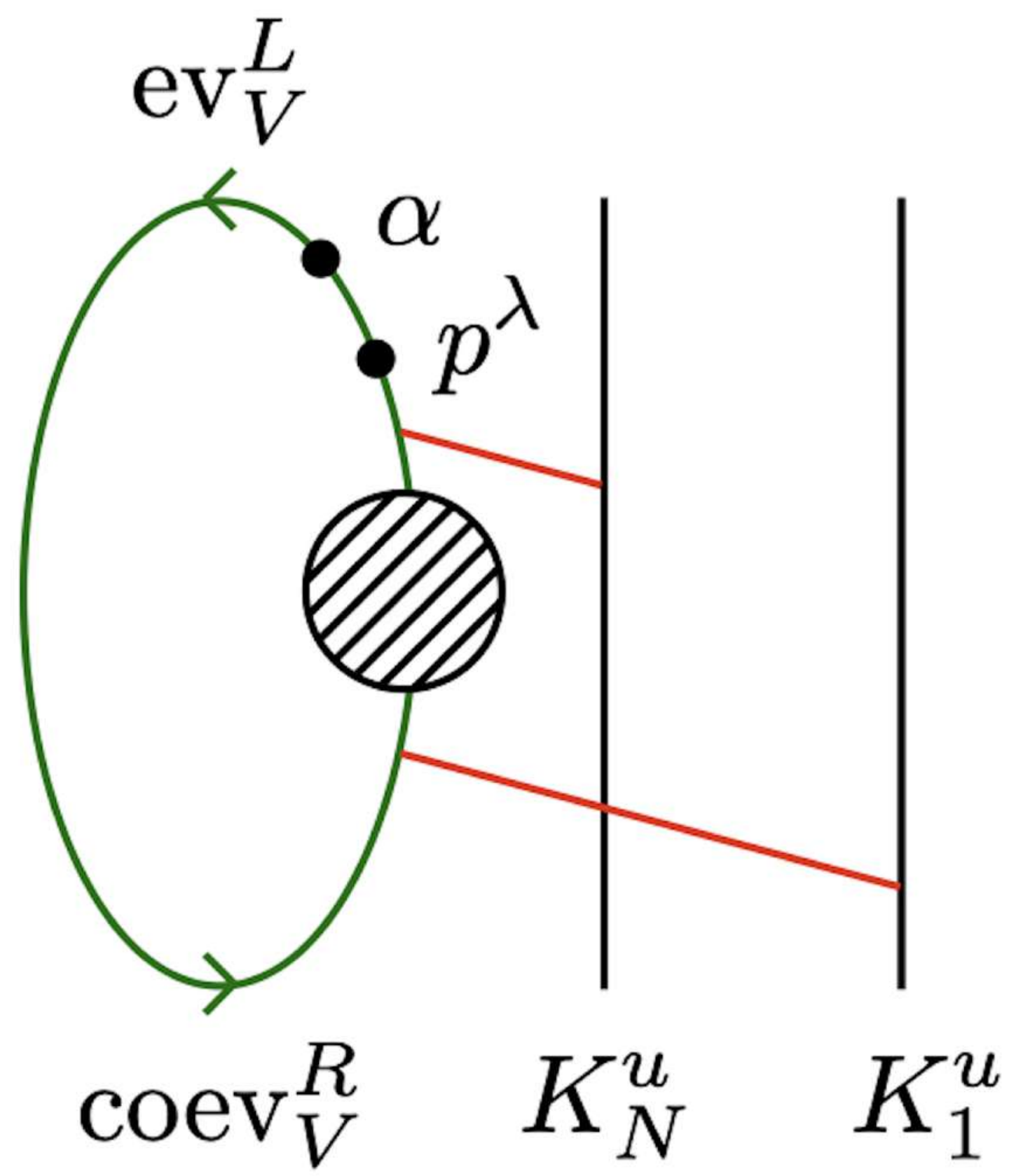} \quad = \quad \adjincludegraphics[valign = c, width = 3cm]{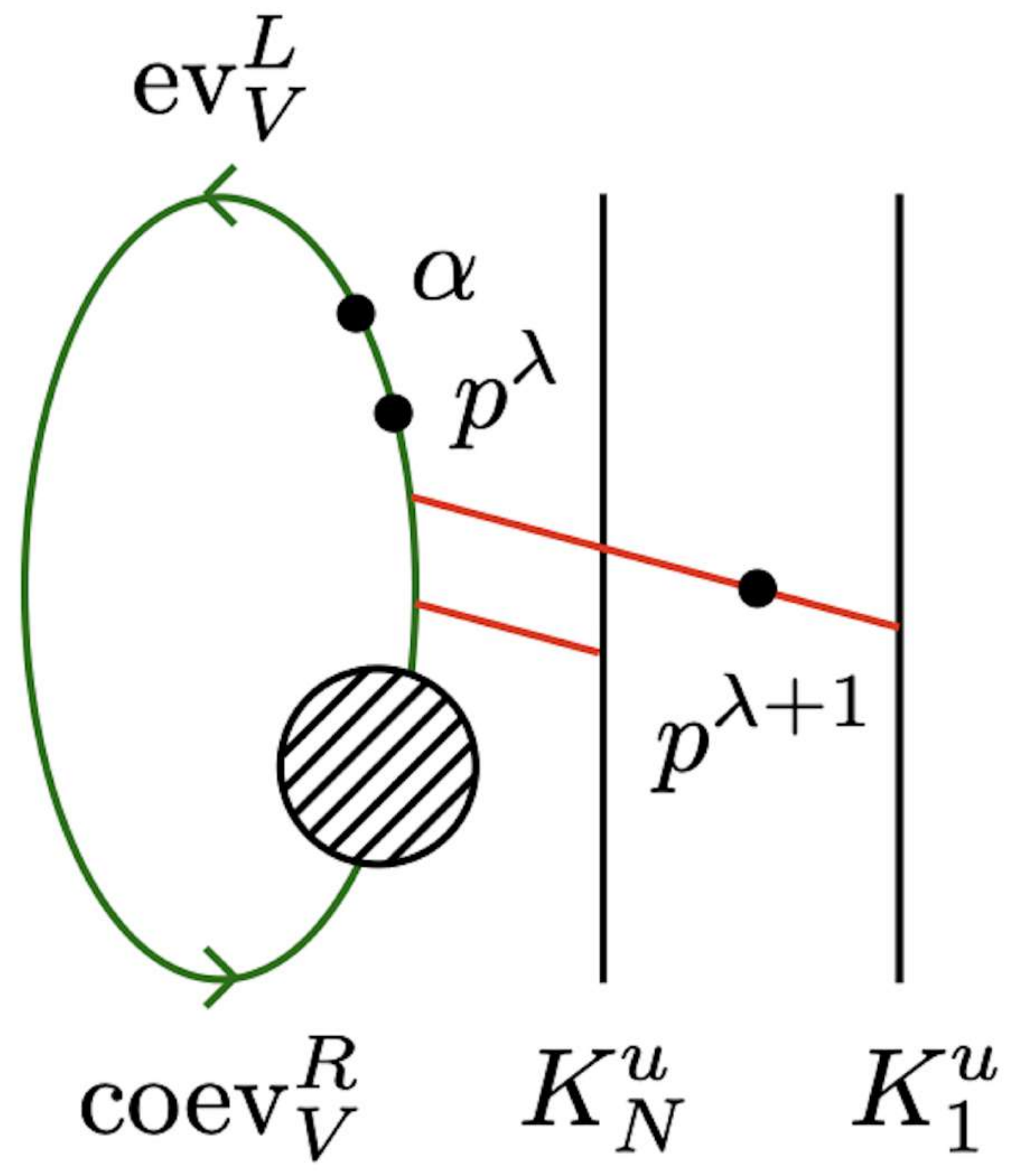},
\end{equation}
which follows from eqs. \eqref{eq: induced right action}, \eqref{eq: zigzag}, and \eqref{eq: left-right relation}.
Therefore, the symmetry action $\widehat{\mathcal{U}}_{V; \alpha}^{\lambda}$ also commutes with $h^{\lambda}_{N, 1}$.
This shows that the Hamiltonian \eqref{eq: Hamiltonian} has a superfusion category symmetry $\mathrm{SRep}(\mathcal{H}^u)$ on the lattice.

Finally, we argue that the action of $\widehat{\mathcal{U}}_{V; \alpha}^{\lambda}$ on the ground states of the Hamiltonian \eqref{eq: Hamiltonian} agrees with the action of the corresponding operator $\mathcal{U}_{V; \alpha}^{\lambda}$ of a fermionic TQFT $\mathcal{T}_f^{K^u}$.
To this end, we compute the actions of $\widehat{\mathcal{U}}_{V; \alpha}^{\lambda}$ and $\mathcal{U}_{V; \alpha}^{\lambda}$ on the dual vector space of the ground state subspace.
States in the dual vector space are denoted by ${}_{\lambda} \bra{N; \gamma}$, where $N$ is a simple right $K^u$-supermodule and $\gamma: N \rightarrow N$ is a right $K^u$-supermodule morphism.
A dual state ${}_{\lambda} \bra{N; \gamma}$ is the transition amplitude on a cylinder whose boundary circles consist of an in-boundary and a topological boundary, where $N$ and $\gamma$ represent a boundary condition and a point-like defect on the topological boundary.
In string diagram notation, a dual state ${}_{\lambda} \bra{N; \gamma}$ can be written as
\begin{equation}
{}_{\lambda} \bra{N; \gamma} := \adjincludegraphics[valign = c, width = 3.5cm]{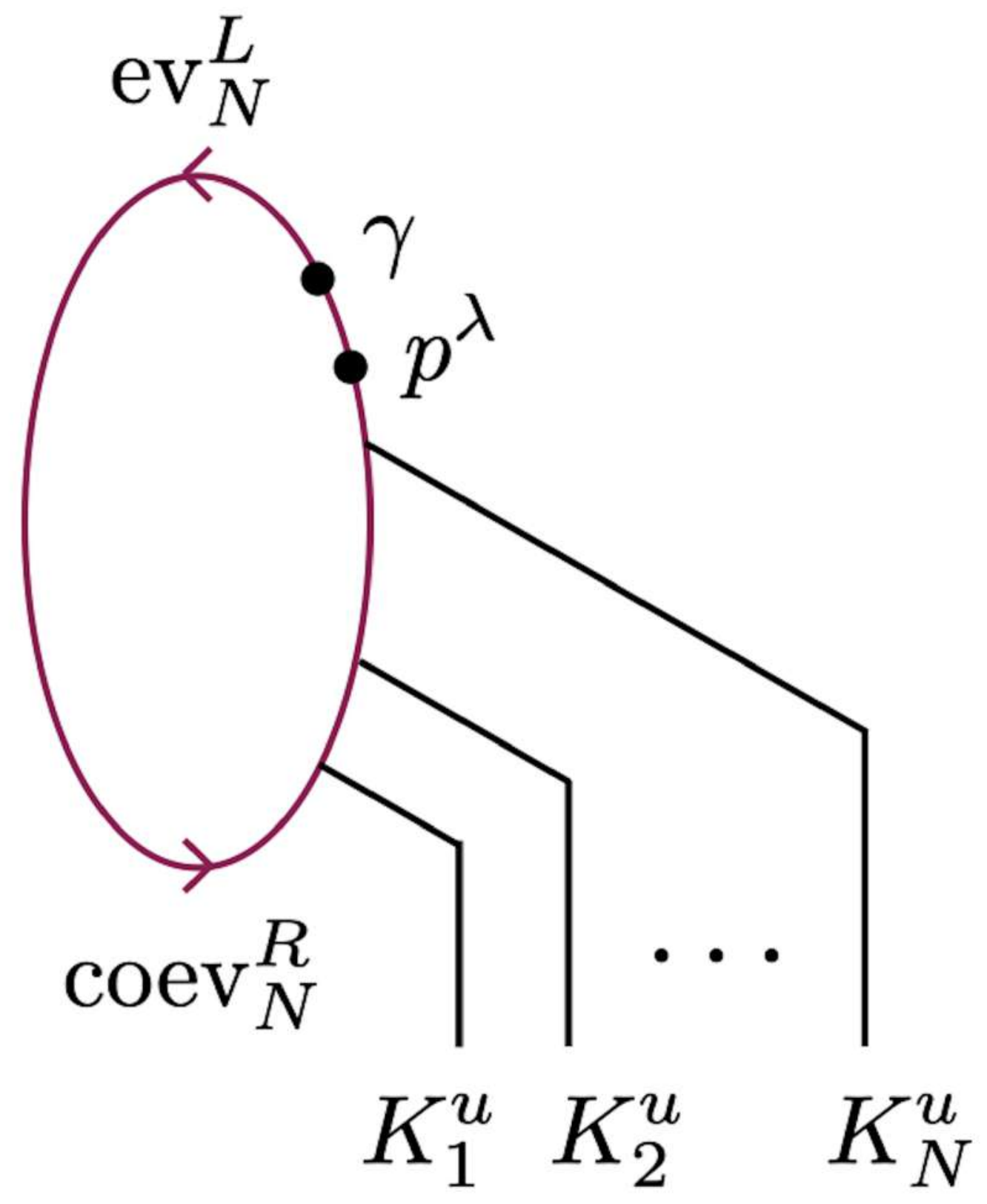},
\label{eq: N}
\end{equation}
which are the wave function of a fermionic matrix product state \cite{KTY2018, BWHV2017}.
Since $N$ is simple, $\gamma$ is the identity morphism $\mathrm{id}$ or a unique $\mathbb{Z}_2$-odd automorphism $f$.
We note that ${}_{\lambda = 0} \bra{N; f} = {}_{\lambda = 1} \bra{N; \mathrm{id}} = 0$ if $N$ has an odd automorphism $f$.
The action of $\widehat{\mathcal{U}}_{V; \alpha}^{\lambda}$ on a dual state \eqref{eq: N} is computed as ${}_{\lambda} \bra{N; \gamma} \widehat{\mathcal{U}}_{V; \alpha}^{\lambda} = {}_{\lambda} \bra{N \otimes V; \gamma \otimes \alpha}$, where the right $K^u$-supermodule structure on $N \otimes V$ is defined via the left $\mathcal{H}^u$-supercomodule action on $K^u$.
On the other hand, the symmetry operator $\mathcal{U}_{V; \alpha}^{\lambda}$ of a fermionic TQFT $\mathcal{T}_f^{K^u}$ acts on a dual state ${}_{\lambda} \bra{N; \gamma}$ as
\begin{equation}
{}_{\lambda} \bra{N; \gamma} \mathcal{U}_{V; \alpha}^{\lambda} = {}_{\lambda} \bra{N \otimes_{K^u} \mathcal{F}_{K^u}(V); \gamma \otimes_{K^u} \mathcal{F}_{K^u}(\alpha)} = {}_{\lambda} \bra{N \otimes V; \gamma \otimes \alpha},
\end{equation}
where we used the isomorphism of right $K^u$-supermodules $N \otimes_{K^u} \mathcal{F}_{K^u}(V) \cong N \otimes_{K^u} (V \otimes K^u) \cong N \otimes_{K^u} (K^u \otimes V) \cong N \otimes V$.\footnote{The second isomorphism follows from $V^* \otimes K^u \cong \mathcal{F}_{K^u}(V^*) \cong \mathcal{F}_{K^u}(V)^* \cong (K^u)^* \otimes V^* \cong K^u \otimes V^*$, where the first and the third isomorphisms follow from the definition of $\mathcal{F}_{K^u}$, the second isomorphism is due to the fact that $\mathcal{F}_{K^u}$ is a supertensor functor, and the last isomorphism is given by the superalgebra isomorphism \eqref{eq: Phi}.}
Therefore, we find that the operators $\widehat{\mathcal{U}}_{V; \alpha}^{\lambda}$ and $\mathcal{U}_{V; \alpha}^{\lambda}$, which are defined in lattice models and TQFTs respectively, acts on the dual of the ground state subspace in the same way.
In particular, we have ${}_{\lambda} \bra{N; \gamma} \widehat{\mathcal{U}}_{V; \alpha}^{\lambda} \ket{M; \beta}_{\lambda} = {}_{\lambda} \bra{N; \gamma} \mathcal{U}_{V; \alpha}^{\lambda} \ket{M; \beta}_{\lambda}$ for any ${}_{\lambda} \bra{N; \gamma}$ and $\ket{M; \beta}_{\lambda}$.
This shows that $\widehat{\mathcal{U}}_{V; \alpha}^{\lambda}$ is equal to $\mathcal{U}_{V; \alpha}^{\lambda}$ when it acts on the ground states.

\section*{Acknowledgments}
The author thanks Ryohei Kobayashi for discussions, and Ryohei Kobayashi and Yunqin Zheng for comments on the manuscript.
The author also thanks anonymous referees for helpful suggestions on the manuscript.
The author is supported by FoPM, WINGS Program, the University of Tokyo, and also by JSPS Research Fellowship for Young Scientists.

\appendix

\section{Weak Hopf superalgebra structure on $\mathcal{H}^u$}
\label{sec: weak Hopf superalgebra}
In this appendix, we show that the super vector space $\mathcal{H}^u$ equipped with the structure maps defined in section \ref{sec: Fermionization of anomalous symmetries} is a weak Hopf superalgebra.
Throughout this section, we will use the string diagram notation introduced in section \ref{sec: Fermionization of fusion category symmetries}.
In string diagrams, the symmetric braiding $c_{\mathrm{super}}$ of super vector spaces will be represented by a crossing with a black dot, whereas the trivial braiding $c_{\mathrm{triv}}$ will be represented by a crossing without a black dot.

We first recall the definition of $\mathcal{H}^u$.
Let $H$ be a finite dimensional semisimple weak Hopf algebra with structure maps $(m, \eta, \Delta, \epsilon, S)$, and let $u \in H^*$ be a $\mathbb{Z}_2$ group-like element of the dual weak Hopf algebra $H^*$.
A $\mathbb{Z}_2$ group-like element $u \in H^*$ is a non-zero element of $H^*$ that satisfies the following properties:
\begin{equation}
\adjincludegraphics[valign = c, width = 1.3cm]{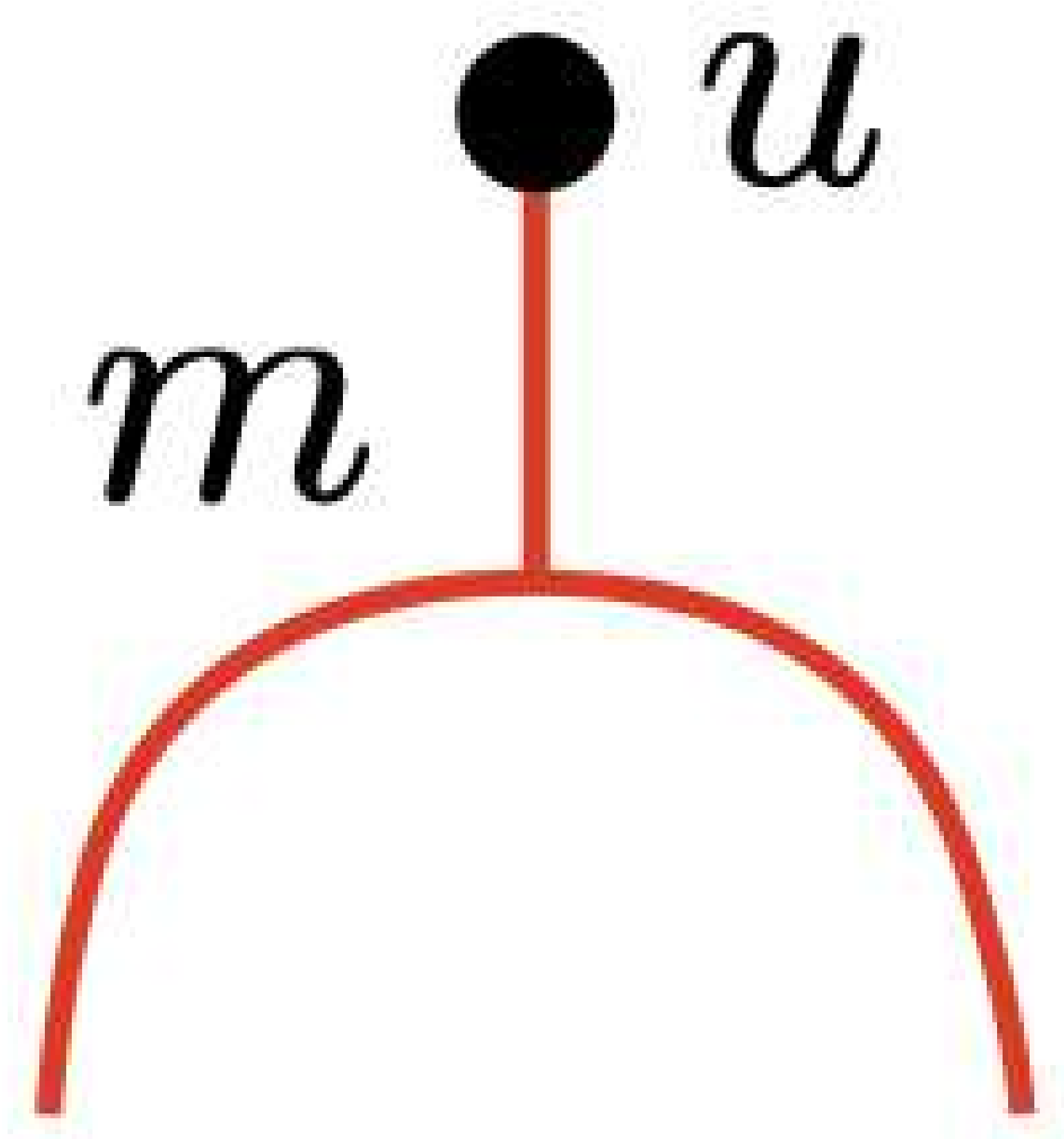} =
\adjincludegraphics[valign = c, width = 3cm]{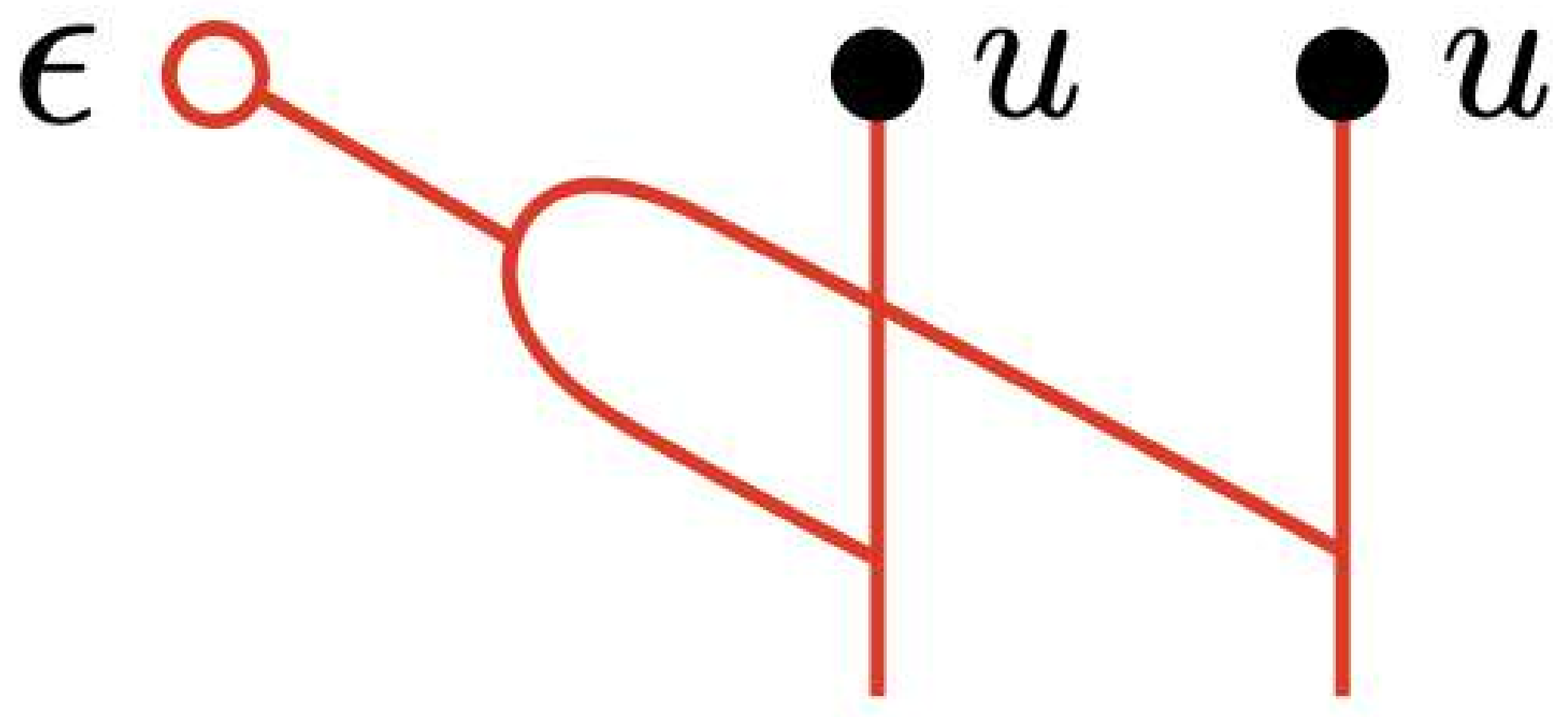} =
\adjincludegraphics[valign = c, width = 3cm]{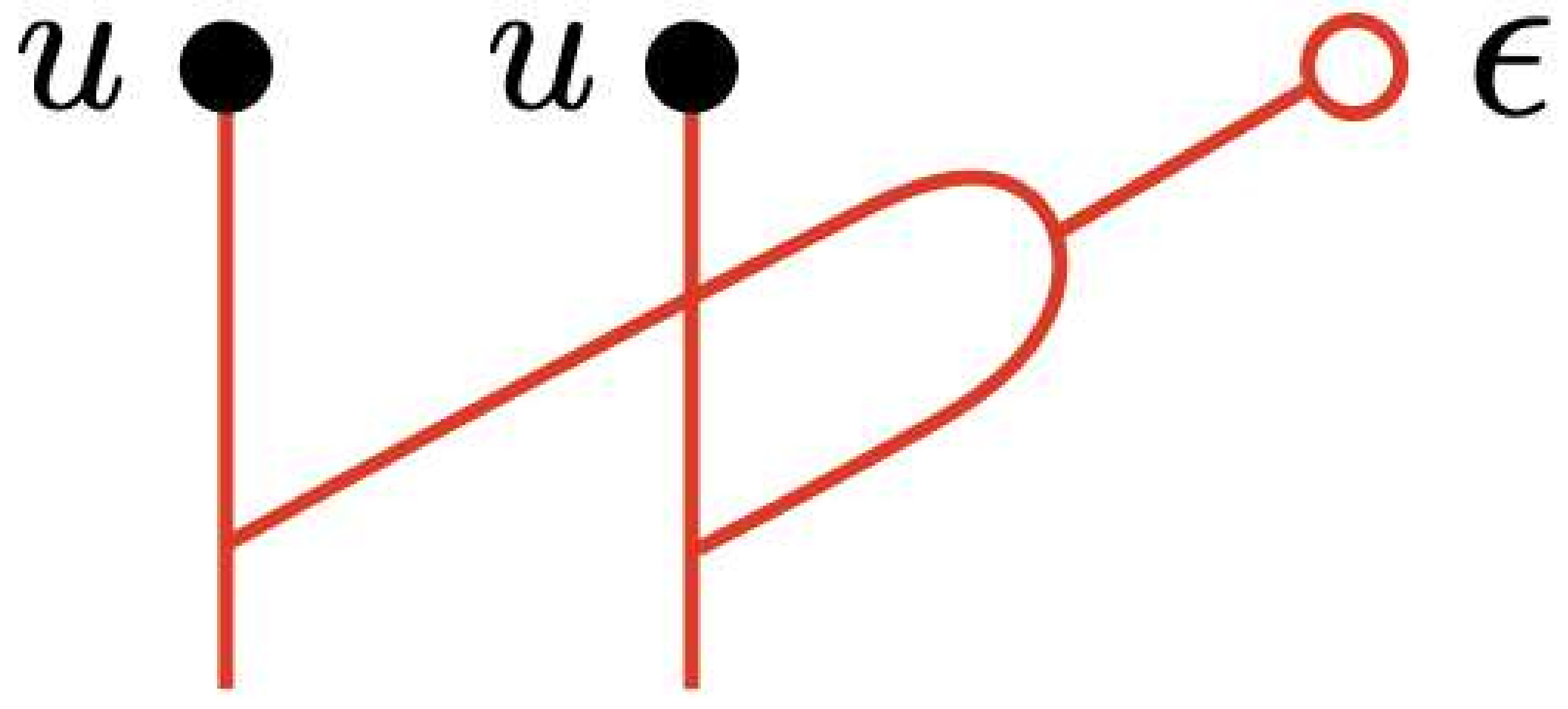}, \qquad
\adjincludegraphics[valign = c, width = 1.8cm]{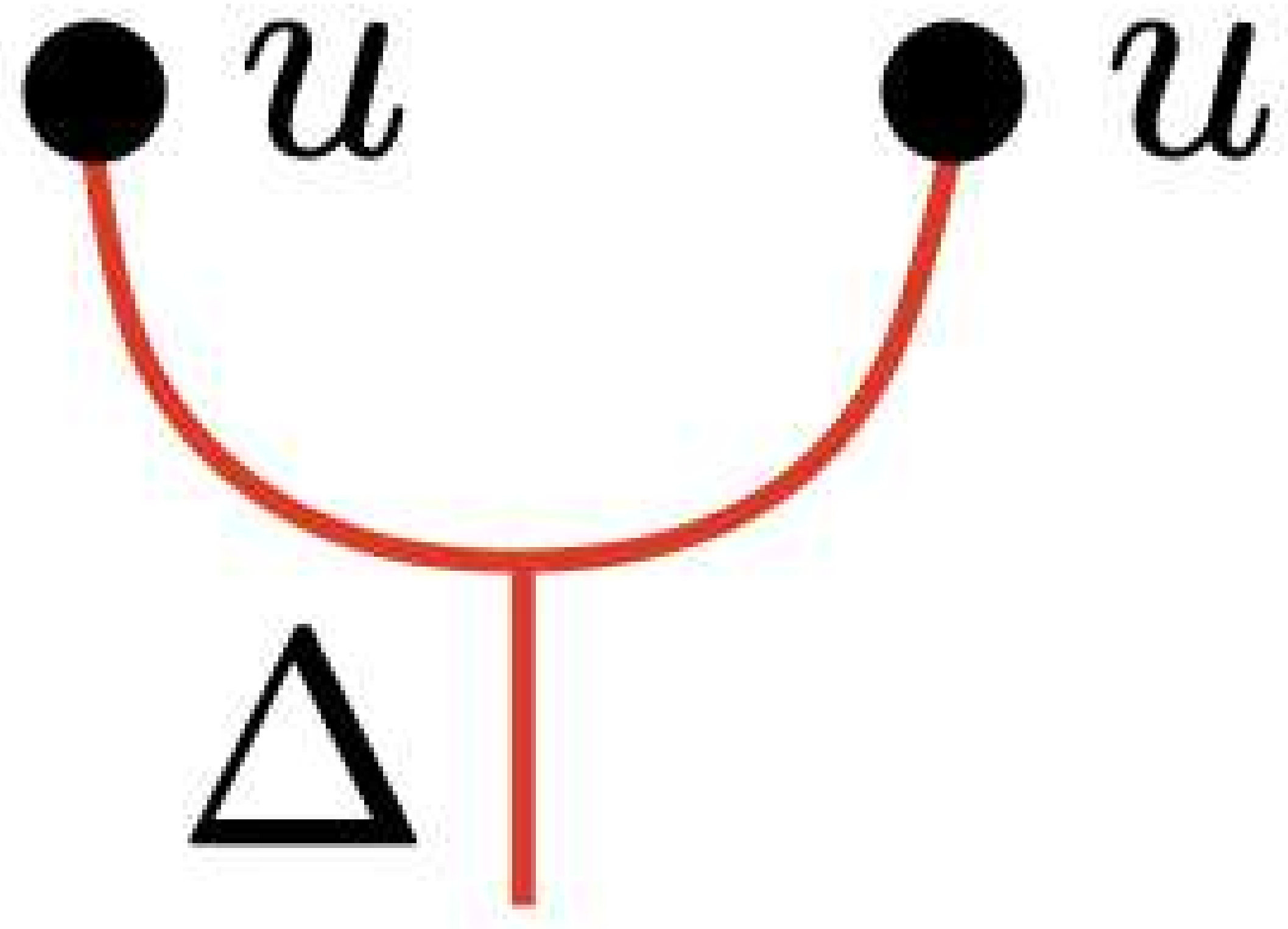} =
\adjincludegraphics[valign = c, width = 0.7cm]{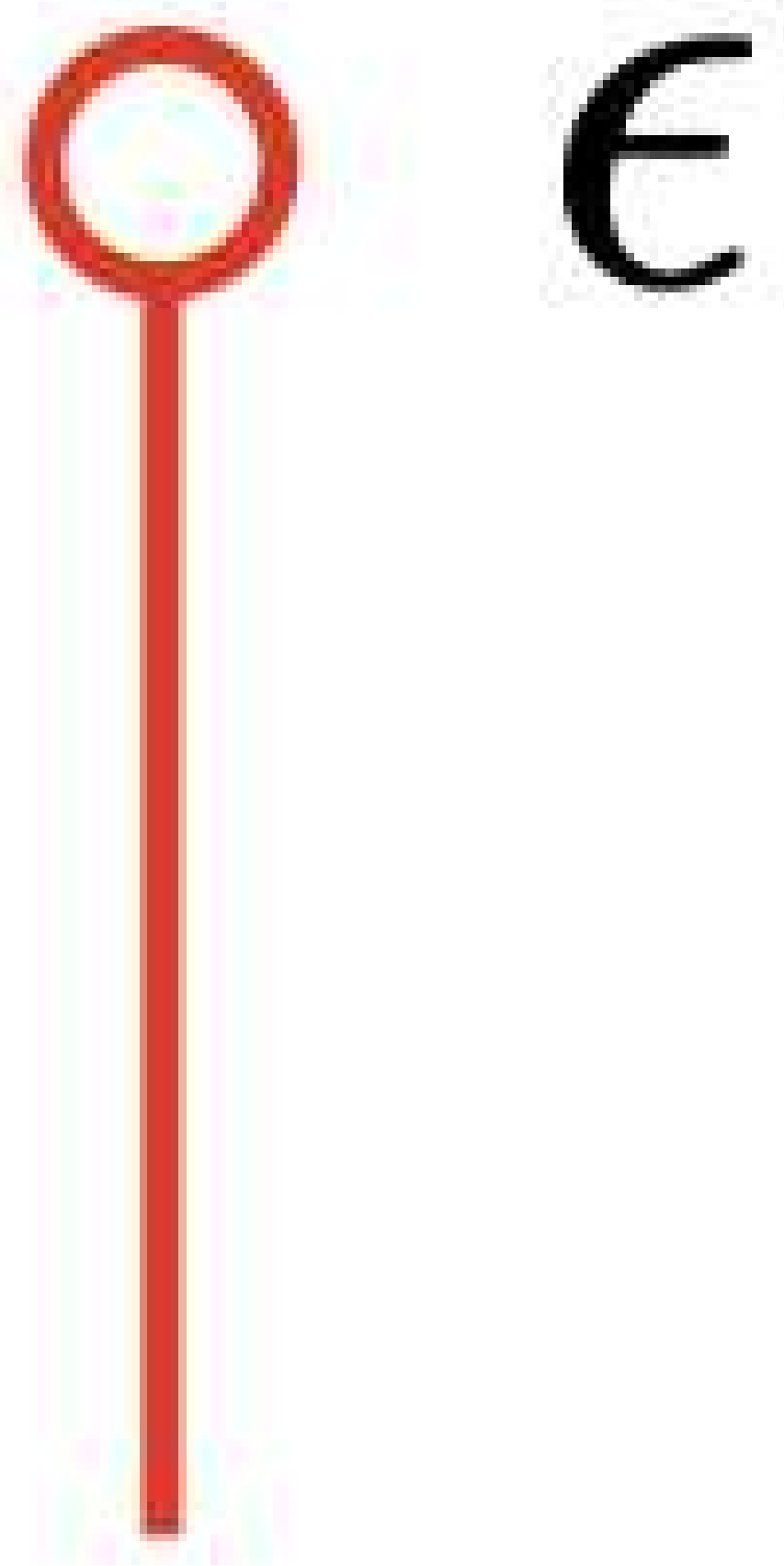}.
\label{eq: Z2 group-like}
\end{equation}
As described in section \ref{sec: Fermionization of anomalous symmetries}, the weak Hopf superalgebra $\mathcal{H}^u$ and its structure maps are defined as follows:
\begin{itemize}
\item The underlying vector space of $\mathcal{H}^u$ is $H$.
\item The $\mathbb{Z}_2$-grading automorphism $p_u: \mathcal{H}^u \rightarrow \mathcal{H}^u$ is defined by eq. \eqref{eq: grading of Hopf superalgebra}:
\begin{equation}
\adjincludegraphics[valign = c, width = 0.8cm]{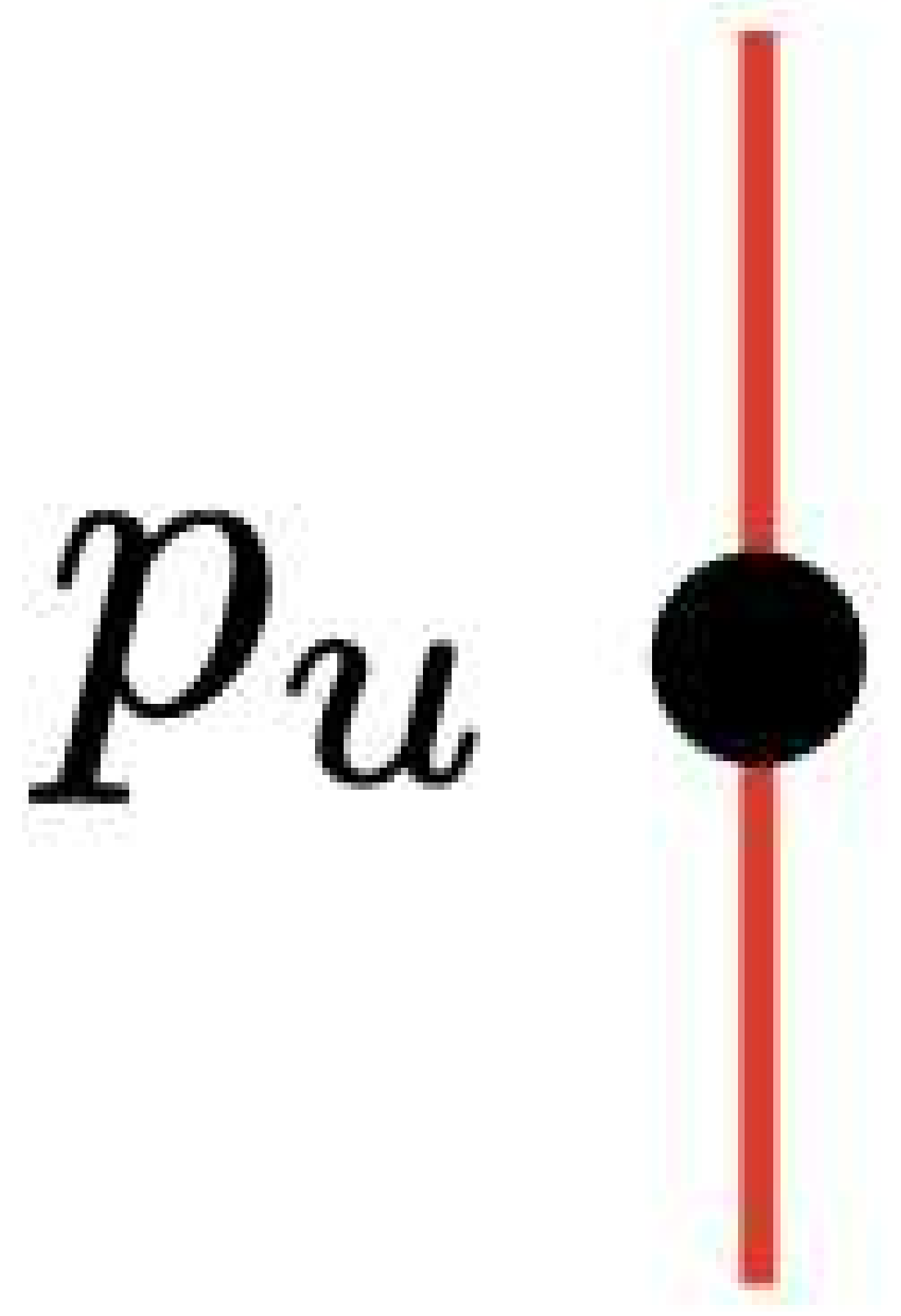} ~ = ~ 
\adjincludegraphics[valign = c, width = 2.1cm]{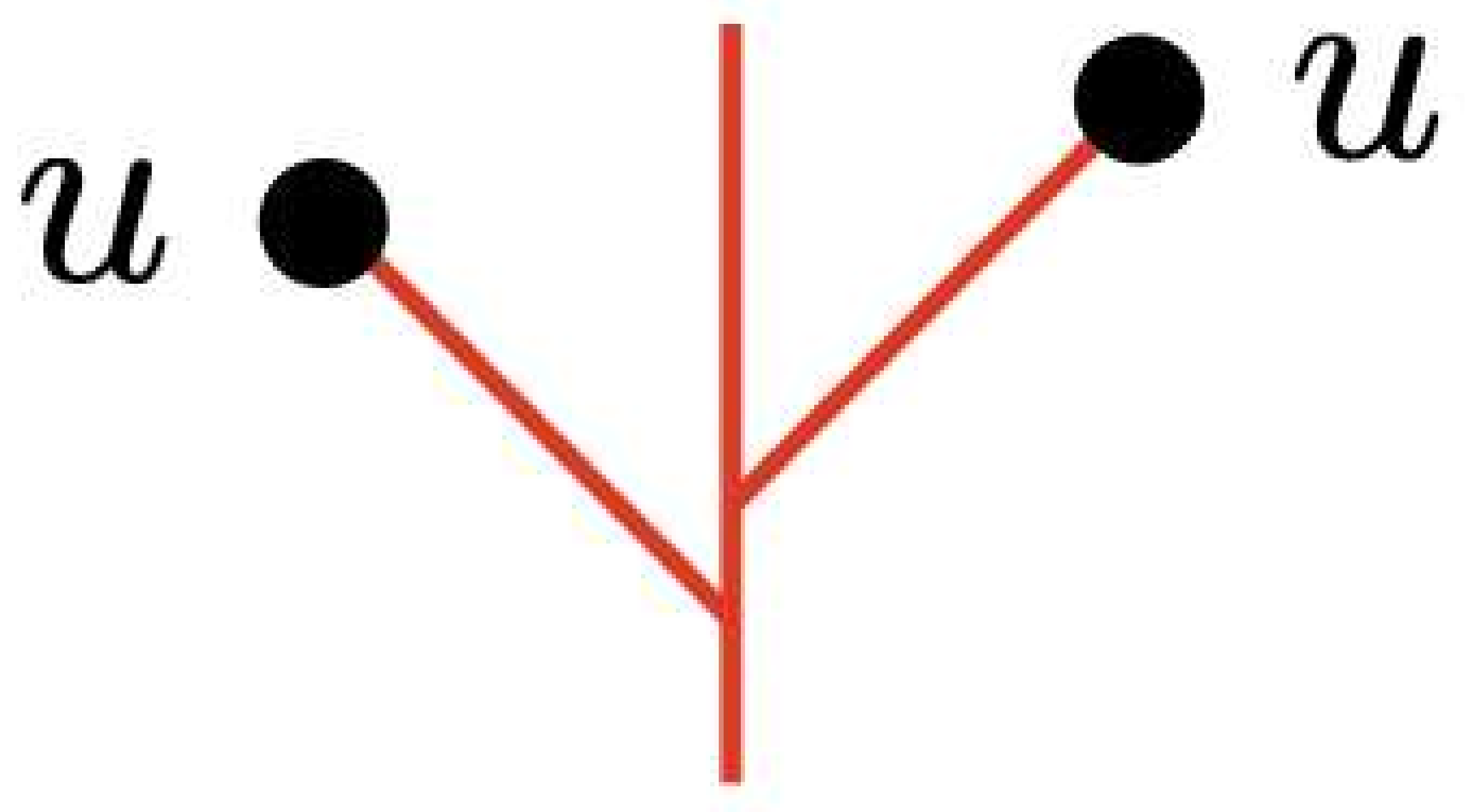}.
\label{eq: p_u}
\end{equation}
\item The multiplication $m_u: \mathcal{H}^u \otimes \mathcal{H}^u \rightarrow \mathcal{H}^u$ is defined by eq. \eqref{eq: multiplication of Hopf superalgebra}:
\begin{equation}
\adjincludegraphics[valign = c, width = 1.5cm]{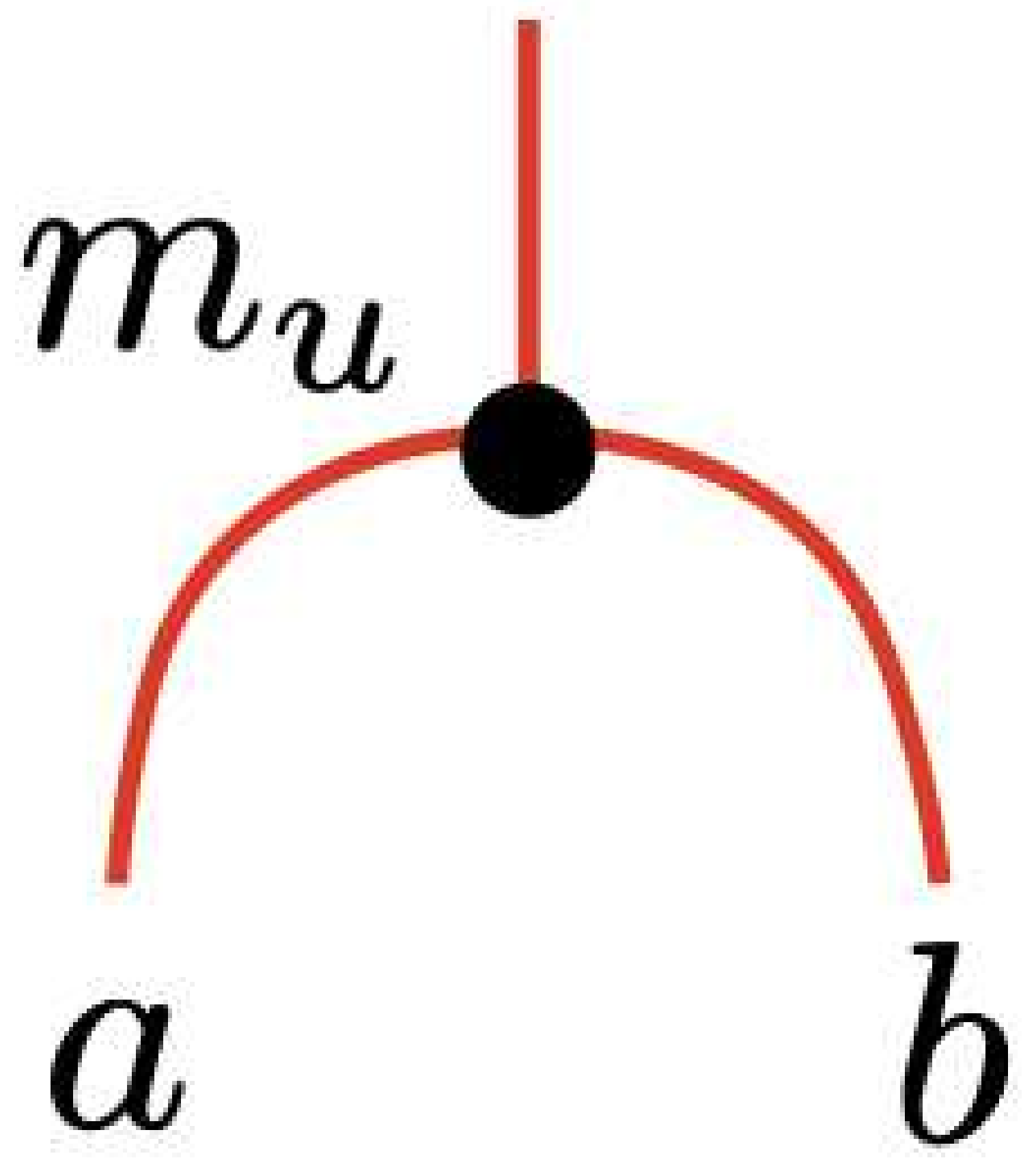} ~ = ~ 
\adjincludegraphics[valign = c, width = 1.8cm]{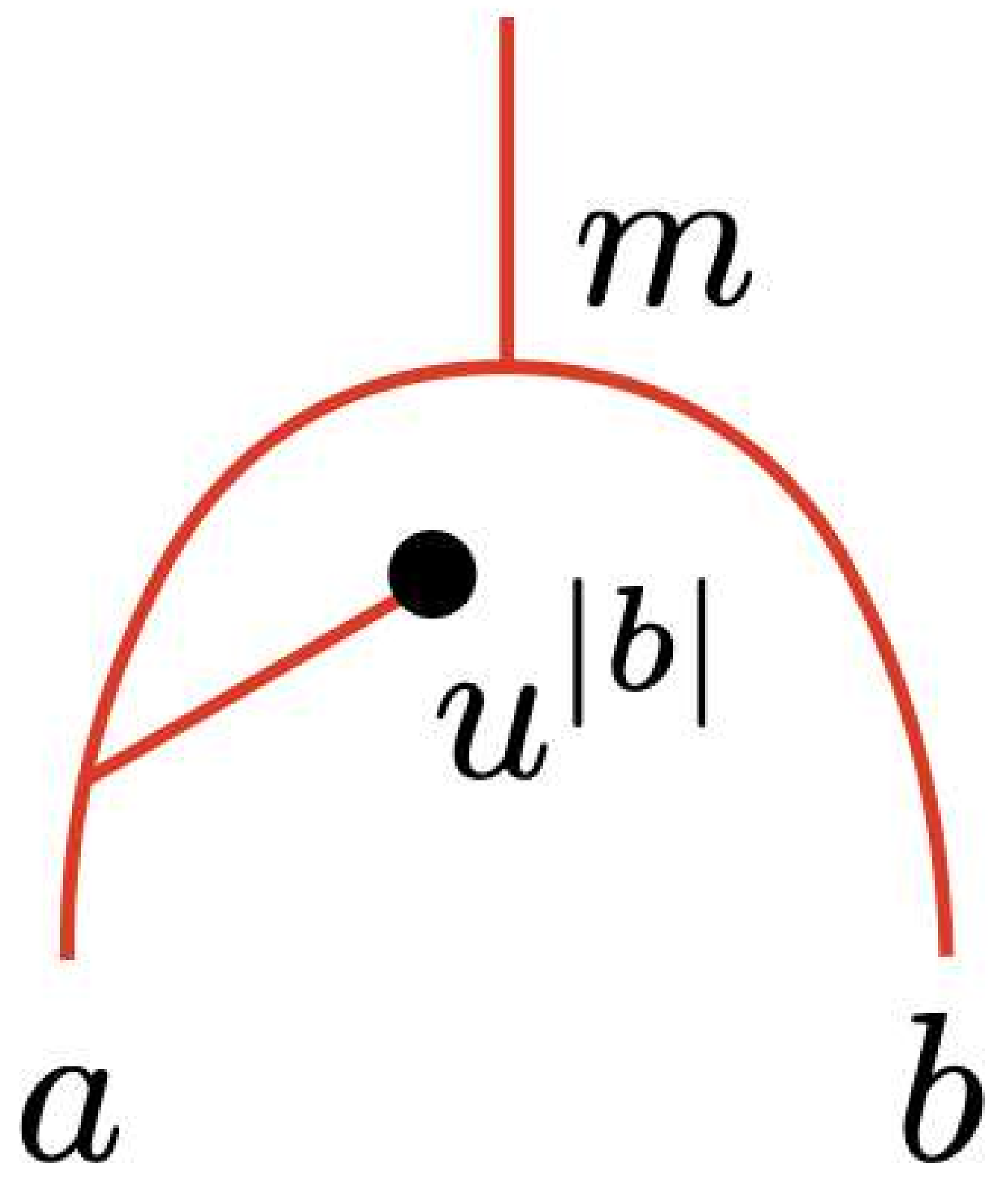}.
\end{equation}
\item The antipode $S_u: \mathcal{H}^u \rightarrow \mathcal{H}^u$ is defined by eq. \eqref{eq: antipode of Hopf superalgebra}:
\begin{equation}
\adjincludegraphics[valign = c, width = 0.8cm]{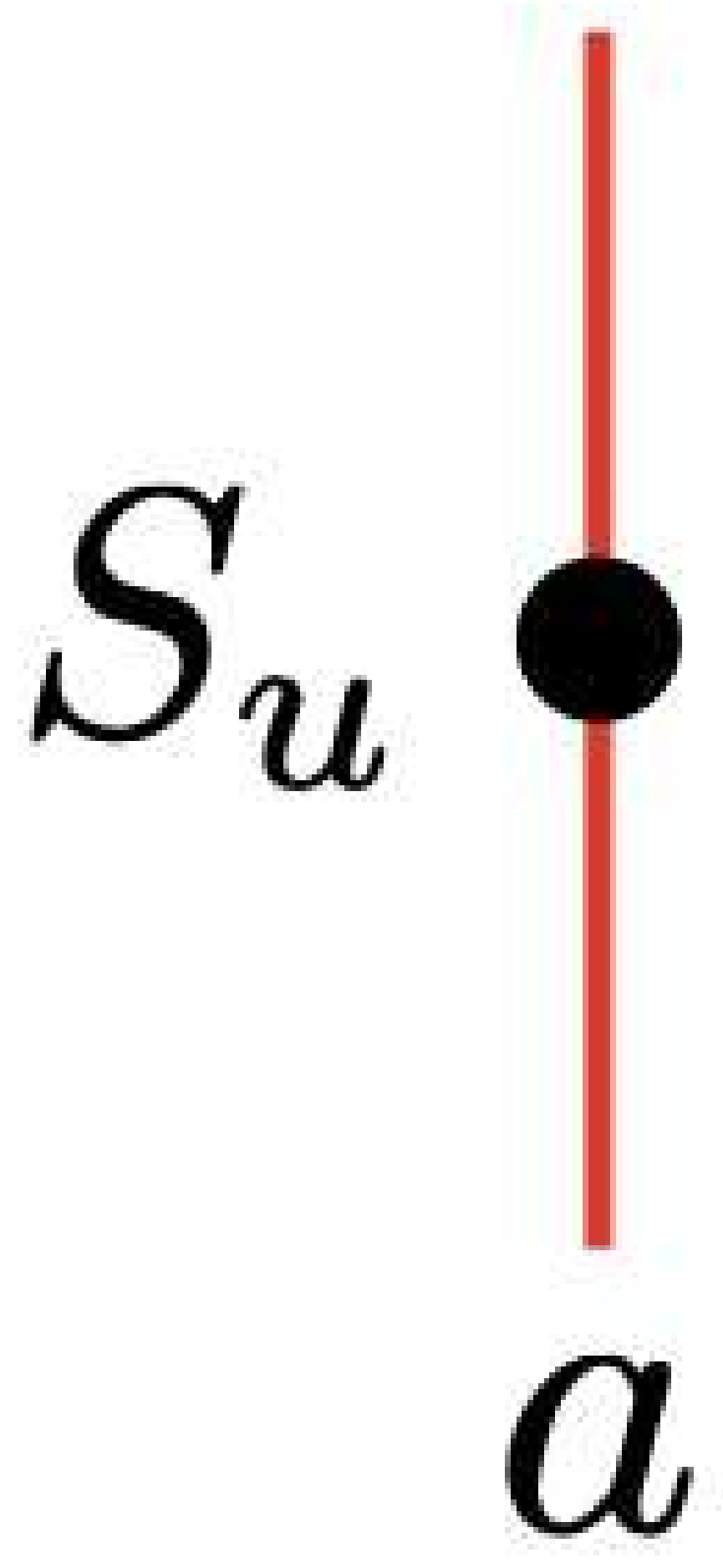} ~ = ~ 
\adjincludegraphics[valign = c, width = 1.9cm]{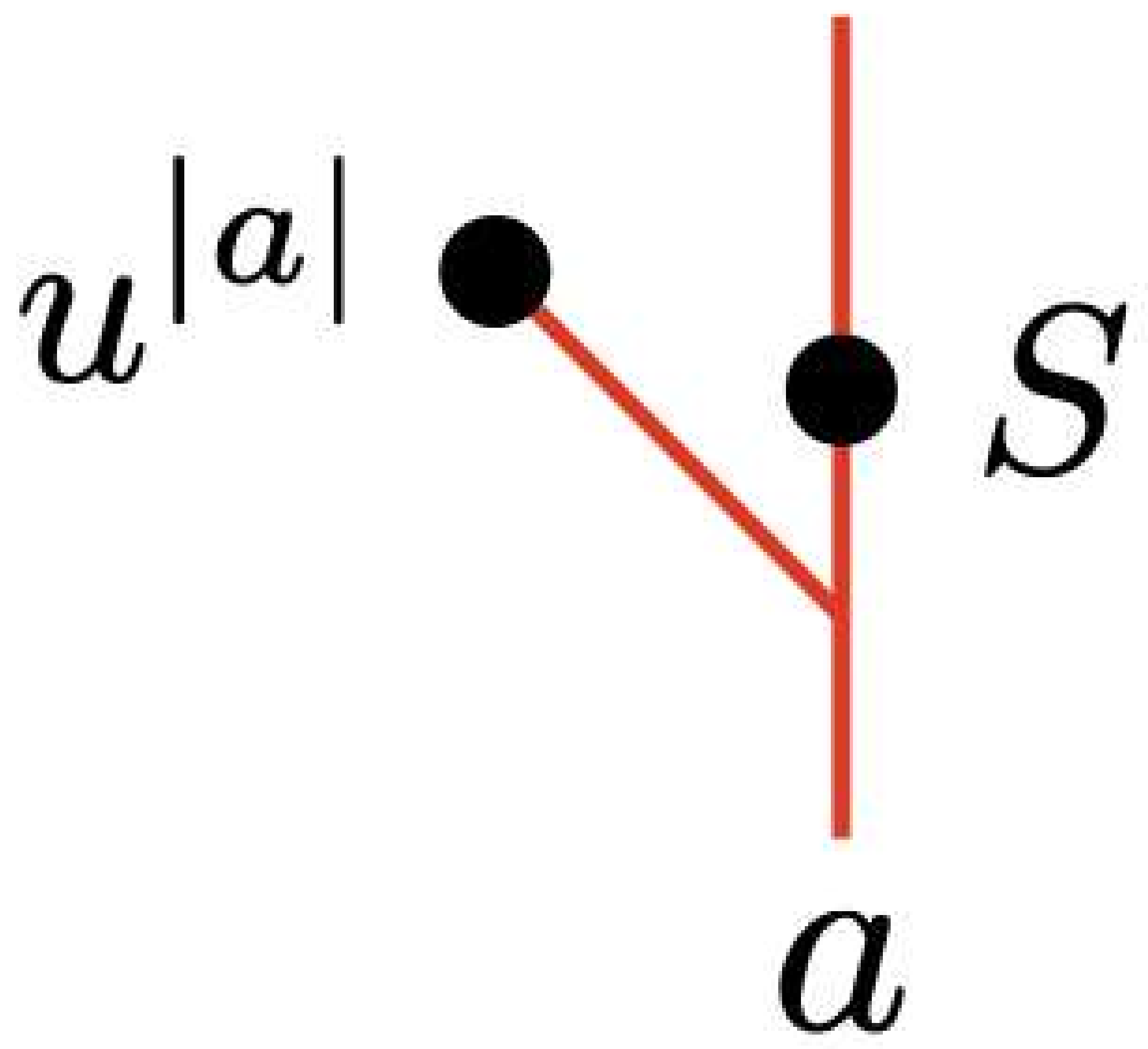}.
\end{equation}
\item The comultiplication $\Delta$, counit $\epsilon$, and unit $\eta$ are the same as those of $H$.
\end{itemize}

Now, we show that $\mathcal{H}^u$ is a weak Hopf superalgebra.
\paragraph{Even structure maps.}
To begin with, we show that the structure maps of $\mathcal{H}^u$ are even with respect to the $\mathbb{Z}_2$-grading \eqref{eq: p_u}.
It suffices to show that the $\mathbb{Z}_2$ group-like element $u \in H^*$ and the structure maps of the original weak Hopf algebra $H$ are even.
The element $u$ is even due to the second equation in eq. \eqref{eq: Z2 group-like}.
The multiplication $m$ is even because
\begin{equation}
\adjincludegraphics[valign = c, width = 1.5cm]{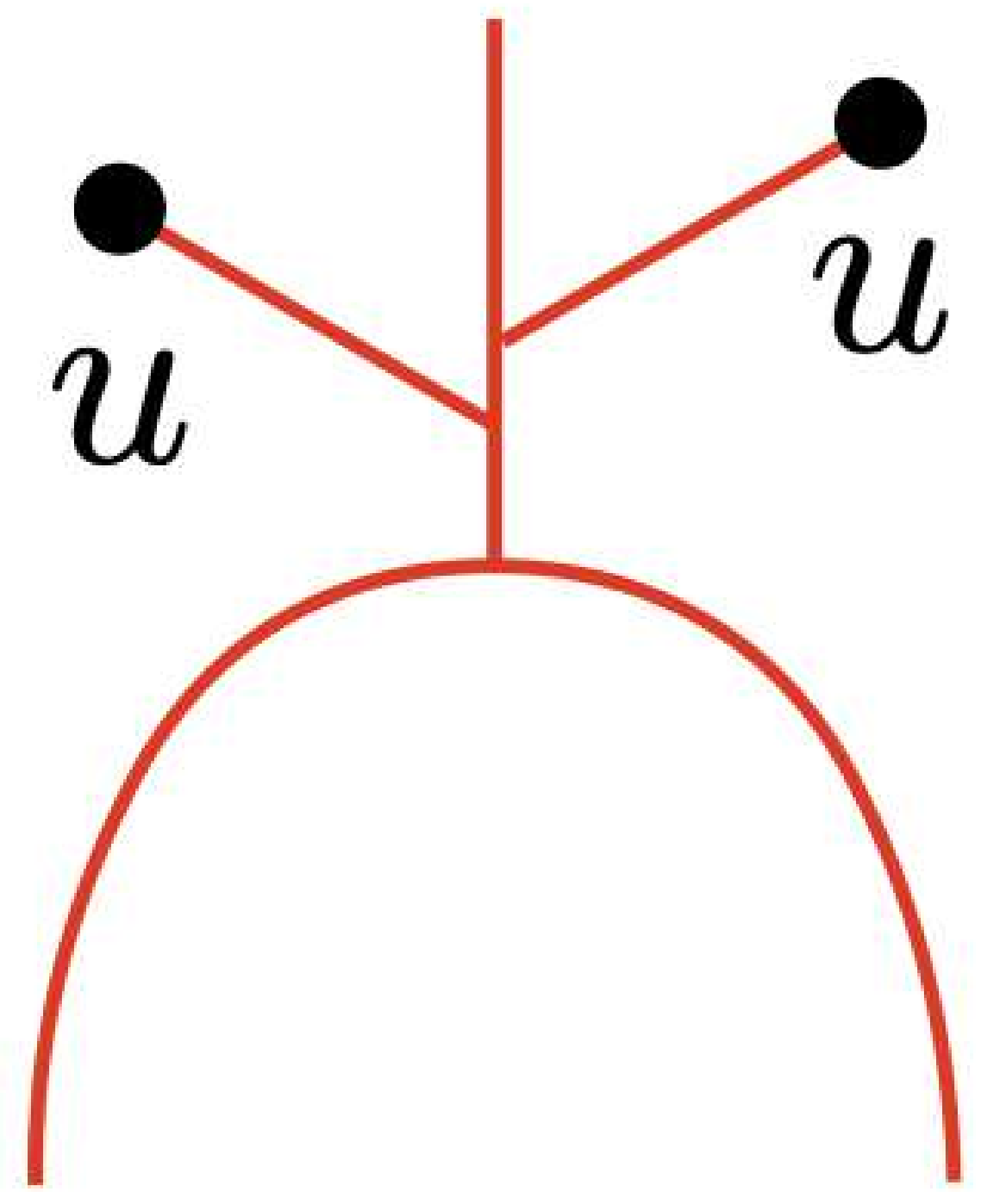} ~ = ~ 
\adjincludegraphics[valign = c, width = 2.2cm]{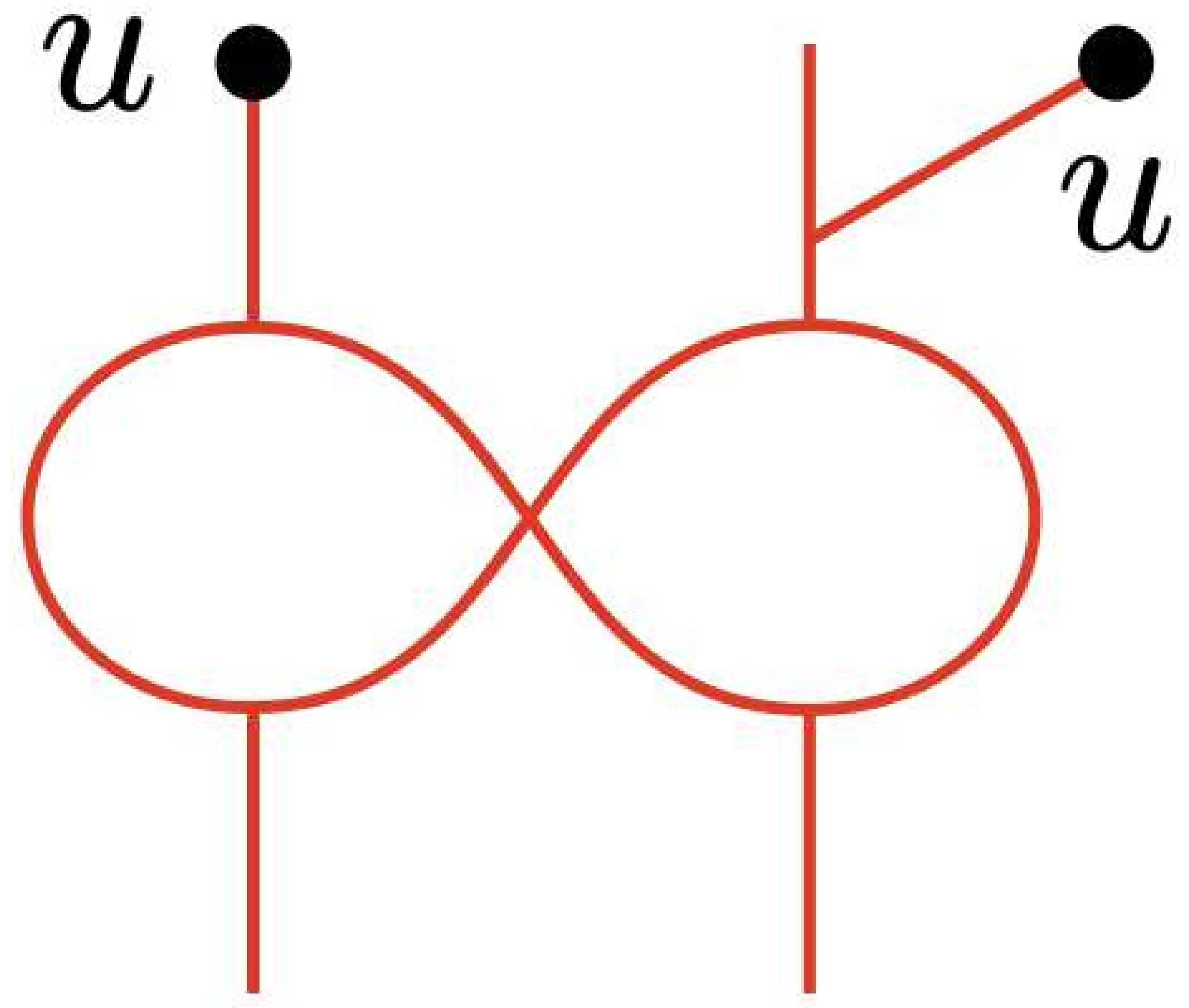} ~ = ~
\adjincludegraphics[valign = c, width = 2.5cm]{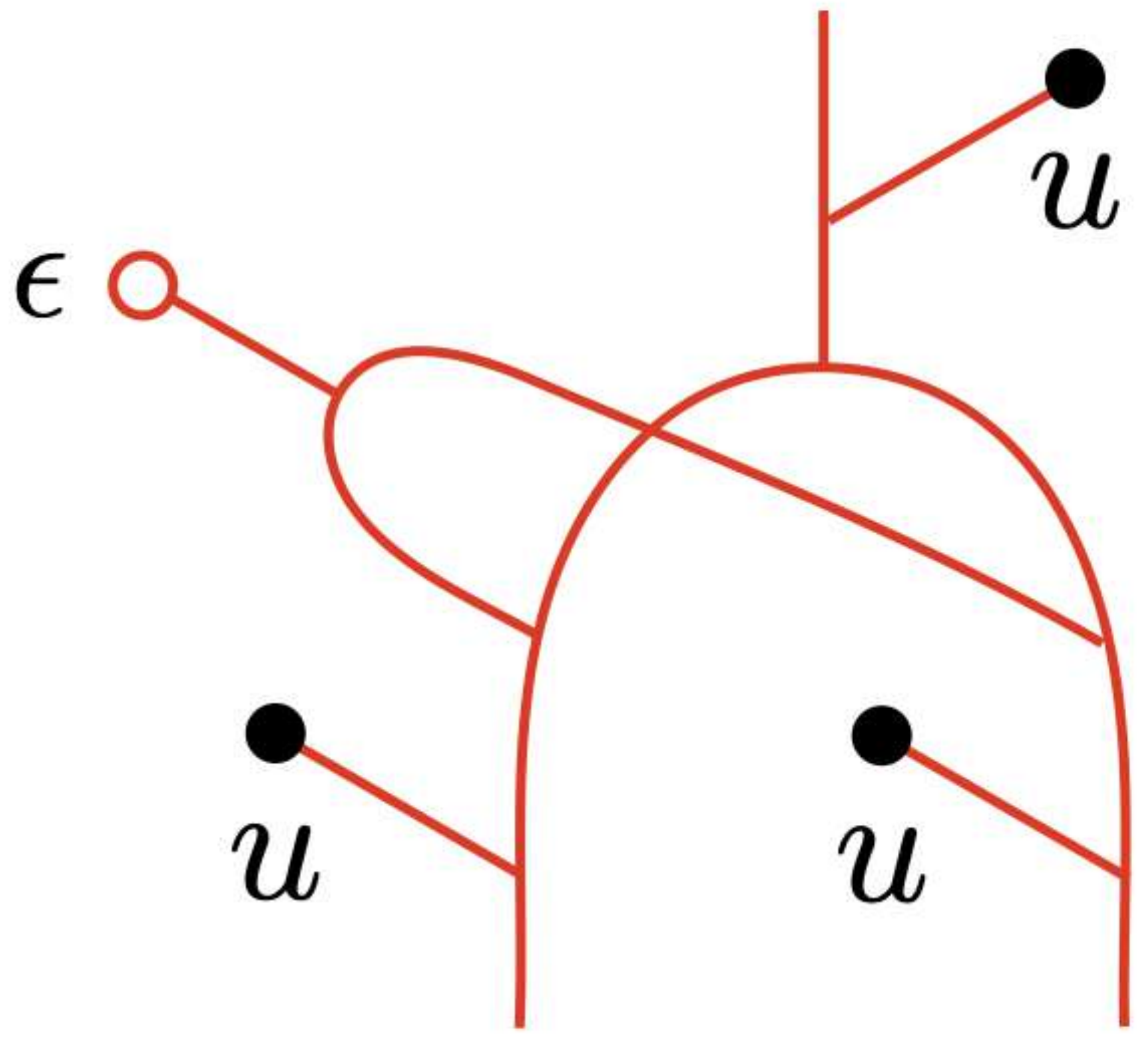} ~ = ~
\adjincludegraphics[valign = c, width = 2.2cm]{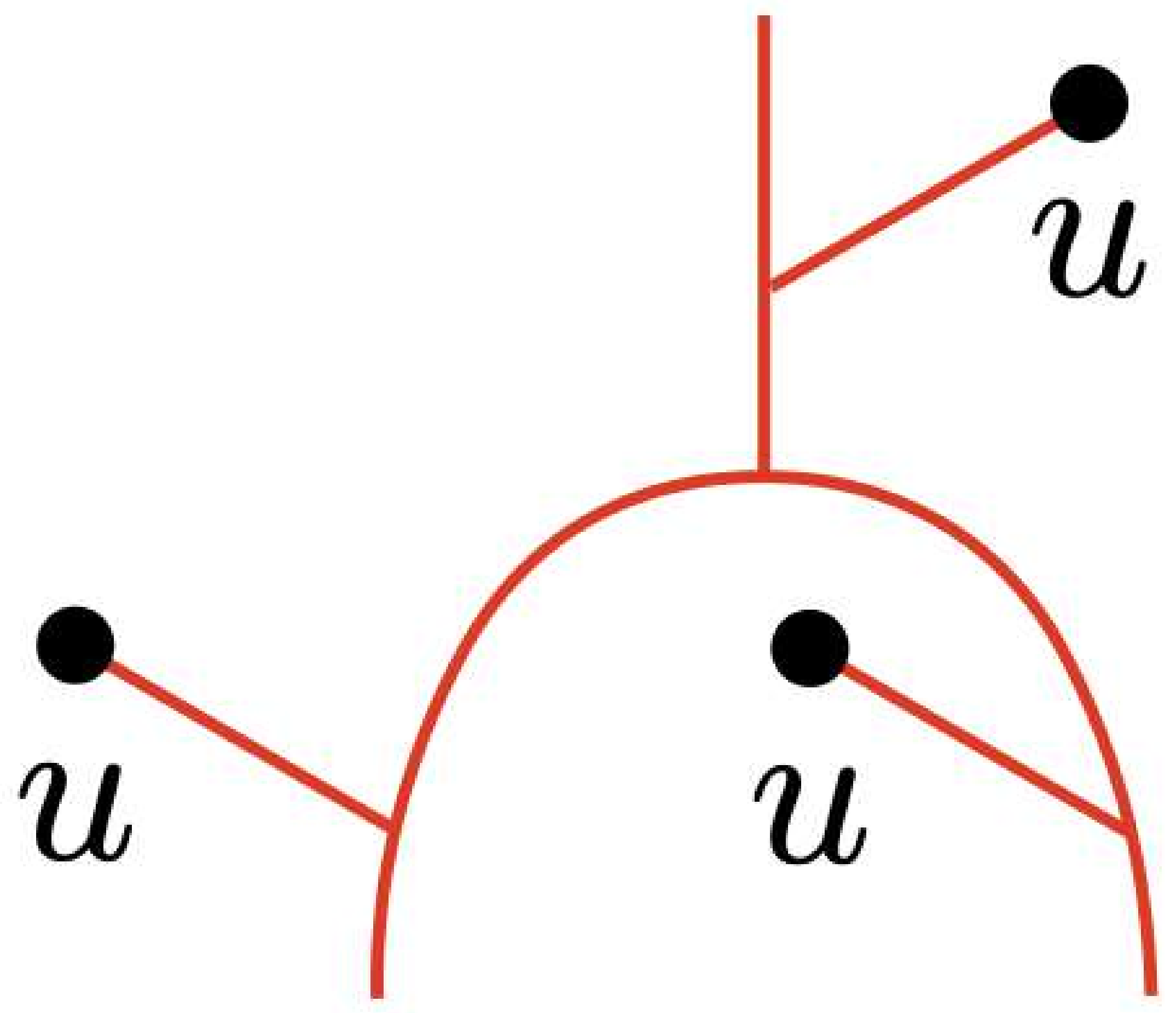} ~ = ~
\adjincludegraphics[valign = c, width = 2.4cm]{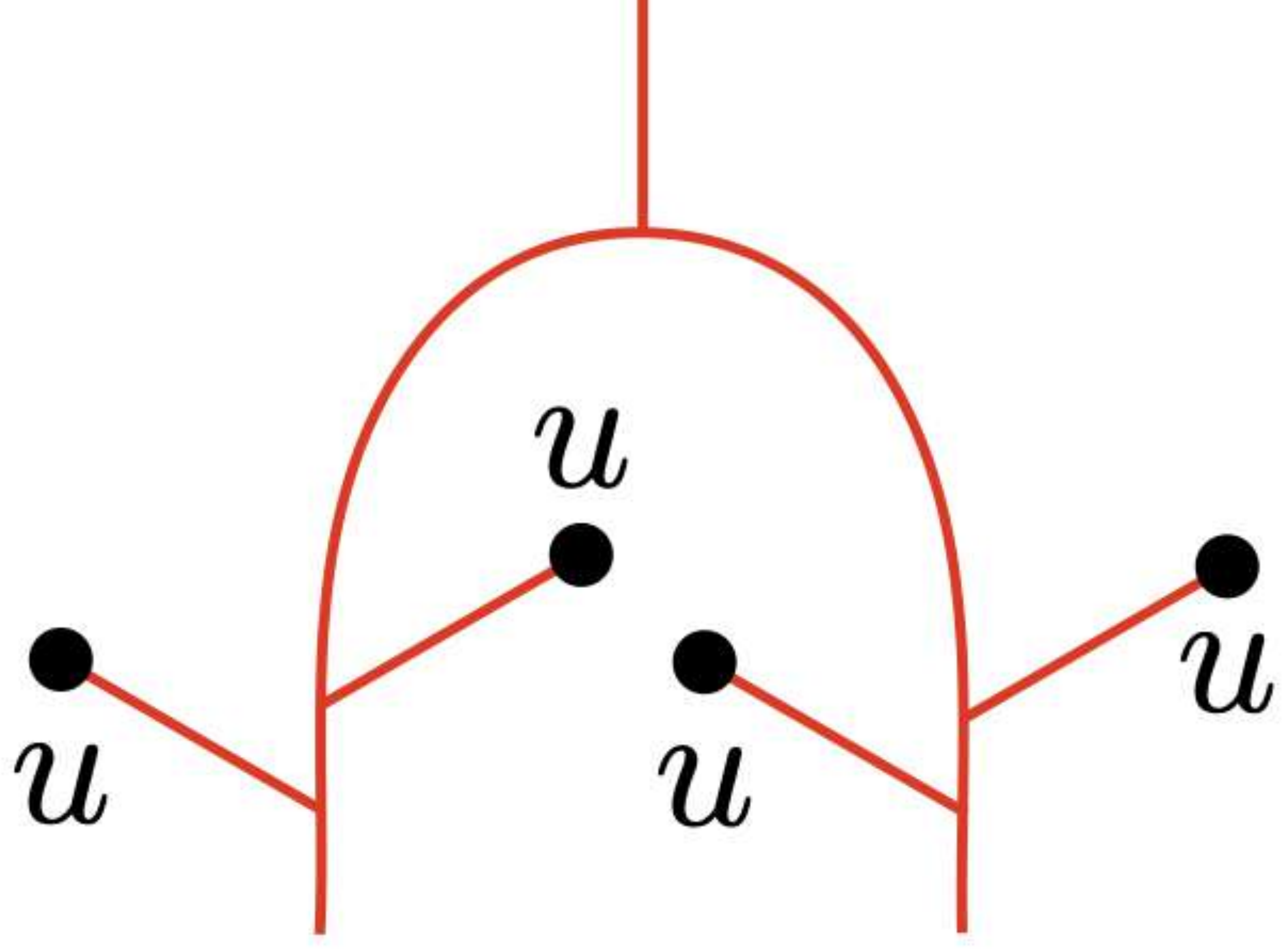}.
\end{equation}
The comultiplication $\Delta$ is even due to the coassociativity of $\Delta$ and the second equation in eq. \eqref{eq: Z2 group-like}:
\begin{equation}
\adjincludegraphics[valign = c, width = 2.4cm]{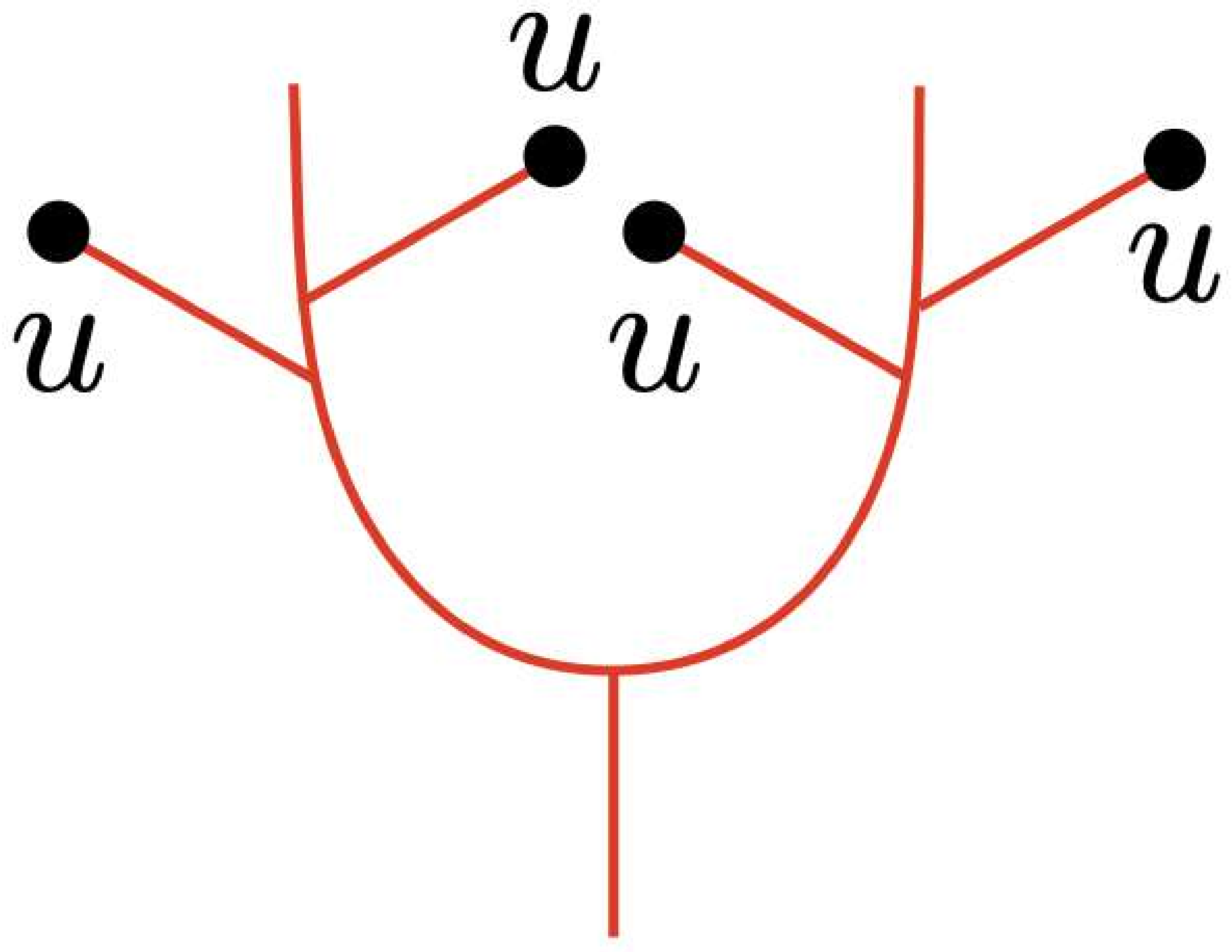} ~ = ~
\adjincludegraphics[valign = c, width = 1.5cm]{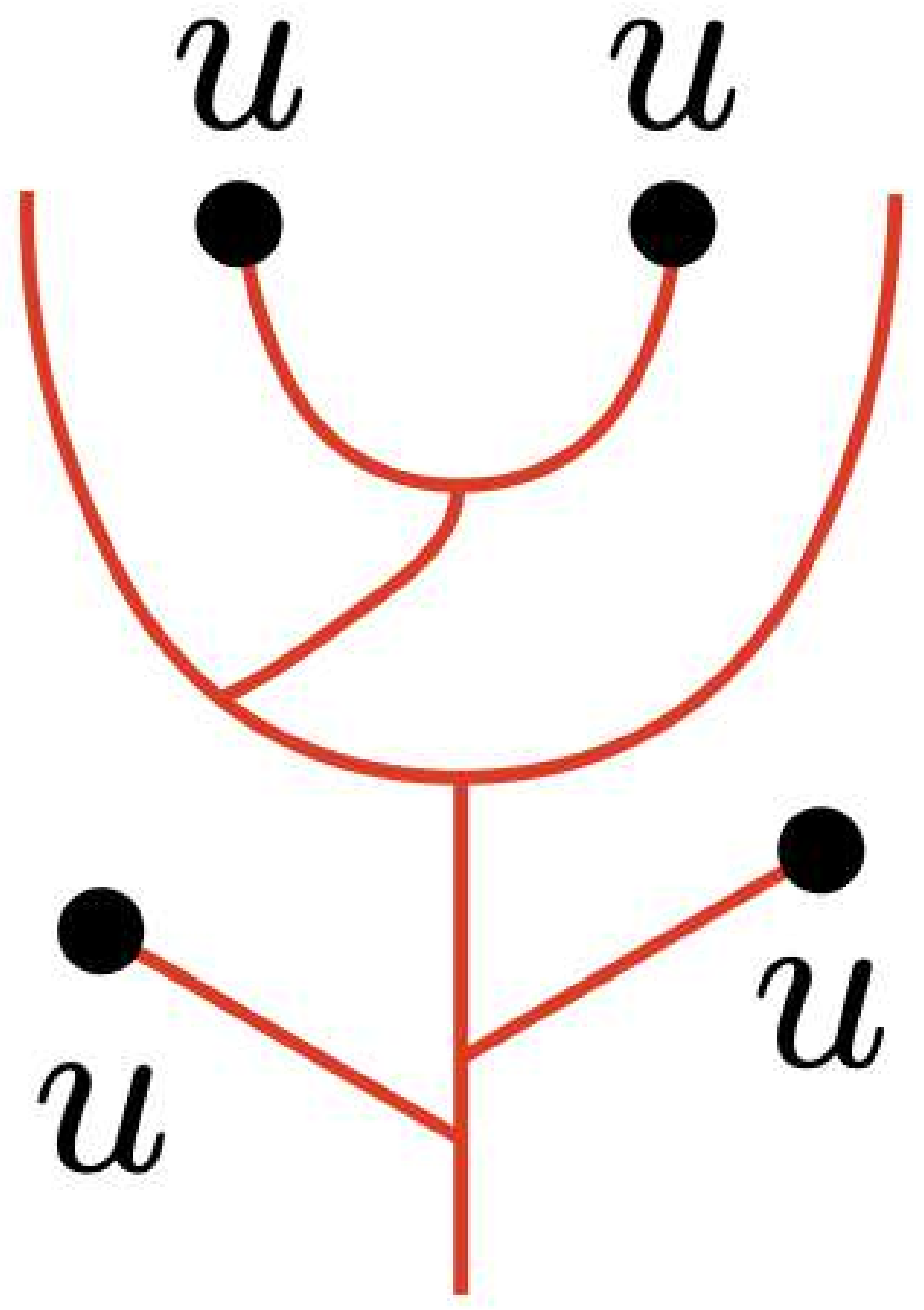} ~ = ~
\adjincludegraphics[valign = c, width = 1.5cm]{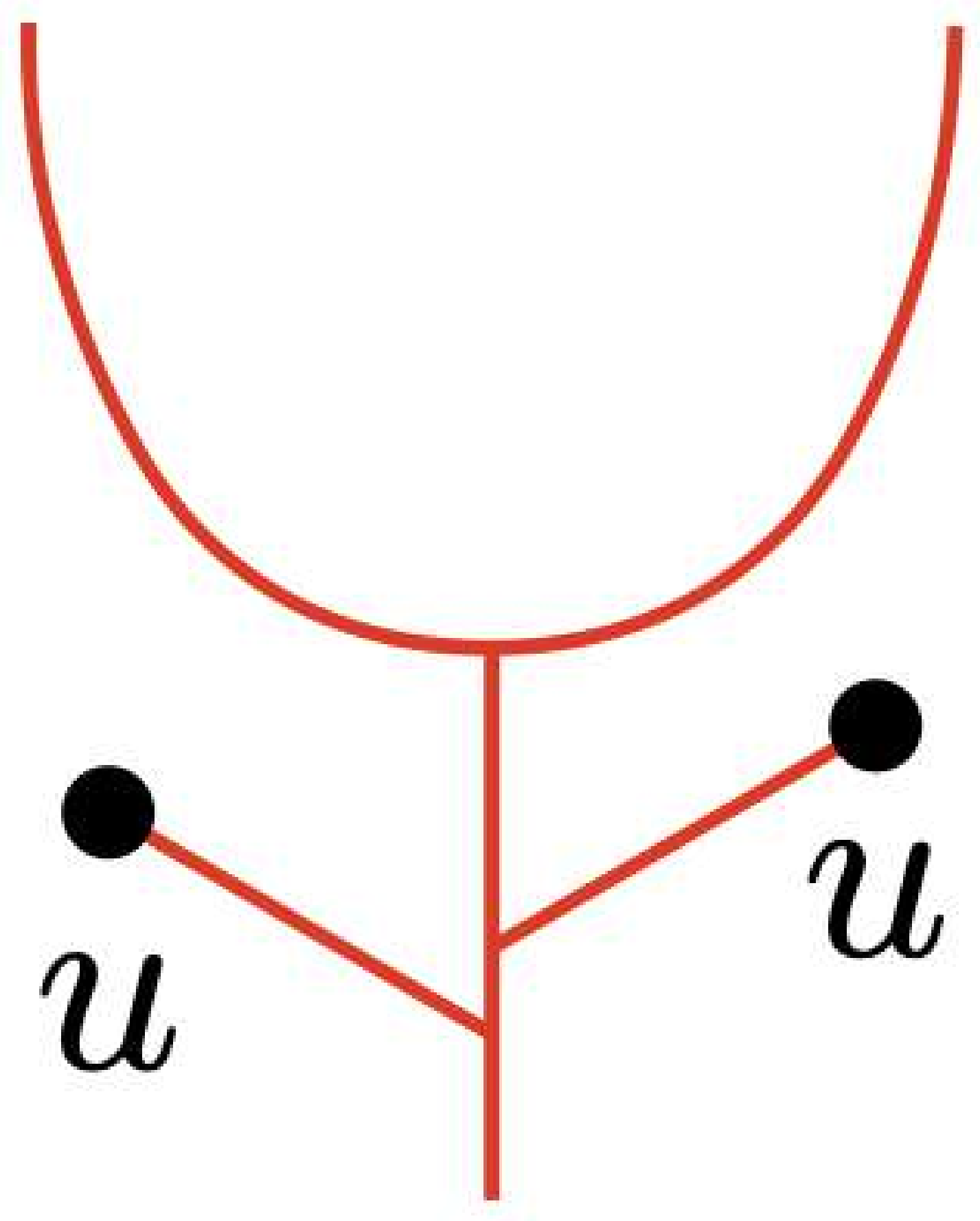}.
\end{equation}
The antipode $S$ is also even, which follows from the fact that $S$ is an algebra and coalgebra homomorphism:
\begin{equation}
\adjincludegraphics[valign = c, width = 1.5cm]{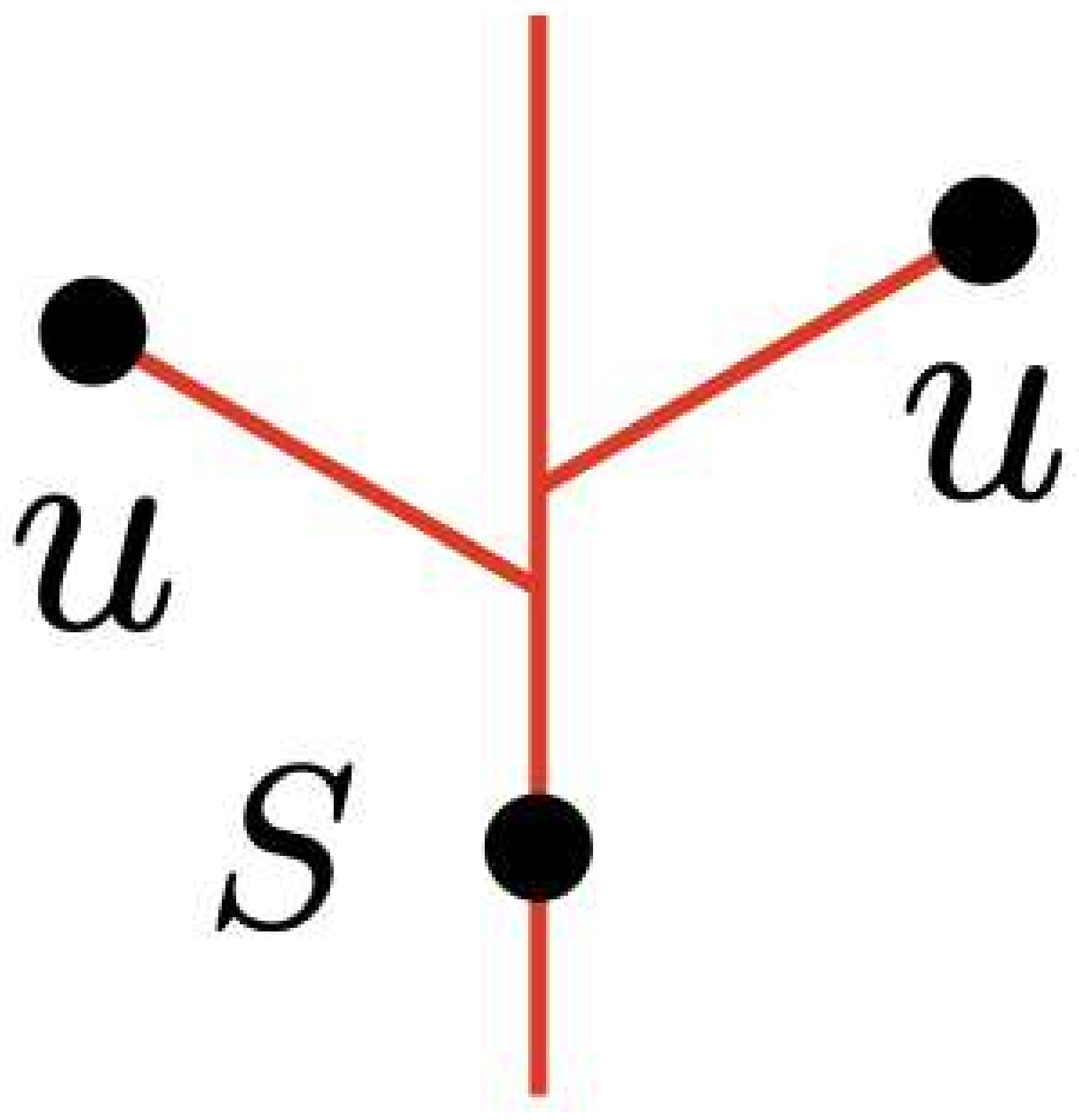} ~ = ~
\adjincludegraphics[valign = c, width = 1.2cm]{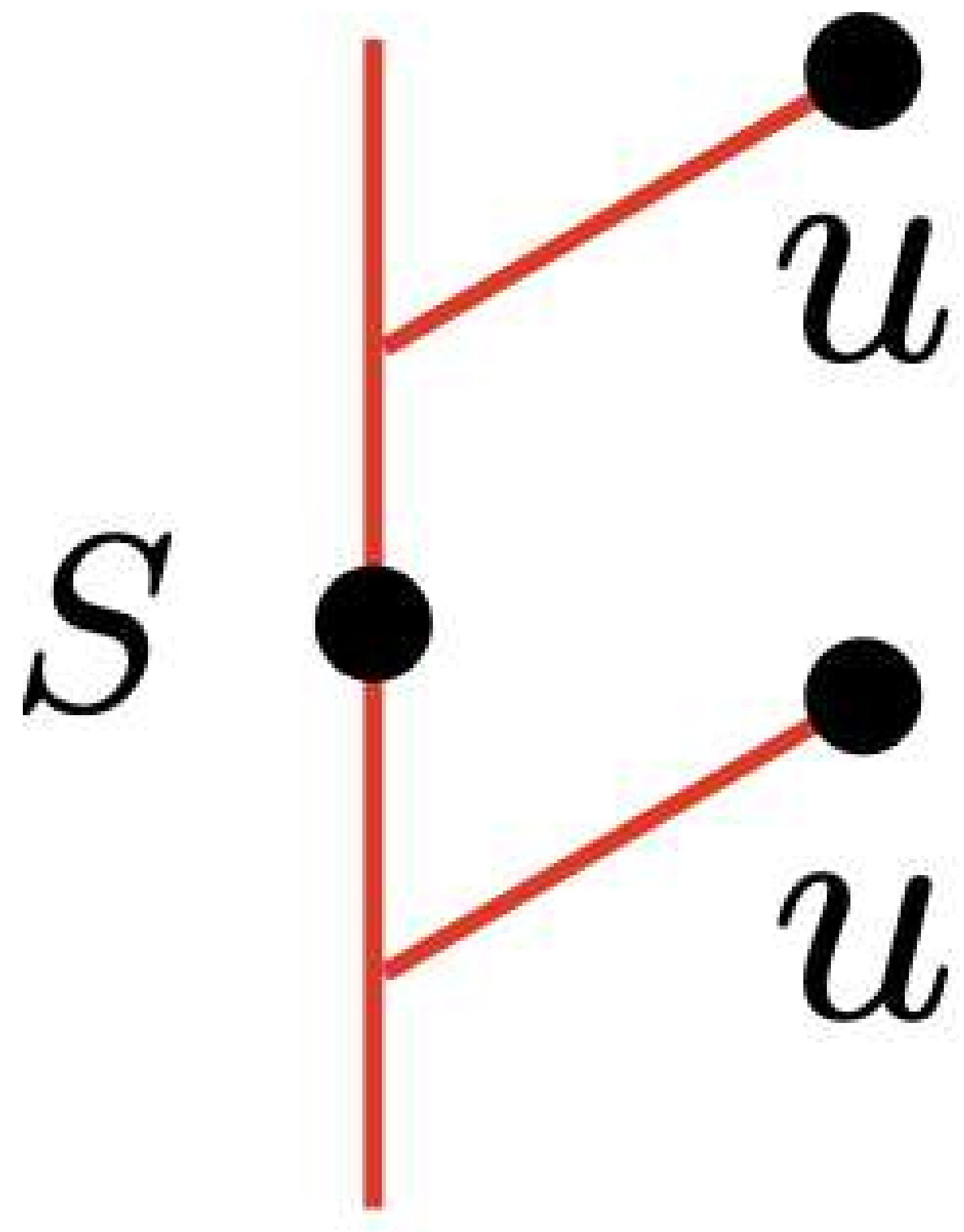} ~ = ~
\adjincludegraphics[valign = c, width = 1.5cm]{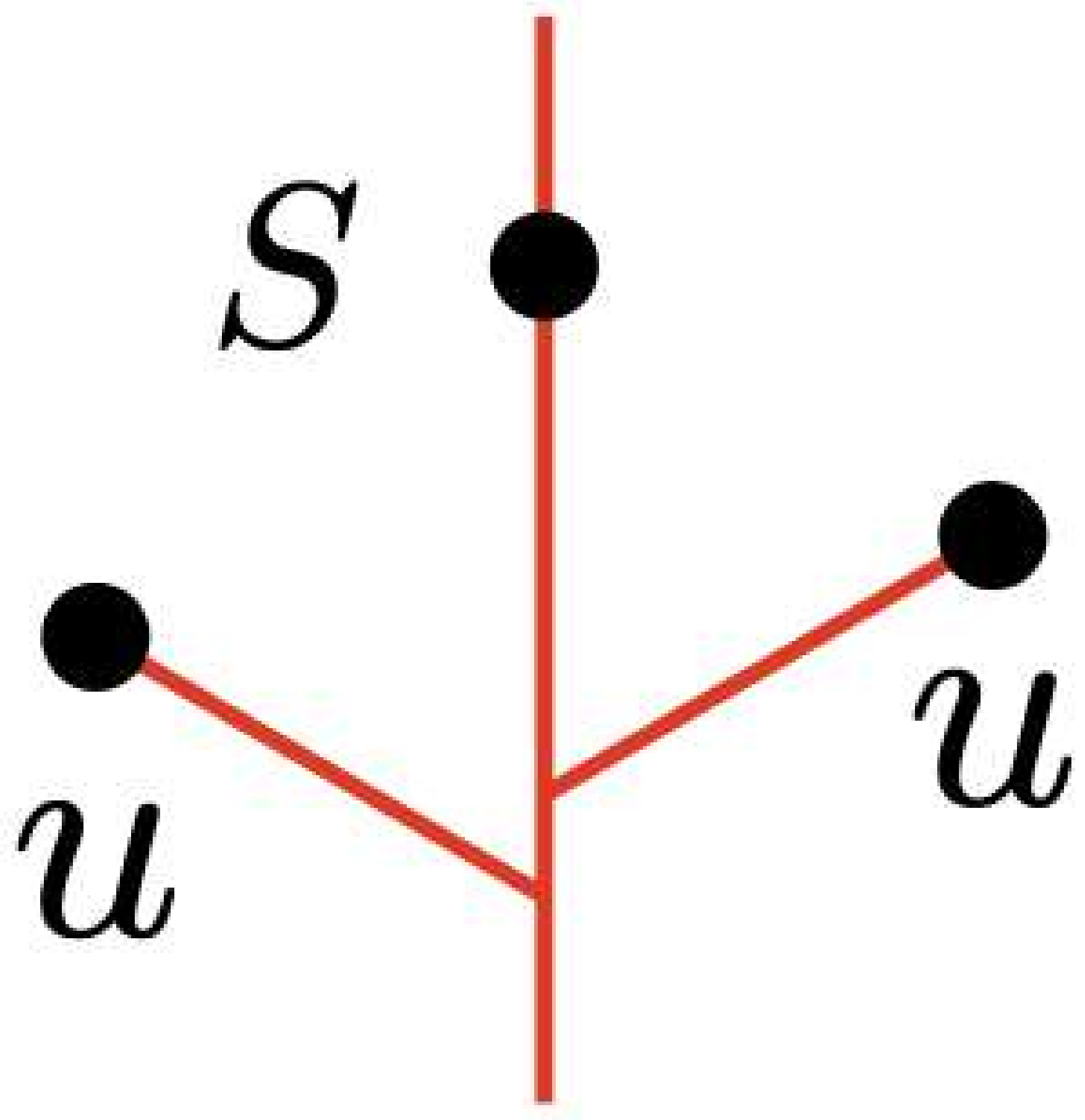}.
\end{equation}
The unit $\eta$ and counit $\epsilon$ are automatically even because the multiplication $m$ and comultiplication $\Delta$ are even.
Therefore, the structure maps of $\mathcal{H}^u$ are all even.

\paragraph{Superalgebra and supercoalgebra structures.}
Next, we show that $(\mathcal{H}^u, m_u, \eta)$ is an associative unital superalgebra and $(\mathcal{H}^u, \Delta, \epsilon)$ is a coassociative counital supercoalgebra.
The associativity of the multiplication $m_u$ is an immediate consequence of the definition of $m_u$:
\begin{equation*}
\adjincludegraphics[valign = c, width = 2.6cm]{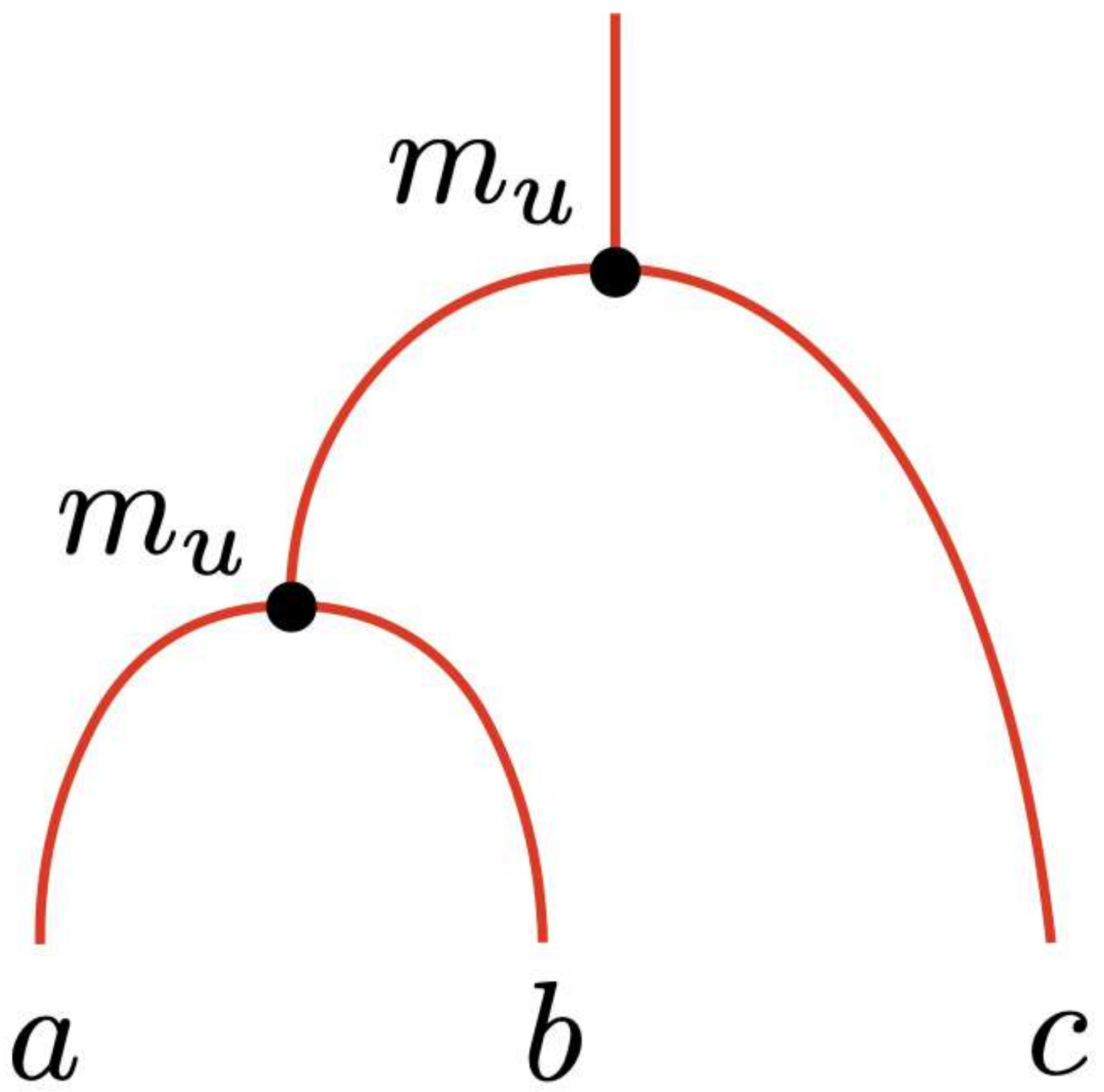} ~ = ~
\adjincludegraphics[valign = c, width = 2.6cm]{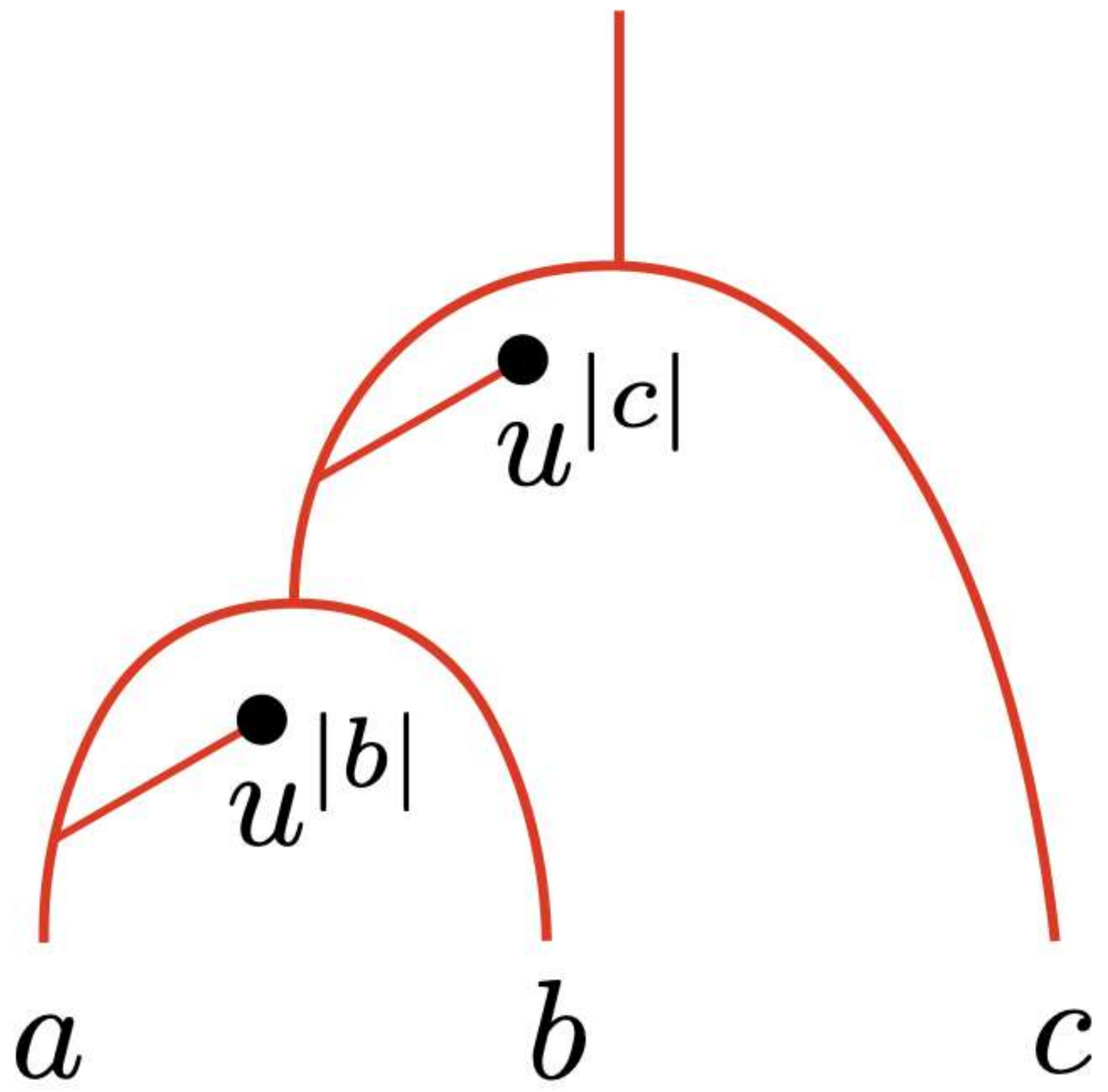} ~ = ~
\adjincludegraphics[valign = c, width = 2.6cm]{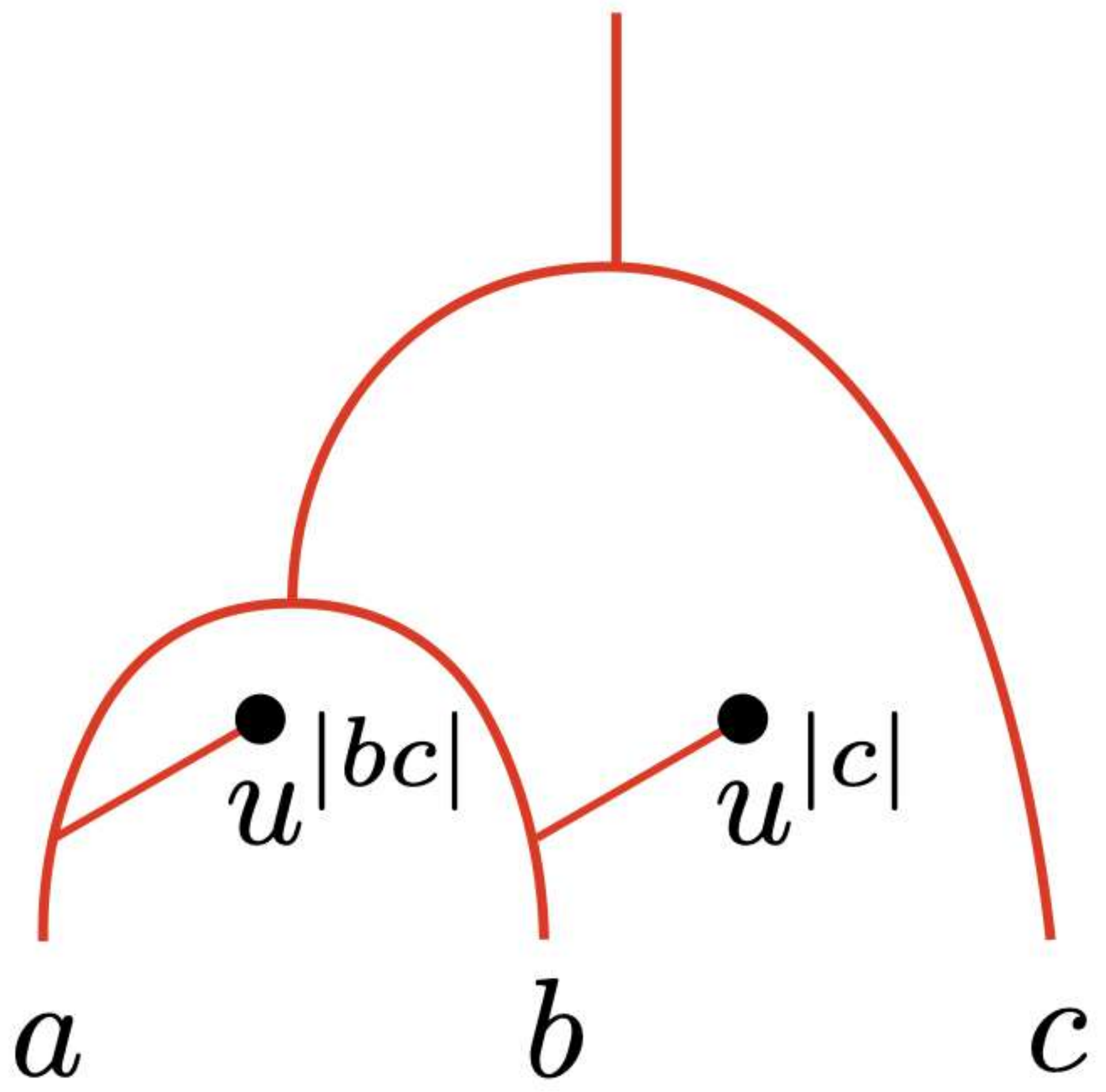} ~ = ~
\adjincludegraphics[valign = c, width = 2.6cm]{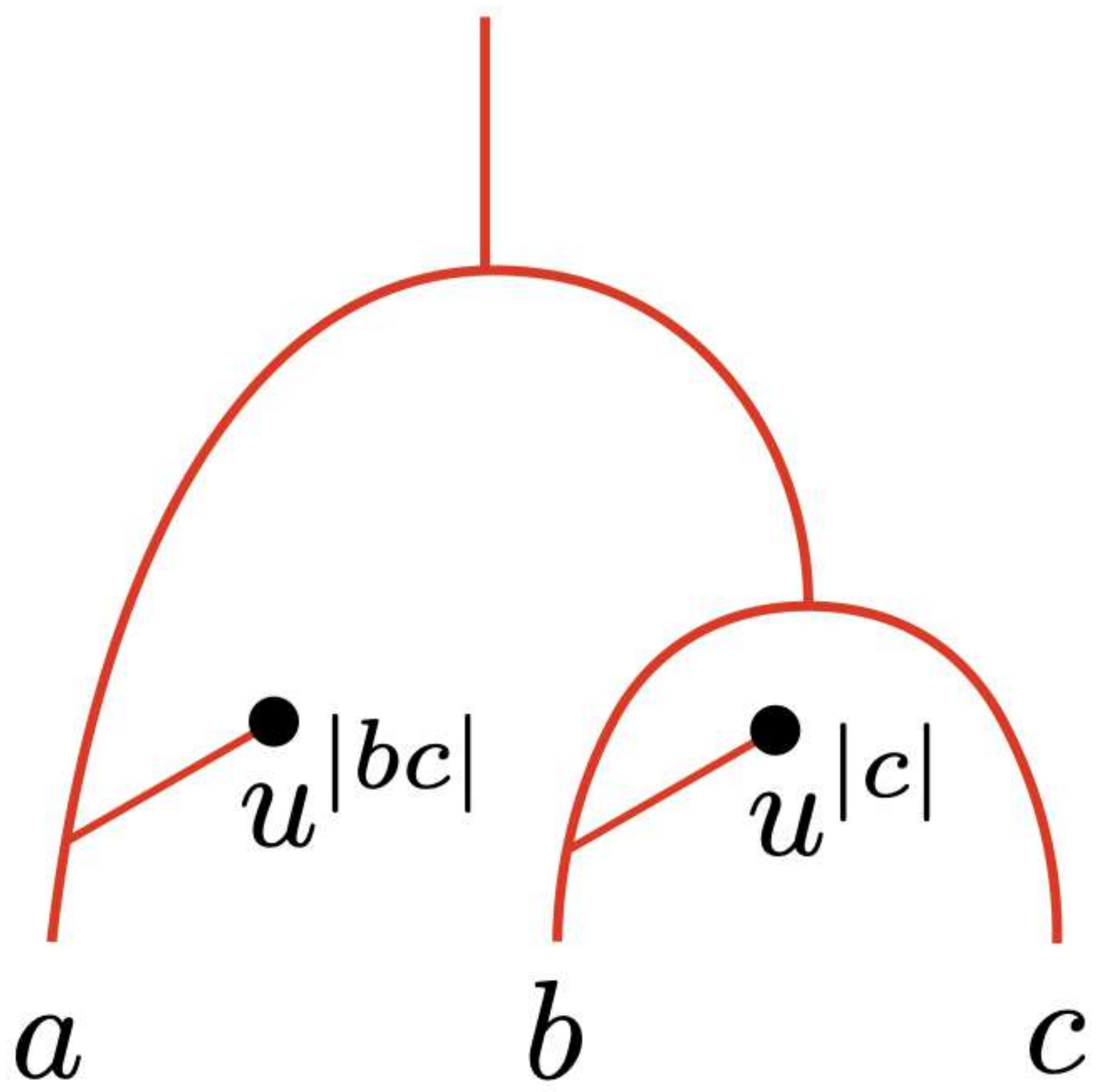} ~ = ~
\adjincludegraphics[valign = c, width = 2.6cm]{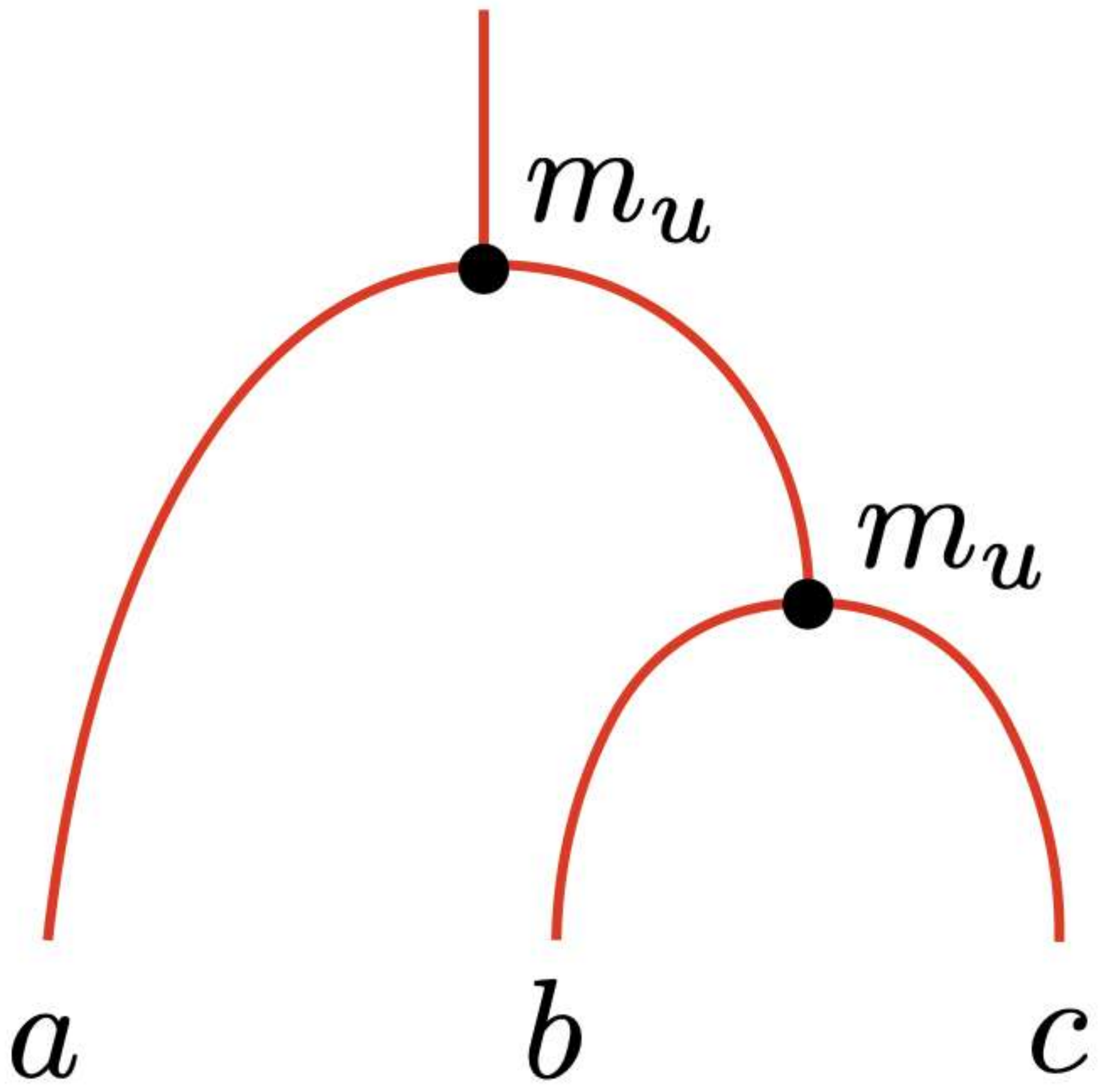}.
\end{equation*}
The unit $\eta$ of the original weak Hopf algebra behaves also as the unit of $\mathcal{H}^u$:
\begin{equation}
\adjincludegraphics[valign = c, width = 1.8cm]{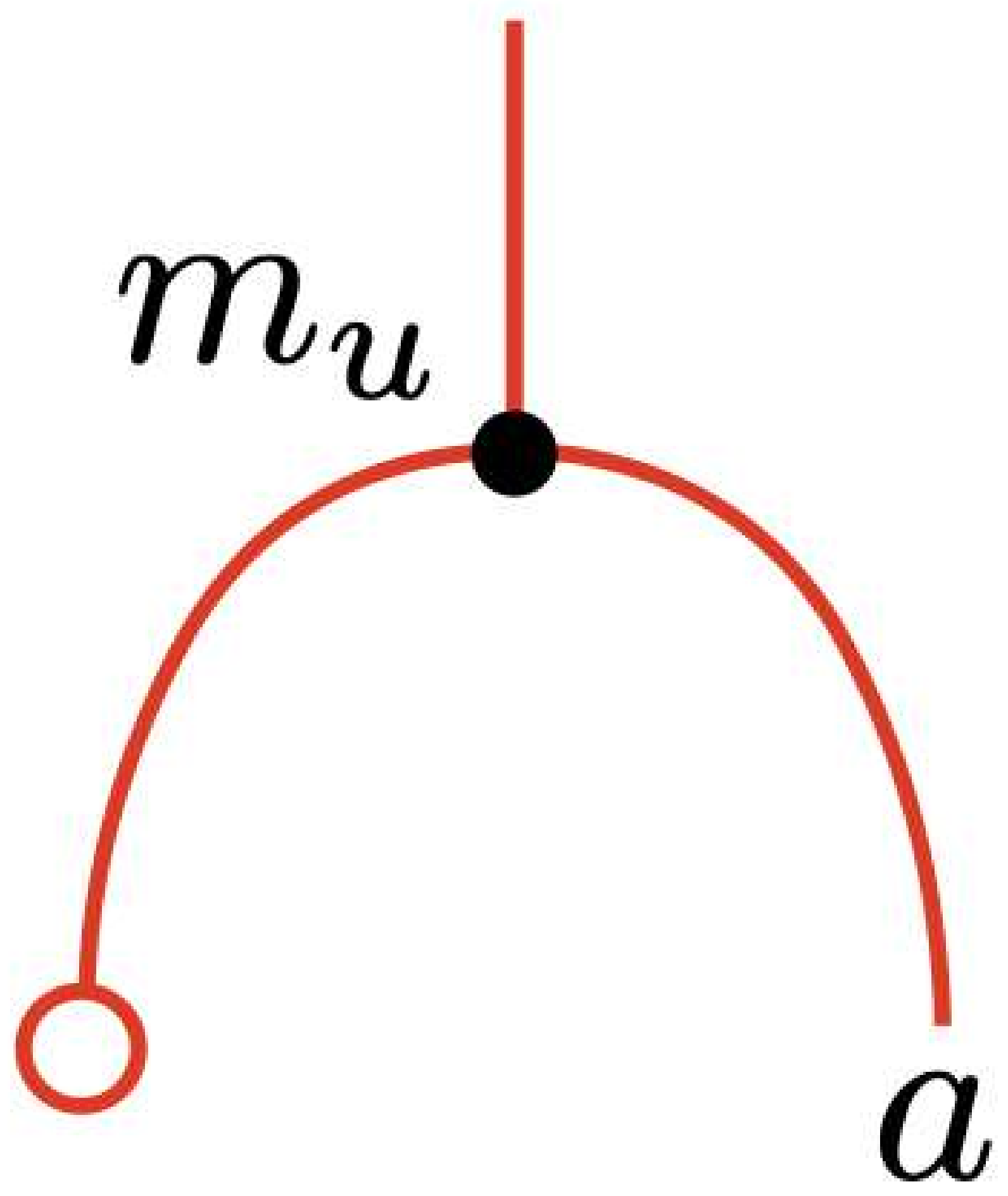} ~ = ~
\adjincludegraphics[valign = c, width = 1.8cm]{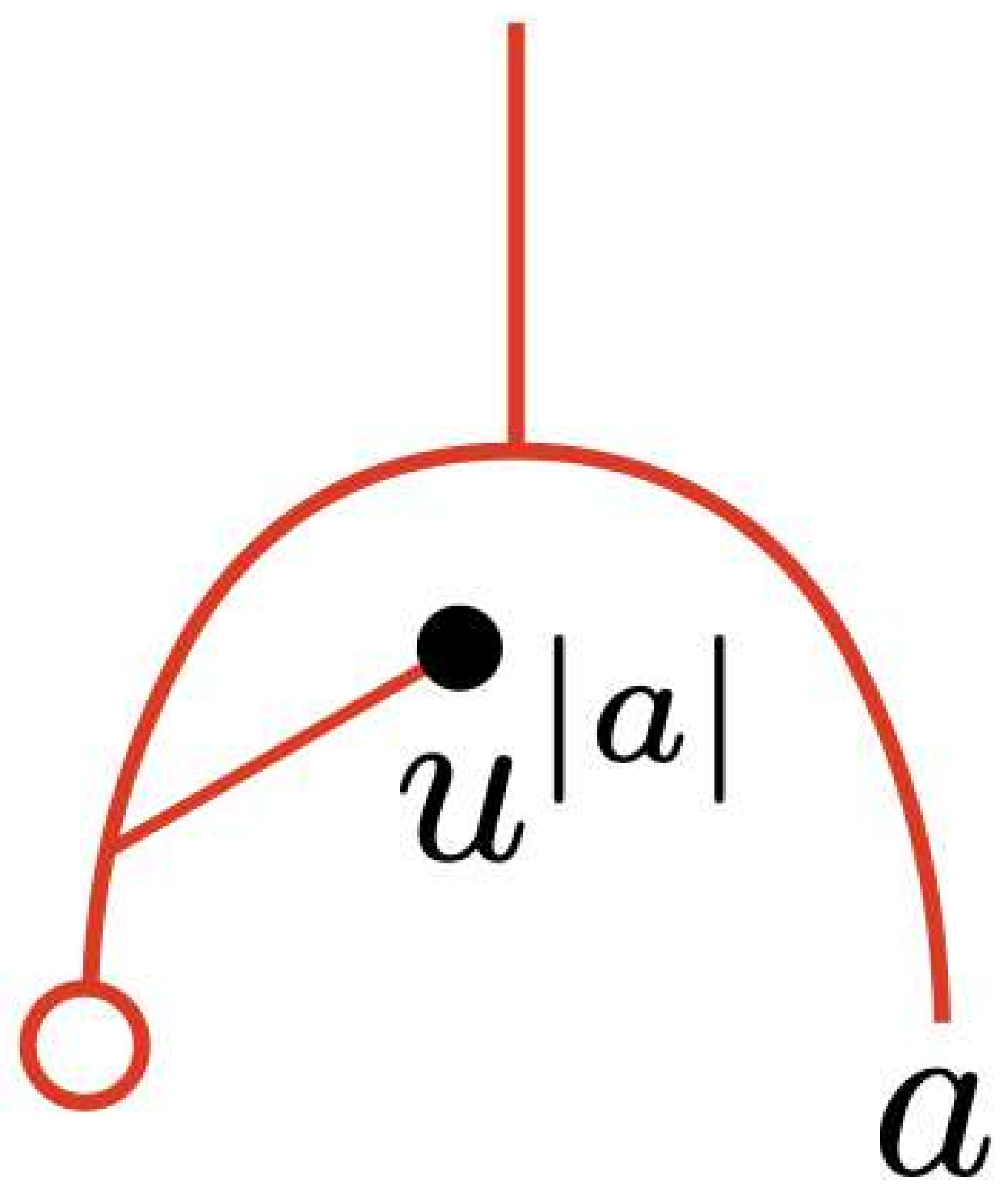} ~ = ~
\adjincludegraphics[valign = c, width = 0.25cm]{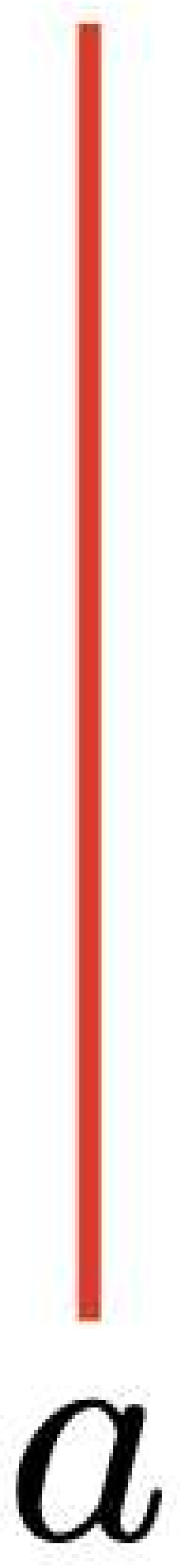} ~ = ~
\adjincludegraphics[valign = c, width = 1.8cm]{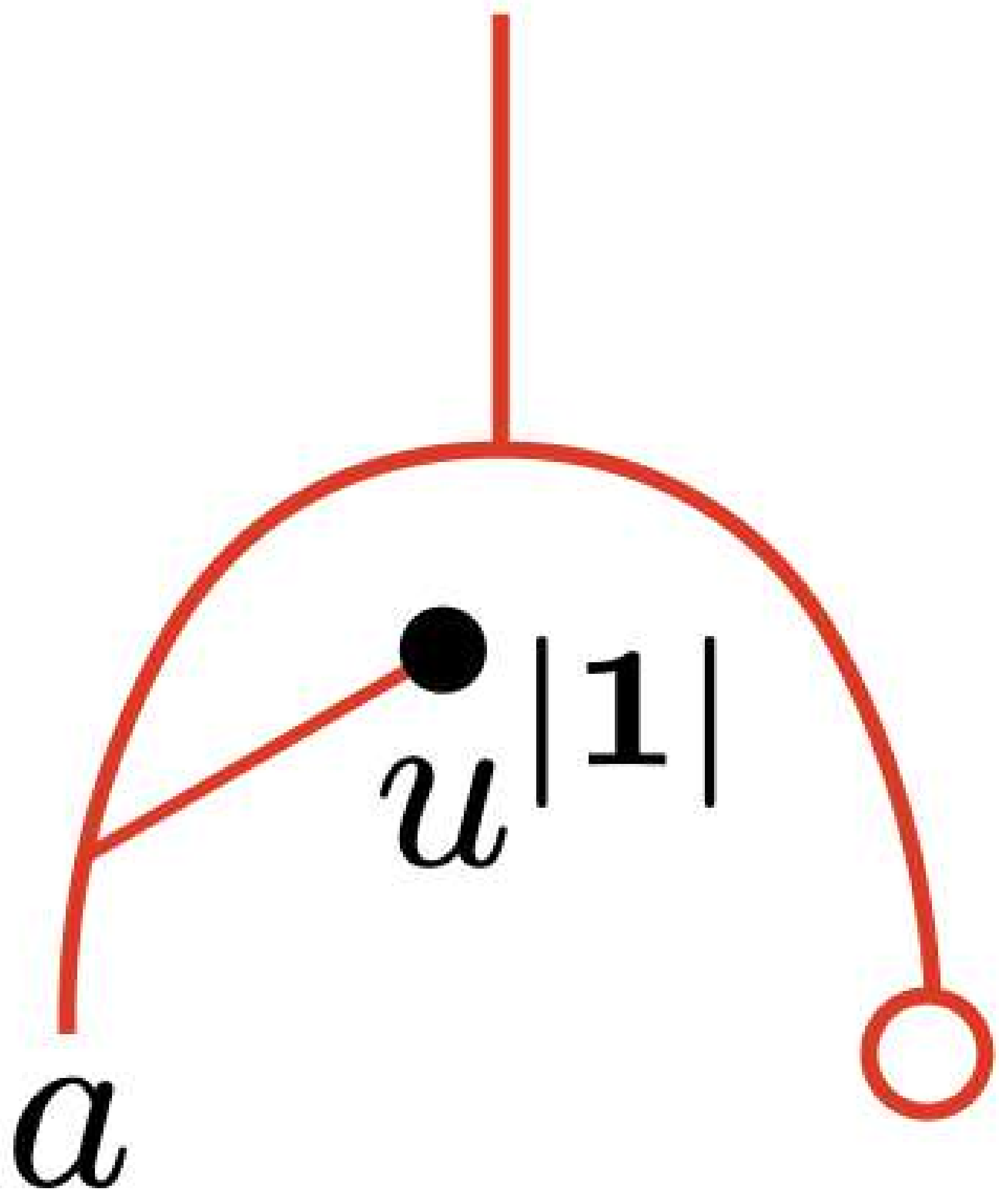} ~ = ~
\adjincludegraphics[valign = c, width = 1.8cm]{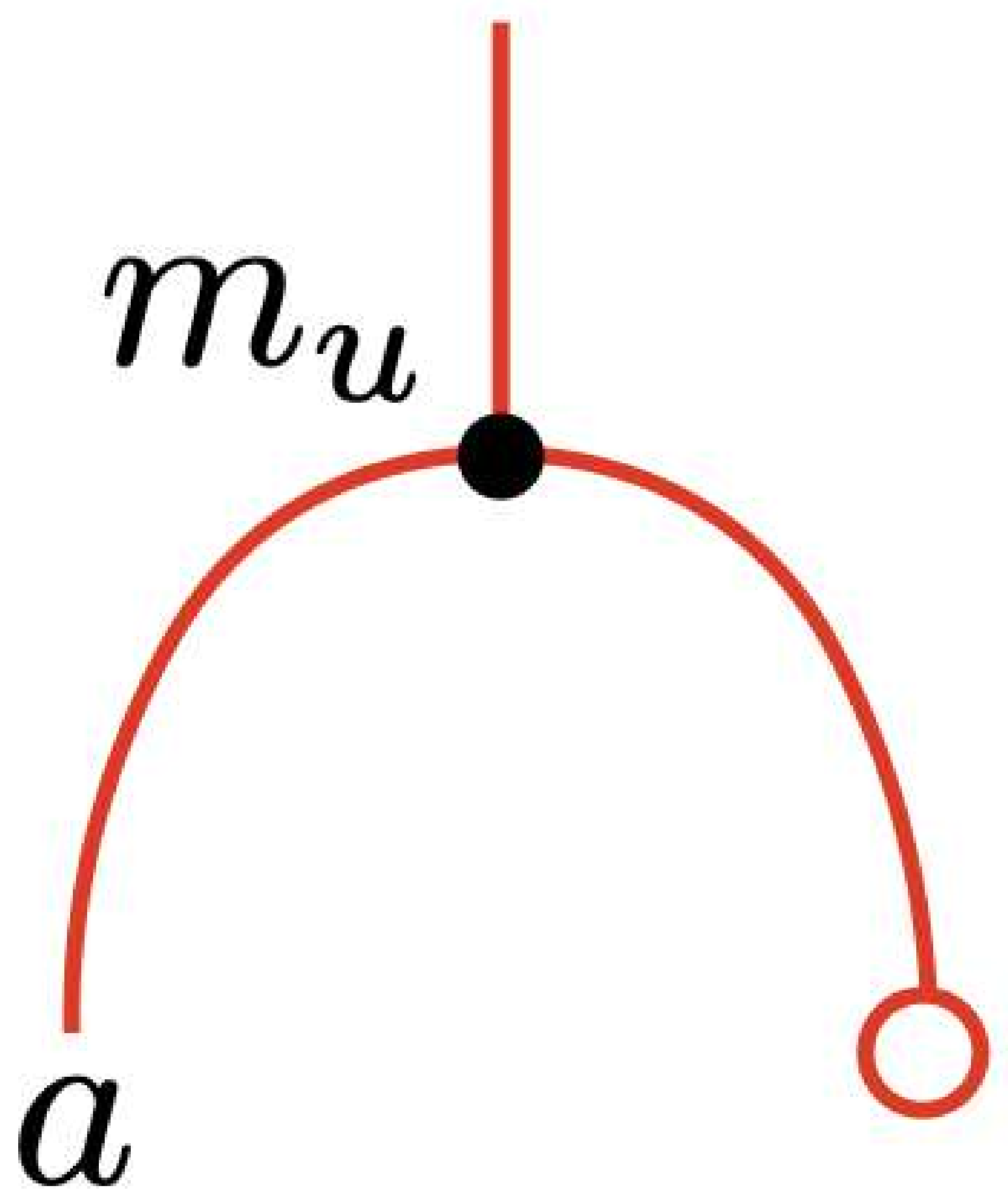},
\end{equation}
where $\mathbf{1} := \eta(1) \in \mathcal{H}^u$.
In the second equality of the above equation, we used the relation $(\mathrm{id} \otimes u) \circ \Delta \circ \eta = \eta$, which can be derived as follows:
\begin{equation}
\adjincludegraphics[valign = c, width = 0.9cm]{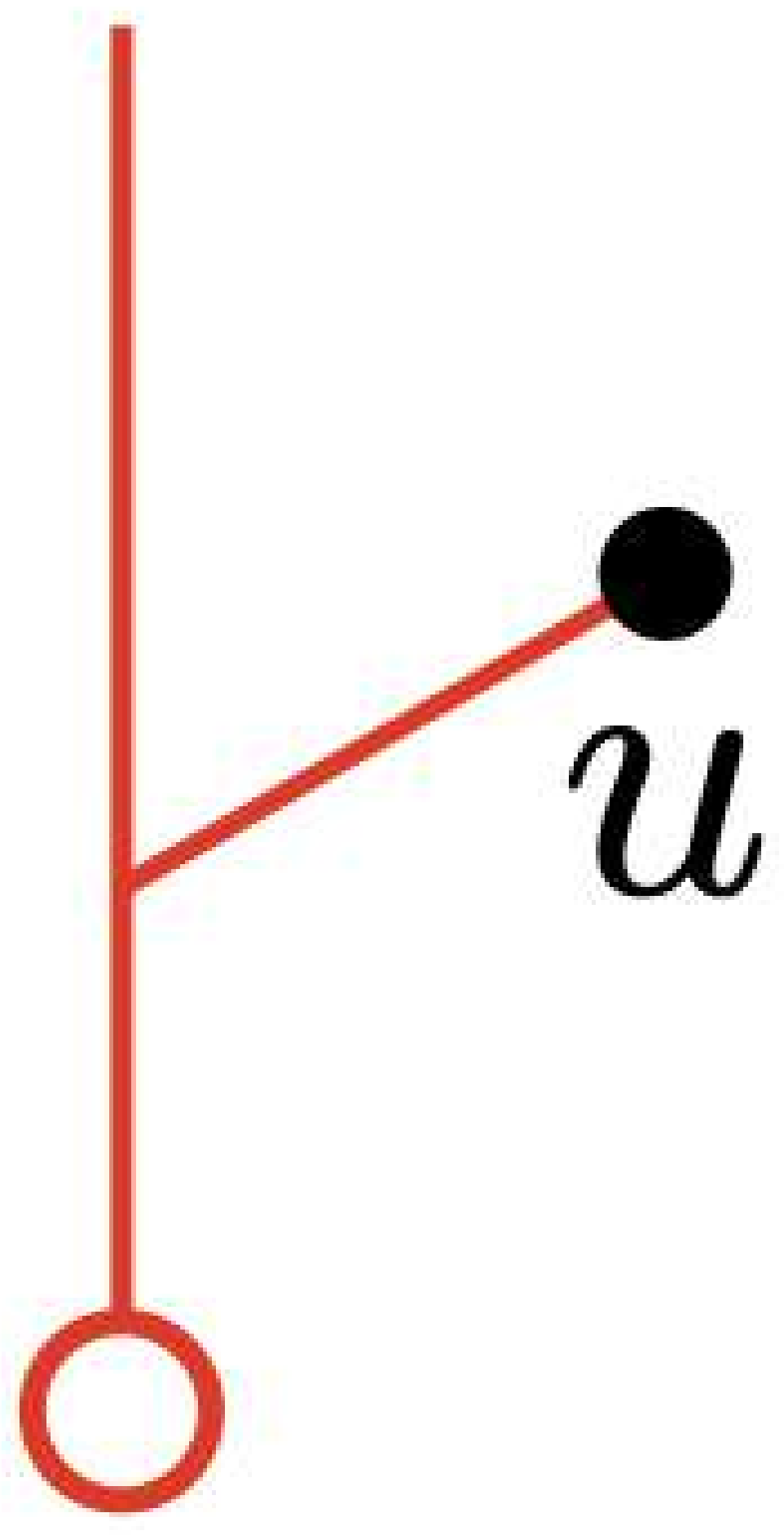} ~ = ~
\adjincludegraphics[valign = c, width = 2.2cm]{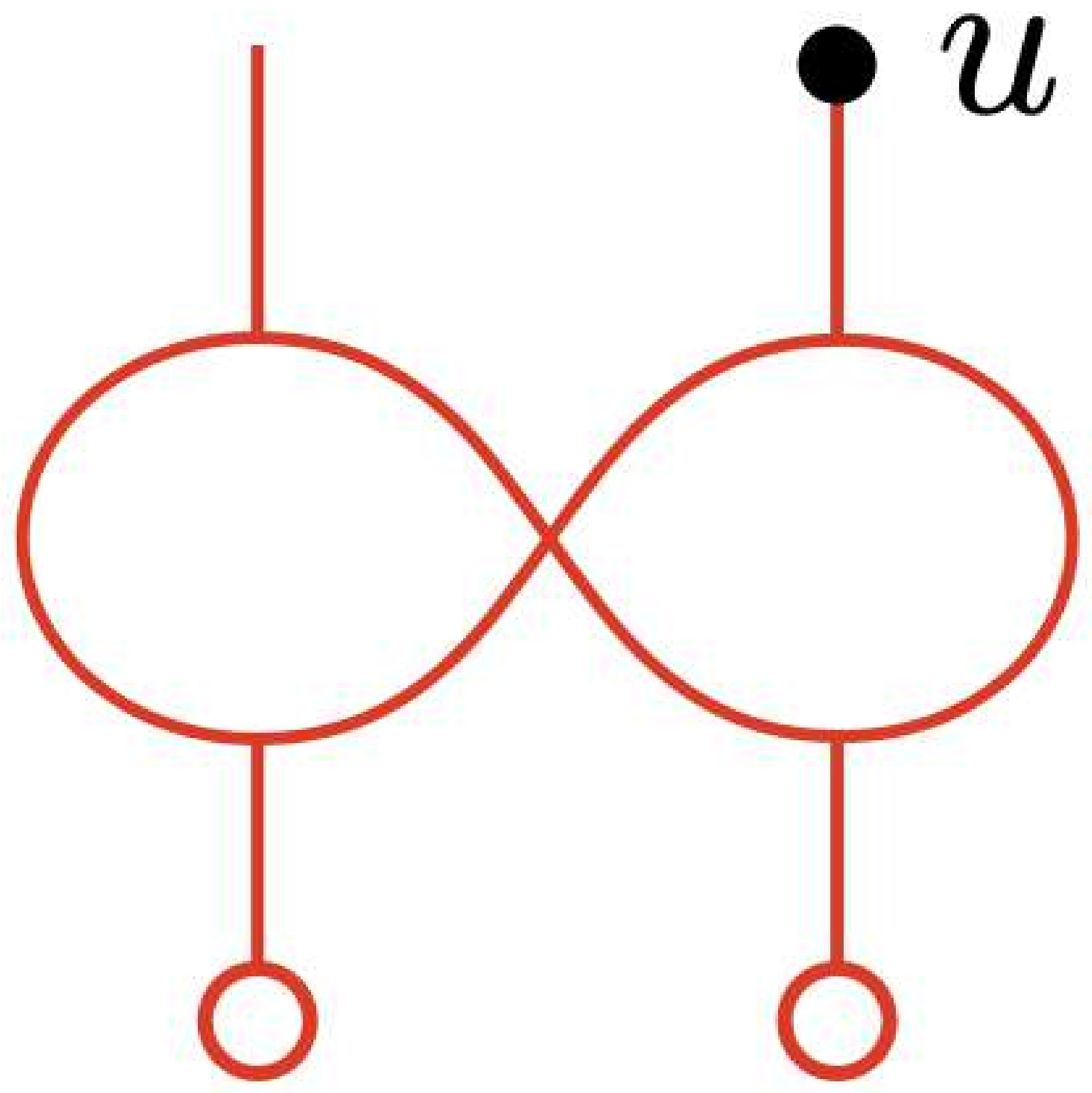} ~ = ~
\adjincludegraphics[valign = c, width = 2.1cm]{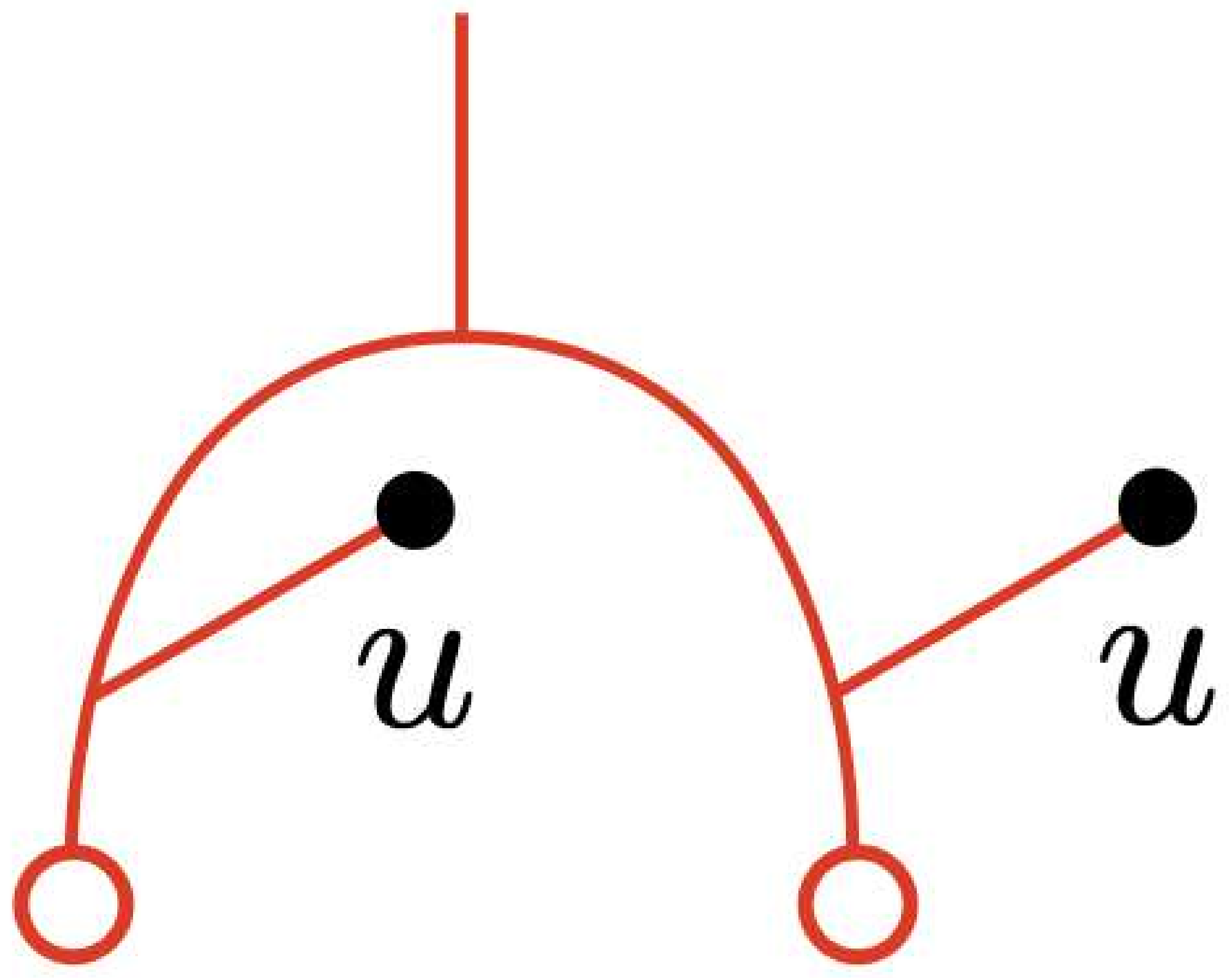} ~ = ~
\adjincludegraphics[valign = c, width = 2.5cm]{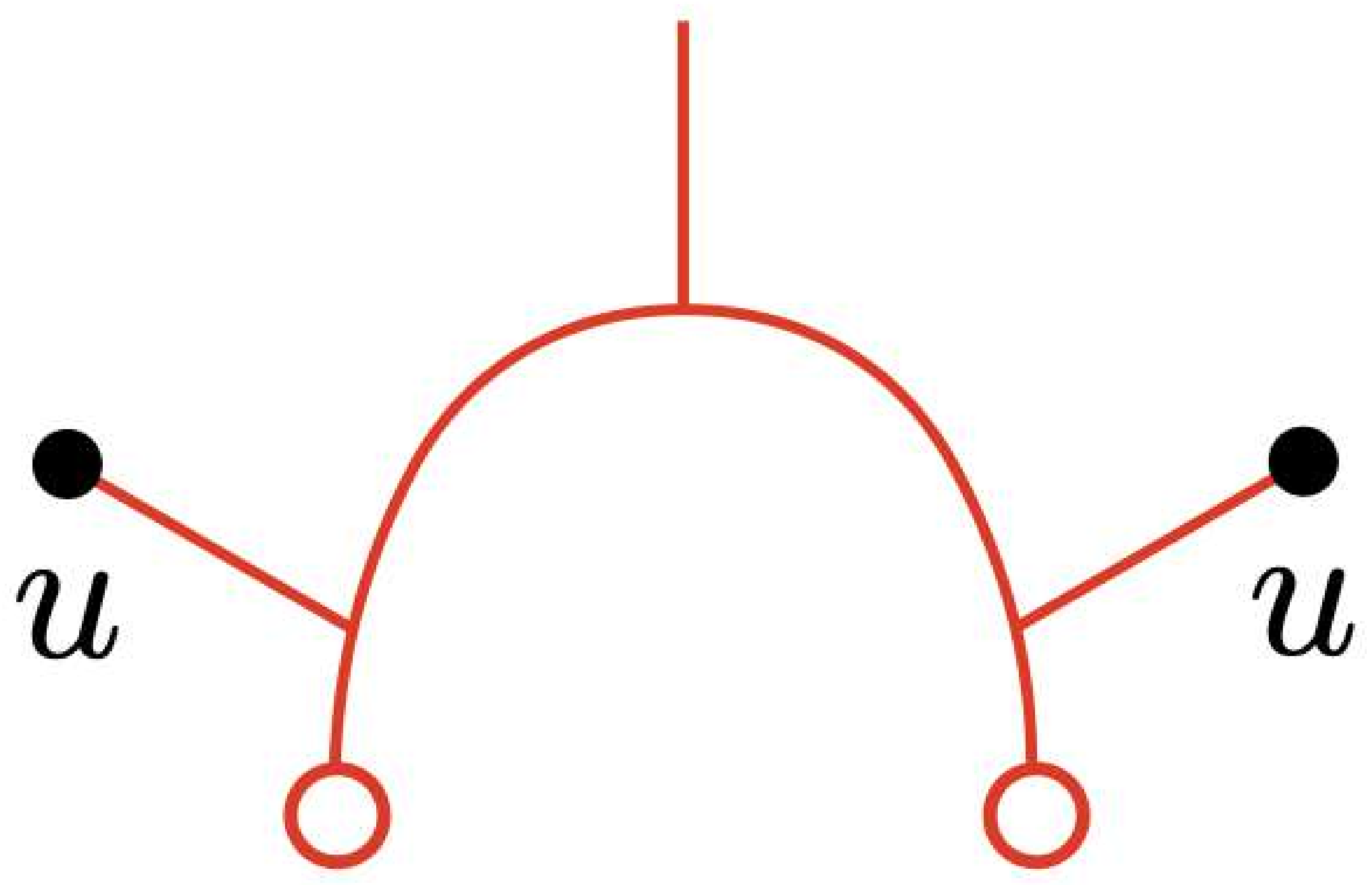} ~ = ~
\adjincludegraphics[valign = c, width = 1.4cm]{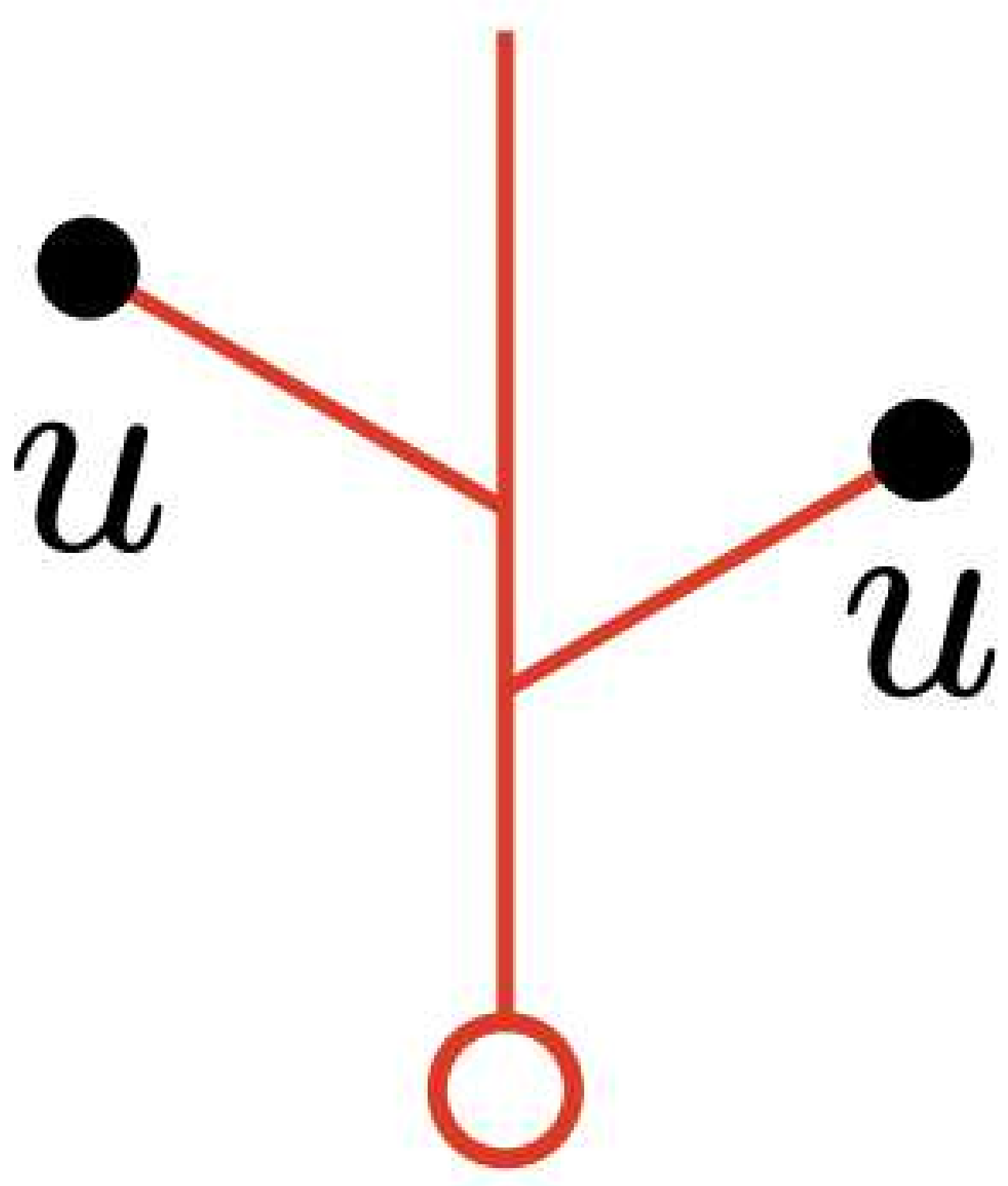} ~ = ~
\adjincludegraphics[valign = c, width = 0.25cm]{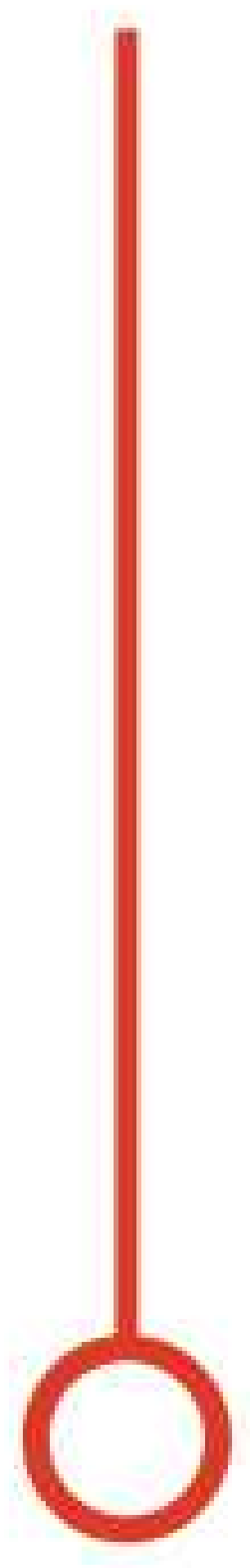}.
\end{equation}
The third and last equalities follow from the fact that the unit $\eta$ is even.
The above computations show that $(\mathcal{H}^u, m_u, \eta)$ is an associative unital superalgebra.
It is obvious that $(\mathcal{H}^u, \Delta, \epsilon)$ is a coassociative counital supercoalgebra because the comultiplication $\Delta$ and counit $\epsilon$ are the same as those of the original weak Hopf algebra $H$.

\paragraph{Weak Hopf superalgebra structure.}
Finally, we show that the structure maps $(m_u, \eta, \Delta, \epsilon, S_u)$ satisfy the defining properties of a weak Hopf superalgebra, namely, eqs. \eqref{eq: weak Hopf comultiplication}--\eqref{eq: weak Hopf antipode} where the trivial braiding $c_{\mathrm{triv}}$ is replaced by the symmetric braiding $c_{\mathrm{super}}$.
Equation \eqref{eq: weak Hopf comultiplication} can be checked by the same computation as in the case of Hopf algebra, cf. eq. \eqref{eq: multiplicative Delta of Hopf superalgebra}.
Equation \eqref{eq: weak Hopf counit} is shown by a direct computation as follows:
\begin{equation*}
\adjincludegraphics[valign = c, width = 2.4cm]{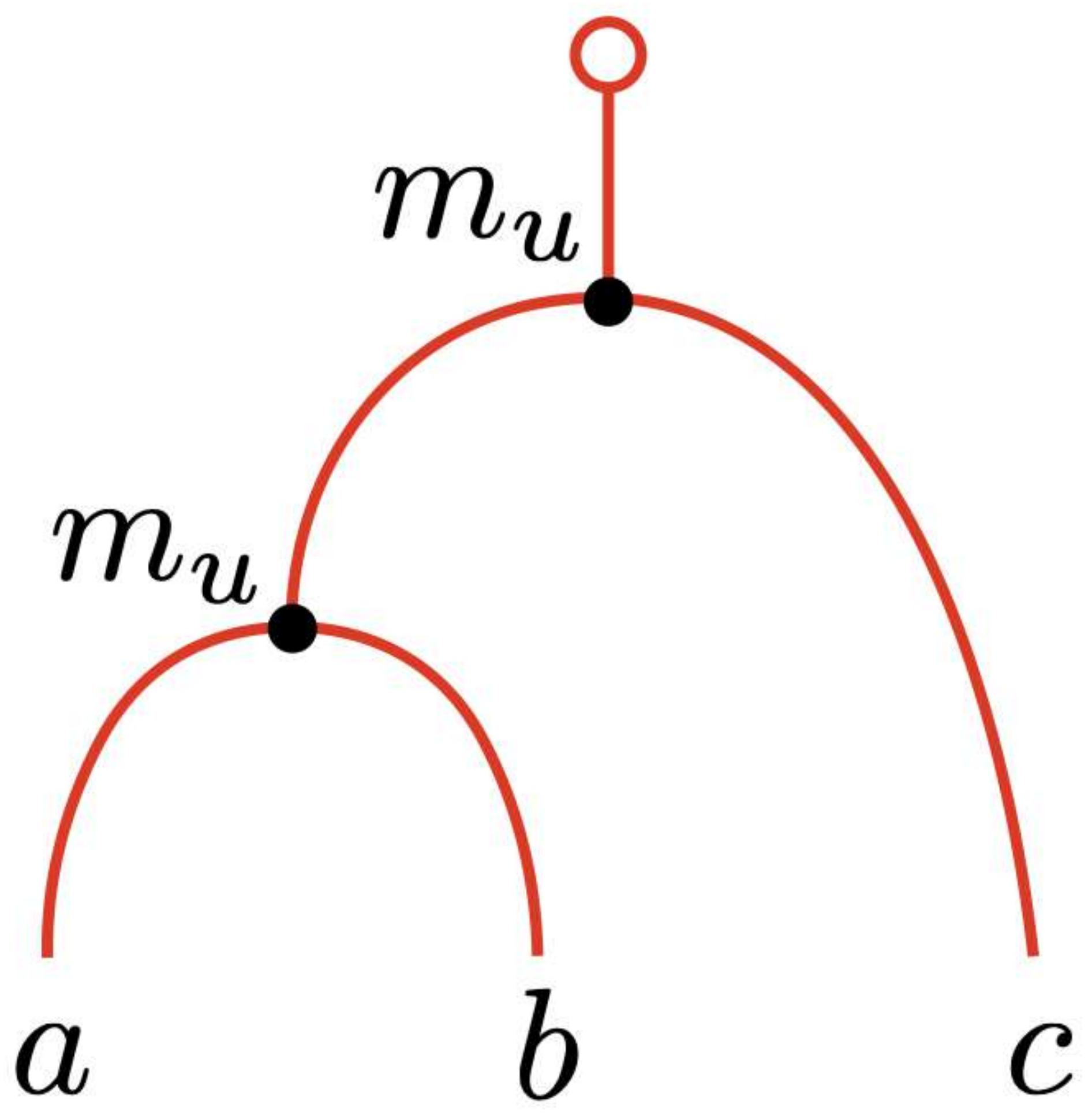} ~ = ~
\adjincludegraphics[valign = c, width = 2.4cm]{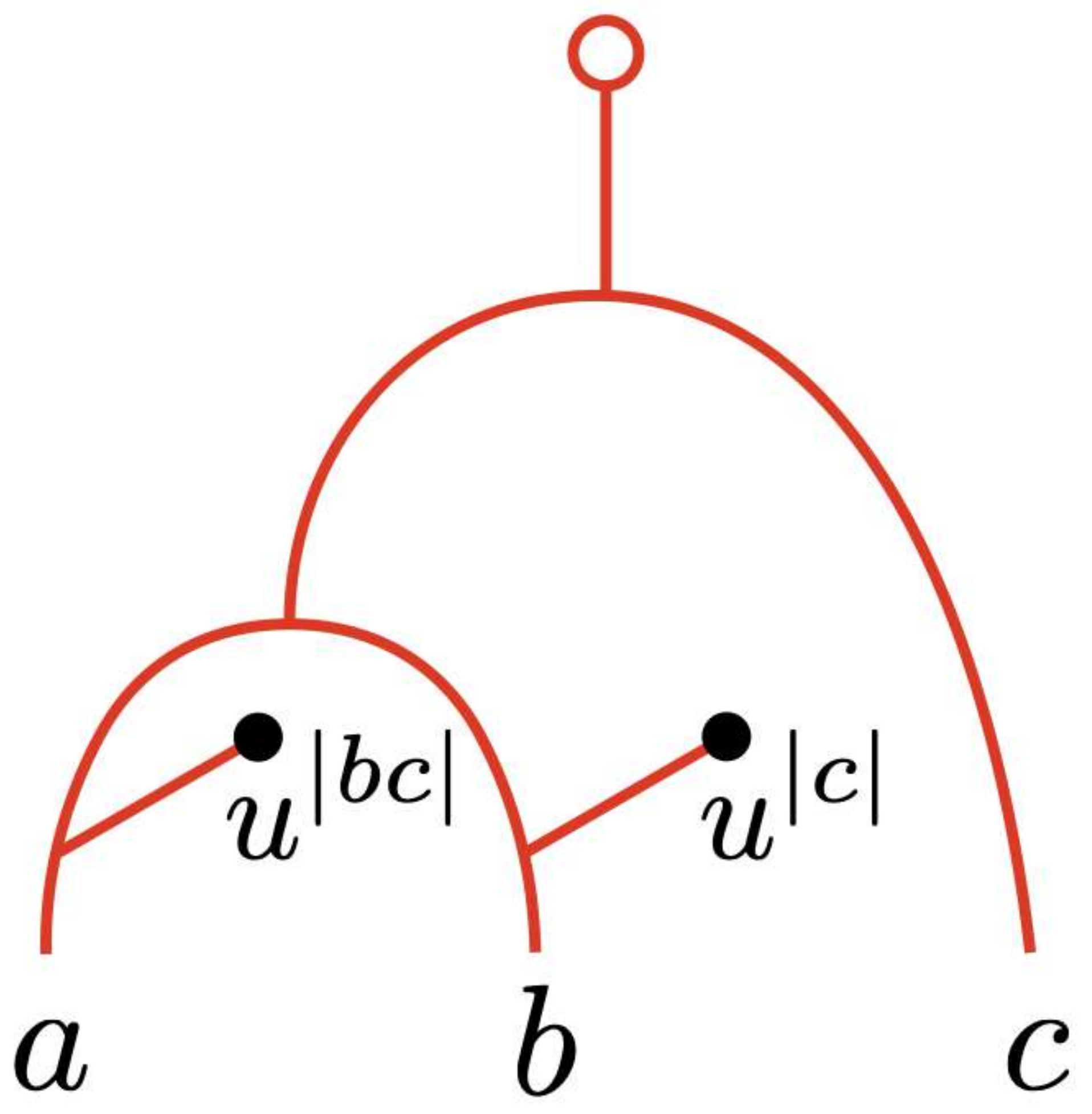} ~ = ~
\adjincludegraphics[valign = c, width = 2.4cm]{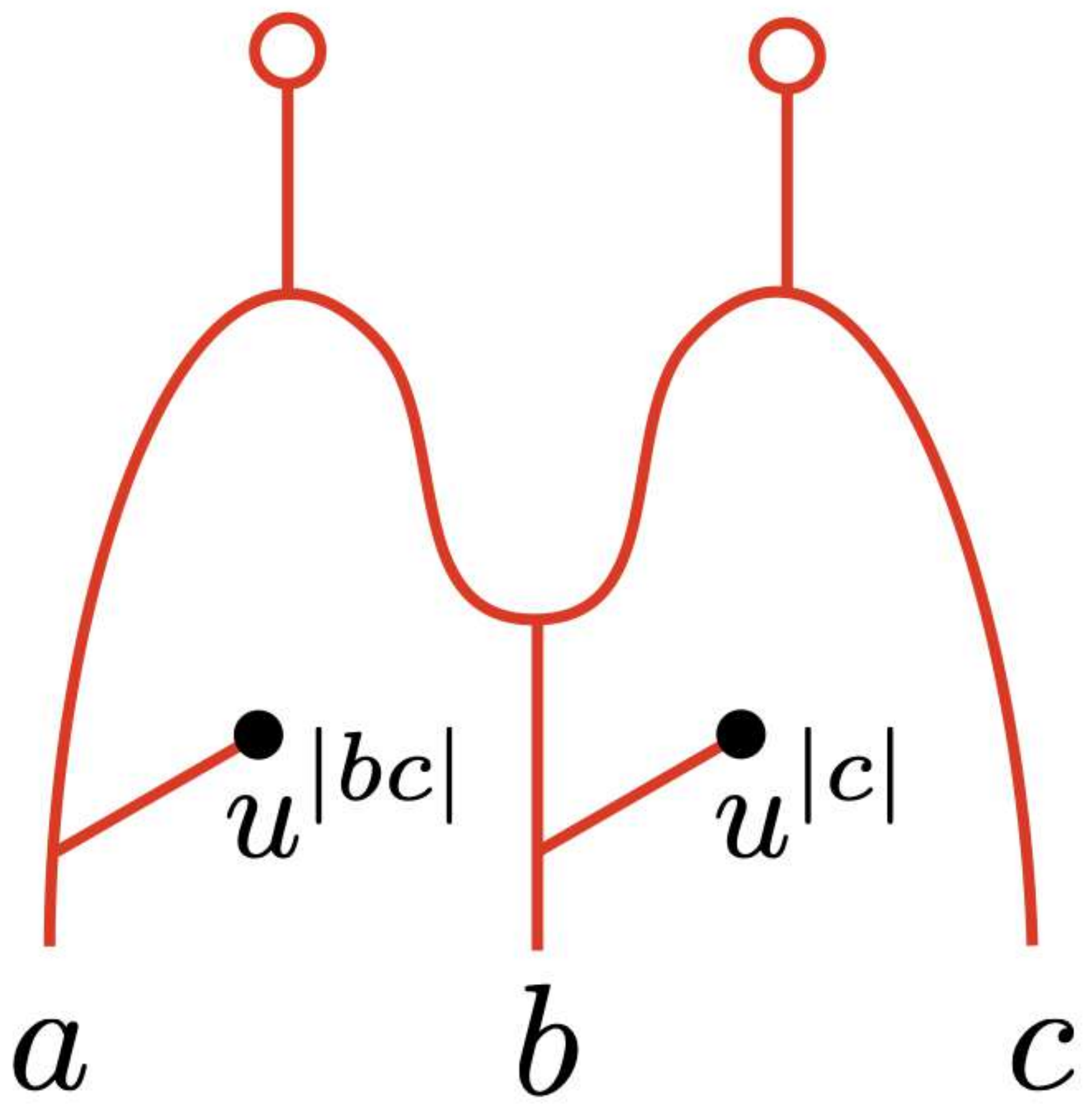} ~ = ~
\adjincludegraphics[valign = c, width = 2.5cm]{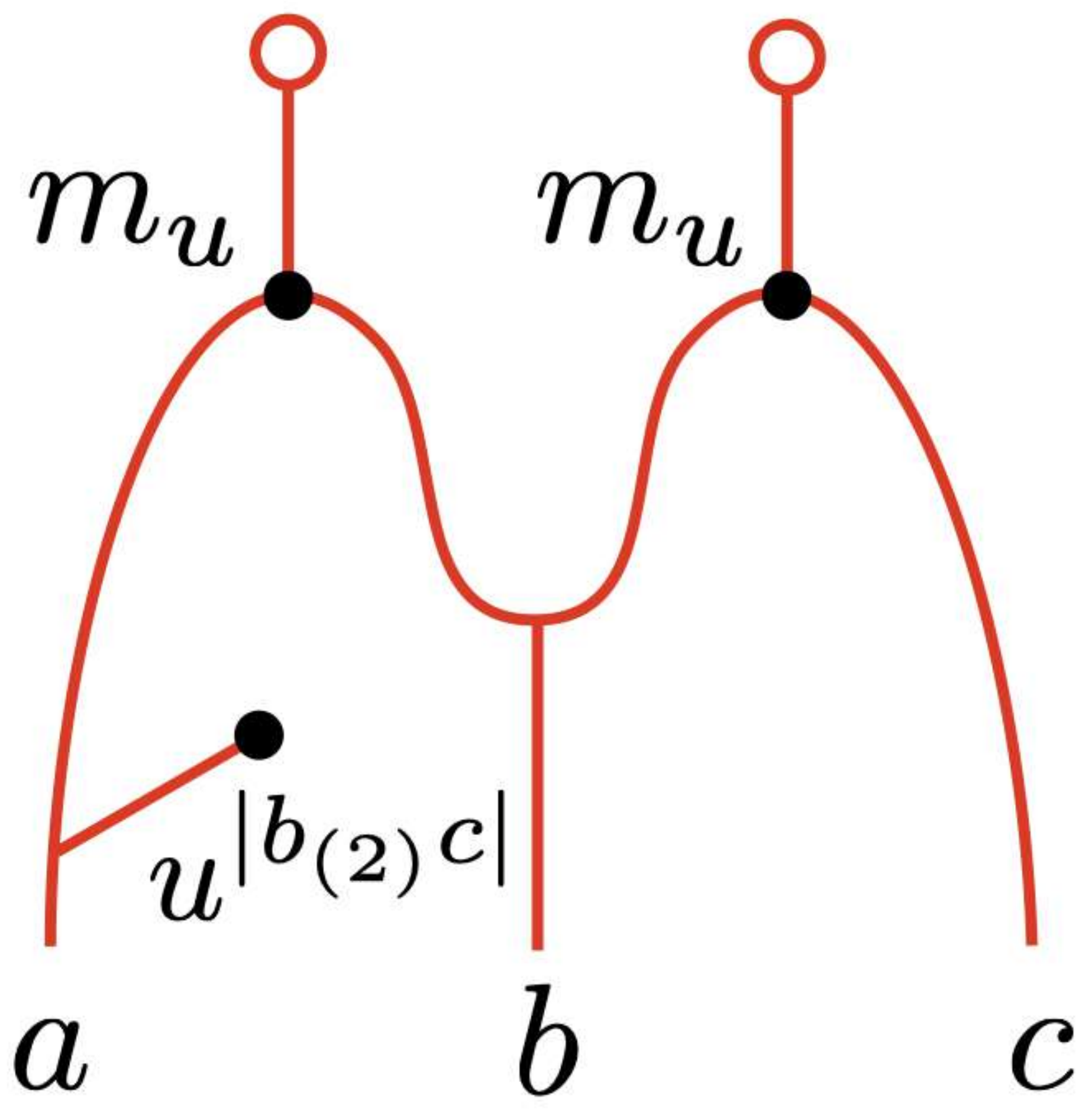} ~ = ~
\adjincludegraphics[valign = c, width = 2.4cm]{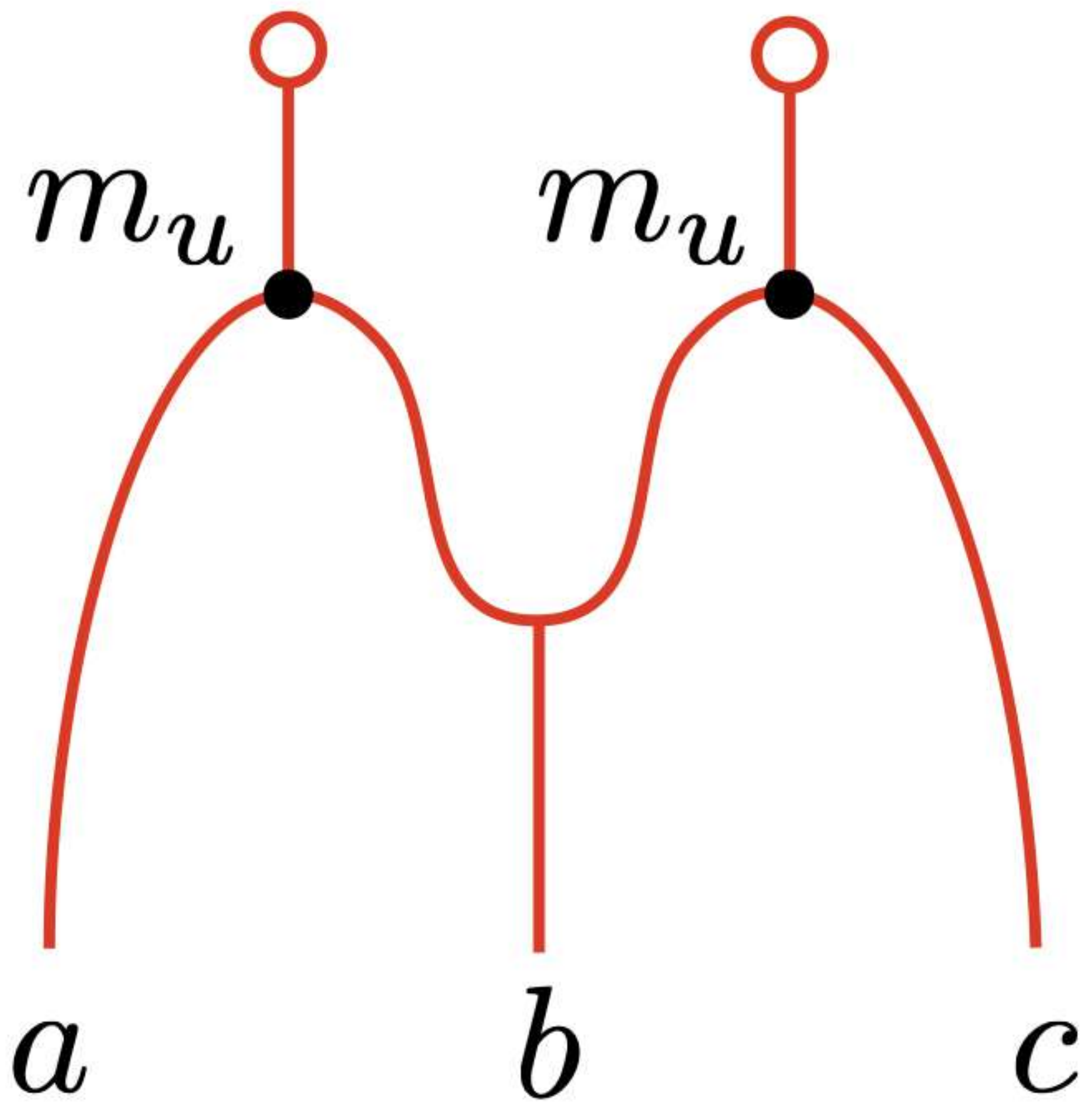},
\end{equation*}
\begin{equation*}
\adjincludegraphics[valign = c, width = 2.4cm]{counit_u1.pdf} ~ = ~
\adjincludegraphics[valign = c, width = 2.4cm]{counit_u2.pdf} ~ = ~
\adjincludegraphics[valign = c, width = 2.4cm]{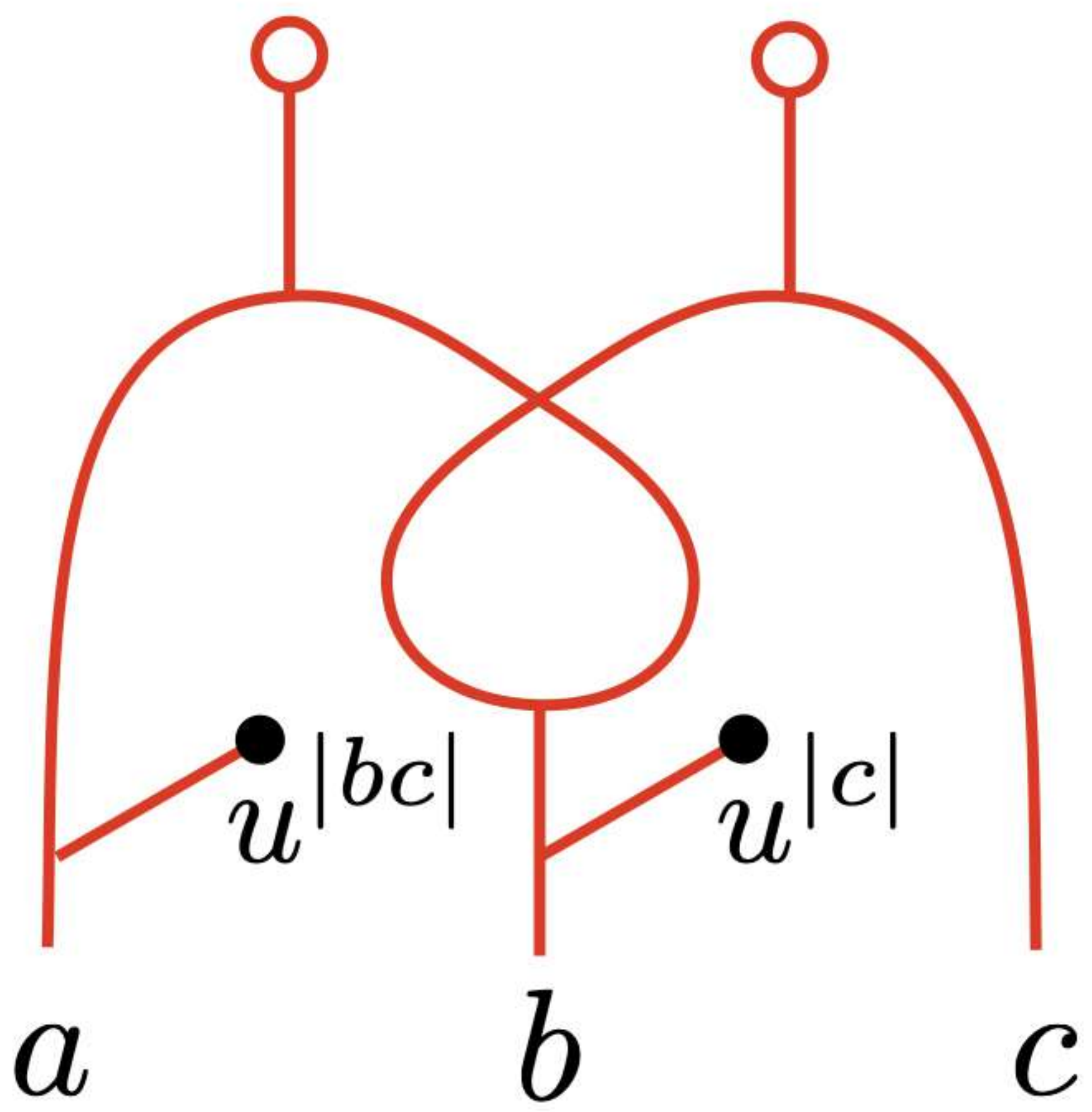} ~ = ~
\adjincludegraphics[valign = c, width = 2.5cm]{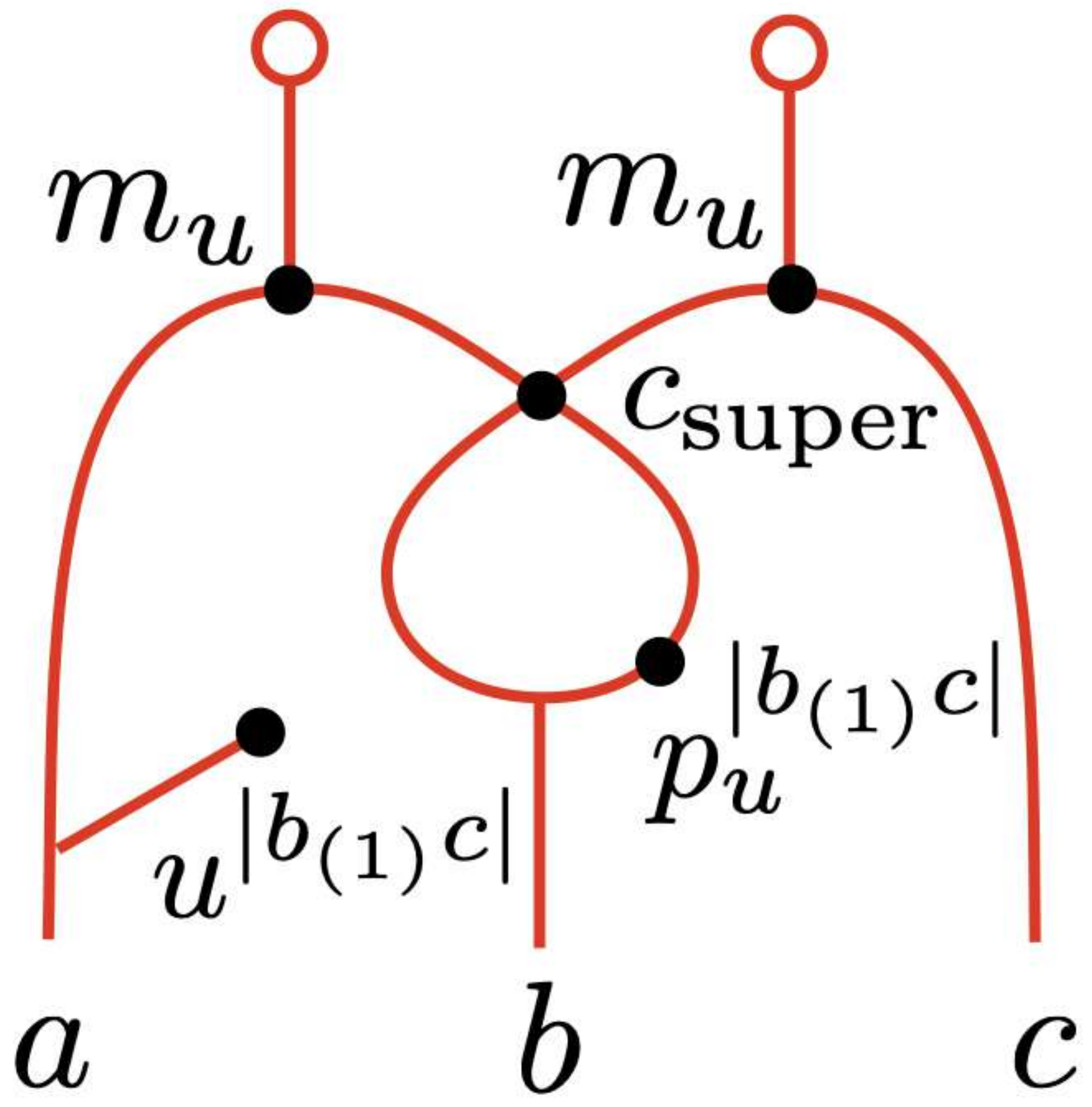} ~ = ~
\adjincludegraphics[valign = c, width = 2.4cm]{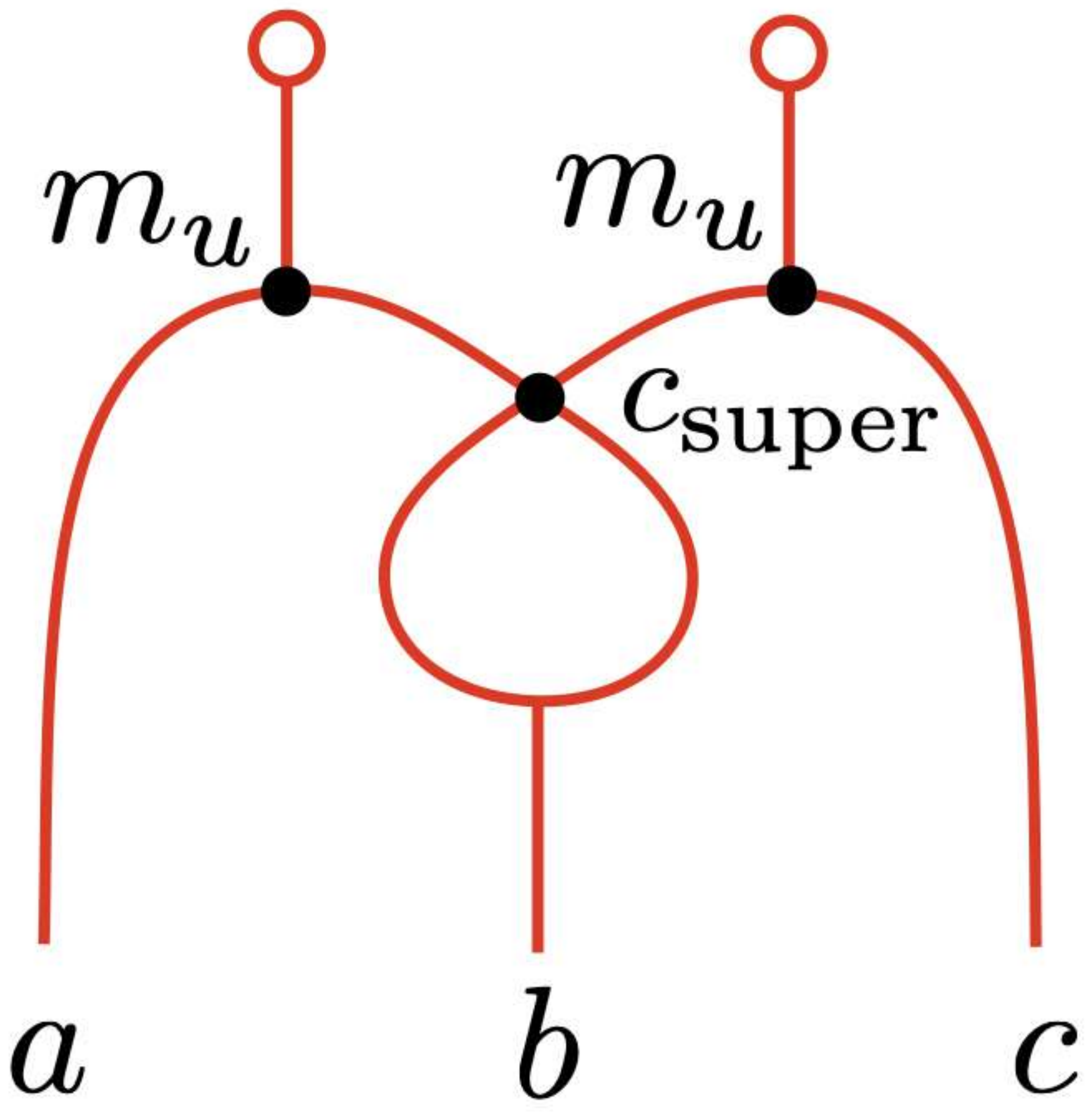}.
\end{equation*}
Here, we used the Sweedler notation $\Delta(b) = b_{(1)} \otimes b_{(2)}$.
The last equality of the first line is due to the fact that $\epsilon \circ m_u (b_{(2)} \otimes c)$ vanishes unless $|b_{(2)}c| = 0$.
Similarly, the last equality of the second line is because $\epsilon \circ m_u (b_{(1)} \otimes c)$ vanishes unless $|b_{(1)}c| = 0$.
We can also show eq. \eqref{eq: weak Hopf unit} analogously.
Equation \eqref{eq: weak Hopf antipode} can be derived as follows:
\begin{equation*}
\adjincludegraphics[valign = c, width = 1.7cm]{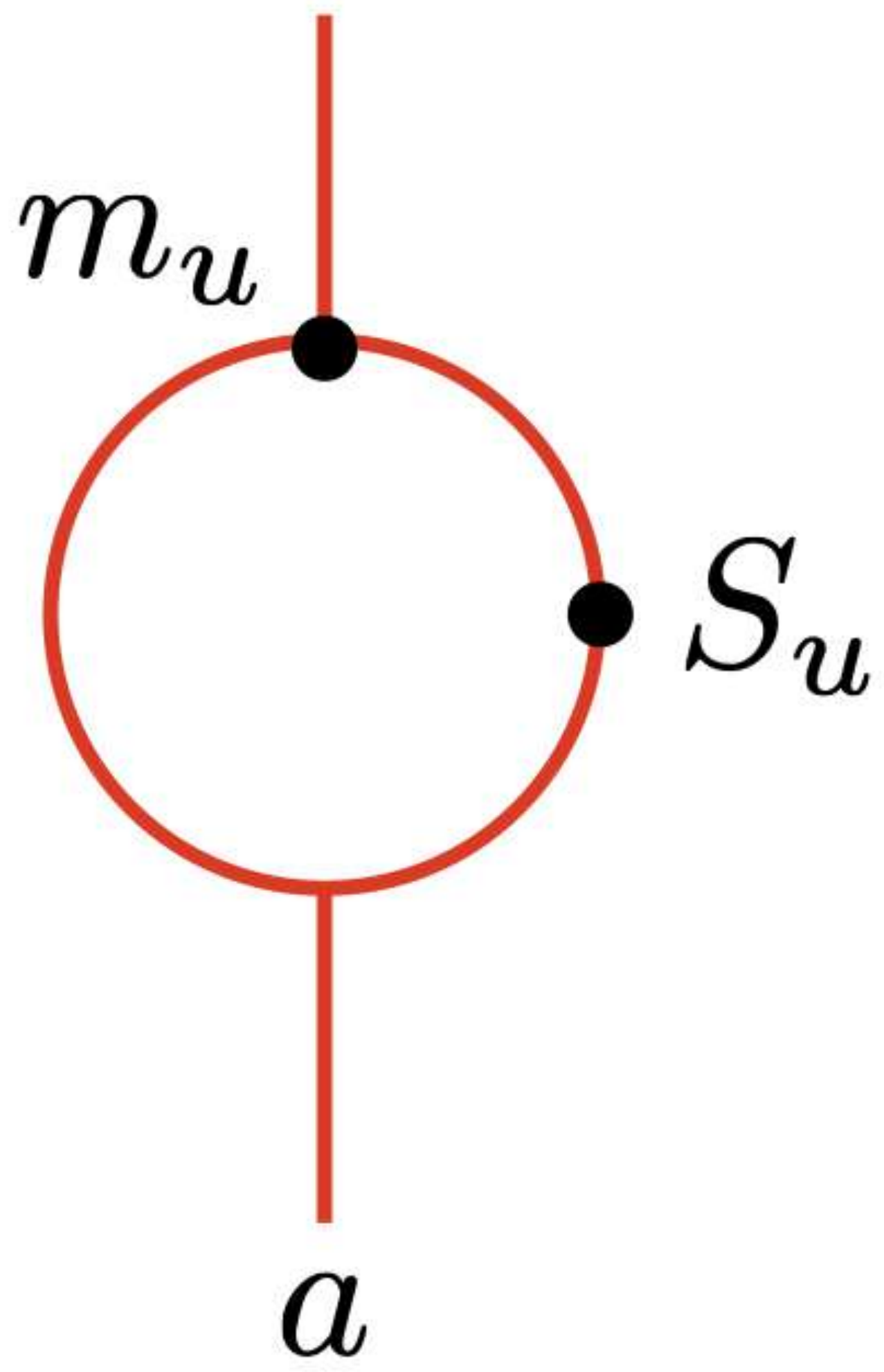} ~ = ~ 
\adjincludegraphics[valign = c, width = 2.2cm]{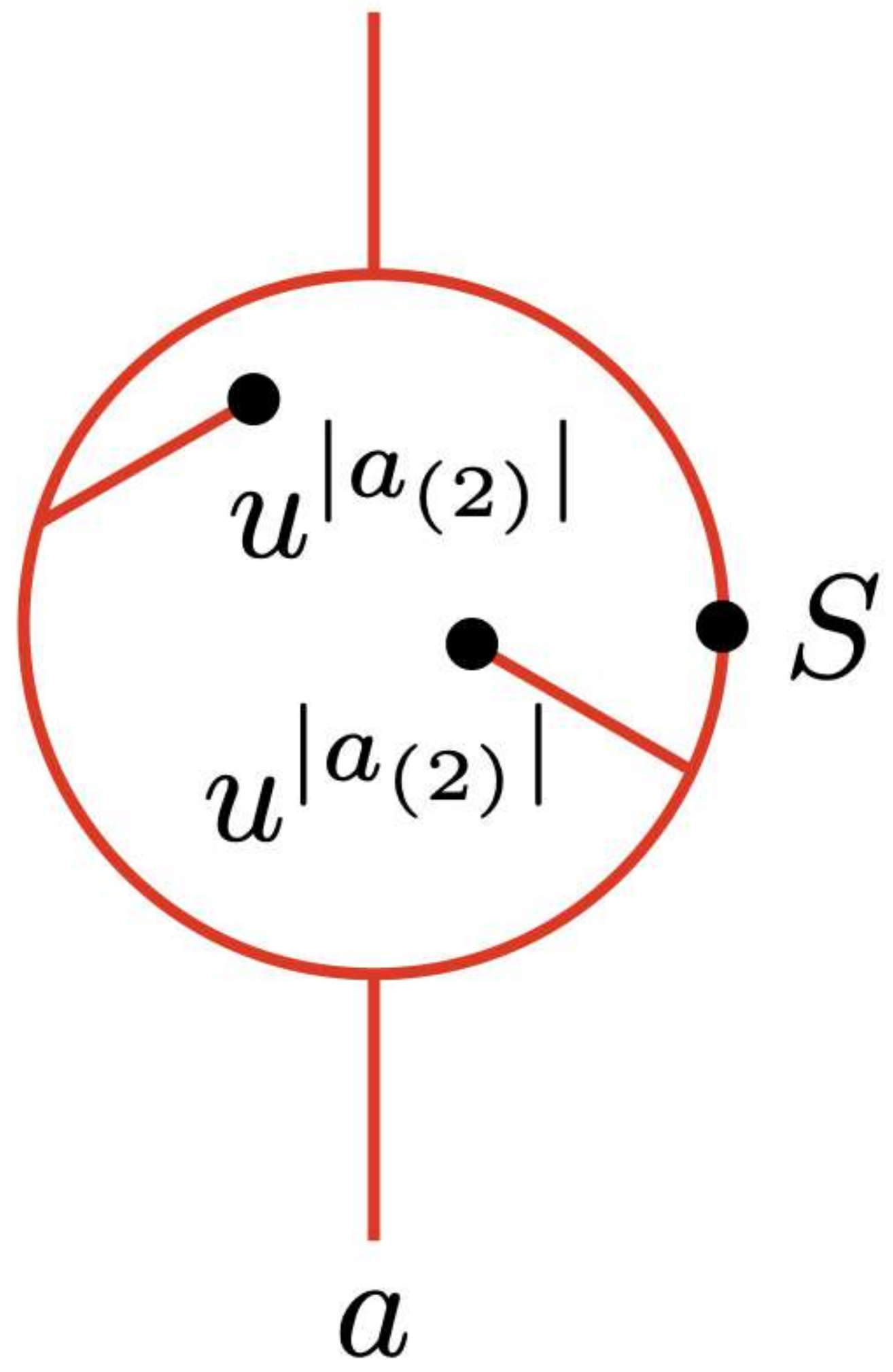} ~ = ~ 
\adjincludegraphics[valign = c, width = 1.5cm]{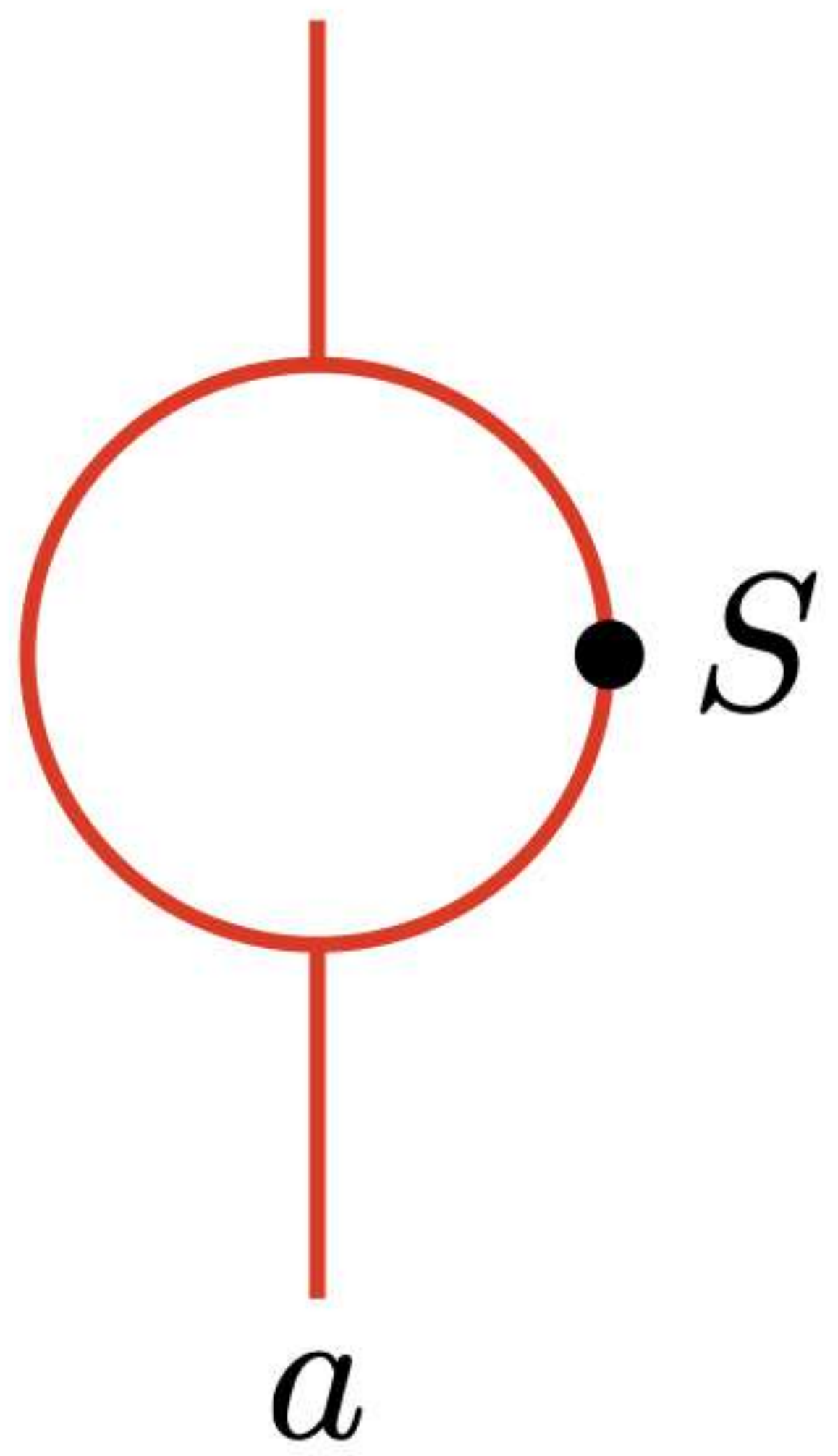} ~ = ~ 
\adjincludegraphics[valign = c, width = 1.2cm]{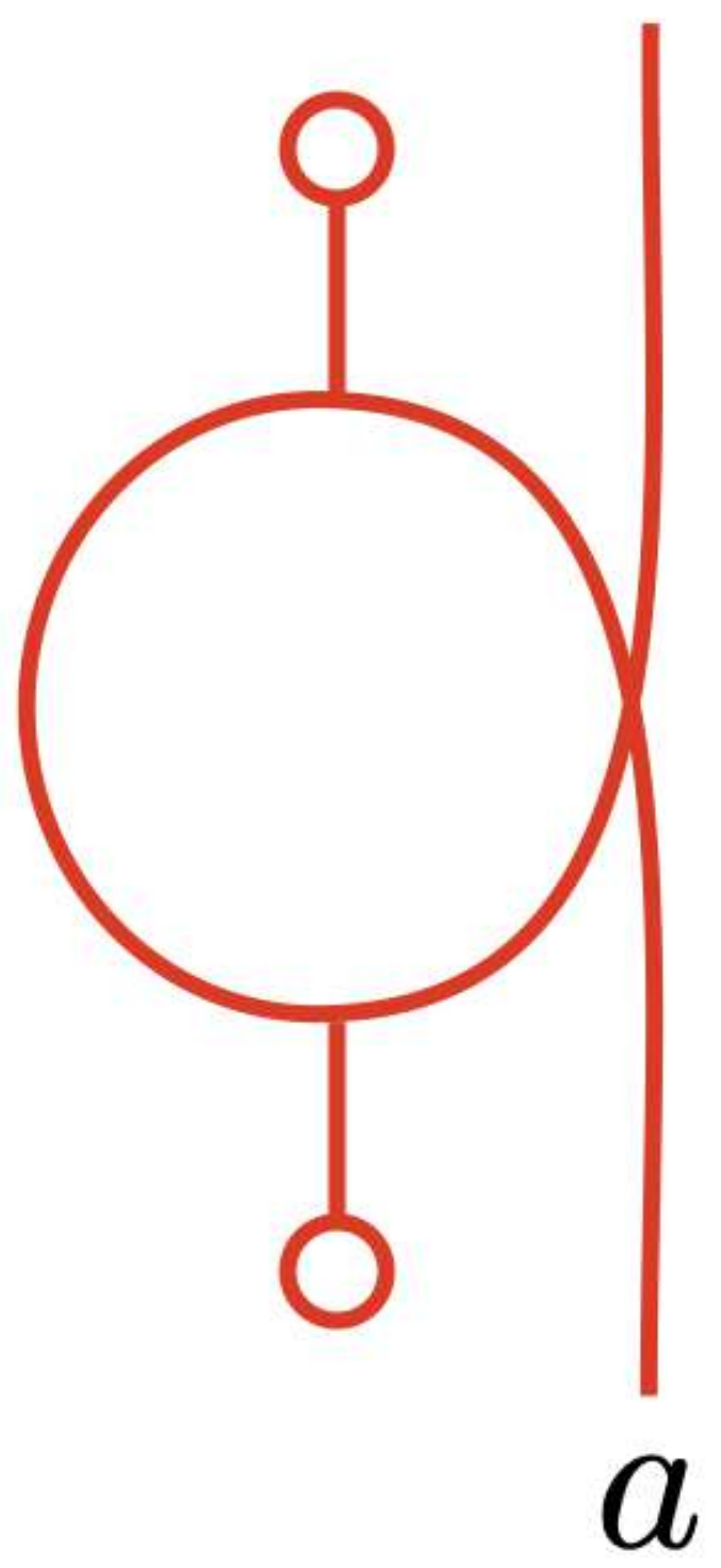} ~ = ~ 
\adjincludegraphics[valign = c, width = 2.1cm]{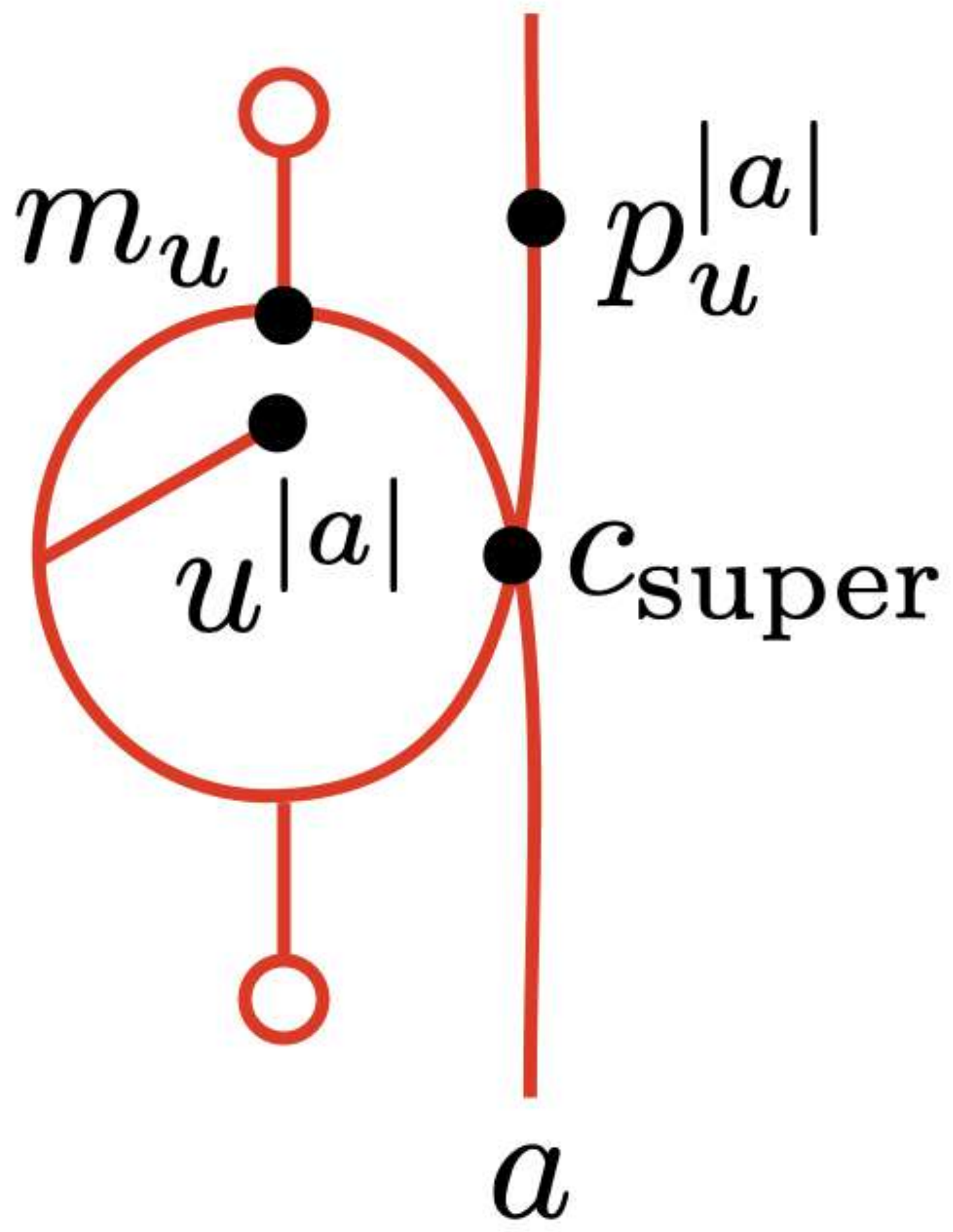} ~ = ~ 
\adjincludegraphics[valign = c, width = 2.1cm]{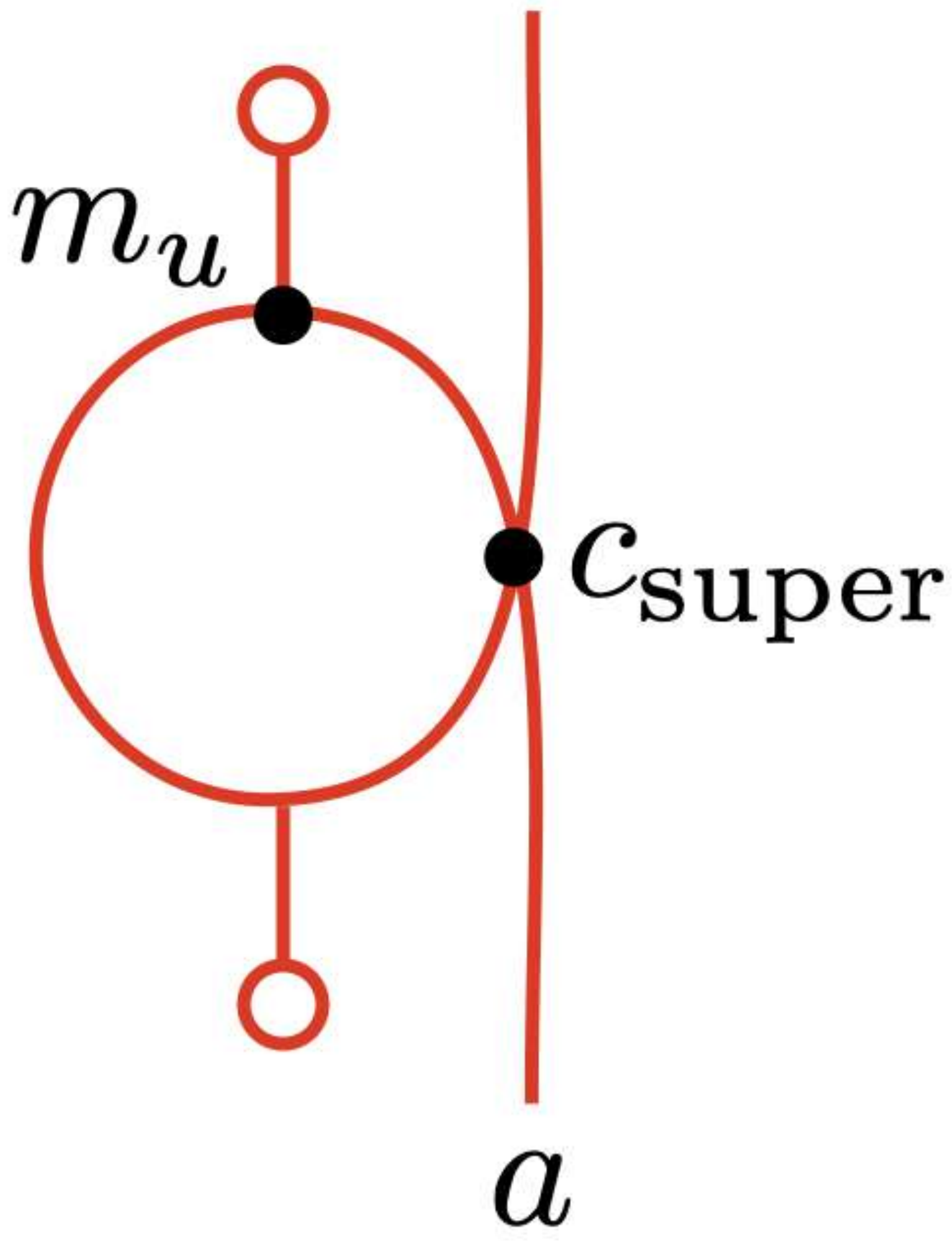},
\end{equation*}
\begin{equation*}
\adjincludegraphics[valign = c, width = 1.7cm]{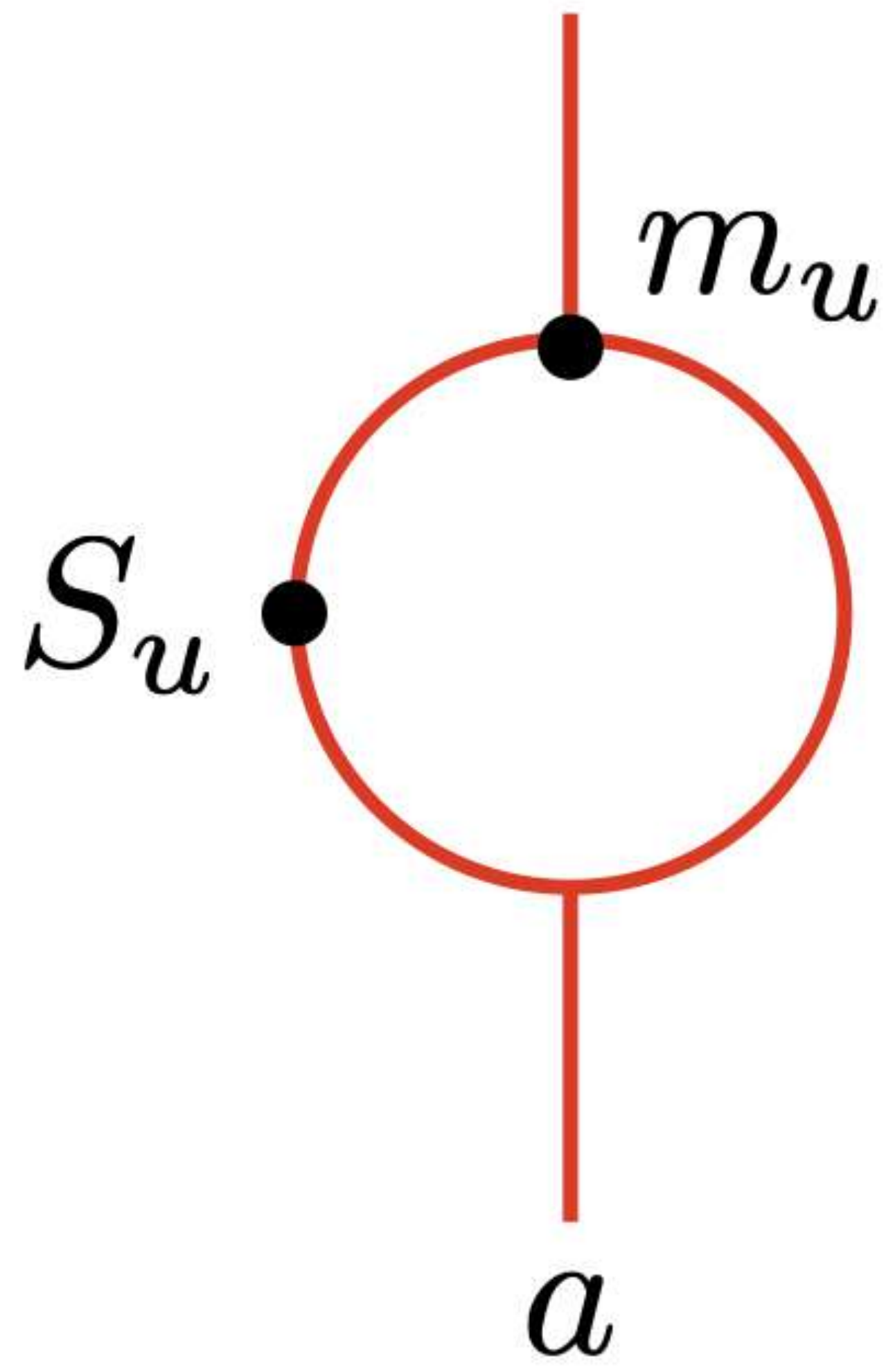} ~ = ~ 
\adjincludegraphics[valign = c, width = 2.4cm]{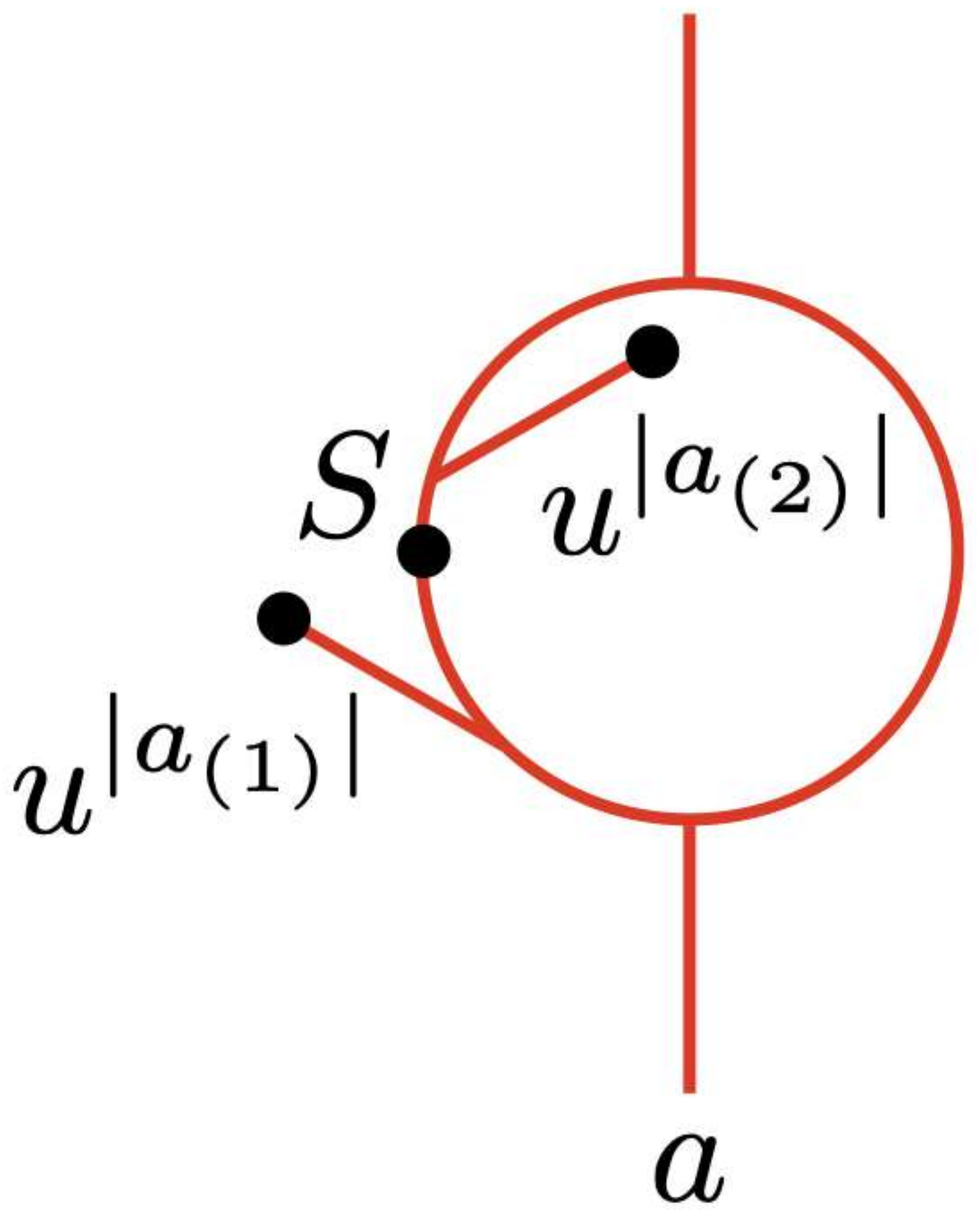} ~ = ~ 
\adjincludegraphics[valign = c, width = 1.6cm]{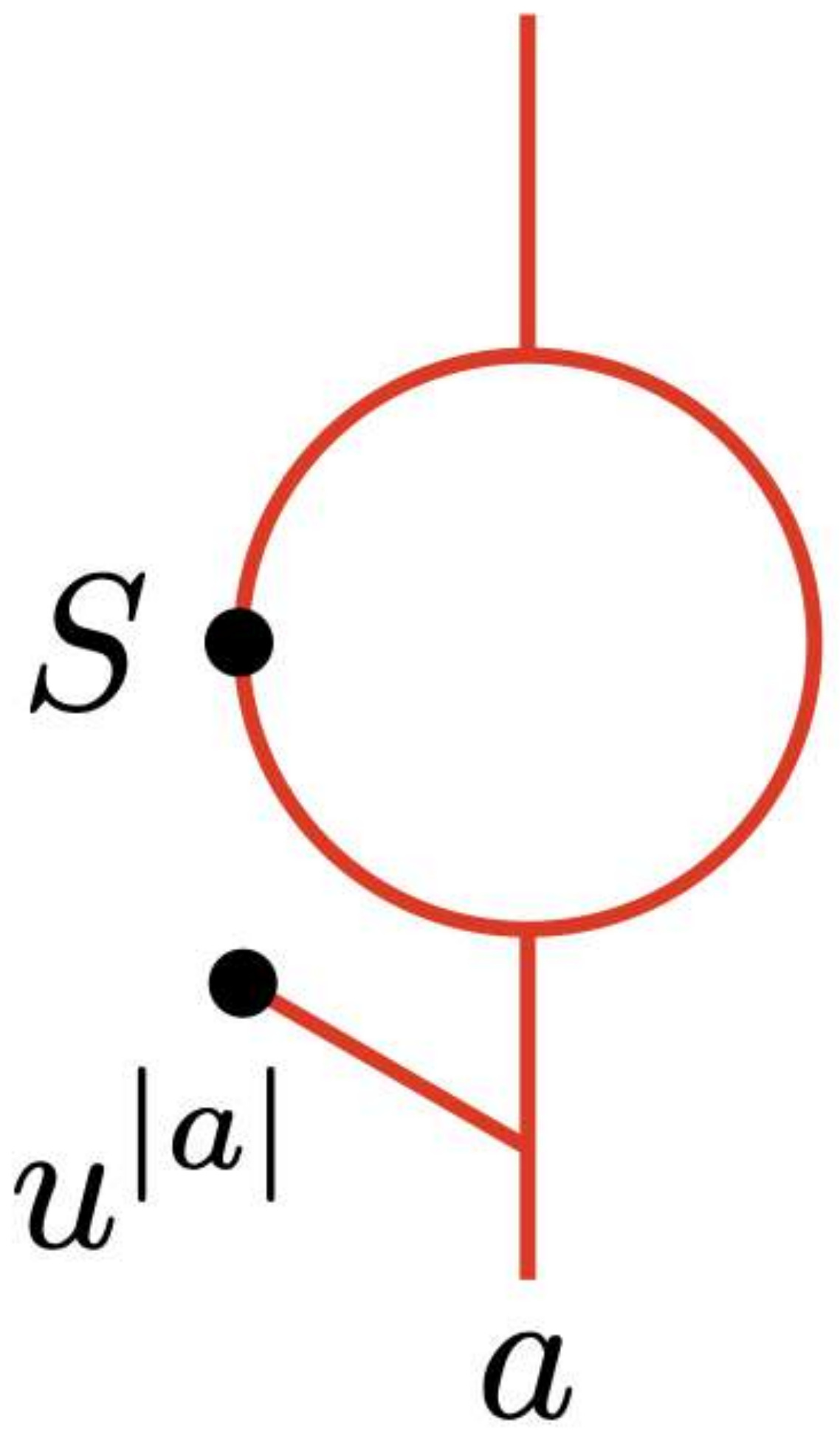} ~ = ~ 
\adjincludegraphics[valign = c, width = 2cm]{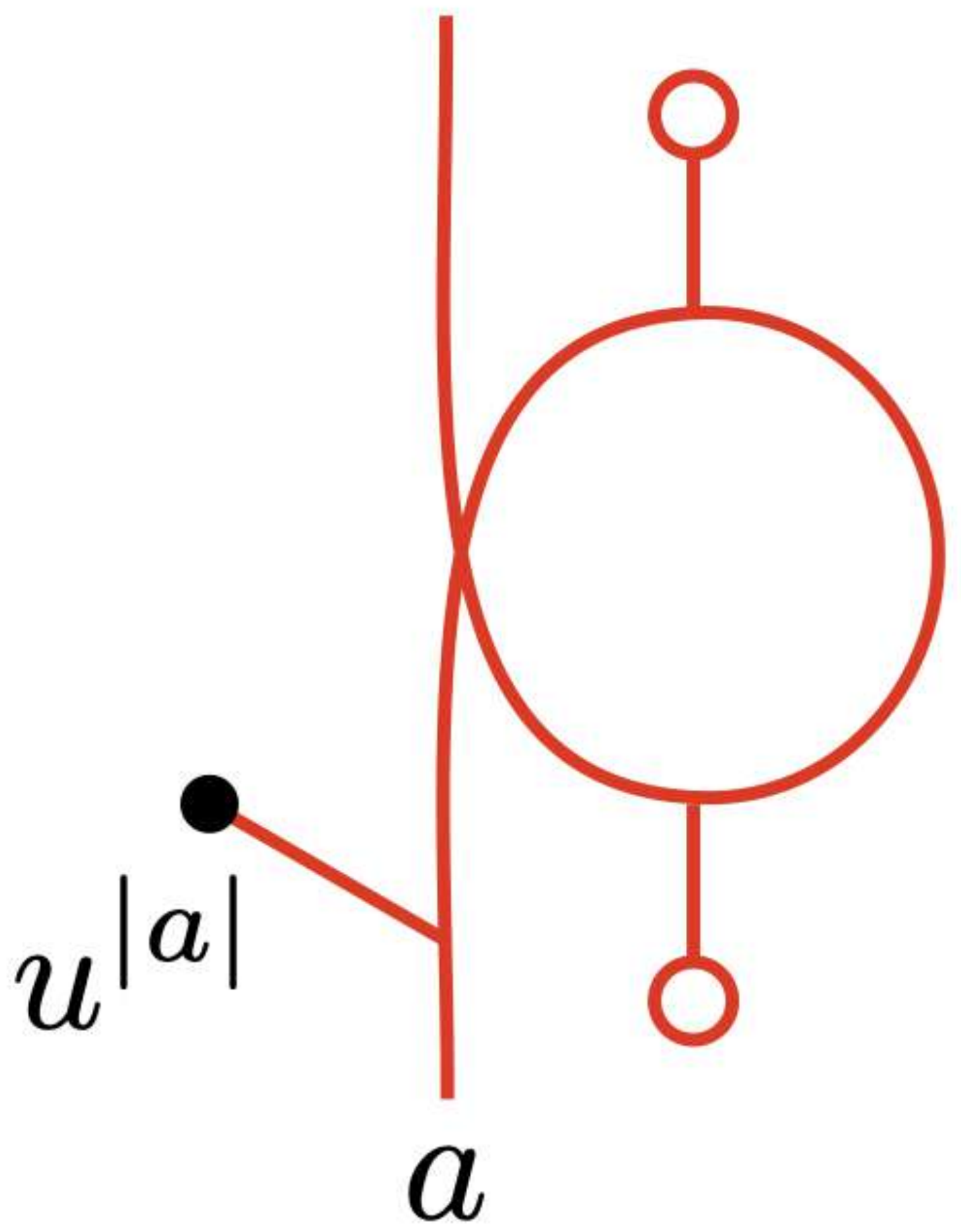} ~ = ~ 
\adjincludegraphics[valign = c, width = 2.1cm]{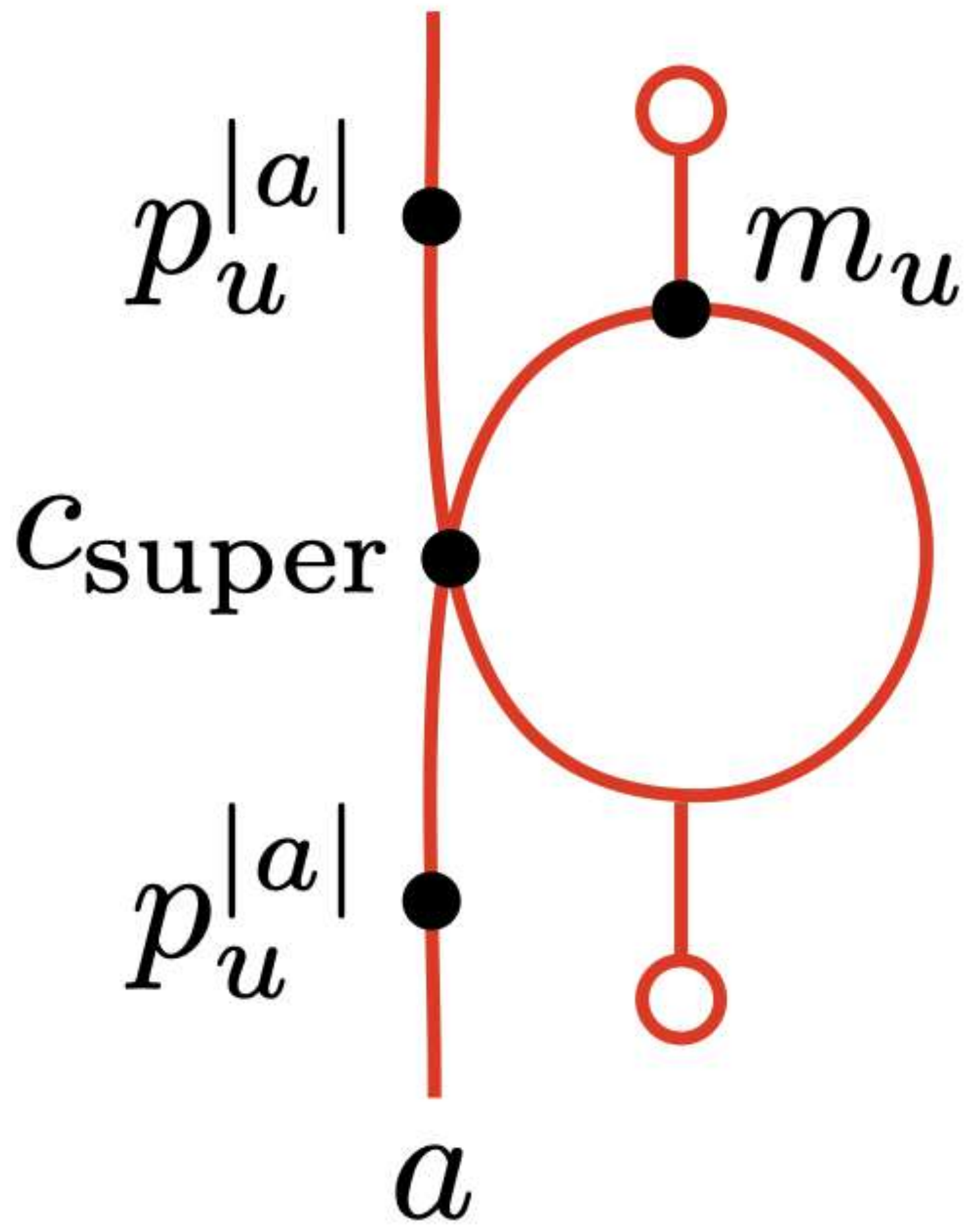} ~ = ~ 
\adjincludegraphics[valign = c, width = 2.1cm]{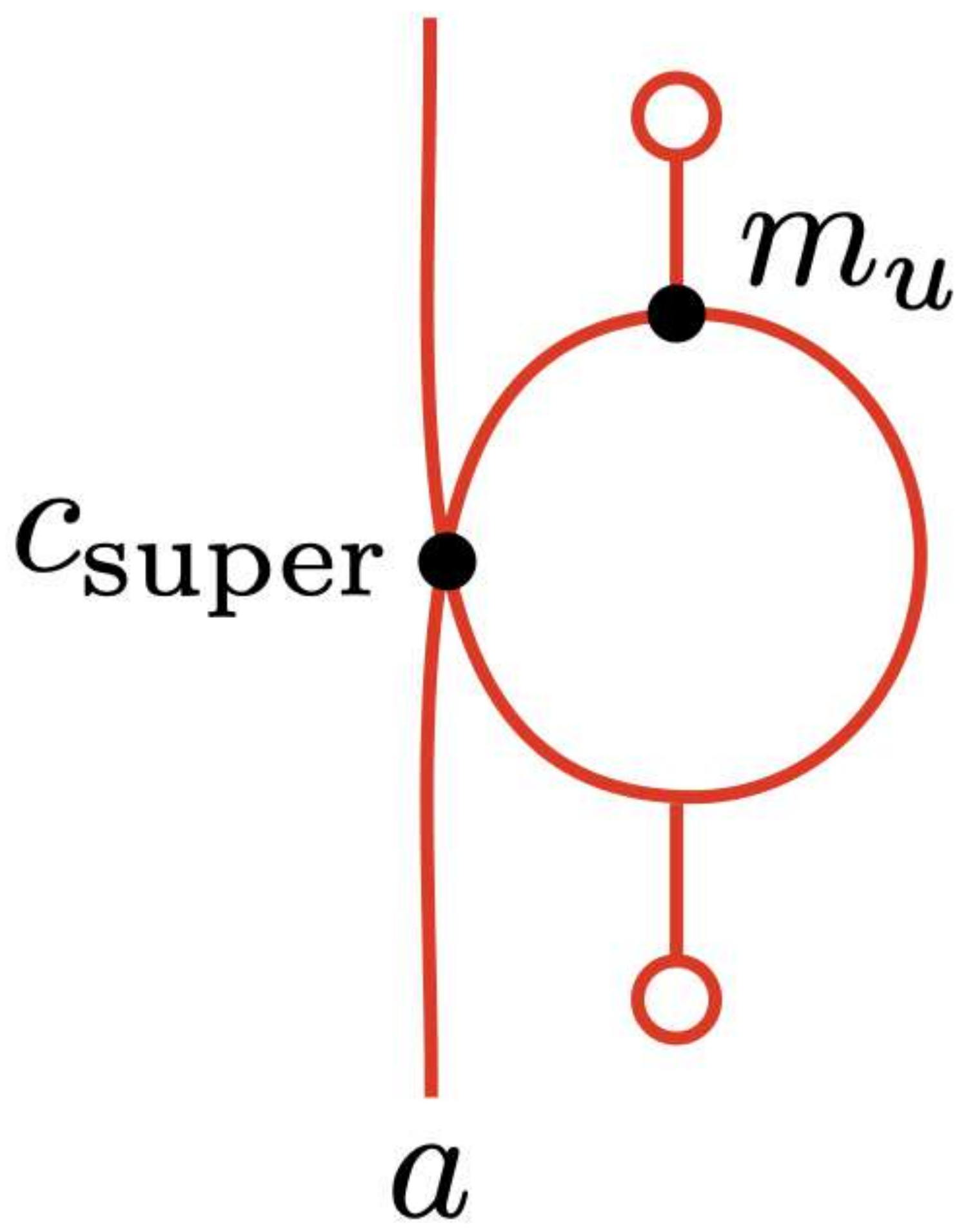},
\end{equation*}
\begin{equation*}
\adjincludegraphics[valign = c, width = 2.5cm]{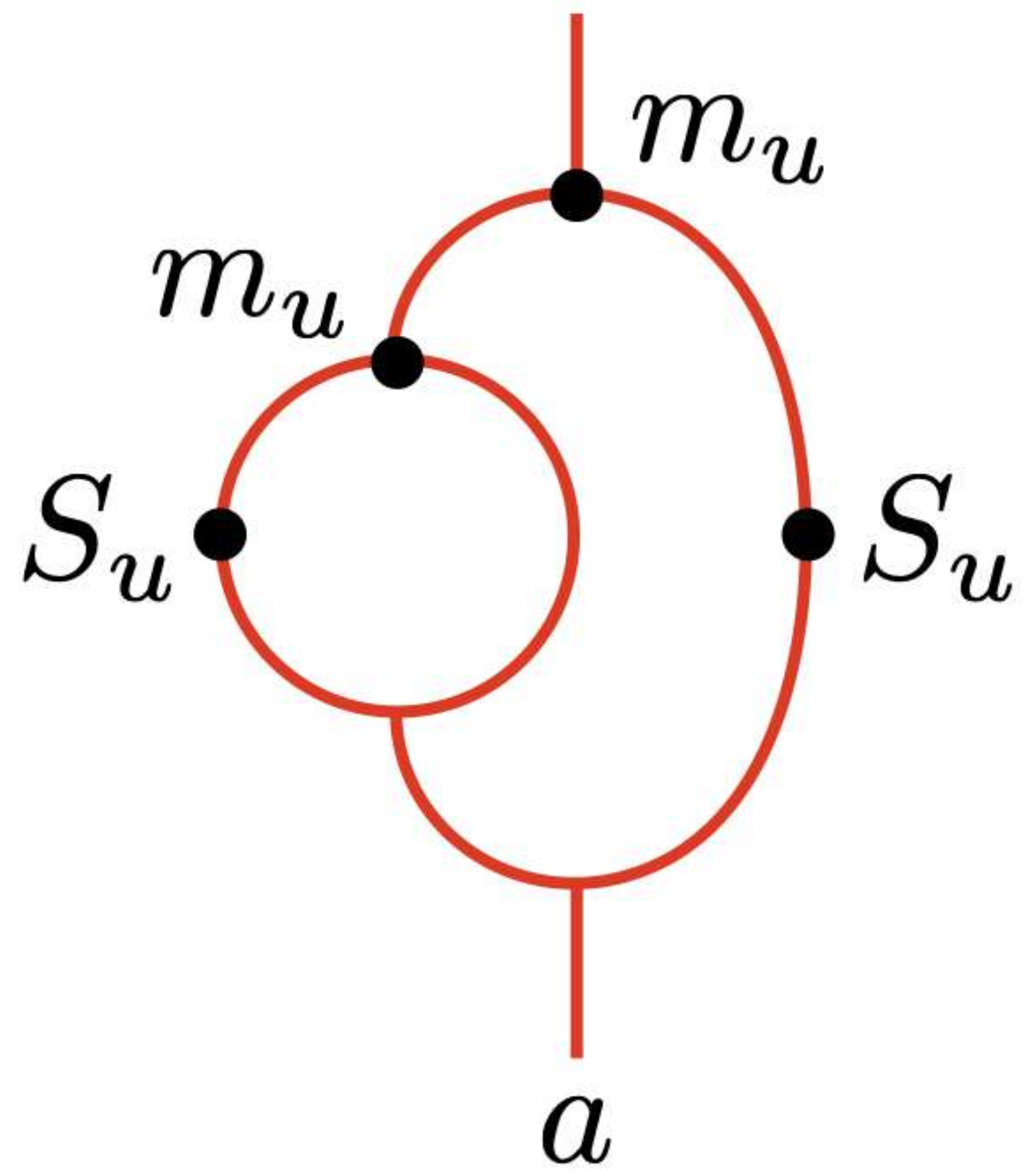} ~ = ~ 
\adjincludegraphics[valign = c, width = 2.6cm]{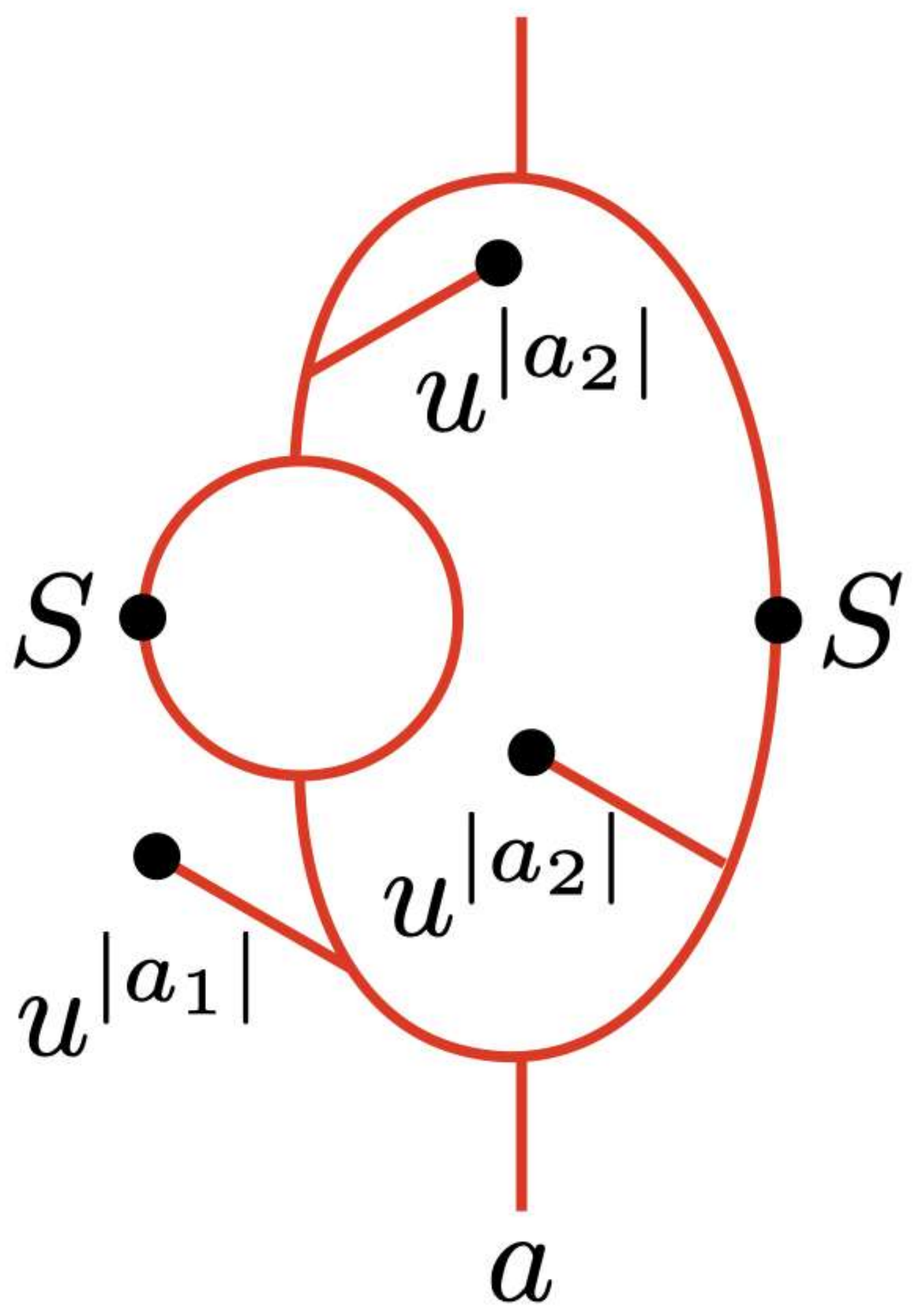} ~ = ~ 
\adjincludegraphics[valign = c, width = 2.4cm]{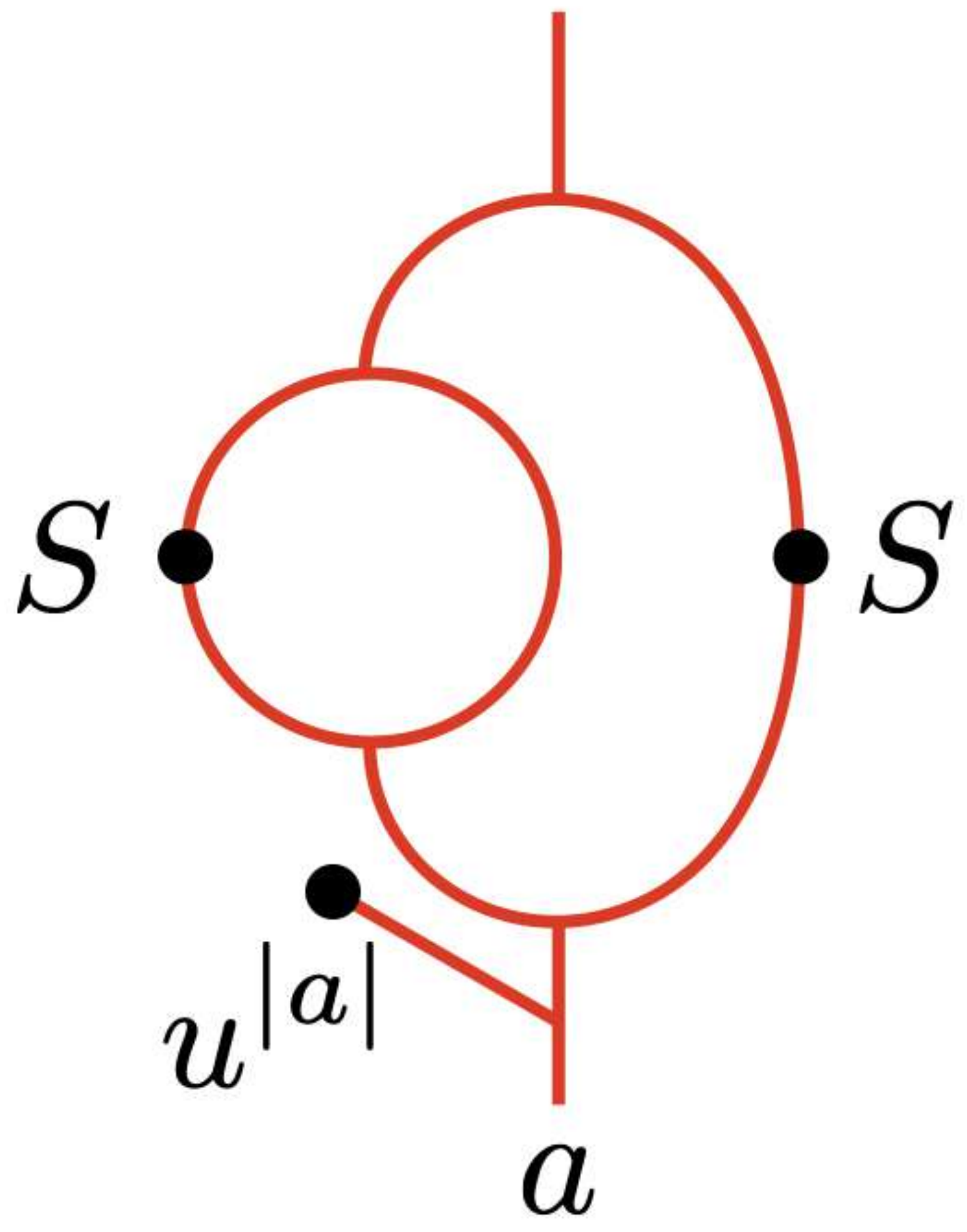} ~ = ~ 
\adjincludegraphics[valign = c, width = 1.15cm]{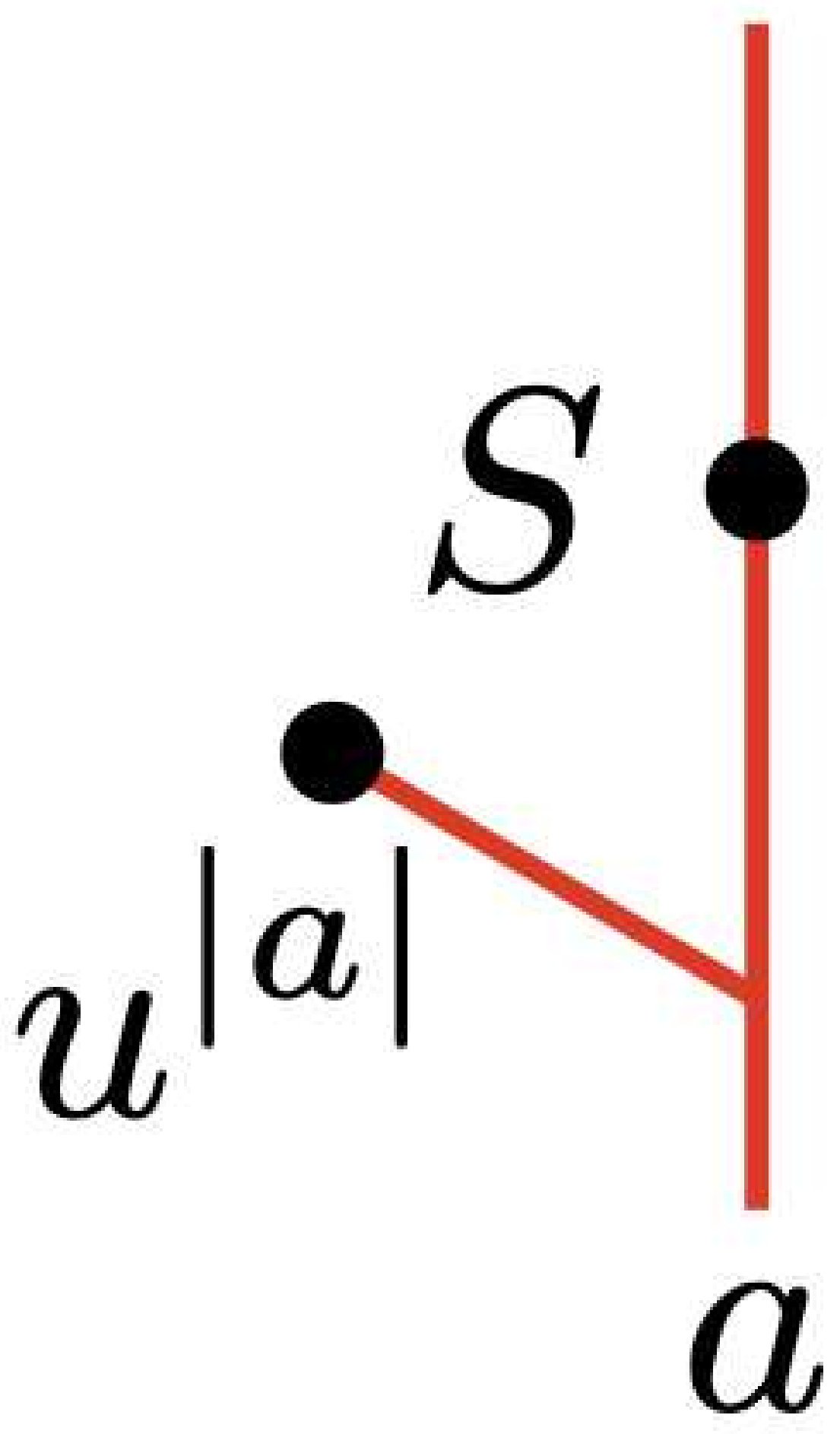} ~ = ~ 
\adjincludegraphics[valign = c, width = 0.7cm]{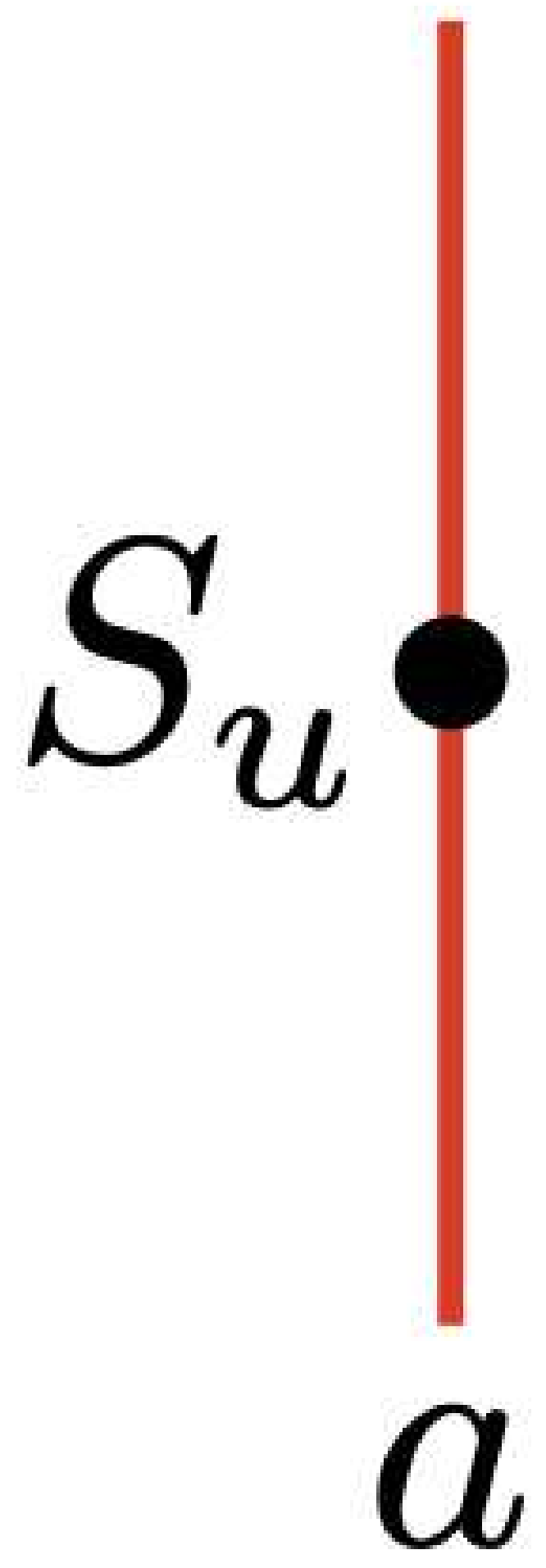}.
\end{equation*}
This completes the proof of the weak Hopf superalgebra structure on $\mathcal{H}^u$.

\section{Supercomodule algebra structure on $K^u$}
\label{sec: supercomodule algebra}
In this appendix, we show that the semisimple superalgebra $K^u$ defined in section \ref{sec: Fermionic TQFTs with SRep symmetry} is an $\mathcal{H}^u$-supercomodule algebra, where $\mathcal{H}^u$ is a weak Hopf superalgebra defined in section \ref{sec: Fermionization of anomalous symmetries}.
As in appendix \ref{sec: weak Hopf superalgebra}, the symmetric braiding $c_{\mathrm{super}}$ and trivial braiding $c_{\mathrm{triv}}$ are represented by a crossing with and without a black dot respectively in string diagrams.

We first recall the definition of $K^u$. 
Let $K$ be a left $H$-comodule algebra, where $H$ is a finite dimensional semisimple weak Hopf algebra.
The multiplication and unit of $K$ are denoted by $m_K$ and $\eta_K$ respectively, and the left $H$-comodule structure on $K$ is denoted by $\delta_K^H: K \rightarrow H \otimes K$.
We define a $\mathbb{Z}_2$-grading $p: K \rightarrow K$ using a $\mathbb{Z}_2$ group-like element $u \in H^*$ as in eq. \eqref{eq: grading on H-comodule algebra K}.
This $\mathbb{Z}_2$-grading is represented by the following string diagram:
\begin{equation}
\adjincludegraphics[valign = c, width = 0.55cm]{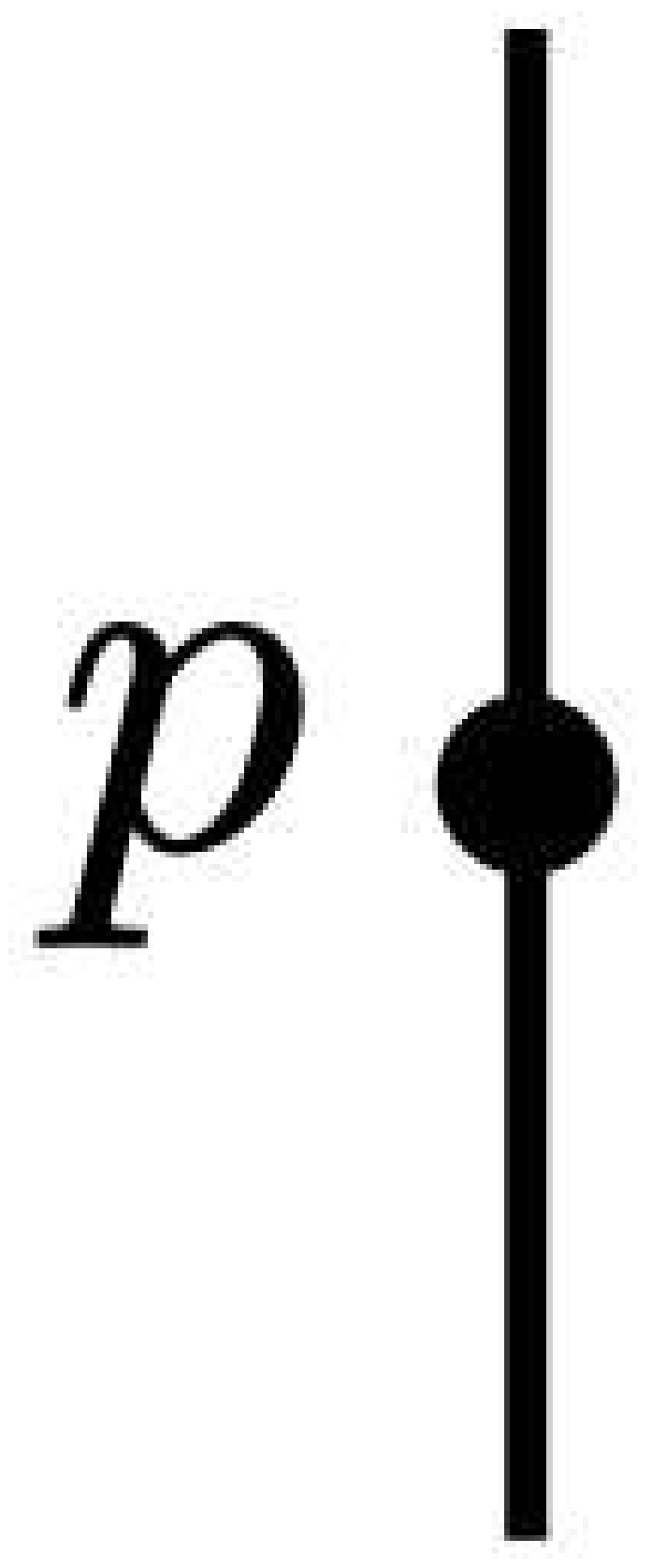} ~ = ~
\adjincludegraphics[valign = c, width = 1.4cm]{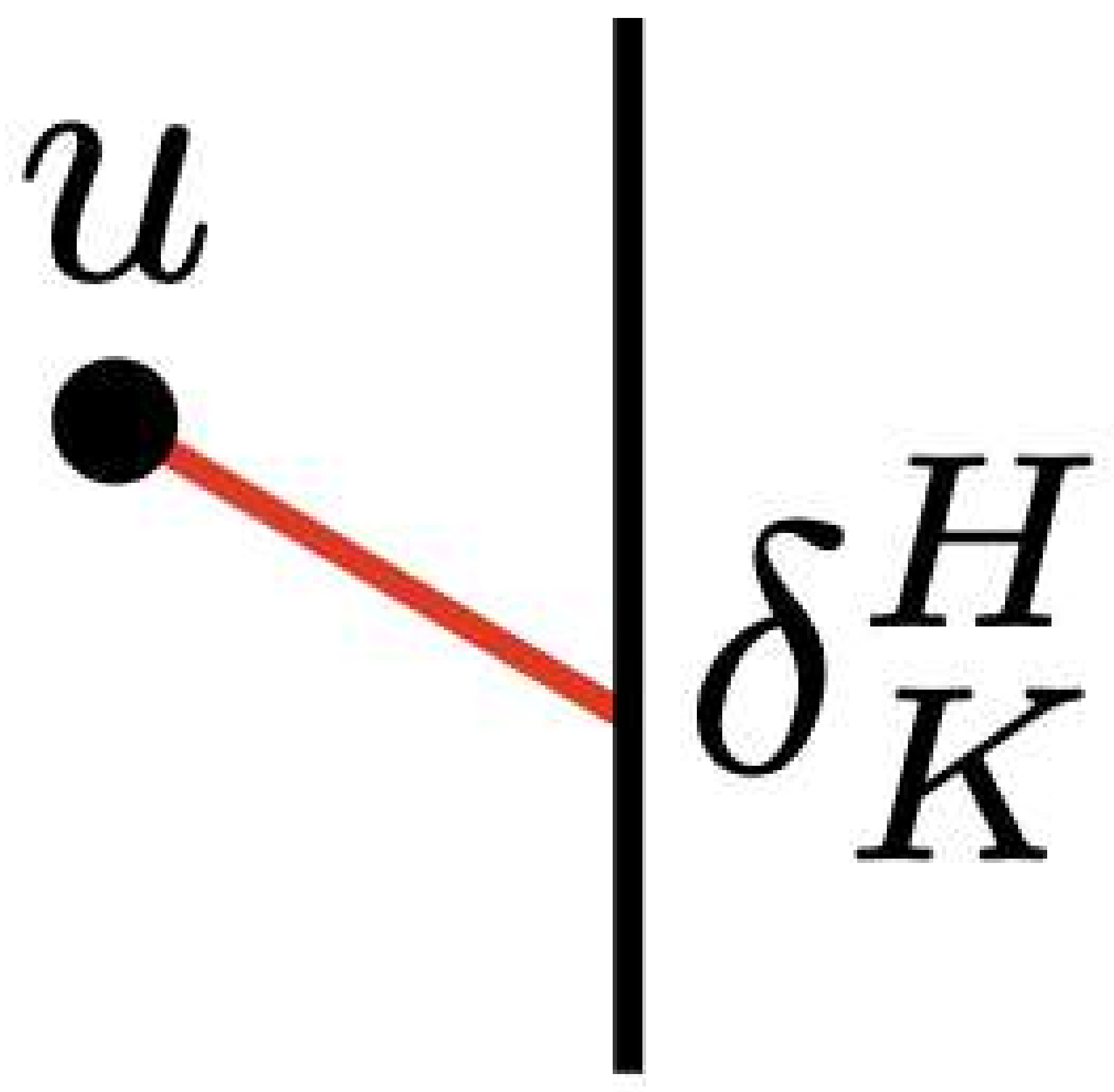}~,
\label{eq: K grading}
\end{equation}
where the black strand corresponds to $K$ and the red strand corresponds to $H$.
An algebra $(K, m_K, \eta_K)$ equipped with the above $\mathbb{Z}_2$-grading $p$ is denoted by $K^u$.
We define a left $\mathcal{H}^u$-coaction on $K^u$ by $\delta_K^H$.\footnote{This is possible because $\mathcal{H}^u$ is $H$ as a coalgebra and $K^u$ is $K$ as a vector space.}

\paragraph{Even structure maps.}
The structure maps $(m_K, \eta_K, \delta_K^H)$ are even with respect to the $\mathbb{Z}_2$-grading \eqref{eq: K grading}.
Indeed, the following computation shows that the multiplication $m_K$ and left $\mathcal{H}^u$-coaction $\delta_K^H$ are even:
\begin{equation}
\adjincludegraphics[valign = c, width = 1.6cm]{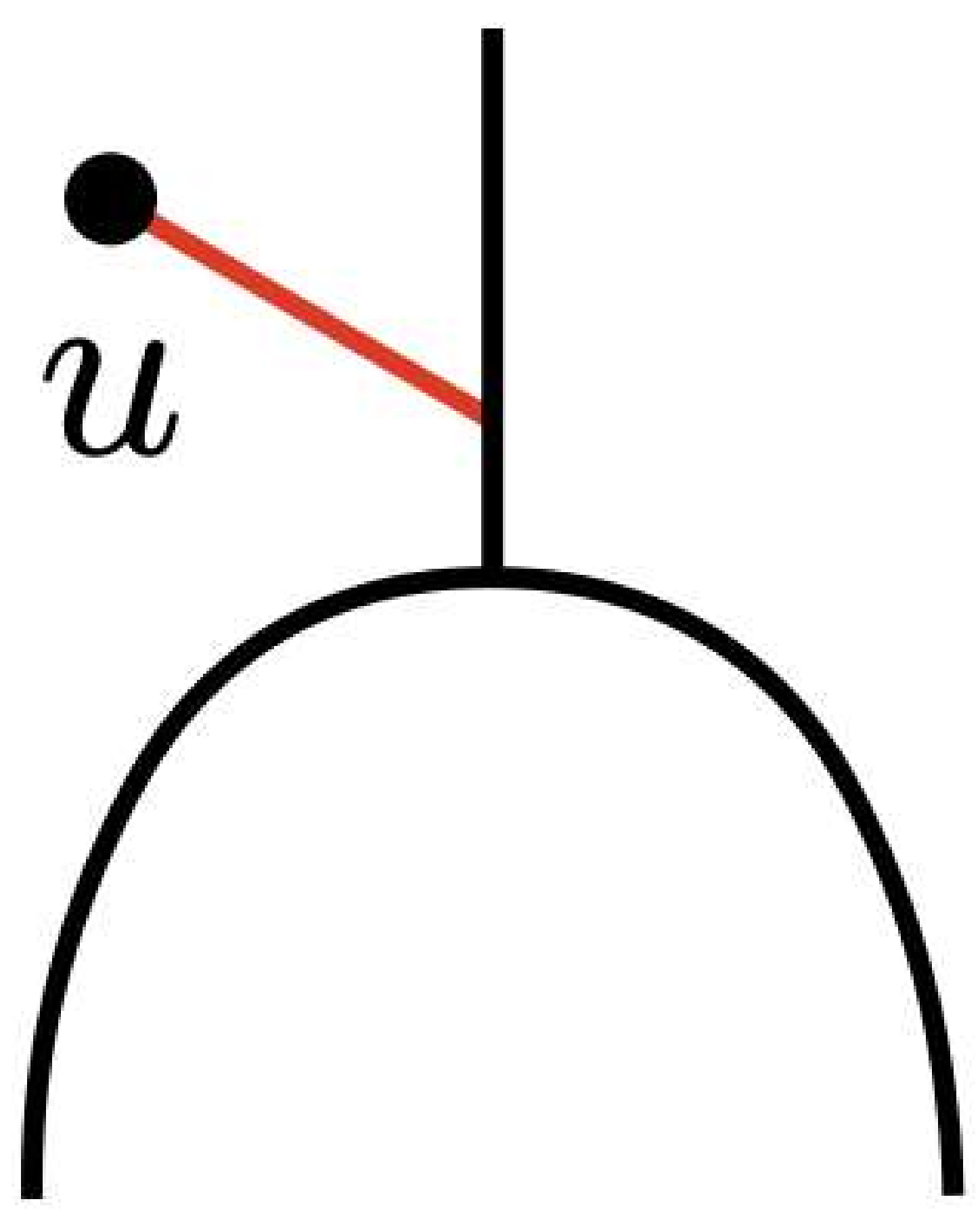} ~ = ~
\adjincludegraphics[valign = c, width = 2.5cm]{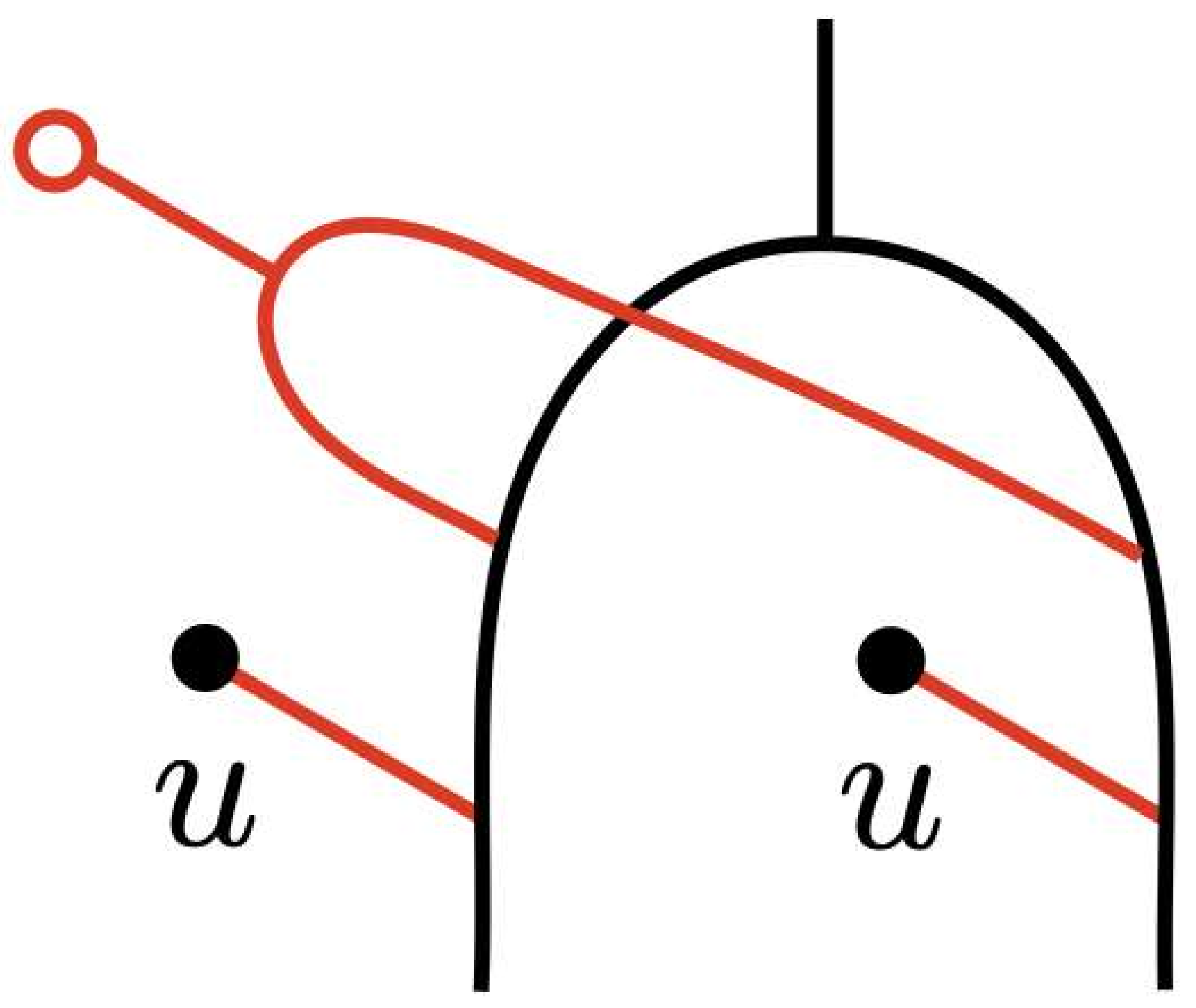} ~ = ~
\adjincludegraphics[valign = c, width = 2cm]{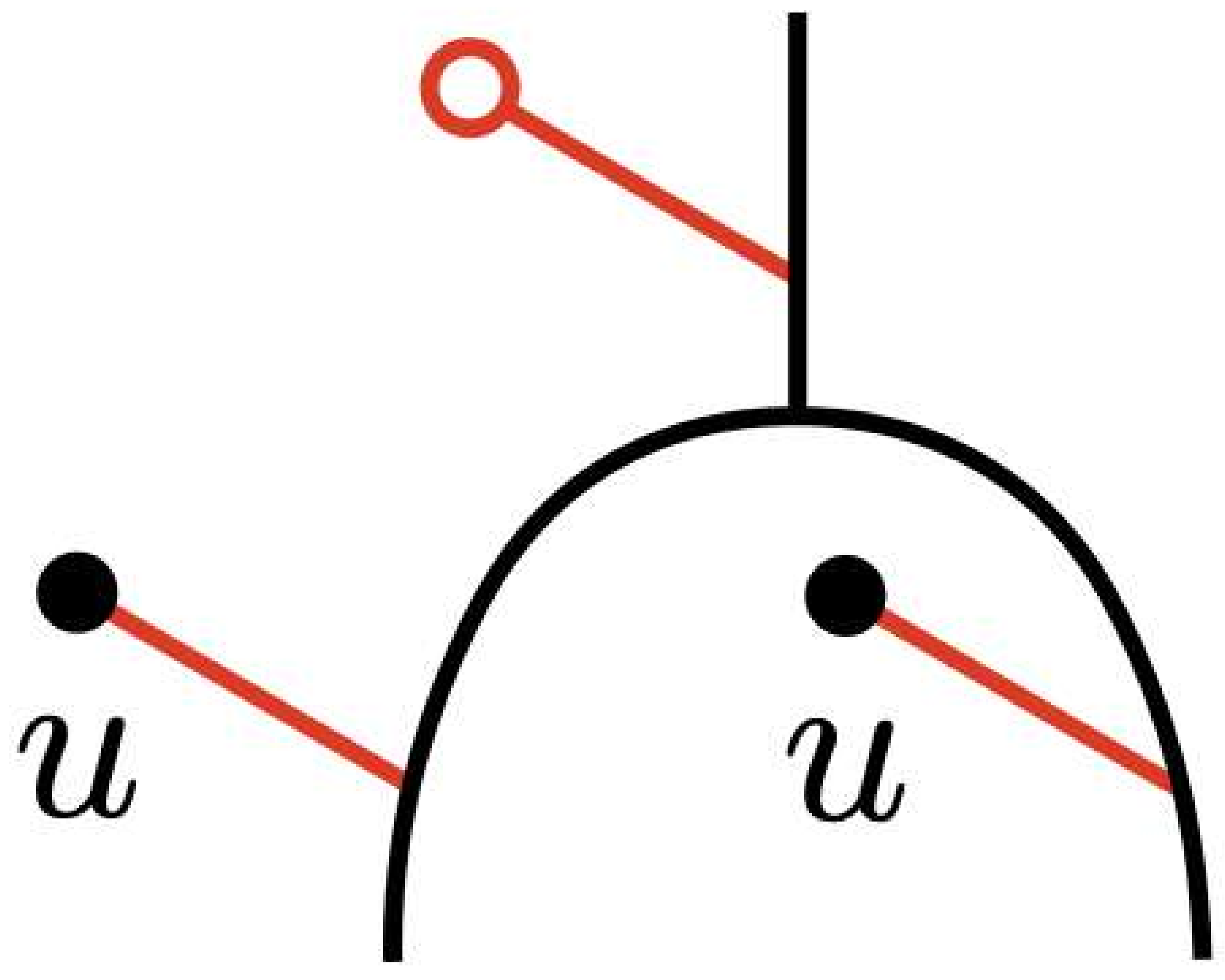} ~ = ~
\adjincludegraphics[valign = c, width = 2cm]{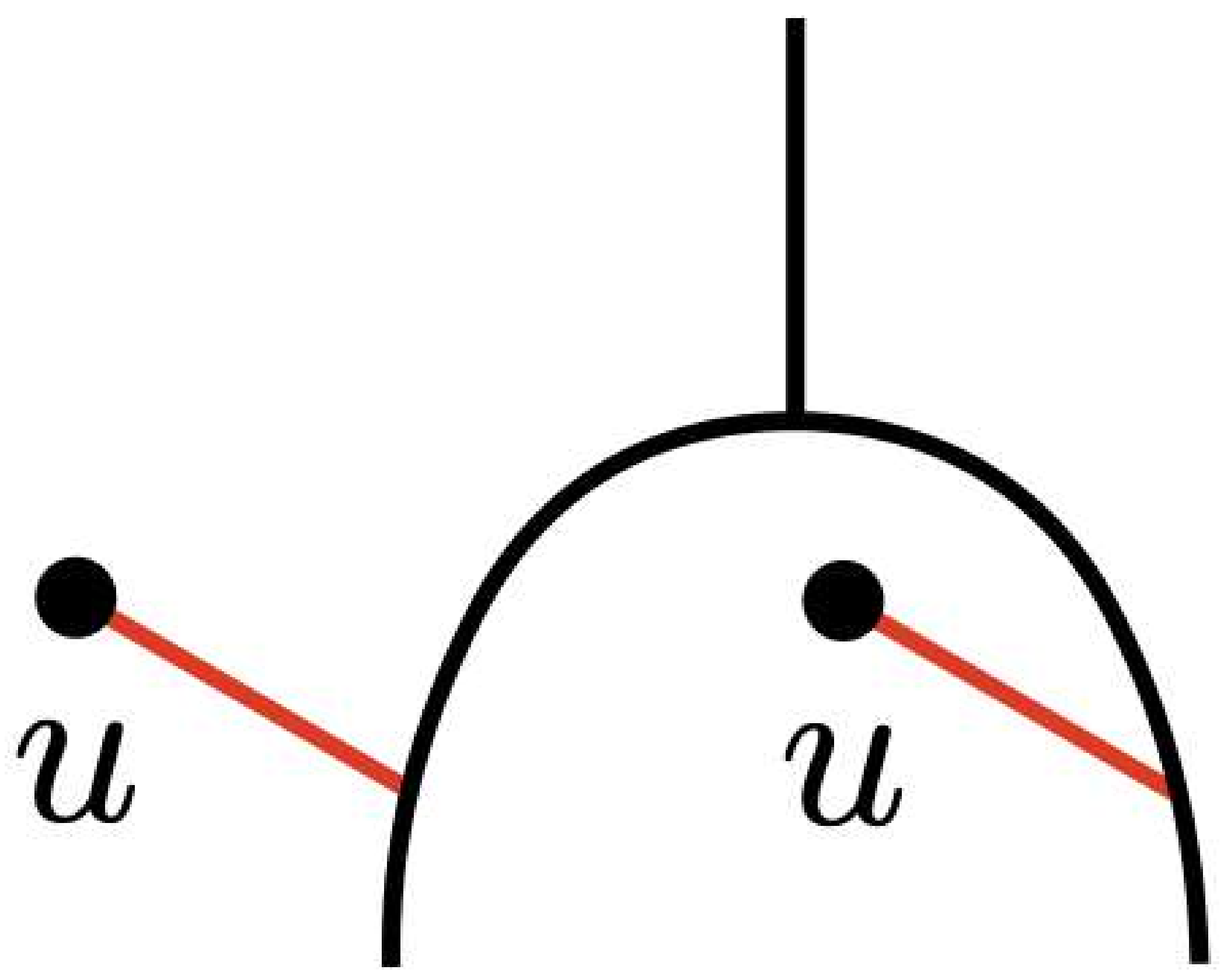},
\end{equation}
\begin{equation}
\adjincludegraphics[valign = c, width = 1.4cm]{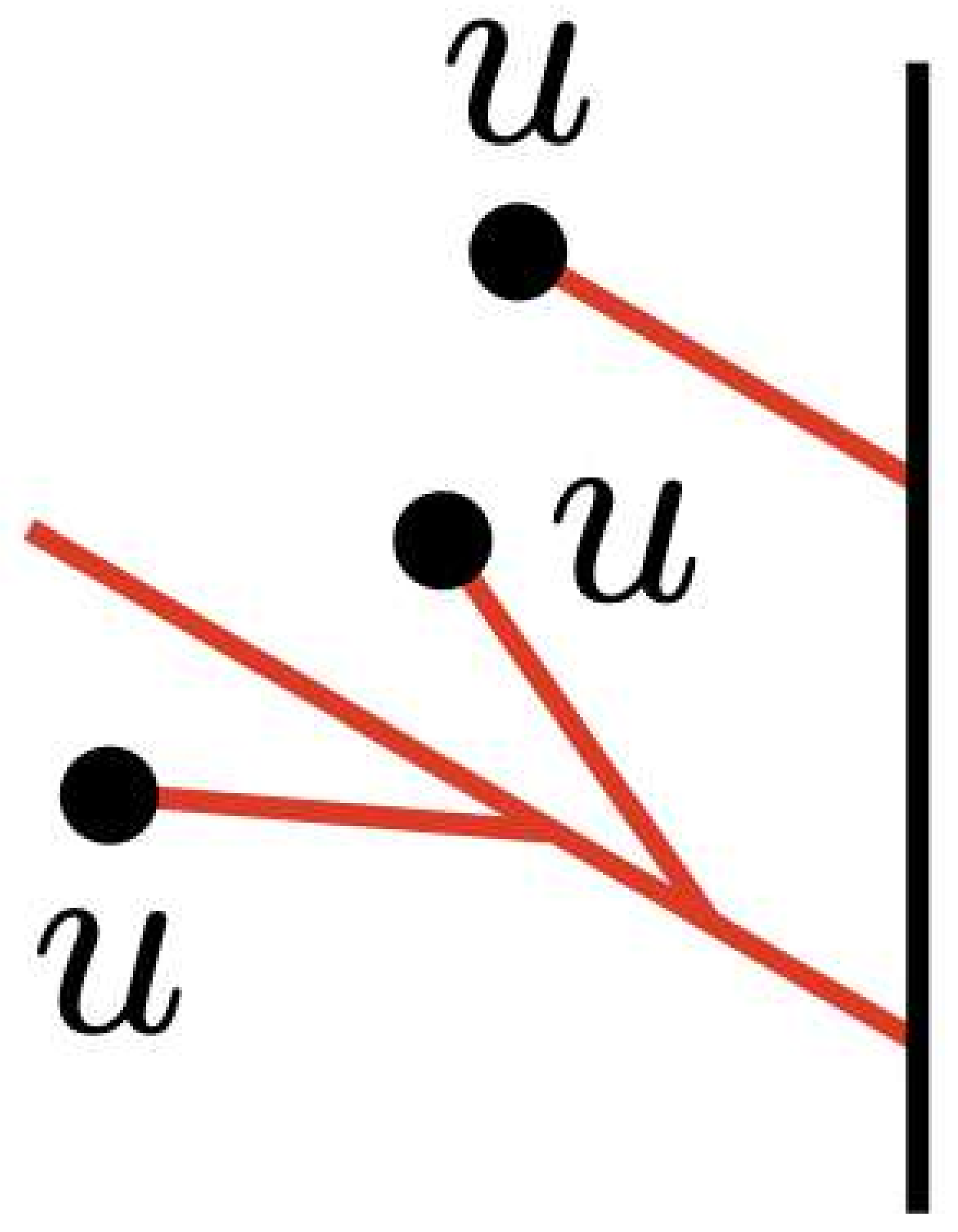} ~ =
\adjincludegraphics[valign = c, width = 1.2cm]{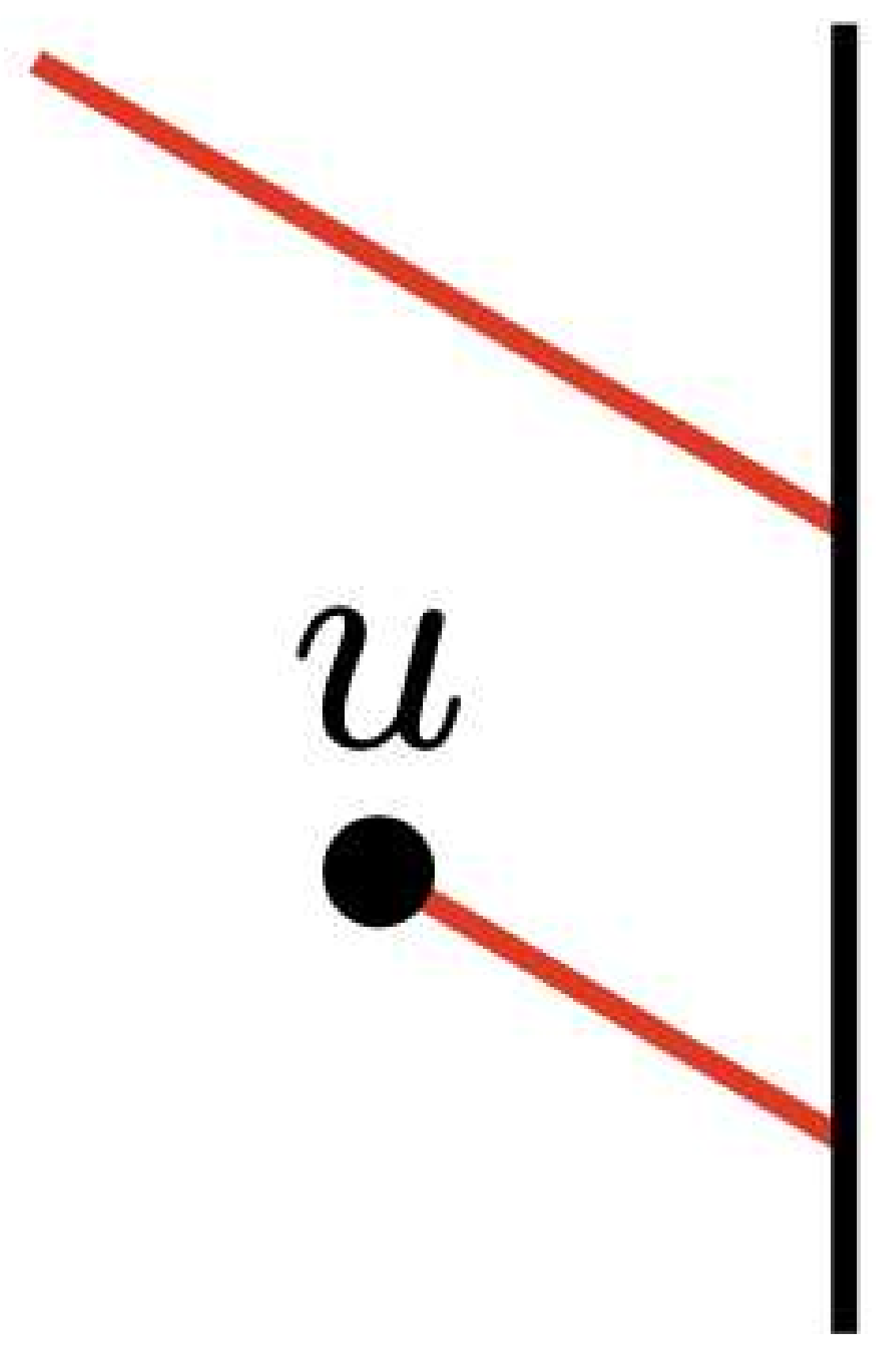}.
\end{equation}
The unit $\eta_K$ is automatically even because the multiplication $m_K$ is even.

\paragraph{Supercomodule algebra structure.}
Since the structure maps are even, the algebra $(K^u, m_K, \eta_K)$ is a superalgebra and $(K^u, \delta_K^H)$ is an $\mathcal{H}^u$-supercomodule.
Therefore, to prove that $K^u$ is an $\mathcal{H}^u$-supercomodule algebra, it suffices to show that the algebra structure and the comodule structure are compatible with each other in the sense of eq. \eqref{eq: weak comodule algebra}.\footnote{Precisely, the trivial braiding $c_{\mathrm{triv}}$ in eq. \eqref{eq: weak comodule algebra} needs to be replaced by the symmetric braiding $c_{\mathrm{super}}$.}
The compatibility between the multiplication $m_K$ and coaction $\delta_K^H$ can be seen from the following equation:
\begin{equation}
\adjincludegraphics[valign = c, width = 1.5cm]{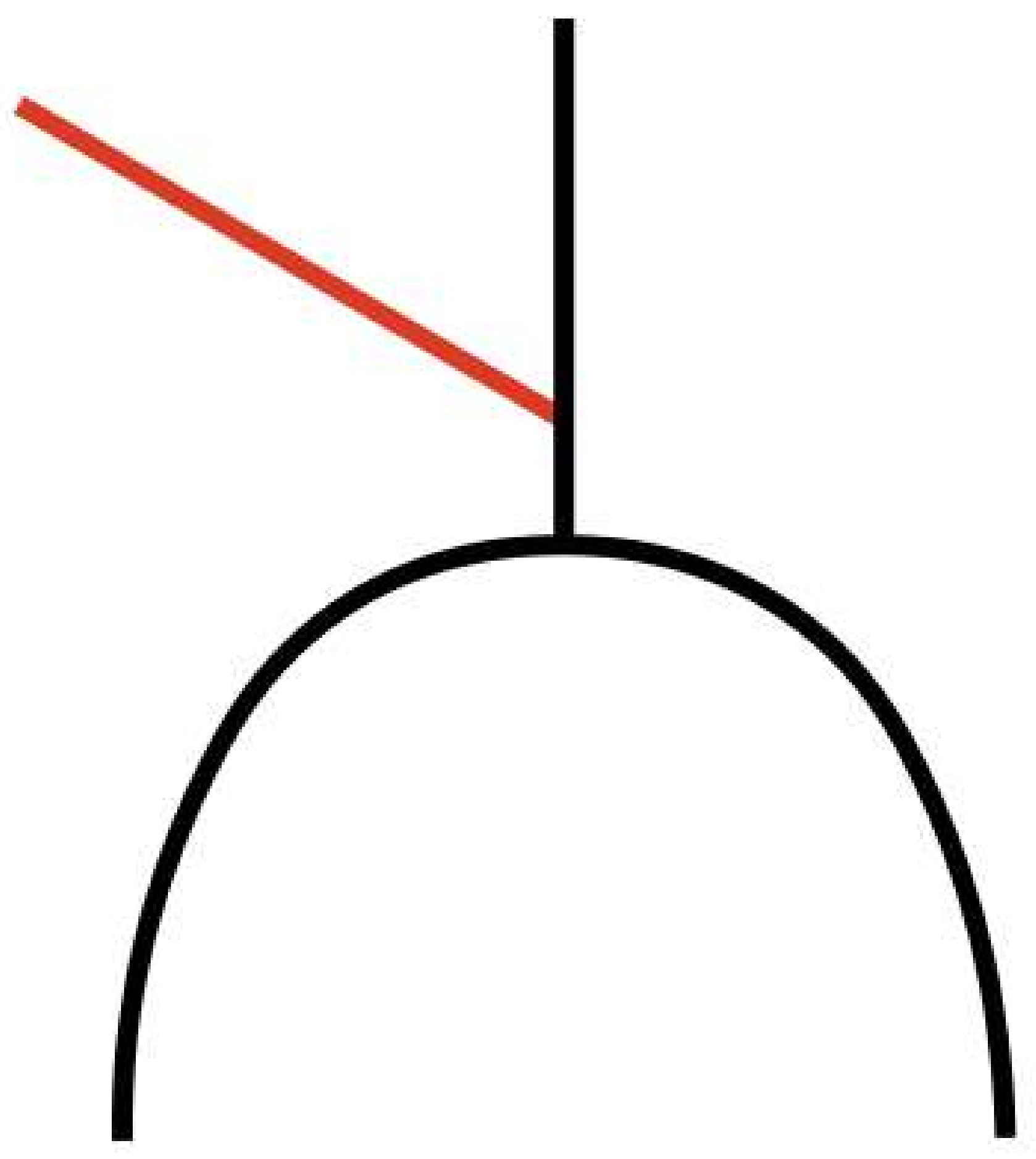} ~ = ~
\adjincludegraphics[valign = c, width = 2.4cm]{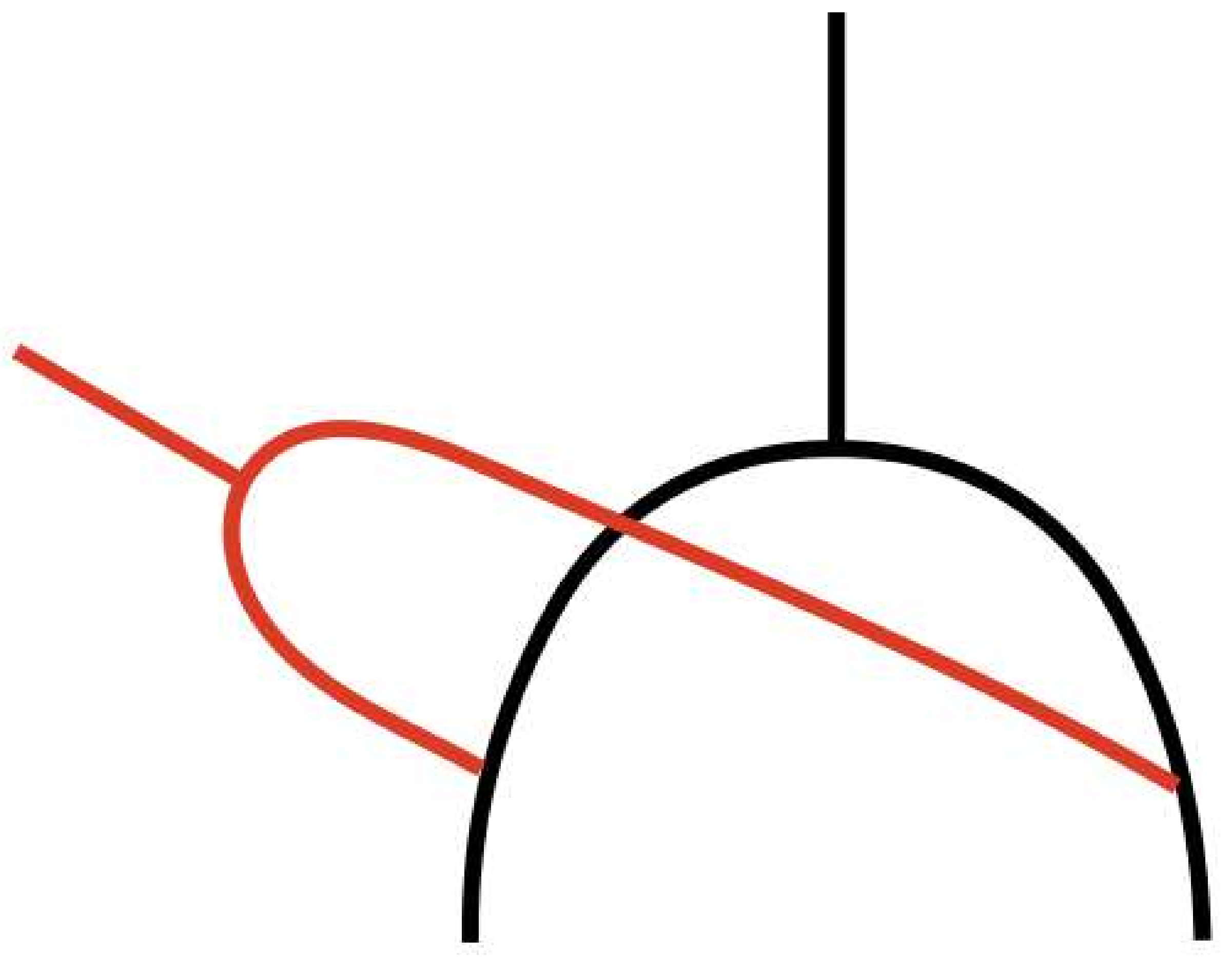} ~ = ~
\adjincludegraphics[valign = c, width = 2.5cm]{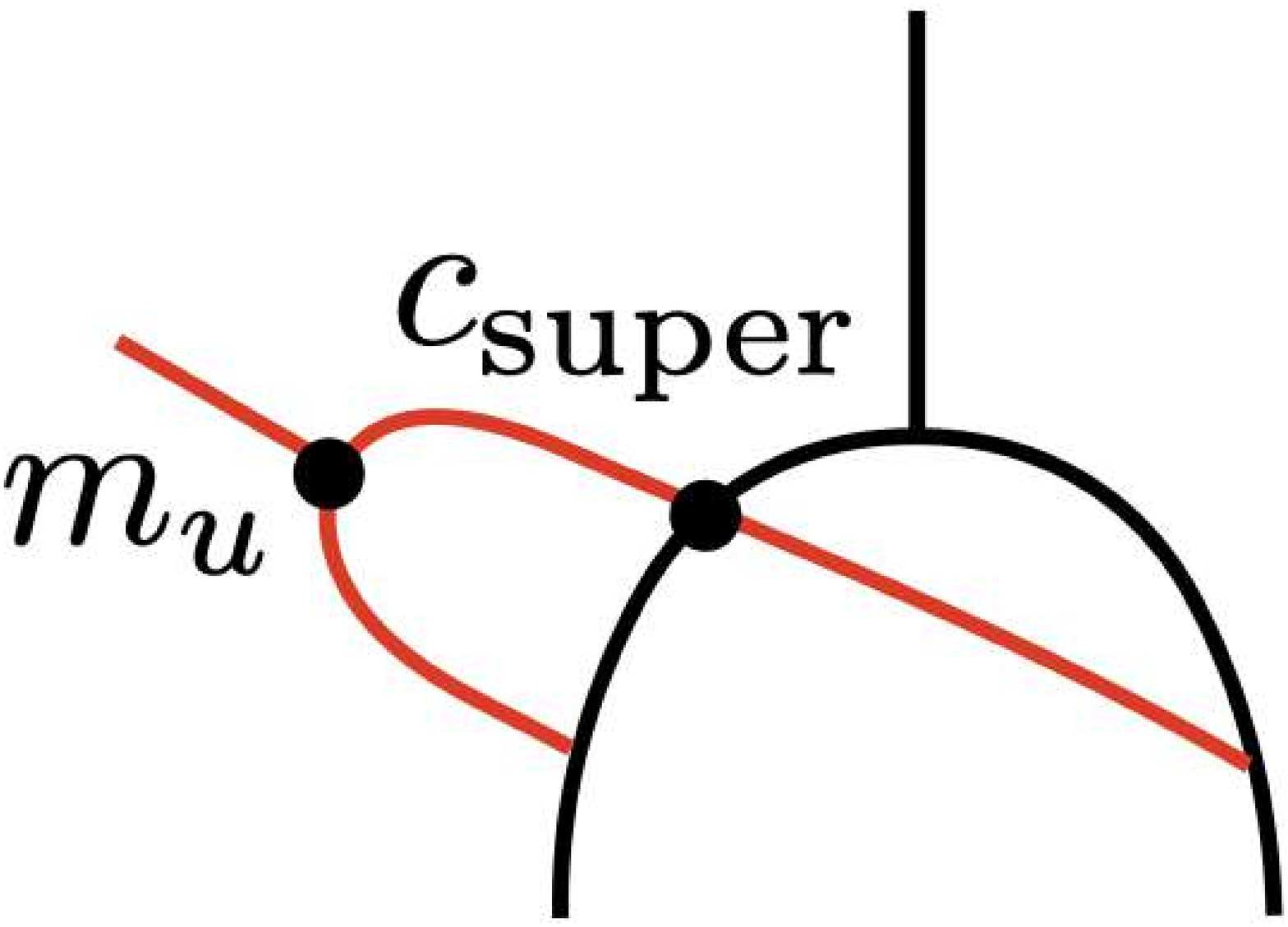}.
\label{eq: compatibility mK}
\end{equation}
The unit $\eta_K$ is also compatible with the coaction $\delta_K^H$ because we have
\begin{equation}
\adjincludegraphics[valign = c, width = 1.15cm]{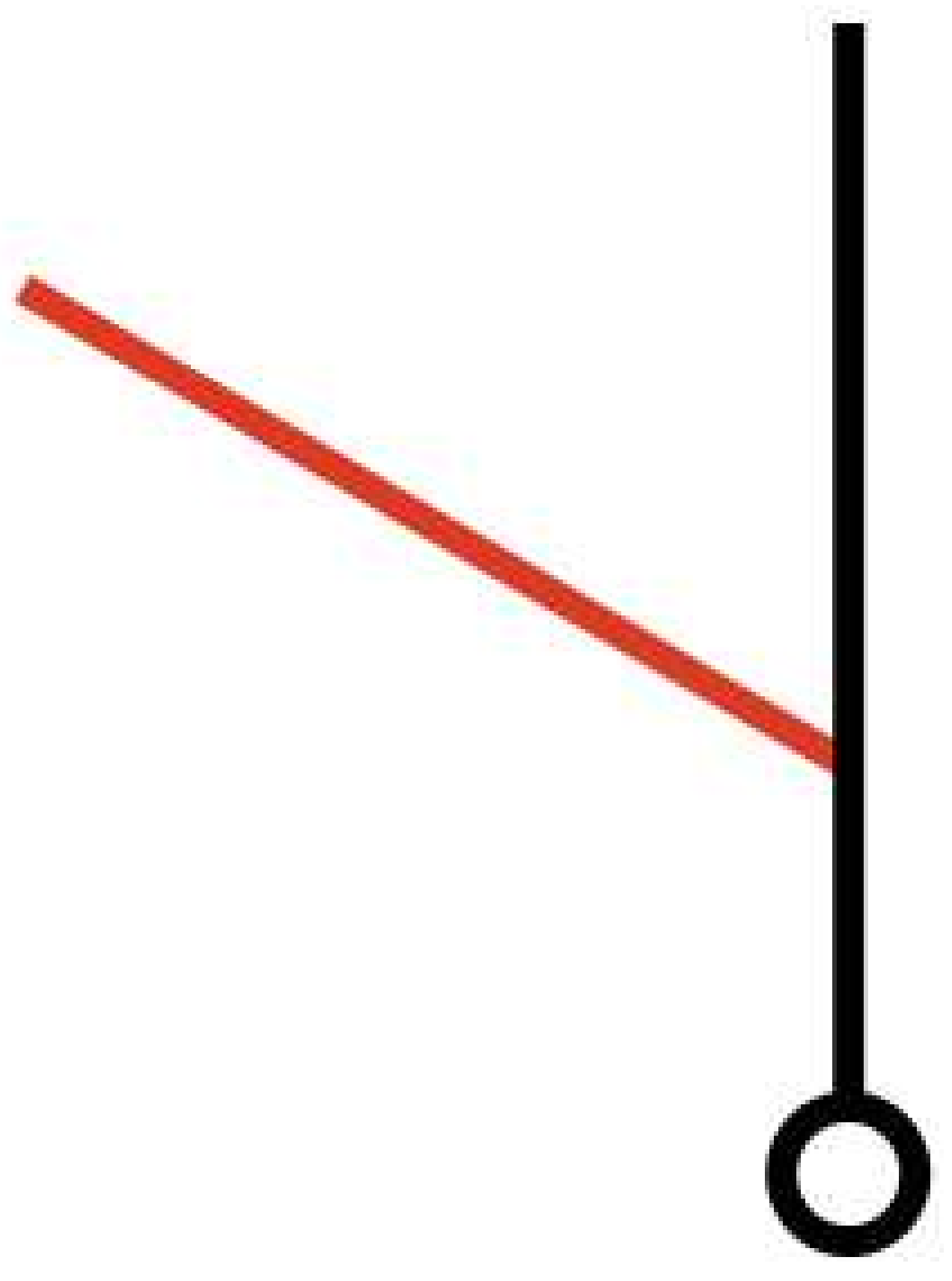} ~ = ~
\adjincludegraphics[valign = c, width = 1.8cm]{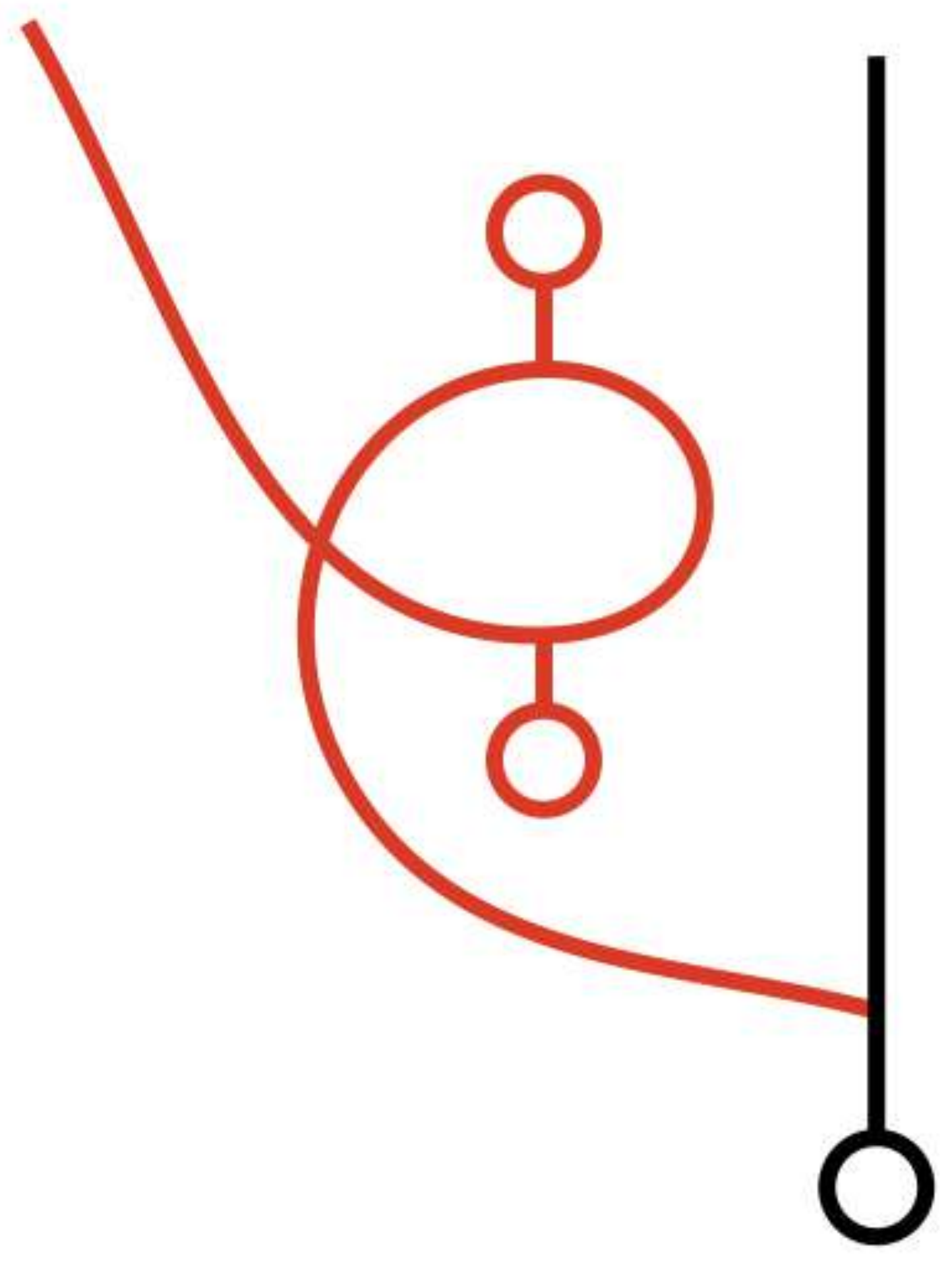} ~ = ~
\adjincludegraphics[valign = c, width = 3.6cm]{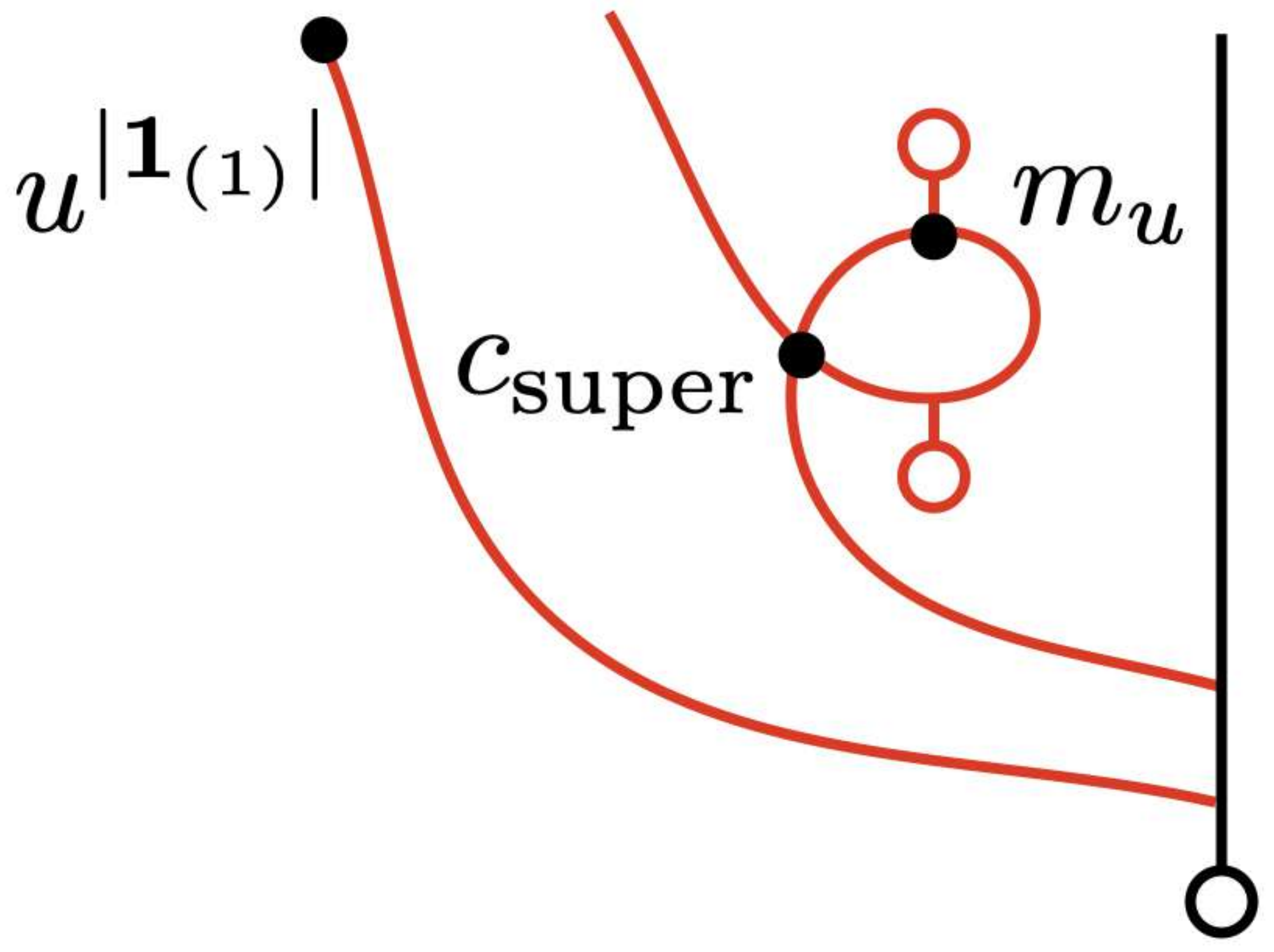} ~ = ~
\adjincludegraphics[valign = c, width = 2.4cm]{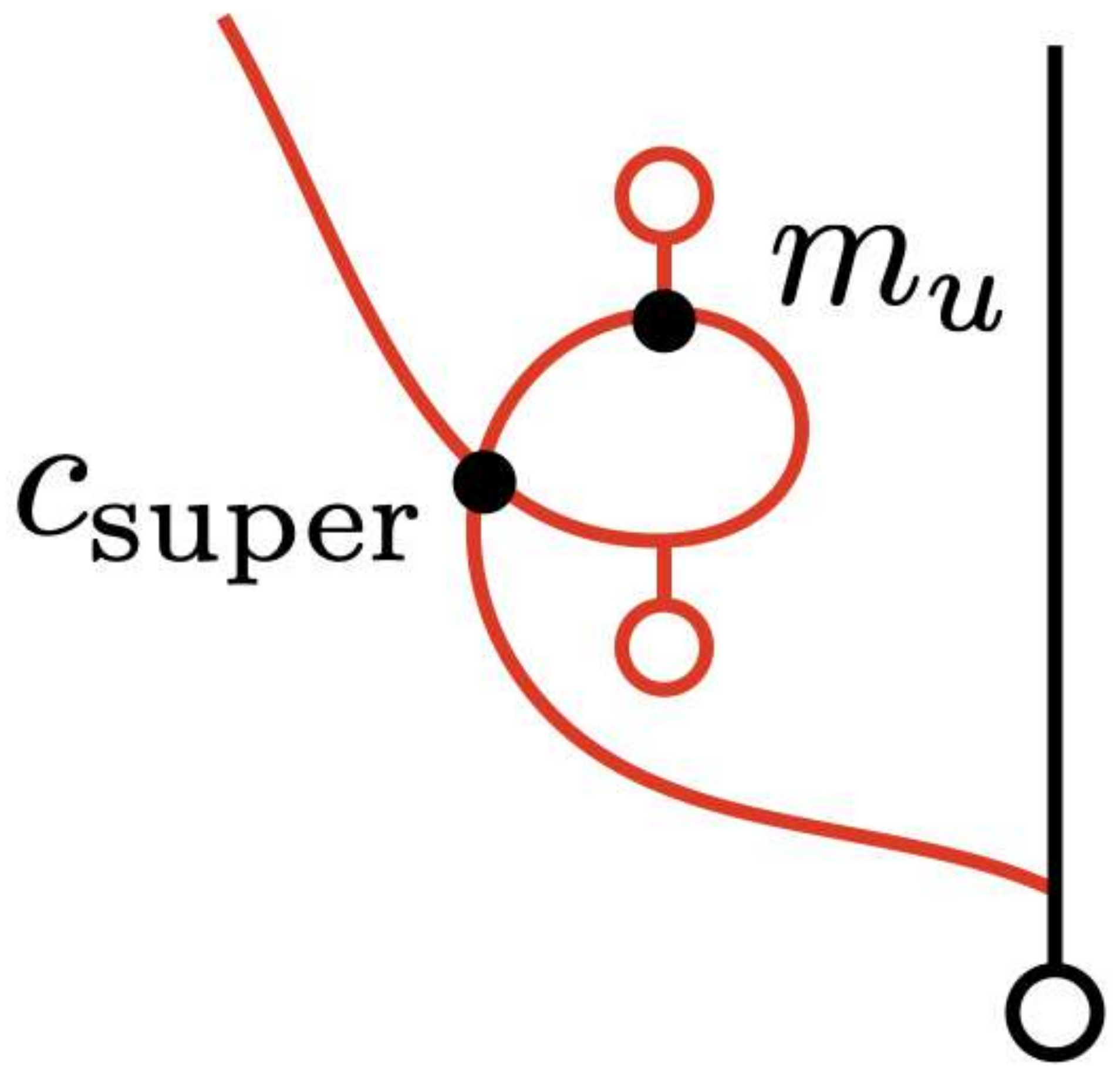},
\label{eq: compatibility etaK}
\end{equation}
where $\mathbf{1} := \eta(1) \in \mathcal{H}^u$.
The first equality follows from the definition of a left $H$-comodule algebra, and the last equality is due to the fact that the unit $\eta_K$ is $\mathbb{Z}_2$-even.
Equations \eqref{eq: compatibility mK} and \eqref{eq: compatibility etaK} show that $K^u$ is an $\mathcal{H}^u$-supercomodule algebra.

\section{Fusion rules of superfusion category symmetries}
\label{sec: Fusion rules of superfusion category symmetries}
In this appendix, we derive the fusion rules of the superfusion categories discussed in section \ref{sec: Examples of fermionization}.

\subsection{Fusion rules of $\mathrm{SRep}(\mathcal{H}^u_{\mathbb{C}[G]^*})$}
We recall that the Hopf superalgebra $\mathcal{H}^u_{\mathbb{C}[G]^*}$ has one-dimensional super representations $V_g \cong \mathbb{C}^{1|0}$ for $g \in C_G(u)$ and two-dimensional super representations $W_g$ for $g \notin C_G(u)$.
The two-dimensional super representation $W_g$ is evenly isomorphic to $W_{ugu}$ and oddly isomorphic to $W_{gu}$ and $W_{ug}$, namely, we have even isomorphisms $W_g \cong W_{ugu} \cong \Pi W_{gu} \cong \Pi W_{ug}$.
In the following, we will compute the fusion rules of these irreducible super representations.
We will follow the definitions and the notations used in section \ref{sec: Fermionization of finite group symmetries}.

\paragraph*{The fusion rule of $V_g$ and $V_h$.}
Let us first compute the fusion rule of one-dimensional super representations $V_g$ and $V_h$.
The action of $\widehat{l} \in \mathcal{H}^u_{\mathbb{C}[G]^*}$ on the tensor product representation $V_g \otimes V_h$ is computed as
\begin{equation}
\Delta ( \widehat{l} ) \cdot v_g \otimes v_h = \delta_{l, gh} v_g \otimes v_h.
\end{equation}
This implies that $V_g \otimes V_h$ is evenly isomorphic to $V_{gh}$, i.e. we have $V_g \otimes V_h \cong V_{gh}$.

\paragraph*{The fusion rule of $V_g$ and $W_h$.}
The action of $\widehat{l} \in \mathcal{H}^u_{\mathbb{C}[G]^*}$ on $V_g \otimes W_h$ is given by
\begin{equation}
\Delta(\widehat{l}) \cdot v_g \otimes w_h = \delta_{l, gh} v_g \otimes \widehat{h} w_h + \delta_{l, ughu} v_g \otimes \widehat{uhu} w_h + \delta_{l, ghu} v_g \otimes \widehat{hu} w_h + \delta_{l, ugh} v_g \otimes \widehat{uh} w_h,
\end{equation}
where we used the fact that $g \in C_G(u)$ commutes with $u$.
This equation implies that the linear map $v_g \otimes w_h \mapsto w_{gh}$ gives an even isomorphism between super representations $V_g \otimes W_h$ and $W_{gh}$.
Therefore, we have $V_g \otimes W_h \cong W_{gh}$.
Similarly, the action of $\widehat{l} \in \mathcal{H}^u_{\mathbb{C}[G]^*}$ on $W_g \otimes V_h$ is given by
\begin{equation}
\Delta(\widehat{l}) \cdot w_g \otimes v_h = \delta_{l, gh} \widehat{g} w_g \otimes v_h + \delta_{l, ughu} \widehat{ugu} w_g \otimes v_h + \delta_{l, ghu} \widehat{gu} w_g \otimes v_h + \delta_{l, ugh} \widehat{ug} w_g \otimes v_h,
\end{equation}
which implies that the linear map $w_g \otimes v_h \mapsto w_{gh}$ gives an even isomorphism $W_g \otimes V_h \cong W_{gh}$.

\paragraph*{The fusion rule of $W_g$ and $W_h$.}
The action of $\widehat{l} \in \mathcal{H}^u_{\mathbb{C}[G]^*}$ on $W_g \otimes W_h$ is given by
\begin{equation}
\begin{aligned}
\Delta(\widehat{l}) \cdot (w_g)_s \otimes (w_h)_t & = \frac{1}{4} (\widehat{lh^{-1}} + \widehat{luh^{-1}u} + \widehat{luh^{-1}} + \widehat{lh^{-1}u}) (w_g)_s \otimes (w_h)_t \\
& \quad + \frac{t}{4} (\widehat{lh^{-1}} + \widehat{luh^{-1}u} - \widehat{luh^{-1}} - \widehat{lh^{-1}u}) (w_g)_s \otimes (w_h)_t \\
& \quad + \frac{s}{4} (\widehat{lh^{-1}} - \widehat{luh^{-1}u} + \widehat{luh^{-1}} - \widehat{lh^{-1}u}) (w_g)_s \otimes (w_h)_{-t} \\
& \quad + \frac{st}{4} (\widehat{lh^{-1}} - \widehat{luh^{-1}u} - \widehat{luh^{-1}} + \widehat{lh^{-1}u}) (w_g)_s \otimes (w_h)_{-t},
\end{aligned}
\end{equation}
where $s, t = \pm 1$ represents the $\mathbb{Z}_2$-grading of $(w_g)_s$ and $(w_h)_t$.
We note that an element $\widehat{l}$ acts non-trivially on $W_g \otimes W_h$ only when $l$ is in a set $\{gh, ughu, ghu, ugh\} \sqcup \{uguh, guhu, uguhu, guh\}$.
This set depends on whether $gh$ and $uguh$ commute with $u$.
Accordingly, the fusion rule of $W_g$ and $W_h$ is divided into four cases.
Let us investigate these cases one by one.

When both $gh$ and $uguh$ commute with $u$, the subalgebra of $\mathcal{H}^u_{\mathbb{C}[G]^*}$ that acts non-trivially on $W_g \otimes W_h$ is spanned by $\{\widehat{gh}, \widehat{ghu}\} \sqcup \{\widehat{uguh}, \widehat{guh}\}$.
Since this subalgebra only has one-dimensional irreducible super representations, the four-dimensional super representation $W_g \otimes W_h$ can be decomposed into the direct sum of four one-dimensional super representations associated with $\widehat{gh}$, $\widehat{ghu}$, $\widehat{uguh}$, and $\widehat{guh}$.
More specifically, the non-trivial action of $\widehat{l} \in \mathcal{H}^u_{\mathbb{C}[G]^*}$ is summarized as
\begin{equation}
\begin{aligned}
\Delta(\widehat{gh}) \cdot [(w_g)_+ \otimes (w_h)_+ + (w_g)_- \otimes (w_h)_-] & = (w_g)_+ \otimes (w_h)_+ + (w_g)_- \otimes (w_h)_-, \\
\Delta(\widehat{ghu}) \cdot [(w_g)_+ \otimes (w_h)_- + (w_g)_- \otimes (w_h)_+] & = (w_g)_+ \otimes (w_h)_- + (w_g)_- \otimes (w_h)_+, \\
\Delta(\widehat{uguh}) \cdot [(w_g)_+ \otimes (w_h)_+ - (w_g)_- \otimes (w_h)_-] & = (w_g)_+ \otimes (w_h)_+ - (w_g)_- \otimes (w_h)_-, \\
\Delta(\widehat{guh}) \cdot [(w_g)_+ \otimes (w_h)_- - (w_g)_- \otimes (w_h)_+] & = (w_g)_+ \otimes (w_h)_- - (w_g)_- \otimes (w_h)_+.
\end{aligned}
\label{eq: gh and uguh commute with u}
\end{equation}
This shows that we have an even isomorphism of super representations
\begin{equation}
W_g \otimes W_h \cong V_{gh} \oplus V_{uguh} \oplus \Pi V_{ghu} \oplus \Pi V_{guh}.
\end{equation}

When only $gh$ commutes with $u$, the subalgebra of $\mathcal{H}^u_{\mathbb{C}[G]^*}$ that acts non-trivially on $W_g \otimes W_h$ is spanned by $\{\widehat{gh}, \widehat{ghu}\} \sqcup \{\widehat{uguh}, \widehat{guhu}, \widehat{uguhu}, \widehat{guh}\}$.
The action of $\widehat{gh}$ and $\widehat{ghu}$ are given by the first two equalities in eq. \eqref{eq: gh and uguh commute with u}, which implies that the tensor product representation $W_g \otimes W_h$ contains one-dimensional super representations $V_{gh}$ and $\Pi V_{ghu}$.
The remaining two-dimensional subspace spanned by $(w_g)_+ \otimes (w_h)_+ - (w_g)_- \otimes (w_h)_-$ and $(w_g)_+ \otimes (w_h)_- - (w_g)_- \otimes (w_h)_+$ becomes an irreducible super representation $W_{uguh}$, on which $\mathcal{H}^u_{\mathbb{C}[G]^*}$ acts as
\begin{equation}
\begin{aligned}
\Gamma_{uguh} [(w_g)_+ \otimes (w_h)_+ - (w_g)_- \otimes (w_h)_-] & = (w_g)_+ \otimes (w_h)_- - (w_g)_- \otimes (w_h)_+, \\
\Gamma_{uguh} [(w_g)_+ \otimes (w_h)_- - (w_g)_- \otimes (w_h)_+] & = (w_g)_+ \otimes (w_h)_+ - (w_g)_- \otimes (w_h)_-, \\
\Gamma_{uguh}^{\prime} [(w_g)_+ \otimes (w_h)_+ - (w_g)_- \otimes (w_h)_- ] & = (w_g)_+ \otimes (w_h)_- - (w_g)_- \otimes (w_h)_+, \\
\Gamma_{uguh}^{\prime} [(w_g)_+ \otimes (w_h)_- - (w_g)_- \otimes (w_h)_+] & = - [(w_g)_+ \otimes (w_h)_+ - (w_g)_- \otimes (w_h)_-].
\end{aligned}
\label{eq: gh commutes with u}
\end{equation}
Therefore, we have an even isomorphism of super representations
\begin{equation}
W_g \otimes W_h \cong V_{gh} \oplus \Pi V_{ghu} \oplus W_{uguh}.
\end{equation}

When only $uguh$ commutes with $u$, the subalgebra of $\mathcal{H}^u_{\mathbb{C}[G]^*}$ that acts non-trivially on $W_g \otimes W_h$ is spanned by $\{\widehat{gh}, \widehat{ughu}, \widehat{ghu}, \widehat{ugh}\} \sqcup \{\widehat{uguh}, \widehat{guh}\}$.
The action of $\widehat{uguh}$ and $\widehat{guh}$ are given by the last two equalities in eq. \eqref{eq: gh and uguh commute with u}, which implies that the tensor product representation $W_g \otimes W_h$ contains one-dimensional super representations $V_{uguh}$ and $\Pi V_{guh}$.
The remaining two-dimensional subspace spanned by $(w_g)_+ \otimes (w_h)_+ + (w_g)_- \otimes (w_h)_-$ and $(w_g)_+ \otimes (w_h)_- + (w_g)_- \otimes (w_h)_+$ is an irreducible super representation $W_{gh}$, on which $\mathcal{H}^u_{\mathbb{C}[G]^*}$ acts as
\begin{equation}
\begin{aligned}
\Gamma_{gh} [(w_g)_+ \otimes (w_h)_+ + (w_g)_- \otimes (w_h)_-] & = (w_g)_+ \otimes (w_h)_- + (w_g)_- \otimes (w_h)_+, \\
\Gamma_{gh} [(w_g)_+ \otimes (w_h)_- + (w_g)_- \otimes (w_h)_+] & = (w_g)_+ \otimes (w_h)_+ + (w_g)_- \otimes (w_h)_-, \\
\Gamma_{gh}^{\prime} [(w_g)_+ \otimes (w_h)_+ + (w_g)_- \otimes (w_h)_- ] & = (w_g)_+ \otimes (w_h)_- + (w_g)_- \otimes (w_h)_+, \\
\Gamma_{gh}^{\prime} [(w_g)_+ \otimes (w_h)_- + (w_g)_- \otimes (w_h)_+] & = - [(w_g)_+ \otimes (w_h)_+ + (w_g)_- \otimes (w_h)_-].
\end{aligned}
\label{eq: uguh commutes with u}
\end{equation}
Therefore, we have an even isomorphism of super representations
\begin{equation}
W_g \otimes W_h \cong W_{gh} \oplus V_{uguh} \oplus \Pi V_{guh}.
\end{equation}

Finally, when both $gh$ and $uguh$ do not commute with $u$, the subalgebra of $\mathcal{H}^u_{\mathbb{C}[G]^*}$ that acts non-trivially on $W_g \otimes W_h$ is spanned by $\{\widehat{gh}, \widehat{ughu}, \widehat{ghu}, \widehat{ugh}\} \sqcup \{\widehat{uguh}, \widehat{guhu}, \widehat{uguhu}, \widehat{guh}\}$.
The action of this subalgebra on $W_g \otimes W_h$ is given by eqs. \eqref{eq: gh commutes with u} and \eqref{eq: uguh commutes with u}.
Therefore, we have an even isomorphism of super representations
\begin{equation}
W_g \otimes W_h \cong W_{gh} \oplus W_{uguh}.
\end{equation}

\subsection{Fusion rules of $\mathrm{SRep}(\mathcal{H}^u_8)$}
The Hopf superalgebra $\mathcal{H}^u_8$ has four one-dimensional super representations $V_i$ labeled by $1\leq i \leq 4$ and two two-dimensional super representations $W_1$ and $W_2$.
The one-dimensional super representation $V_i$ is a super vector space $\mathbb{C}^{1|0}$ on which the idempotent $e_i \in \mathcal{H}^u_8$ acts as the identity and $e_j \in \mathcal{H}^u_8$ acts as zero if $j \neq i$.
The two-dimensional super representation $W_1$ is a super vector space $\mathbb{C}^{1|1}$ on which the subalgebra of $\mathcal{H}^u_8$ spanned by $e_5$ and $e_7$ acts non-trivially and $e_{j \neq 5, 7}$ acts as zero.
The action of $e_5$ and $e_7$ on $W_1$ is given by
\begin{equation}
e_5 \cdot w_{\pm} = w_{\pm}, \quad e_7 \cdot w_{\pm} = w_{\mp},
\label{eq: W1}
\end{equation}
where $w_+$ and $w_-$ are $\mathbb{Z}_2$-even and $\mathbb{Z}_2$-odd elements of $W_1$.
Similarly, the two-dimensional super representation $W_2$ is a super vector space $\mathbb{C}^{1|1}$ on which the subalgebra of $\mathcal{H}^u_8$ spanned by $e_6$ and $e_8$ acts analogously to eq. \eqref{eq: W1} and $e_{j \neq 6, 8}$ acts as zero.

In order to derive the fusion rules, we first compute the comultiplication of $\mathcal{H}^u_8$.
A direct computation shows that the comultiplication is given by 
\begin{equation}
\begin{aligned}
\Delta(e_1) & = e_1 \otimes e_1 + e_2 \otimes e_2 + e_3 \otimes e_3 + e_4 \otimes e_4 + \frac{1}{2} (e_5 \otimes e_5 - u(z) e_7 \otimes e_7) + \frac{1}{2} (e_6 \otimes e_6 + u(z) e_8 \otimes e_8), \\
\Delta(e_2) & = e_1 \otimes e_2 + e_2 \otimes e_1 + e_3 \otimes e_4 + e_4 \otimes e_3 + \frac{1}{2} (e_5 \otimes e_5 + u(z) e_7 \otimes e_7) + \frac{1}{2} (e_6 \otimes e_6 - u(z) e_8 \otimes e_8), \\
\Delta(e_3) & = e_1 \otimes e_3 + e_2 \otimes e_4 + e_3 \otimes e_1 + e_4 \otimes e_2 + \frac{1}{2} (e_5 \otimes e_6 + e_6 \otimes e_5 + u(z) e_7 \otimes e_8 - u(z) e_8 \otimes e_7), \\
\Delta(e_4) & = e_1 \otimes e_4 + e_2 \otimes e_3 + e_3 \otimes e_2 + e_4 \otimes e_1 + \frac{1}{2} (e_5 \otimes e_6 + e_6 \otimes e_5 - u(z) e_7 \otimes e_8 + u(z) e_8 \otimes e_7), \\
\Delta(e_5) & = (e_1 + e_2) \otimes e_5 + (e_3 + e_4) \otimes e_6 + e_5 \otimes (e_1 + e_2) + e_6 \otimes (e_3 + e_4), \\
\Delta(e_6) & = (e_1 + e_2) \otimes e_6 + (e_3 + e_4) \otimes e_5 + e_5 \otimes (e_3 + e_4) + e_6 \otimes (e_1 + e_2), \\
\Delta(e_7) & = (e_1 - e_2) \otimes e_7 + (e_3 - e_4) \otimes e_8 + e_7 \otimes (e_1 - e_2) - e_8 \otimes (e_3 - e_4), \\
\Delta(e_8) & = (e_1 - e_2) \otimes e_8 + (e_3 - e_4) \otimes e_7 - e_7 \otimes (e_3 - e_4) + e_8 \otimes (e_1 - e_2).
\end{aligned}
\label{eq: comultiplication of H8}
\end{equation}
Let us derive the fusion rules of irreducible super representations of $\mathcal{H}^u_8$ based on the above expression of the comultiplication.

\paragraph*{The fusion rule of $V_i$ and $V_j$.}
An element $e_k \in \mathcal{H}^u_8$ acts non-trivially on the tensor product representation $V_i \otimes V_j$ only if the comultiplication $\Delta(e_k)$ contains $e_i \otimes e_j$.
When $e_k$ acts non-trivially on $V_i \otimes V_j$, this is the unique element (up to scalar multiplication) that acts non-trivially on $V_i \otimes V_j$ because the tensor product of one-dimensional super representations $V_i$ and $V_j$ is also one-dimensional.
This gives an even isomorphism of super representations $V_i \otimes V_j \cong V_k$.
More specifically, we have the following fusion rules:
\begin{equation}
V_1 \otimes V_i \cong V_i \otimes V_1 \cong V_i, \quad V_2 \otimes V_2 \cong V_3 \otimes V_3 \cong V_4 \otimes V_4 \cong V_1, \quad V_2 \otimes V_3 \cong V_3 \otimes V_2 \cong V_4.
\end{equation}

\paragraph*{The fusion rule of $V_i$ and $W_j$.}
An element $e_k \in \mathcal{H}^u_8$ acts non-trivially on $V_i \otimes W_1$ only if the comultiplication $\Delta(e_k)$ contains $e_i \otimes e_5$ or $e_i \otimes e_7$.
Similarly, an element $e_k \in \mathcal{H}^u_8$ acts non-trivially on $W_1 \otimes V_i$ only if the comultiplication $\Delta(e_k)$ contains $e_5 \otimes e_i$ or $e_7 \otimes e_i$.
These facts completely determine the fusion rules of $V_i$ and $W_1$ as follows:
\begin{equation}
\begin{aligned}
V_1 \otimes W_1 & \cong V_2 \otimes W_1 \cong W_1 \cong W_1 \otimes V_1 \cong W_1 \otimes V_2, \\
V_3 \otimes W_1 & \cong V_4 \otimes W_1 \cong W_2 \cong W_1 \otimes V_3 \cong W_1 \otimes V_4.
\end{aligned}
\end{equation}
The fusion rules of $V_i$ and $W_2$ are uniquely determined by the associativity of the fusion rules.

\paragraph*{The fusion rule of $W_i$ and $W_j$.}
We only need to consider the fusion of $W_1$ with itself because the associativity of the fusion rules uniquely determines the other fusion rules.
An element $e_k \in \mathcal{H}^u_8$ acts non-trivially on $W_1 \otimes W_1$ only if the comultiplication $\Delta(e_k)$ contains at least one of $e_5 \otimes e_5$, $e_5 \otimes e_7$, $e_7 \otimes e_5$, and $e_7 \otimes e_7$.
As we can see from eq. \eqref{eq: comultiplication of H8}, the elements that satisfy this condition are $e_1$ and $e_2$.
Therefore, the tensor product representation $W_1 \otimes W_1$ only contains one-dimensional super representations $V_1$, $V_2$, and their oddly isomorphic variants $\Pi V_1$ and $\Pi V_2$.
More specifically, the non-trivial action of $e_1$ and $e_2$ on $W_1 \otimes W_1$ is given by
\begin{equation}
\begin{aligned}
\Delta(e_1) \cdot [w_+ \otimes w_+ - u(z) w_- \otimes w_-] & = w_+ \otimes w_+ - u(z) w_- \otimes w_-, \\
\Delta(e_1) \cdot [w_+ \otimes w_- - u(z) w_- \otimes w_+] & = w_+ \otimes w_- - u(z) w_- \otimes w_+, \\
\Delta(e_2) \cdot [w_+ \otimes w_+ + u(z) w_- \otimes w_-] & = w_+ \otimes w_+ + u(z) w_- \otimes w_-, \\
\Delta(e_2) \cdot [w_+ \otimes w_- + u(z) w_- \otimes w_+] & = w_+ \otimes w_- + u(z) w_- \otimes w_+.
\end{aligned}
\end{equation}
This implies that we have an even isomorphism of super representations
\begin{equation}
W_1 \otimes W_1 \cong V_1 \oplus V_2 \oplus \Pi V_1 \oplus \Pi V_2.
\end{equation}

\subsection{Fusion rules of $\mathrm{SRep}(\mathcal{H}^u_{\mathbb{Z}_2, \chi, \epsilon})$}
A weak Hopf superalgebra $\mathcal{H}_{\mathbb{Z}_2, \chi, \epsilon}^u$ has two three-dimensional super representations $V_g \cong \mathbb{C}^{2|1}$ labeled by $g \in \mathbb{Z}_2$ and two four-dimensional super representations $W_s \cong \mathbb{C}^{2|2}$ labeled by $s = \pm 1$.
The actions of $\mathcal{H}_{\mathbb{Z}_2, \chi, \epsilon}^u$ on these super representations are defined by eqs. \eqref{eq: Z2 TY action on Vg} and \eqref{eq: Z2 TY action on Ws}.
In the following, we compute the fusion rules of these super representations.

\paragraph*{The fusion rule of $V_g$ and $V_h$}
The tensor product representation $V_g \boxtimes V_h$ is obtained by projecting the vector space $V_g \otimes V_h$ to the image of the action of the unit element $\eta(1) \in \mathcal{H}_{\mathbb{Z}_2, \chi, \epsilon}^u$.
Based on the definitions of the unit and the comultiplication of $\mathcal{H}_{\mathbb{Z}_2, \chi, \epsilon}^u$ given in section \ref{sec: Fermionization of Z2 Tambara-Yamagami symmetries}, we can compute the action of $\eta(1)$ on $V_g \otimes V_h$ as
\begin{equation}
\begin{aligned}
& \quad \Delta(\eta(1)) \cdot \sum_{\alpha, \beta} c_{\alpha \beta} v^g_{\alpha} \otimes v^h_{\beta} \\
& = \sum_{a \in \mathbb{Z}_2}\frac{c_{1a} + (-1)^{\delta_{h, u}}c_{u, au}}{2} [v^g_1 \otimes v^h_a + (-1)^{\delta_{h, u}} v^g_u \otimes v^h_{au}] + c_{mm} v^g_m \otimes v^h_m,
\end{aligned}
\end{equation}
where $c_{\alpha \beta}$ is an arbitrary complex number.
The above equation implies that the tensor product representation $V_g \boxtimes V_h$ is spanned by two $\mathbb{Z}_2$-even elements $v^g_1 \otimes v^h_1 + (-1)^{\delta_{h, u}} v^g_u \otimes v^h_u$ and $v^g_m \otimes v^h_m$ and a $\mathbb{Z}_2$-odd element $v^g_1 \otimes v^h_u + (-1)^{\delta_{h, u}} v^g_u \otimes v^h_1$.
Therefore, $V_g \boxtimes V_h \cong \mathbb{C}^{2|1}$ is a three-dimensional super representation, which is evenly isomorphic to either $V_1$ or $V_u$.
In order to identify this super representation, we compute the action of $x^l_{mm} \in \mathcal{H}^u_{\mathbb{Z}_2, \chi, \epsilon}$ on $v^g_m \otimes v^h_m \in V_g \boxtimes V_h$:
\begin{equation}
\Delta(x^l_{mm}) \cdot v^g_m \otimes v^h_m = \delta_{l, gh} v^g_m \otimes v^h_m.
\end{equation}
This indicates that $V_g \boxtimes V_h$ contains a three-dimensional super representation $V_{gh}$.
Since $V_g \boxtimes V_h$ itself is three-dimensional, we have an even isomorphism of super representations
\begin{equation}
V_g \boxtimes V_h \cong V_{gh}.
\end{equation}

\paragraph*{The fusion rule of $V_g$ and $W_s$}
We first consider the tensor product representation $V_g \boxtimes W_s$.
The action of the unit $\eta(1)$ on a general element of $V_g \otimes W_s$ is computed as
\begin{equation}
\begin{aligned}
& \quad \Delta(\eta(1)) \cdot \sum_{\alpha, \beta, p} c_{\alpha \beta}^p v^g_{\alpha} \otimes (w^s_{\beta})_p\\
& = \sum_p \left[ \frac{c_{11}^p + c_{u1}^{-p}}{2} [v_1^g \otimes (w_1^s)_p + v_u^g \otimes (w_1^s)_{-p}] + c_{m \overline{1}}^p v_m^g \otimes (w_{\overline{1}}^s)_p \right],
\end{aligned}
\end{equation}
where $c_{\alpha \beta}^p$ is an arbitrary complex number.
The above equation implies that $V_g \boxtimes W_s$ is a four-dimensional super representation spanned by $\{v^g_1 \otimes (w^s_1)_p + v^g_u \otimes (w^s_1)_{-p}, v^g_m \otimes (w^s_{\overline{1}})_p \mid p = \pm 1 \}$.
Hence, $V_g \boxtimes W_s$ is isomorphic to either $W_+$ or $W_-$.
Furthermore, the action of $(x^{m, t}_{\overline{1} \overline{1}})_+ \in \mathcal{H}_{\mathbb{Z}_2, \chi, \epsilon}^u$ on $v^g_m \otimes (w^s_{\overline{1}})_p \in V_g \boxtimes W_s$ is computed as
\begin{equation}
\Delta((x^{m, t}_{\overline{1} \overline{1}})_+) \cdot v^g_m \otimes (w^s_{\overline{1}})_p = \frac{1 + st (-1)^{\delta_{g, u}}}{2} v^g_m \otimes (w^s_{\overline{1}})_p,
\end{equation}
which indicates that $V_g \boxtimes W_s$ contains a four-dimensional super representation $W_{s (-1)^{\delta_{g, u}}}$.
Therefore, we have isomorphisms of super representations
\begin{equation}
V_1 \boxtimes W_s \cong W_{s}, \quad V_u \boxtimes W_s \cong W_{-s}.
\end{equation}
We can also compute the tensor product representation $W_s \boxtimes V_g$ similarly.
The action of the unit $\eta(1)$ on a general element of $W_s \otimes V_g$ is given by
\begin{equation}
\begin{aligned}
& \quad \Delta(\eta(1)) \cdot \sum_{\alpha, \beta, p} c_{\alpha \beta}^p (w^s_{\alpha})_p \otimes v^g_{\beta} \\
& = \sum_p \left[ \frac{c_{\overline{1} 1}^p + isp(-1)^{\delta_{g, u}}c_{\overline{1} u}^{-p}}{2} [(w^s_{\overline{1}})_p \otimes v^g_1 - isp(-1)^{\delta_{g, u}} (w^s_{\overline{1}})_{-p} \otimes v^g_u] + c_{1m}^p (w^s_1)_p \otimes v^g_m \right],
\end{aligned}
\end{equation}
which implies that $W_s \boxtimes V_g$ is a four-dimensional super representation spanned by $\{(w^s_{\overline{1}})_p \otimes v^g_1 - isp(-1)^{\delta_{g, u}} (w^s_{\overline{1}})_{-p} \otimes v^g_{u}, (w^s_{1})_p \otimes v^g_m \mid p = \pm 1\}$.
Furthermore, $W_s \boxtimes V_g$ contains a four-dimensional super representation $W_{s(-1)^{\delta_{g, u}}}$ because $(x^{m, t}_{11})_+$ acts on $(w^s_{1})_p \otimes v^g_m \in W_s \boxtimes V_g$ as 
\begin{equation}
\Delta((x^{m, t}_{11})_+) \cdot (w^s_1)_p \otimes v^g_m = \frac{1 + st(-1)^{\delta_{g, u}}}{2} (w^s_1)_p \otimes v^g_m.
\end{equation}
Therefore, we have isomorphisms of super representations 
\begin{equation}
W_s \boxtimes V_1 \cong W_{s}, \quad W_s \boxtimes V_u \cong W_{-s}.
\end{equation}

\paragraph*{The fusion rule of $W_s$ and $W_t$}
The unit element $\eta(1)$ acts on a general element of $W_s \otimes W_t$ as 
\begin{equation}
\begin{aligned}
& \quad \Delta(\eta(1)) \cdot \sum_{\alpha, \beta, p, q} c_{\alpha \beta}^{pq} (w^s_{\alpha})_p \otimes (w^t_{\beta})_q \\
& = \sum_{pq} c_{1\overline{1}}^{pq} (w^s_1)_p \otimes (w^t_{\overline{1}})_q + \sum_q \frac{c_{\overline{1}1}^{+, q} + is c_{\overline{1}1}^{-, -q}}{2} [(w^s_{\overline{1}})_+ \otimes (w^t_1)_q - is (w^s_{\overline{1}})_- \otimes (w^t_1)_{-q}],
\end{aligned}
\end{equation}
This implies that $W_s \boxtimes W_t$ is a six-dimensional super representation spanned by $(w^s_1)_p \otimes (w^t_{\overline{1}})_q$ and $(w^s_{\overline{1}})_+ \otimes (w^t_1)_q - is (w^s_{\overline{1}})_- \otimes (w^t_1)_{-q}$ for $p, q = \pm 1$.
Therefore, $W_s \boxtimes W_t$ can be decomposed into the direct sum of two three-dimensional super representations.
Since $W_s \boxtimes W_t$ has an odd automorphism due to the fact that $W_s$ and $W_t$ are q-type objects, a six-dimensional super representation $W_s \boxtimes W_t$ is isomorphic to either $V_1 \oplus \Pi V_1$ or $V_u \oplus \Pi V_u$.
This isomorphism is determined by computing the action of $\mathcal{H}_{\mathbb{Z}_2, \chi, \epsilon}^u$ on $W_s \boxtimes W_t$.
Specifically, it turns out that $x^g_{11} \in \mathcal{H}_{\mathbb{Z}_2, \chi, \epsilon}^u$ acts non-trivially on $(w^s_1)_p \otimes (w^t_{\overline{1}})_q \in W_s \boxtimes W_t$ if $st = (-1)^{\delta_{g, u}}$, which indicates that $W_s \boxtimes W_t$ contains $V_1$ if $s = t$ and $V_u$ if $s = -t$.
Therefore, we find the following isomorphisms of super representations:
\begin{equation}
W_s \boxtimes W_s \cong V_1 \oplus \Pi V_1, \quad W_s \boxtimes W_{-s} \cong V_u \oplus \Pi V_u.
\end{equation}

\bibliographystyle{JHEP}
\bibliography{bibliography}

\end{document}